\documentclass[10pt]{article}
\pdfoutput=1

\usepackage[usenames,dvipsnames,table]{xcolor}
\usepackage[utf8]{inputenc}
\usepackage{booktabs}
\usepackage{amsmath}
\usepackage{float}

\usepackage{jheppub} % for details on the use of the package, please
\usepackage{tikz-cd} 
\usepackage{circuitikz}
\usepgfmodule{shapes}
\usetikzlibrary{decorations.pathmorphing}
\usetikzlibrary{decorations.pathreplacing,decorations.markings,snakes}
\usetikzlibrary{backgrounds, arrows,calc,shapes,decorations.pathreplacing, automata,positioning}

\usepackage{cancel} 
\usepackage{xcolor}
\usepackage{enumitem}

\usetikzlibrary{shapes.misc}

\tikzset{cross/.style={cross out, draw=black, minimum size=5*(#1-\pgflinewidth), inner sep=0pt, outer sep=0pt},
cross/.default={2pt}}

\tikzset{snake it/.style={decorate, decoration=snake}}

\tikzset{
  hyper/.style    = { thick, double, 
  double distance = 3pt }}

\pgfdeclarepatternformonly{coarsedots}{\pgfqpoint{-1pt}{-1pt}}{\pgfqpoint{5pt}{5pt}}{\pgfqpoint{12pt}{12pt}}{
    \pgfpathcircle{\pgfqpoint{.5pt}{.5pt}}{.5pt}
    \pgfusepath{fill}}
    
\tikzset{gauge/.style={rounded rectangle, draw=black!100, thick, minimum size=5mm},  gaugeD/.style={rounded rectangle, draw=black!100,double,thick,minimum size=5mm},  empty/.style={rounded rectangle, draw=white!100, thick, minimum size=5mm}, flavor/.style={rectangle, draw=black!100, thick, minimum size=5mm},flavorD/.style={rectangle, draw=black!100, double,thick, minimum size=5mm}}

\definecolor{cobalt}{rgb}{0.0, 0.28, 0.67}
\definecolor{penred}{RGB}{200,20,20}
\definecolor{penblue}{RGB}{20,60,160}

\usepackage{graphicx}
\usepackage{amsmath, amssymb}
\usepackage{youngtab}
\usepackage{changepage}

\newcommand{\be}{\begin{eqnarray}}
\newcommand{\ee}{\end{eqnarray}}
\newcommand{\ba}{\begin{array}}
\newcommand{\ea}{\end{array}}
\newcommand{\bea}{\begin{eqnarray}}
\newcommand{\eea}{\end{eqnarray}}
\newcommand{\bpic}{\begin{tikzpicture}}
\newcommand{\epic}{\end{tikzpicture}}

\newcommand{\nn}{\nonumber}
\newcommand{\bn}{\begin{enumerate}}
\newcommand{\en}{\end{enumerate}}

\newcommand{\dFu}[1]{{}_2F_1\left(#1\right)}
\newcommand{\G}[1]{\Gamma\left(#1\right)}
\newcommand{\qPoc}[2]{\left(#1;#2\right)_\infty}
\newcommand{\qth}[1]{\theta\left(#1\right)}

\usepackage[mathscr]{euscript}

\def\CC{\mathscr{C}}
\def\VV{\mathscr{V}}
\def\FF{\mathscr{F}}
\def\SS{\mathscr{S}}
\def\BB{\mathscr{B}}
\def\TT{\mathscr{T}}
\def\EE{\mathscr{E}}

\title{\begin{center}
    {\Huge\boldmath$2+2=4$}
\\ \vspace{0.3cm}
{\LARGE A 2d/2d unitary/non-unitary correspondence}
\end{center}}

\author[a]{Leonardo Rastelli,}
\author[b]{Brandon C.~Rayhaun,}
\author[c]{Matteo Sacchi,}
\author[d,e]{Gabi Zafrir}

\affiliation[a]{C.N.~Yang Institute for Theoretical Physics, Stony Brook University, Stony Brook, NY 11794, USA}
\affiliation[b]{School of Natural Sciences, Institute for Advanced Study, Princeton, NJ 08540, USA}
\affiliation[c]{Simons Center for Geometry and Physics, Stony Brook University, Stony Brook, NY 11794, USA}
\affiliation[d]{Department of Physics, University of Haifa at Oranim, Kiryat Tivon 36006, Israel}
\affiliation[e]{Haifa Research Center for Theoretical Physics and Astrophysics, University of Haifa,
Haifa 3498838, Israel}
\emailAdd{leonardo.rastelli@stonybrook.edu}\emailAdd{rayhaun@ias.edu}\emailAdd{msacchi@scgp.stonybrook.edu}\emailAdd{gabi.zafrir@oranim.ac.il}

\abstract{
Motivated by the observation that $2+2=4$, we consider four-dimensional $\mathcal{N}=2$ superconformal field theories on $S^2\times\Sigma$, turning on a suitable rigid supergravity background. On the one hand, reduction of a four-dimensional theory ${T}$ on a Riemann surface $\Sigma$ leads to a family $\mathscr{F}[{T}, \Sigma]$ of two-dimensional $(2,2)$ \emph{unitary} SCFTs, a two-dimensional analog of the four-dimensional theories of class $\mathscr{S}$. On the other hand, reduction on $S^2$ yields a \emph{non-unitary} two-dimensional CFT $\mathscr{C}[{T}]$ whose chiral algebra is the same as the one associated to ${T}$ by the standard SCFT/VOA correspondence. This construction upgrades the vertex operator algebra to a full-fledged two-dimensional CFT. What's more, it leads to a novel 2d/2d correspondence, a ``$2+2 = 4$'' analog of the ``$4+2=6$'' AGT correspondence: the $S^2$ partition function of $\mathscr{F}[{T}; \Sigma]$ is computed by correlation functions of $\mathscr{C}[{T}]$ on $\Sigma$. The elliptic genus of $\mathscr{F}[{T}; \Sigma]$ is instead computed by a topological QFT $\mathscr{E}[T]$ on $\Sigma$. A central question is whether one can give a purely two-dimensional presentation of the family $\mathscr{F}[{T}; \Sigma]$ of $(2, 2)$ theories. We propose an algorithm to realize the $(2, 2)$ theories as gauged linear sigma models when ${T}$ is an Argyres--Douglas theory of type $(A_1, A_{2k})$ and $\Sigma$ an $n$-punctured sphere. We perform stringent checks  of our conjecture for $k=1$ and $k=2$. 
}

\begin{document} 

\maketitle

\flushbottom

%%%%%%%%%%%%%%%%%%%%%%%%%%%%%%%%%

%%%%%%%%%%%%%%%%%%%%%%%%%%%%%%%%%
\section{Introduction and summary}

A pervasive phenomenon in the dynamics of quantum field theory (QFT) is \emph{duality}, the fact that a single physical system can admit two seemingly different but ultimately equivalent descriptions. A duality entails a one-to-one map between the observables in each description. The ability to view a QFT from different duality frames enriches our understanding of its physics, and often leads to calculational approaches that would be otherwise hard to come by.
Perhaps less familiar is the related notion of a \emph{correspondence}. A correspondence relates two theories which are genuinely different, but which nevertheless share specific sectors of observables. Often, a correspondence can be derived by viewing a single parent system in two different limits, and arguing that the physics of interest does not depend on which limit one takes. 

We focus here on the correspondence~\cite{Beem:2013sza} relating a protected subsector of ${\cal N}=2$ four-dimensional  superconformal field theories (SCFTs) to vertex operator algebras (VOAs), henceforth the ``SCFT/VOA correspondence''. We can think of it as a canonical map that associates a VOA $\VV[T]$ to any 4d ${\cal N}=2$ SCFT $T$,
\begin{equation}\label{eq:SCFTVOA}
    T\mapsto\VV[T]\,.
\end{equation}
As a vector space, $\VV[T]$ comprises the {\it Schur operators} of $T$, a class of local operators belonging to specific shortened representation of the four-dimensional superconformal algebra. The SCFT/VOA correspondence has proved  extremely powerful in a variety of ways. Let us mention a few. On the pragmatic side, as it is often possible to ``bootstrap'' $\VV[T]$ even when $T$ is a strongly coupled fixed point, one can access precious protected information about ``non-Lagrangian'' theories, see e.g.~\cite{Beem:2014rza, Lemos:2014lua, Lemos:2016xke,  Creutzig:2017qyf,Song:2017oew,  Buican:2017fiq, Choi:2017nur,Beem:2019snk, Xie:2019yds, Xie:2019zlb, Xie:2019vzr}. More conceptually, the correspondence provides an organizing principle for the  whole space of ${\cal N}=2$ four-dimensional SCFTs~\cite{Beem:2018duj, Bonetti:2018fqz, Beem:2019tfp,  Kaidi:2022sng,Rastelli:2023sfk,  Deb:2025cqr, ArabiArdehali:2025fad}, complementary to their exploration via symplectic geometry of the  Coulomb branch, i.e.~Seiberg--Witten theory (see e.g.~\cite{Martone:2020hvy, Argyres:2024uuc} for recent overviews). Finally it has inspired important new developments in the mathematics of vertex operator algebras~\cite{Arakawa:2016hkg, Arakawa:2017aon, Arakawa:2018egx, Arakawa:2023cki}.
However a very natural question remains open:

\begin{center}
\itshape
Can we associate a full-fledged two-dimensional CFT to a four-dimensional  ${\cal N}=2$ SCFT,\\ rather than just a chiral algebra?
\end{center}

\noindent 
We would like to find a prescription to obtain a full two-dimensional CFT $\CC[T]$ such that the chiral algebra of $\CC[T]$ coincides with the VOA $\VV[T]$ of the SCFT/VOA correspondence. This would upgrade the map \eqref{eq:SCFTVOA} to
\begin{equation}
    T\mapsto\CC[T]\,,\qquad \text{chiral algebra}(\CC[T])\cong\VV[T]\,.
\end{equation}
\smallskip
\noindent
This upgraded correspondence is implicit in a compelling geometric picture.

\subsection*{Engineering the full CFT}
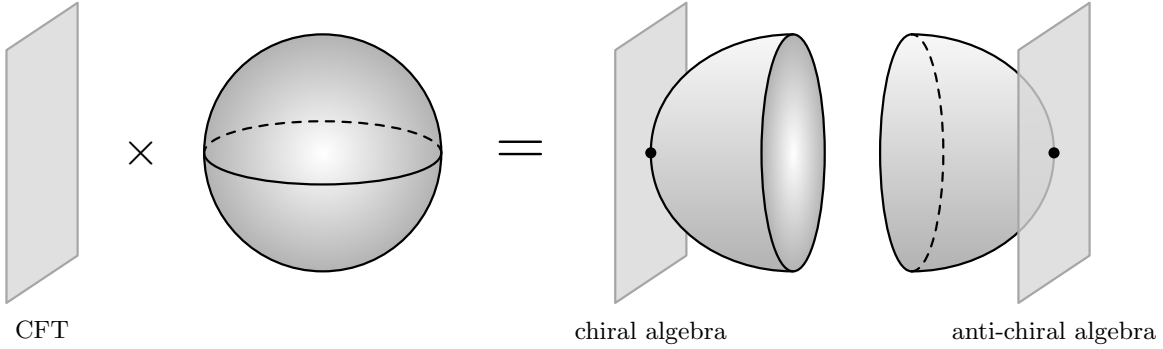
\begin{figure}[t]
\center
\resizebox{\textwidth}{!}{\begin{tikzpicture}[thick, line cap=round, line join=round]
    %----------------------------
    % Parameters
    %----------------------------
    \def\R{1.5}   % Sphere radius
    \def\Rx{0.4}  % Sphere equator vertical radius
    \def\Pr{1.5}  % Piece radius (height)
    \def\Pw{0.4}  % Piece ellipse width
    \def\Pl{1.8}  % Piece length

    % Planes
    \def\PlaneW{0.9}       % SHORT edge length (depth)
    \def\PlaneH{3.2}       % LONG vertical sides
    \def\PlaneSkew{0.60}   % tilt of the short edges (depth impression)
    \def\LabelGap{0.1}     % gap between plane bottom and label

    % Spacing (adjust these to avoid overlaps)
    \def\GapPlaneTimes{0.8}    % gap between LHS plane right edge and \times
    \def\GapTimesSphere{0.8}   % gap between \times and sphere left boundary
    \def\GapSphereEq{1.0}      % gap between sphere right boundary and "="
    \def\GapEqRHS{1.2}         % gap between "=" and start of RHS planes/pieces
    \def\RHSSeparation{1.5}    % separation between the two RHS caps

    %----------------------------
    % Derived x-positions (LHS = plane \times sphere)
    %----------------------------
    % Choose LHS plane center so that its left edge starts at x=0
    \pgfmathsetmacro{\LHSPlaneX}{\PlaneW/2}

    % \times after plane
    \pgfmathsetmacro{\TimesX}{\LHSPlaneX + \PlaneW/2 + \GapPlaneTimes}

    % Sphere after \times
    \pgfmathsetmacro{\SphereX}{\TimesX + \GapTimesSphere + \R}

    % "=" after sphere
    \pgfmathsetmacro{\EqX}{\SphereX + \R + \GapSphereEq}

    % RHS left-piece group position so that the left edge of its plane
    % is GapEqRHS to the right of "="
    \pgfmathsetmacro{\LeftGroupX}{\EqX + \GapEqRHS + \Pl + \PlaneW/2}
    \pgfmathsetmacro{\RightGroupX}{\LeftGroupX + \RHSSeparation}

    %----------------------------
    % Plane centered at (0,0)
    %----------------------------
    \newcommand{\PlaneShape}{%
      \path
        (-\PlaneW/2, -\PlaneH/2 - \PlaneSkew/2) coordinate (A)
        ( \PlaneW/2, -\PlaneH/2 + \PlaneSkew/2) coordinate (B)
        ( \PlaneW/2,  \PlaneH/2 + \PlaneSkew/2) coordinate (C)
        (-\PlaneW/2,  \PlaneH/2 - \PlaneSkew/2) coordinate (D);
      \fill[gray!30, opacity=0.8] (A)--(B)--(C)--(D)--cycle;
      \draw[gray!70] (A)--(B)--(C)--(D)--cycle;
    }

    %----------------------------
    % LHS: Plane \times Sphere
    %----------------------------
    \begin{scope}[shift={(\LHSPlaneX,0)}]
      \PlaneShape
      \node[font=\small, below] at (0, -\PlaneH/2 - \PlaneSkew/2 - \LabelGap) {CFT};
    \end{scope}

    \node at (\TimesX,0) {\LARGE $\times$};

    \begin{scope}[shift={(\SphereX,0)}]
      \shade[inner color=white, outer color=gray!60] (0,0) circle (\R);
      \draw (0,0) circle (\R);
      \draw (-\R,0) arc (180:360:\R cm and \Rx cm);
      \draw[dashed] (\R,0) arc (0:180:\R cm and \Rx cm);
    \end{scope}

    % "="
    \node at (\EqX, 0) {\Huge $=$};

    %----------------------------
    % RHS: two caps with planes
    %----------------------------

    % 3. Left Piece (Right-facing cap)
    \begin{scope}[shift={(\LeftGroupX,0)}]
        % LEFT PLANE (behind the hemisphere), centered at bullet (-\Pl,0)
        \begin{scope}[shift={(-\Pl,0)}]
          \PlaneShape
          \node[font=\small, below] at (0, -\PlaneH/2 - \PlaneSkew/2 - \LabelGap)
            {chiral algebra};
        \end{scope}

        % Body
        \shade[top color=gray!5,bottom color=gray!60]
          (0,\Pr) to[out=180,in=90] (-\Pl,0)
          to[out=270,in=180] (0,-\Pr) -- cycle;
        \shade[inner color=white,outer color=gray!60]
          (0,0) ellipse (\Pw cm and \Pr cm);

        \draw (0,\Pr) to[out=180,in=90] (-\Pl,0)
              to[out=270,in=180] (0,-\Pr);
        \draw (0,0) ellipse (\Pw cm and \Pr cm);

        % Bullet at tip (center of plane)
        \fill (-\Pl,0) circle (2pt);
    \end{scope}

    % 4. Right Piece (Left-facing cap)
    \begin{scope}[shift={(\RightGroupX,0)}]
        % Body
        \shade[top color=gray!5,bottom color=gray!60]
          (0,\Pr) to[out=0,in=90] (\Pl,0)
          to[out=270,in=0] (0,-\Pr) -- cycle;

        \shade[top color=gray!5,bottom color=gray!60]
          (0,0) ellipse (\Pw cm and \Pr cm);

        \draw (0,\Pr) to[out=0,in=90] (\Pl,0)
              to[out=270,in=0] (0,-\Pr);

        \draw (0,\Pr) arc (90:270:\Pw cm and \Pr cm);
        \draw[dashed] (0,-\Pr) arc (-90:90:\Pw cm and \Pr cm);

        % RIGHT PLANE (in front), centered at bullet (\Pl,0)
        \begin{scope}[shift={(\Pl,0)}]
          \PlaneShape
          \node[font=\small, below] at (0, -\PlaneH/2 - \PlaneSkew/2 - \LabelGap)
            {anti-chiral algebra};
        \end{scope}

        % Bullet at tip (center of plane)
        \fill (\Pl,0) circle (2pt);
    \end{scope}
\end{tikzpicture}
}

\caption{A sphere as a gluing of two cigars,  realizing the combination of a chiral algebra with its anti-chiral counterpart to yield a full two-dimensional CFT.}
\label{fig:spheregluing}
\end{figure}

The original construction of~\cite{Beem:2013sza} carved out the VOA from the four-dimensional SCFT by performing a cohomological 
reduction with respect to a certain nilpotent supercharge. For our purposes, 
it is useful to consider an alternative but equivalent viewpoint~\cite{Oh:2019bgz,Jeong:2019pzg, Dedushenko:2023cvd}. 
In a cartoon, one  considers reducing the four-dimensional theory $T$ on an infinitely long two-dimensional cigar or hemisphere,
in such a way that the theory living in the two orthogonal directions at the tip of the cigar is the VOA $\VV[T]$. To get a full two-dimensional CFT, we  wish to 
combine the chiral algebra together with its anti-chiral counterpart. This corresponds to gluing two cigars into a sphere, see Figure \ref{fig:spheregluing}; the gluing amounts to a sum over (modulus squares of) conformal blocks. 
In other words, we expect that reduction of $T$ on $S^2$ 
will lead to a two-dimensional CFT $\CC[T]$, whose chiral algebra coincides with the VOA $\VV[T]$ predicted by the SCFT/VOA correspondence. 

This cartoon can be made more precise. Let us start by recalling the picture of
\cite{Oh:2019bgz,Jeong:2019pzg} for the standard SCFT/VOA correspondence. First, one places $T$ on $\mathbb{R}^2\times\Sigma$, where $\Sigma$ is a Riemann surface,
with a holomorphic-topological (HT) twist~\cite{Kapustin:2006hi}. To wit:
\begin{itemize}
\item twist by $U(1)_r$ on $\mathbb{R}^2$, which become the topological directions;
\item twist by the Cartan $U(1)_R\subset SU(2)_R$ on $\Sigma$, which become the holomorphic directions.
\end{itemize}
Cohomological reduction of the local operator algebra of $T$ with respect to the nilpotent HT supercharge 
yields a {\it commutative} vertex Poisson algebra (VPA) \cite{Oh:2019mcg}. 
The VOA $\VV[T]$
is a 
non-commutative deformation of this VPA. This is realized  by introducing an $\Omega$-deformation~\cite{Nekrasov:2002qd,Nekrasov:2003rj} 
of the topological plane, $\mathbb{R}^2 \rightarrow \mathbb{R}^2_\epsilon$.
The (cohomological) degrees of freedom localize at the origin of $\mathbb{R}^2_\epsilon$, indeed one can think of 
 $\mathbb{R}^2_\epsilon$
 as an infinitely long cigar with the VOA living at its tip \cite{Nekrasov:2010ka, Dedushenko:2023cvd}. This 
  construction corresponds to one half of Figure \ref{fig:spheregluing}. To get a full non-chiral CFT,
  we should then glue two cigars of opposite $U(1)_r$ topological twist into an $S^2$ with no overall twist, and with an additional $SU(2)_R$ twist along $\Sigma$.
A first goal in this paper, accomplished in Section~\ref{sec:kinematics}, is to consolidate this picture
by constructing the precise four-dimensional off-shell supergravity background.

\subsection*{What does the full CFT compute?}
An immediate question is which protected observables are captured by the two-dimensional
CFT~$\CC [T]$. The answer is provided by the following diagram:

\begin{center}
\tikzset{every picture/.style={line width=0.75pt}} %set default line width to 0.75pt        
\begin{tikzpicture}[x=0.75pt,y=0.75pt,yscale=-1,xscale=1,scale=.9]
%uncomment if require: \path (0,219); %set diagram left start at 0, and has height of 219
\path[use as bounding box] (20,10) rectangle (660,180);

%Straight Lines [id:da055433214918889084] 
\draw [line width=1]    (264+30,56) -- (116.66+30,132.62) ;
\draw [shift={(114+30,134)}, rotate = 332.53] [color={rgb, 255:red, 0; green, 0; blue, 0 }  ][line width=1.5]    (14.21,-4.28) .. controls (9.04,-1.82) and (4.3,-0.39) .. (0,0) .. controls (4.3,0.39) and (9.04,1.82) .. (14.21,4.28)   ;
%Straight Lines [id:da6657060278054253] 
\draw [line width=1]    (365+30,55) -- (501.39+30,132.52) ;
\draw [shift={(504+30,134)}, rotate = 209.61] [color={rgb, 255:red, 0; green, 0; blue, 0 }  ][line width=1.5]    (14.21,-4.28) .. controls (9.04,-1.82) and (4.3,-0.39) .. (0,0) .. controls (4.3,0.39) and (9.04,1.82) .. (14.21,4.28)   ;

% Text Node
\draw (240,29) node [anchor=north west][inner sep=0.75pt]   [align=left] {4d $\displaystyle \mathcal{N} =2$ SCFT $T$ on $\displaystyle S^{2} \times \Sigma _{g,n}$};
% Text Node
\draw (20,142) node [anchor=north west][inner sep=0.75pt]   [align=left] {\begin{minipage}[lt]{171.41pt}\setlength\topsep{0pt}
~~~~~~~$S^{2}$ partition function of \\2d   $(2,2)$
 unitary SCFT $\FF[T;\Sigma_{g,n};\dots]$

\end{minipage}};
% Text Node
\draw (430,142) node [anchor=north west][inner sep=0.75pt]   [align=left] {\begin{minipage}[lt]{207.42pt}\setlength\topsep{0pt}

 \hspace{.05in}$\Sigma_{g,n}$ correlator of 
2d non-unitary \\  CFT $\CC[T]$ with chiral algebra $\displaystyle \VV[T]$ 
\end{minipage}};
% Text Node
\draw (68+30,74) node [anchor=north west][inner sep=0.75pt]   [align=left] {reduce on $\displaystyle \Sigma _{g,n}$};
% Text Node
\draw (451+30,74) node [anchor=north west][inner sep=0.75pt]   [align=left] {reduce on $\displaystyle S^{2}$};

\draw [line width=1]    (275+15,160) -- (380+15,160) ;
\draw [shift={(380+15,160)}, rotate = 180] [color={rgb, 255:red, 0; green, 0; blue, 0 }  ][line width=1.5]    (14.21,-4.28) .. controls (9.04,-1.82) and (4.3,-0.39) .. (0,0) .. controls (4.3,0.39) and (9.04,1.82) .. (14.21,4.28)   ;
\draw [shift={(275+15,160)}, rotate = 0] [color={rgb, 255:red, 0; green, 0; blue, 0 }  ][line width=1.5]    (14.21,-4.28) .. controls (9.04,-1.82) and (4.3,-0.39) .. (0,0) .. controls (4.3,0.39) and (9.04,1.82) .. (14.21,4.28)   ;
\draw (292,132) node [anchor=north west][inner sep=0.75pt]   [align=left] {correspondence};
\end{tikzpicture}
\end{center}
\vspace{.2in}

\noindent We propose a new 2d/2d correspondence, heuristically derived by taking different limits of the compactification of $T$ on $S^2 \times \Sigma_{g, n}$,
where  $\Sigma_{g, n}$ is a genus $g$ Riemann surface with $n$ punctures. On the one hand, reduction on $S^2$
yields the non-unitary 2d CFT $\CC[T]$ on $\Sigma_{g, n}$, as we have just argued. The punctures correspond to insertions of local operators  of $\CC[T]$. On the other hand, 
we claim that (with our choice of background)
reduction on $\Sigma_{g, n}$ yields a unitary $\mathcal{N}=(2, 2)$  SCFT on $S^2$, which we dub $\FF[T; \Sigma_{g, n}; \dots]$. The dots are a reminder that we should be more precise about the boundary conditions at the punctures; in general one will have to make certain discrete choices.
The diagram suggests that the $S^2$ partition function of $\FF[T; \Sigma_{g, n}; \dots]$ is computed by
an $n$-point correlator of $\CC[T]$ on $\Sigma_{g, n}$. Agreement of the two reductions can be argued by
recalling that the holomorphic twist on $\Sigma_{g,n}$ eliminates the dependence on its size. As the full 4d partition function of $T$ on $S^2 \times \Sigma_{g,n}$
can only depend on the {\it ratio} of the sizes of $S^2$ and $\Sigma_{g,n}$, it can be equivalently computed by reducing on $S^2$ or reducing on $\Sigma_{g,n}$.
The second goal of this paper will be to make our proposed correspondence precise and to illustrate it in a class of simple examples. 

\medskip

\noindent\textbf{Analogy with AGT.} 
Our  perceptive readers will not have failed to  notice that our new 2d/2d correspondence is structurally analogous to the 4d/2d
AGT correspondence~\cite{Alday:2009aq, Wyllard:2009hg}. According to AGT, the $S^4$ partition function \cite{Pestun:2007rz, Hama:2012bg} of a 4d $\mathcal{N}=2$ superconformal theory of class $\SS$~\cite{Witten:1997sc,Gaiotto:2009hg, Gaiotto:2009we} --- defined by the choice of an ADE Lie algebra $\mathfrak{g}$ and a Riemann surface $\Sigma_{g,n}$, decorated by certain data at its punctures --- is computed by an $n$-point genus-$g$ correlator of the Toda CFT based on $\mathfrak{g}$. The AGT correspondence can be heuristically derived  
by considering the 6d $\mathcal{N}=(2,0)$ SCFT of type $\mathfrak{g}$ on a supersymmetric background with topology $S^4\times \Sigma_{g,n}$. By reducing on $\Sigma_{g,n}$, one goes over to the class $\SS$ side of the correspondence, while reducing on $S^4$ one recovers the Toda side \cite{Cordova:2016cmu}.

\begin{center}
\tikzset{every picture/.style={line width=0.75pt}} %set default line width to 0.75pt     
\begin{tikzpicture}[x=0.75pt,y=0.75pt,yscale=-1,xscale=1,scale=.9]
\path[use as bounding box] (20,10) rectangle (650,180);
\draw [line width=1]    (264+30,56) -- (116.66+30,132.62) ;
\draw [shift={(114+30,134)}, rotate = 332.53] [color={rgb, 255:red, 0; green, 0; blue, 0 }  ][line width=1.5]    (14.21,-4.28) .. controls (9.04,-1.82) and (4.3,-0.39) .. (0,0) .. controls (4.3,0.39) and (9.04,1.82) .. (14.21,4.28)   ;
%Straight Lines [id:da6657060278054253] 
\draw [line width=1]    (365+30,55) -- (501.39+30,132.52) ;
\draw [shift={(504+30,134)}, rotate = 209.61] [color={rgb, 255:red, 0; green, 0; blue, 0 }  ][line width=1.5]    (14.21,-4.28) .. controls (9.04,-1.82) and (4.3,-0.39) .. (0,0) .. controls (4.3,0.39) and (9.04,1.82) .. (14.21,4.28)   ;

% Text Node
\draw (230,29) node [anchor=north west][inner sep=0.75pt]   [align=left] {6d $\displaystyle \mathcal{N} =(2,0)$ SCFT on $\displaystyle S^{4} \times \Sigma _{g,n}$};
% Text Node
\draw (40,142) node [anchor=north west][inner sep=0.75pt]   [align=left] {\begin{minipage}[lt]{171.41pt}\setlength\topsep{0pt}
~~~~~~~~$S^{4}$ partition function \\
of 4d $\mathcal{N} =2$ SCFT $\SS[\mathfrak{g};\Sigma_{g,n};\dots]$

\end{minipage}};
% Text Node
\draw (470,142) node [anchor=north west][inner sep=0.75pt]   [align=left] {\begin{minipage}[lt]{207.42pt}\setlength\topsep{0pt}

 $\Sigma_{g,n}$ correlator of 
2d \\ Toda CFT of type $\mathfrak{g}$ 
\end{minipage}};
% Text Node
\draw (68+30,74) node [anchor=north west][inner sep=0.75pt]   [align=left] {reduce on $\displaystyle \Sigma _{g,n}$};
% Text Node
\draw (451+30,74) node [anchor=north west][inner sep=0.75pt]   [align=left] {reduce on $\displaystyle S^{4}$};

\draw [line width=1]    (275+15,160) -- (380+15,160) ;
\draw [shift={(380+15,160)}, rotate = 180] [color={rgb, 255:red, 0; green, 0; blue, 0 }  ][line width=1.5]    (14.21,-4.28) .. controls (9.04,-1.82) and (4.3,-0.39) .. (0,0) .. controls (4.3,0.39) and (9.04,1.82) .. (14.21,4.28)   ;
\draw [shift={(275+15,160)}, rotate = 0] [color={rgb, 255:red, 0; green, 0; blue, 0 }  ][line width=1.5]    (14.21,-4.28) .. controls (9.04,-1.82) and (4.3,-0.39) .. (0,0) .. controls (4.3,0.39) and (9.04,1.82) .. (14.21,4.28)   ;
\draw (292,132) node [anchor=north west][inner sep=0.75pt]   [align=left] {correspondence};
\end{tikzpicture}
\end{center}
\vspace{.1in}

\noindent
The obvious parallel between this ``$4+2 = 6$'' diagram and our previous ``$2+2 = 4$'' diagram is made sharper by recalling that the AGT correspondence can  also be justified 
in terms of gluing two (four-dimensional) hemispheres,
corresponding respectively to a chiral algebra and an anti-chiral algebra living in the  two transverse dimensions. 
The chiral algebra of the Toda theory of type $\mathfrak{g}$ is of course the W-algebra $W_{\mathfrak{g}}$. This is precisely the VOA that arises by cohomological reduction~\cite{Beem:2014kka} of  the $(2, 0)$ theory of type $\mathfrak{g}$ by a direct 6d/2d analog of the 4d/2d SCFT/VOA correspondence~\cite{Beem:2013sza}.
What's more, in complete analogy with the $2+2 = 4$ case, one can understand~\cite{Bobev:2020vhe} this VOA as arising from a
holomorphic-topological twist of the 6d theory on $\mathbb{R}^4 \times \Sigma_{g,n}$, followed by an $\Omega$-deformation $\mathbb{R}^4 \rightarrow \mathbb{R}^4_{\epsilon_1, \epsilon_2}$.
Finally, it is well-known that the $S^4$ partition 
of an ${\cal N}=2$ theory (in this case, the class $\SS$ theory obtained by reduction on $\Sigma_{g,n}$)
decomposes into a sum over (modulus squares of) Nekrasov partition functions on $\mathbb{R}^4_{\epsilon_1,\epsilon_2}$ \cite{Nekrasov:2002qd,Nekrasov:2003rj}, which indeed coincide with the Toda conformal blocks.
All in all, the 6d construction of AGT involving reduction on $S^4$ gives a prescription to extract a full 2d CFT whose chiral algebra is the VOA expected by the SCFT/VOA correspondence, in accordance with the picture of Figure \ref{fig:spheregluing}, where the sphere is now understood as being four-dimensional.

In summary, compactification of  a 6d $(2,0)$ SCFT  on $S^4\times\Sigma_{g,n}$ motivates three related developments:
\begin{enumerate}[label=\arabic*)]
    \item[{(i)}] It provides the family $\SS[\mathfrak{g}; \Sigma_{g, n};\dots]$
    of 4d $\mathcal{N}=2$  SCFTs  of class $\SS$. 
    \item[{(ii)}]  It upgrades the 6d version of the SCFT/VOA correspondence by giving a full 2d CFT (the Toda CFT) for each 6d $(2, 0)$ theory, rather than just a chiral algebra.
    \item[{(iii)}]  It relates 
    correlation functions of this full 2d CFT to the $S^4$ partition functions of  class     $\SS$ theories, via the AGT correspondence.
\end{enumerate}
All  three points have direct analogs in our $2+2 = 4$ story. Let us briefly elaborate on each of them.

\medskip

\noindent\textbf{Theories of class \boldmath$\FF$.} When compactifying a 4d $\mathcal{N}=2$ SCFT on $\mathbb{R}^2 \times \Sigma_{g,n}$, we can preserve part of the supersymmetry by performing a twist
on the Riemann surface. There are two possibilities: twisting by $U(1)_r$ or by the Cartan of $SU(2)_R$. The second option is the one relevant for our construction, as we wish to treat $\Sigma_{g,n}$ as the holomorphic directions of the HT twist.
Twisting by the Cartan of $SU(2)_R$ preserves
$(2,2)$ supersymmetry in the transverse $\mathbb{R}^2$.\footnote{Some aspects of this twist, including its connection to the standard Donaldson twist \cite{Witten:1988ze}, were investigated in \cite{Gukov:2017zao}.}
Reduction  on $\Sigma$ gives rise to a unitary SQFT
which is expected to flow to a superconformal field theory in the IR.
We thus obtain a family of two-dimensional $(2,2)$ SCFTs labelled by $\Sigma_{g,n}$ and, of course, by the choice of the 4d $\mathcal{N}=2$ SCFT $T$ that we started with. We baptize these theories as \emph{theories of class $\FF$}
and  denote them by 
\begin{equation}\label{eq:classF}
    \FF[T;\Sigma_{g,n};\dots]\,,
\end{equation}
where the dots stand for possible additional discrete data to be specified at each puncture. As in class~$\SS$, 
the conformal structure moduli of $\Sigma_{g, n}$ are interpreted as the conformal manifold (the space of exactly marginal couplings) of the 2d theory of class $\FF$. (More precisely, as we shall see, it is the {\it chiral} conformal manifold of the $(2, 2)$ theory.) On the the other hand, the K\"ahler structure moduli of  $\Sigma_{g, n}$ are expected to be irrelevant in the RG sense. In analogy with class $\SS$, we anticipate that different pairs-of-pants decompositions of $\Sigma_{g, n}$ (equivalenty, different degeneration limits) should correspond to different duality frames for the $(2, 2)$ SCFT.

\medskip

\noindent\textbf{2d non-unitary CFT.} Reduction on $S^2$ allows us to extract from the parent 4d $\mathcal{N}=2$ SCFT $T$ a full 2d CFT $\CC[T]$, whose chiral algebra coincides with the protected VOA $\VV[T]$. Unitarity of the 4d SCFT implies that the VOA is instead non-unitary \cite{Beem:2013sza}, thus the full non-chiral CFT will be non-unitary as well. From the compactification of the 4d theory on $S^2\times\Sigma_{g,n}$ we will in particular obtain the correlation function of this CFT on $\Sigma_{g,n}$, where the punctures correspond to the insertion of primary operators in the correlator. Different choices of operator insertions in the CFT correspond to different boundary conditions at the punctures in the compactification of the 4d SCFT $T$ on the Riemann surface and lead to different 2d theories of class $\FF$.

\medskip

\noindent\textbf{2d/2d unitary/non-unitary correspondence.} For every 4d $\mathcal{N}=2$ SCFT $T$, there exists a correspondence of the form
\begin{align}\label{eqn:corr}
    Z_{S_B^2}[\FF[T;\Sigma_{g,n};\mathcal{O}_1,\dots,\mathcal{O}_n]] \sim \langle \mathcal{O}_1\cdots\mathcal{O}_n\rangle_{\Sigma_{g,n}}^{\CC[T]}\,.
\end{align}
On the left hand side, we have the $S^2$ partition function of the unitary two-dimensional $(2,2)$ class $\FF$ theory obtained from $T$ by dimensional reduction on a Riemann surface $\Sigma_{g,n}$ of genus $g$ with $n$ punctures. As we shall review below, there is a choice between the $SU(2|1)_A$ and $SU(2|1)_B$ backgrounds in defining the $S^2$ partition function -- we will use the shorthands $S^2_A$ and $S^2_B$ when we need to emphasize this distinction. The correct choice for the 2d/2d unitary/non-unitary correspondence is $S^2_B$. (The A-type background motivates a {\it different} 2d/2d correspondence, which we briefly sketch in Section \ref{sec:A-type}.) 
On the right-hand side, we have an $n$-point genus-$g$ correlator of the 2d non-unitary CFT $\CC[T]$. It will be important for a dynamical check of this correspondence that the  $S^2$ partition function  of $(2, 2)$ SCFT
can be evaluated exactly by supersymmetric localization~\cite{Benini:2012ui,Doroud:2012xw} when the theory admits a  UV Lagrangian description.

\subsection*{A TQFT for the elliptic genus}
There is another observable of $(2, 2)$ SCFTs that is amenable to exact computation, namely their elliptic genus~\cite{Witten:1986bf, Witten:1993jg, Gadde:2013wq,Gadde:2013dda,Benini:2013nda,Benini:2013xpa}. The elliptic genus of a class $\FF$ theory can  be understood as a four-dimensional partition function of the parent theory $T$, this time on the background $T^2\times\Sigma_{g,n}$, turning on a holonomy for the R-symmetry on the Euclidean time cycle of $T^2$.
Consider again reducing in two different orders: 

\begin{center}
\tikzset{every picture/.style={line width=0.75pt}} %set default line width to 0.75pt        
\begin{tikzpicture}[x=0.75pt,y=0.75pt,yscale=-1,xscale=1,scale=.9]
\path[use as bounding box] (20,10) rectangle (660,180);

%Straight Lines [id:da055433214918889084] 
\draw [line width=1]    (264+30,56) -- (116.66+30,132.62) ;
\draw [shift={(114+30,134)}, rotate = 332.53] [color={rgb, 255:red, 0; green, 0; blue, 0 }  ][line width=1.5]    (14.21,-4.28) .. controls (9.04,-1.82) and (4.3,-0.39) .. (0,0) .. controls (4.3,0.39) and (9.04,1.82) .. (14.21,4.28)   ;
%Straight Lines [id:da6657060278054253] 
\draw [line width=1]    (365+30,55) -- (501.39+30,132.52) ;
\draw [shift={(504+30,134)}, rotate = 209.61] [color={rgb, 255:red, 0; green, 0; blue, 0 }  ][line width=1.5]    (14.21,-4.28) .. controls (9.04,-1.82) and (4.3,-0.39) .. (0,0) .. controls (4.3,0.39) and (9.04,1.82) .. (14.21,4.28)   ;

% Text Node
\draw (240,29) node [anchor=north west][inner sep=0.75pt]   [align=left] {4d $\displaystyle \mathcal{N} =2$ SCFT $T$ on $\displaystyle T^{2} \times \Sigma _{g,n}$};
% Text Node
\draw (25,142) node [anchor=north west][inner sep=0.75pt]   [align=left] {\begin{minipage}[lt]{171.41pt}\setlength\topsep{0pt}
~~~~~~~~~~Elliptic genus of 2d   \\$( 2,2)$
 unitary SCFT $\FF[T;\Sigma_{g, n};\dots]$

\end{minipage}};
% Text Node
\draw (430,152) node [anchor=north west][inner sep=0.75pt]   [align=left] {\begin{minipage}[lt]{207.42pt}\setlength\topsep{0pt}

 \hspace{.05in}$\Sigma_{g,n}$ correlator of 
2d TQFT $\EE[T]$
\end{minipage}};
% Text Node
\draw (68+30,74) node [anchor=north west][inner sep=0.75pt]   [align=left] {reduce on $\displaystyle \Sigma _{g,n}$};
% Text Node
\draw (451+30,74) node [anchor=north west][inner sep=0.75pt]   [align=left] {reduce on $\displaystyle T^{2}$};

\draw [line width=1]    (275+15,160) -- (380+15,160) ;
\draw [shift={(380+15,160)}, rotate = 180] [color={rgb, 255:red, 0; green, 0; blue, 0 }  ][line width=1.5]    (14.21,-4.28) .. controls (9.04,-1.82) and (4.3,-0.39) .. (0,0) .. controls (4.3,0.39) and (9.04,1.82) .. (14.21,4.28)   ;
\draw [shift={(275+15,160)}, rotate = 0] [color={rgb, 255:red, 0; green, 0; blue, 0 }  ][line width=1.5]    (14.21,-4.28) .. controls (9.04,-1.82) and (4.3,-0.39) .. (0,0) .. controls (4.3,0.39) and (9.04,1.82) .. (14.21,4.28)   ;
\draw (292,132) node [anchor=north west][inner sep=0.75pt]   [align=left] {correspondence};
\end{tikzpicture}
\end{center}
\vspace{.2in}

\noindent The dependence on the complex structure moduli of $\Sigma_{g,n}$ is lost here, because the elliptic genus does not depend on  continuous  deformations.
We are led at once to the statement that the elliptic genus of  $\FF[T; \Sigma_{g, n}; \dots]$  is computed by an $n$-point  TQFT correlator on a genus-$g$ surface. The TQFT
depends only on the parent four-dimensional theory $T$; we denote it by $\EE[T]$.  Its local operators are in one-to-one correspondence with the different types of punctures. 
The consistency of this TQFT imposes stringent constraints on the dimensional reductions of $T$, which we are going to check are satisfied in our examples.

Once again, this second ``$2+2 = 4$'' correspondence has a direct ``$4+2=6$" counterpart. 
The 4d analog of the elliptic genus is the superconformal index, namely the partition function on $S^3  \times S^1$ with suitable holonomies turned on.
In class $\SS$ theories, the superconformal index
 is computed by correlation functions of a 2d TQFT~\cite{Gadde:2009kb,Gadde:2011ik,Gadde:2011uv, Gaiotto:2012xa, Rastelli:2014jja}, which turns out to be a certain three-parameter generalization of $q$-deformed $\mathfrak{g}$-type Yang--Mills theory in the zero-area limit.
The superconformal index/TQFT correspondence
can be heuristically derived by
considering the 6d $(2,0)$ SCFT of type $\mathfrak{g}$ on  $S^3\times S^1\times\Sigma_{g,n}$:

\begin{center}
\tikzset{every picture/.style={line width=0.75pt}} %set default line width to 0.75pt     
\begin{tikzpicture}[x=0.75pt,y=0.75pt,yscale=-1,xscale=1,scale=.9]
\path[use as bounding box] (20,10) rectangle (690,180);
\draw [line width=1]    (264+30,56) -- (116.66+30,132.62) ;
\draw [shift={(114+30,134)}, rotate = 332.53] [color={rgb, 255:red, 0; green, 0; blue, 0 }  ][line width=1.5]    (14.21,-4.28) .. controls (9.04,-1.82) and (4.3,-0.39) .. (0,0) .. controls (4.3,0.39) and (9.04,1.82) .. (14.21,4.28)   ;
%Straight Lines [id:da6657060278054253] 
\draw [line width=1]    (365+30,55) -- (501.39+30,132.52) ;
\draw [shift={(504+30,134)}, rotate = 209.61] [color={rgb, 255:red, 0; green, 0; blue, 0 }  ][line width=1.5]    (14.21,-4.28) .. controls (9.04,-1.82) and (4.3,-0.39) .. (0,0) .. controls (4.3,0.39) and (9.04,1.82) .. (14.21,4.28)   ;

% Text Node
\draw (215,29) node [anchor=north west][inner sep=0.75pt]   [align=left] {6d $\displaystyle \mathcal{N} =(2,0)$ SCFT on $\displaystyle S^{3}\times S^1 \times \Sigma _{g,n}$};
% Text Node
\draw (40,142) node [anchor=north west][inner sep=0.75pt]   [align=left] {\begin{minipage}[lt]{171.41pt}\setlength\topsep{0pt}
~~~Superconformal index \\
of 4d $\mathcal{N} =2$ class $\SS$ SCFT

\end{minipage}};
% Text Node
\draw (470,142) node [anchor=north west][inner sep=0.75pt]   [align=left] {\begin{minipage}[lt]{207.42pt}\setlength\topsep{0pt}

 ~~~~~$\Sigma_{g,n}$ correlator of 
2d \\ Yang--Mills TQFT of type $\mathfrak{g}$ 
\end{minipage}};
% Text Node
\draw (68+30,74) node [anchor=north west][inner sep=0.75pt]   [align=left] {reduce on $\displaystyle \Sigma _{g,n}$};
% Text Node
\draw (451+30,74) node [anchor=north west][inner sep=0.75pt]   [align=left] {reduce on $\displaystyle S^{3}\times S^1$};

\draw [line width=1]    (275+15,160) -- (380+15,160) ;
\draw [shift={(380+15,160)}, rotate = 180] [color={rgb, 255:red, 0; green, 0; blue, 0 }  ][line width=1.5]    (14.21,-4.28) .. controls (9.04,-1.82) and (4.3,-0.39) .. (0,0) .. controls (4.3,0.39) and (9.04,1.82) .. (14.21,4.28)   ;
\draw [shift={(275+15,160)}, rotate = 0] [color={rgb, 255:red, 0; green, 0; blue, 0 }  ][line width=1.5]    (14.21,-4.28) .. controls (9.04,-1.82) and (4.3,-0.39) .. (0,0) .. controls (4.3,0.39) and (9.04,1.82) .. (14.21,4.28)   ;
\draw (292,132) node [anchor=north west][inner sep=0.75pt]   [align=left] {correspondence};
\end{tikzpicture}
\end{center}
\vspace{.1in}

\subsection*{The 2d/2d unitary/non-unitary correspondence in more detail}

We should first be precise about the off-shell supergravity background that needs to be turned on in order to define an
$S^2\times \Sigma$  compactification of $T$ that preserves the requisite supersymmetry. 
Consider the Riemann surface first. Since the four-dimensional
$\mathcal{N}=2$ superconformal algebra contains an $SU(2)_R\times U(1)_r$ R-symmetry, we can twist by either
$U(1)_r$ or by the Cartan of $SU(2)_R$ (see e.g.~\cite{Benini:2013cda,Benini:2012cz,Putrov:2015jpa,Gadde:2015wta,Gukov:2017zao,Bobev:2017uzs}). The HT construction of the VOA compels us to make the latter choice.

Let us open a brief parenthesis about the physics associated to the opposite choice, a
twist by $U(1)_r$ on $\Sigma$. This is only possible if the four-dimensional  theory does not have operators with fractional $U(1)_r$ charges.\footnote{More precisely, if the local operators of $T$ have $U(1)_r$ charges taking values in $\mathbb{Z}/p$, then the twist is allowed only when the $U(1)_r$ flux $2(1-g)-n$ is an integer multiple of $p$.} 
Considering the 4d theory $T$ on $\mathbb{R}^2 \times \Sigma$, this twist preserves $(0, 4)$ supersymmetry on the transverse  $\mathbb{R}^2$. Several interesting examples
of this construction have been studied in~\cite{Putrov:2015jpa,Nawata:2023aoq, Cui:2025eep}. 
It appears that these $(0, 4)$ theories are independent of both the complex structure moduli of $\Sigma$ and of the 4d conformal manifold.
One could then proceed to compactify this 2d
$(0, 4)$ theory on a second Riemann surface $\widetilde \Sigma$, twisting by the Cartan of $SU(2)_R$ on
$\widetilde \Sigma$. In other terms, we are doing a HT twist on $\widetilde \Sigma \times \Sigma$, but where now $\Sigma$ comprises the topological directions and
$\widetilde \Sigma$ the holomorphic ones.
This is a rich story, which is however
quite orthogonal to the main narrative of this paper.

Back to regular programming, we choose once and for all to twist by the Cartan of $SU(2)_R$ along the Riemann surface.
As we have already emphasized, reduction on $\Sigma$ leads to the family $\FF[T;\Sigma_{g, n};\dots  ]$ 
of  $(2, 2)$ superconformal field theories. 
A structural feature 
of $(2, 2)$ SCFTs is that their conformal manifold (locally) factorizes into chiral and twisted chiral exactly marginal 
deformations. It is important for our correspondence that the dependence of class $\FF$ theories on the complex structure moduli of $\Sigma_{g, n}$ resides in the {\it chiral} conformal manifold of the class $\FF$ theory, while the dependence on the 4d conformal manifold of $T$ (if any) resides in the {\it twisted chiral} conformal manifold.

Next, let us consider placing a class $\FF$
theory on $S^2$.
It would appear  that any $(2, 2)$ SCFT
can be placed supersymmetrically on $S^2$  by
simply performing a Weyl transformation. This is however misleading,
as the $S^2$ partition function is divergent and one is forced to choose a UV regulator that breaks conformal invariance. The upshot is that one must choose a ``massive'' subalgebra of the full $(2, 2)$ superalgebra, e.g.~a subalgebra that closes on the isometries of $S^2$ (as opposed to the full conformal isometries)~\cite{Benini:2012ui,Doroud:2012xw, Closset:2014pda}. There are two possible choices, each corresponding to a different off-shell supergravity background: the massive $SU(2|1)_A$ algebra, which contains the $U(1)_V$ vector R-symmetry of the $(2, 2)$ theory,
and the massive $SU(2|1)_B$ algebra, which contains the $U(1)_A$ axial R-symmetry. 
With either choice, the $S^2$ partition function can be computed by localization techniques, whenever the $(2, 2)$ SCFT can be realized as the IR fixed point of a Lagrangian UV theory.

\begingroup
\renewcommand{\arraystretch}{1.3}
\newcolumntype{L}{>{\centering\arraybackslash}m{.3\textwidth}}
\begin{table}
\begin{center}
    \begin{tabular}{L|L|L}
    4d & 2d unitary & 2d non-unitary \\\toprule
    4d $\mathcal{N}=2$ SCFT $T$ & 2d $(2,2)$ class $\FF$ theory & non-unitary CFT $\CC[T]$ with chiral algebra $\VV[T]$\\\hline
    $S_B^2\times\Sigma_{g,n}$ partition function & $S_B^2$ partition function & $n$-point genus $g$ correlator  \\\hline
    $\mathbb{R}^2_\epsilon\times \Sigma_{g,n}$ partition function & vortex partition function & chiral conformal block of $\VV[T]$ \\\hline
complex structure deformations of $\Sigma_{g,n}$ & chiral exactly marginal parameters & complex structure deformations of spacetime \\\hline 
    conformal manifold & twisted chiral exactly marginal parameters & --- \\\bottomrule
    \end{tabular}
    \caption{Basic dictionary of the 2d/2d unitary/non-unitary correspondence. }\label{tab:corroverview}
    \end{center}
\end{table}
\endgroup

We claim that the background relevant for our correspondence is the B-type background. A top-down argument 
uses the picture of Figure \ref{fig:spheregluing}. We realize 
 the VOA  via  HT twist (with $\Omega$-deformation)
of the 4d $\mathcal{N}=2$ SCFT, which requires 
performing a topological twist by $U(1)_r$ on the cigar. The four-dimensional $U(1)_r$ symmetry is mapped to the two-dimensional $U(1)_A$ axial symmetry after reduction on $\Sigma_{g,n}$. 
The claim then  follows by recalling that
the B-type $S^2$ partition function of a 2d $(2,2)$ theory does indeed factorize into hemispheres partition functions with opposite topological twist for $U(1)_A$~\cite{Benini:2012ui,Doroud:2012xw,Gomis:2012wy}.  (By contrast, of course, the A-type $S^2$
partition function factorizes into hemispheres partition functions\footnote{In both the A and the B cases, this is the old story of topological/antitopological fusion~\cite{Cecotti:1991me}. 
} with opposite topological twist for $U(1)_V$.) 
A bottom-up argument proceeds by insisting that both sides of our correspondence (\ref{eqn:corr}) must depend on the same
set of continuous parameters. Indeed, 
the B-type $S^2$ partition function depends on {\it chiral} exactly marginal deformations \cite{Gerchkovitz:2014gta}, which as we have 
mentioned descend from the complex structure moduli of $\Sigma_{g, n}$.

We have written a squiggle rather than an equal sign in equation \eqref{eqn:corr} because the $S^2$ partition function of a 2d $(2,2)$ theory is well-known to suffer from counterterm ambiguities (reflecting the ambiguity of the Kähler potential on the conformal manifold), which imply that it is only well-defined up to multiplication by 
the modulus square of a  holomorphic function of the exactly marginal parameters. Equation \eqref{eqn:corr} should thus be understood as an equality up to such holomorphically factorizing contributions. However, as we shall explain in Section \ref{subsec:counterterm}, this ambiguity is only present in two dimensions, and is resolved if one computes the full $S_B^2\times \Sigma_{g,n}$ partition function of the parent  4d $\mathcal{N}=2$ SCFT $T$. 

So far we have phrased the description of the off-shell supergravity background by considering separately the two two-dimensional factors, $\Sigma$ and $S^2$. This is sufficient for heuristic purposes but a rigorous  treatment  requires to  determine the full {\it four-dimensional} off-shell supergravity background.
This will be our first technical achievement in Section~\ref{sec:kinematics}. 
       
Factorization of the $S_B^2$ partition function of the class $\FF$ theory can be uplifted to factorization of the full 4d partition, which we denote by $Z[T, S_B^2\times\Sigma_{g,n}]$. Specifically, 
$Z[T, S_B^2\times \Sigma_{g,n}]$
can be factorized into two copies of the partition function on $(\mathbb{R}^2_\epsilon)_B \times\Sigma_{g,n}$ with opposite $U(1)_r$ twist on $\mathbb{R}^2_\epsilon$. This mirrors the conformal blocks decomposition of the CFT correlator, since the $Z[T, (\mathbb{R}^2_\epsilon)_B \times\Sigma_{g,n}]$ engineers an $n$-point genus-$g$ chiral conformal block of the associated VOA $\VV[T]$. 

The $\mathbb{R}^2_\epsilon \times\Sigma$ background
is nearly identical to the one studied by Nekrasov-Shatashvili~(NS)~\cite{Nekrasov:2009rc,Nekrasov:2009ui,Nekrasov:2009uh,Nekrasov:2014xaa}, but with the crucial difference that along the cigar directions we are twisting by $U(1)_r$ (we have indicated this choice by the subscript $B$ in $(\mathbb{R}^2_\epsilon)_B$), while in NS one is twisting by the Cartan of  $SU(2)_R$ (we  denote that choice by $(\mathbb{R}^2_\epsilon)_A$). The resulting physics is very different. To wit, with the choice of type-B background $(\mathbb{R}^2_\epsilon)_B \times\Sigma$ we retain dependence on the {\it conformal structure} of $\Sigma$ (while we lose dependence on the 4d conformal manifold), indeed the partition function yields a conformal block of the VOA $\VV[T]$. By contrast, with the NS choice~\cite{Nekrasov:2009rc,Nekrasov:2014xaa} of A-type  background $(\mathbb{R}^2_\epsilon)_A \times\Sigma$ the partition function depends only on the {\it topology} of $\Sigma$ but retains dependence on the 4d conformal manifold. Another obvious notable difference is that the B-type background is only available for four-dimensional {\it conformal} ${\cal N}=2$ theories, while the A-type can be defined for any  ${\cal N}=2$ QFT. As we have mentioned, there is in fact a distinct 2d/2d correspondence associated with the A-type background. We preview some of its aspects in Section~\ref{sec:A-type} but leave a detailed analysis to future work.

\subsection*{A minimal example}

By now there are large classes of $\mathcal{N}=2$ SCFTs for which the associated chiral algebras have been identified. The vast majority of them are non-rational, though they are expected to enjoy a finiteness condition known as ``quasi-lisse'' \cite{Arakawa:2016hkg}. Indeed, a central conjecture~\cite{Beem:2017ooy}
identifies the so-called ``associated variety''
of $\VV[T]$ with the Higgs branch of $T$. 
Having a trivial associated variety is equivalent to $\VV[T]$ being $C_2$-cofinite, which is necessary for $\VV[T]$ to be (strongly) rational.\footnote{What physicists refer to as ``rational chiral algebras'' are axiomatized by \emph{strongly} rational VOAs in the mathematics literature, see e.g.~\cite{Creutzig:2016fms} for a review. In a strongly rational VOA, one imposes, among other things, both semisimplicity of the representation theory (rationality) as well as a technical condition called $C_2$-cofiniteness. The conjecture of \cite{Beem:2017ooy} implies that Higgsless 4d SCFTs  give rise to VOAs that are only guaranteed to enjoy the second of these two properties -- in the language of \cite{Creutzig:2016fms} one expects such Higgsless VOAs to be strongly finite. It is a further ``experimental'' fact that almost all presently known Higgless VOAs happen to be strongly rational, see the vast set  arising from Argyres-Douglas theories catalogued in~\cite{Xie:2019vzr}. There are however a few examples of Higgless  VOAs arising from {\it Lagrangian} 4d theories that are strongly finite but not rational~\cite{LagrangianC2cofinite}.  } Thus, only 4d theories without a Higgs branch can give rise to (strongly) rational VOAs.
More generally, the expectation that the Higgs branch of $T$ possesses only finitely many symplectic leaves translates to the condition that $\VV[T]$ is quasi-lisse.

A central open problem (also in the mathematical literature) is to upgrade any of these interesting quasi-lisse VOAs to full-fledged two-dimensional CFTs.\footnote{See \cite{Hofer:2025oif} for recent progress on this problem for $C_2$-cofinite VOAs, generalizing the analogous story for (strongly) rational VOAs in \cite{Felder:1999mq,Fuchs:2002cm,Fjelstad:2005ua}. As noted in the previous paragraph, $C_2$-cofinite VOAs include all VOAs which descend from 4d SCFTs without a Higgs branch.} We believe that our new correspondence will give fundamental clues for accomplishing this, but we leave this exciting direction for future work. 

In this paper, we instead choose to consider cases where the associated VOA is strongly rational. This will  make the identification of the non-chiral CFT immediate, the only potential ambiguity being a choice of modular invariant gluing of left-movers with right-movers, but even that will be absent in the cases that we study. However, in order for the correspondence to be a concrete rather than a formal statement, we wish to give an explicit, genuinely two-dimensional description of the class $\FF$ theories on the unitary side. This will be one of our principal tasks.

We shall focus on the simplest example, taking $T$ to be the $(A_1,A_2)$ Argyres--Douglas theory \cite{Argyres:1995jj,Argyres:1995xn}, sometimes also denoted by $H_0$. The $(A_1,A_2)$ model is the ``minimal'' parent theory that one can choose as input for our construction. To wit, it has the smallest central charge $c_{\rm 4d}=\tfrac{11}{30}$ of any interacting 4d $\mathcal{N}=2$ SCFT \cite{Liendo:2015ofa}. This minimality is mirrored on the non-unitary side of the correspondence. Indeed, it is known \cite{Cordova:2015nma} that the associated VOA $\VV[(A_1,A_2)]$ is the (simple quotient of the) $c_{2\mathrm{d}}=-\tfrac{22}{5}$ Virasoro algebra, which is the chiral algebra of the simplest minimal model, the Lee-Yang model $\mathcal{M}(2,5)$. The Lee-Yang model is of course the  unique non-chiral CFT with this VOA, so we 
immediately conclude
$\CC[(A_1,A_2)]=\mathcal{M}(2,5)$.

One of our main results is a  proposal for a concrete UV Lagrangian description of $\FF[(A_1,A_2), \Sigma_{0, n}] $.
\begin{table}
  \centering
  \begin{tabular}{c|cccccc}
    & $\Phi_1$ & $\Phi_2$ & $\tilde{\Phi}_1$ & $\tilde{\Phi}_2$ & $F_1$ & $F_2$ \\\hline
    $U(1)_{\text{gauge}}$ & 1 & 1 & $-1$ & $-1$ & 0 & 0 \\
    $U(1)_R$ & $\frac{1}{5}$ & $\frac{2}{5}$ & $\frac{1}{5}$ & $\frac{2}{5}$ & $\frac{3}{5}$ & $\frac{1}{5}$
  \end{tabular}
  \caption{Matter content of the  UV Lagrangian description of $\FF[(A_1, A_2),\Sigma_{0, 4}] $.}\label{tab:4ptH0}
\end{table}
The simplest example is the reduction of $(A_1,A_2)$ on a four-punctured sphere. We conjecture that $\FF[(A_1,A_2), \Sigma_{0, 4}] $ is the IR fixed point of
a two-dimensional  $(2,2)$ gauged linear sigma model (GLSM) comprising six {\it twisted chiral} superfields and a (twisted)
$U(1)$ gauge field  (whose field strength is a chiral superfield), 
with charge assignments as in Table \ref{tab:4ptH0} and superpotential given by 
\begin{align}
    \mathcal{W} = \Phi_1^2\tilde{\Phi}_1\tilde{\Phi}_2+\Phi_1\Phi_2\tilde{\Phi}_1^2+F_1\Phi_1\tilde{\Phi}_1+F_2\Phi_2\tilde{\Phi}_2+F_2^5\,.
\end{align}
This theory possesses an exactly marginal complexified Fayet--Iliopoulos parameter $\xi$ \cite{Fayet:1974jb}. Since all the fields are taken to be twisted, this is a {\it chiral} deformation, which we identify with the complex structure modulus $z$ of $\Sigma_{0,4}$ via $z=e^{-2\pi \xi}$.
Indeed, one can check, using formulae obtained via supersymmetric localization, that the B-type $S^2$ partition function precisely reproduces the four-point function of the unique non-trivial primary $\phi_{(2,1)}(z,\bar z)$ of dimension $h=\bar h=-\tfrac{1}{5}$ in the Lee--Yang minimal model (up to counterterm ambiguities described further in Section~\ref{subsec:counterterm}):
\begin{align}
\begin{split}
    Z_{S^2}[\FF[(A_1,A_2),\Sigma_{0,4}]](\xi,\bar{\xi}) &\sim |z|^{\frac{4}{5}}|1-z|^{\frac{4}{5}}\Big|\dFu{\frac{3}{5},\frac{4}{5};\frac{6}{5};z}\Big|^2\nn
    +C_{\phi\phi\phi}^2|z|^{\frac{2}{5}}|1-z|^{\frac{4}{5}}\Big|\dFu{\frac{2}{5},\frac{3}{5};\frac{4}{5};z}\Big|^2\\[1ex]
    &=\langle \phi_{(2,1)}(0)\phi_{(2,1)}(1)\phi_{(2,1)}(\infty)\phi_{(2,1)}(z,\bar z)\rangle_{\Sigma_{0,4}}^{\mathcal{M}(2,5)}\, ,
\end{split}
\end{align}
with $C_{\phi\phi\phi}$ the OPE coefficient given in equation \eqref{eq:YLCppp}. 
We extend this analysis to five- and higher-point functions. In this case, we use known Coulomb gas integral expressions for the conformal blocks to engineer GLSMs whose hemisphere partition functions reproduce them. We again interpret these GLSMs as describing the compactifications of $(A_1,A_2)$ on punctured spheres with $SU(2)_R$ twist. 

We subject this proposal to a variety of stringent consistency checks. 
Leveraging our knowledge of various properties of the parent 4d $\mathcal{N}=2$ SCFT and of the details of the compactification, we make various predictions for the resulting 2d $(2,2)$ theory, which we can independently  reproduce in our model -- e.g.~by computing  central charges and the elliptic genus. Perhaps our most striking result regards the TQFT structure of the elliptic genus. 
The conjectured GLSM for $\FF[(A_1,A_2),\Sigma_{0,4}]$, anomaly considerations, and associativity of the TQFT are sufficient to completely determine the propagators and structure constants of the TQFT $\EE[(A_1,A_2)]$, see \eqref{TQFTsummary}. We can then for example check that the TQFT prediction for the elliptic genus of $\FF[(A_1,A_2),\Sigma_{0,5}]$ agrees with the direct localization calculation in the GLSM. In fact, by this route we are able to 
determine the elliptic genus of $\FF[(A_1,A_2),\Sigma_{g,n}]$ for arbitrary genus and number of punctures, even if we currently a lack a genuine two-dimensional Lagrangian description of the $(2, 2)$ theories with $g>0$.

The minimal example $(A_1, A_2)$ is the first of an infinite sequence $(A_1,A_{2k})$ of Argyres--Douglas theories, for which $\CC [(A_1,A_{2k})]$ is the non-unitary minimal model CFT $\mathcal{M}(2, 2k+3)$.
Our results generalize straightforwardly to the whole sequence, even if the story becomes more intricate because the number of distinct types of punctures increases with $k$. We study in some detail the correspondence for the second instance, $\CC [(A_1,A_{4})] = {\cal M}(2, 7)$.

\bigskip

\noindent
The detailed organization of the paper is best apprehended from the Table of Contents. In Section~\ref{sec:kinematics} we construct the $S^2\times\Sigma$ rigid supergravity background and address some kinematical aspects of the correspondence from the four-dimensional perspective. In Section~\ref{sec:gencomp} we instead discuss some general two-dimensional features of the correspondence, both on the unitary and the non-unitary sides. In Section~\ref{sec:dimredA1A2} we consider our first interacting example, the $(A_1,A_2)$ minimal Argyres--Douglas SCFT, focusing on its compactifications on punctured spheres. In Section~\ref{sec:EGTQFT} we investigate the TQFT structure of the elliptic genera of the $\FF[(A_1,A_2),\Sigma]$ theories. In Section~\ref{sec:dimredA1A4} we extend our analysis to the $(A_1,A_4)$ Argyres--Douglas SCFT. We conclude in Section~\ref{sec:discussion} with a brief discussion of  open questions and directions for future research. The main text is supplemented by several appendices with background material and additional technical details.

\subsection*{Notations and conventions}

We summarize here the meaning of the main symbols that we will use throughout the paper.
\begin{itemize}
    \item $T$: a generic 4d $\mathcal{N}=2$ SCFT.
    \item $\Sigma_{g,n}$: a Riemann surface of genus $g$ with $n$ punctures.
    \item $\SS[\mathfrak{g};\Sigma_{g,n};\dots]$: the 4d $\mathcal{N}=2$ class $\SS$ theory from reduction of the 6d $(2,0)$ $\mathfrak{g}$-type SCFT 
    on~$\Sigma_{g,n}$.
    \item $\FF[T;\Sigma_{g,n};\dots]$: the 2d $(2,2)$ class $\FF$ theory obtained by reducing $T$ on $\Sigma_{g,n}$ with $SU(2)_R$ twist.
    \item $\VV[T]$: the VOA associated to $T$ via the SCFT/VOA correspondence. We can think of it as arising from reducing $T$ on $(\mathbb{R}^2_\epsilon)_B$ (the $\Omega$-deformed cigar with $U(1)_r$ twist).
    \item $\CC[T]$: the 2d CFT (whose chiral algebra is  $\VV[T]$) obtained by reducing $T$ on the B-type $S^2$.
    \item $\EE[T]$: the 2d TQFT that computes the elliptic genus of class $\FF$, obtained by reducing $T$ on $T^2$.
    \item $\BB[T]$: the relative 2d TQFT obtained from 
    A-twisting the $(2, 2)$ theory that arises by reduction of $T$ on $(\mathbb{R}^2_\epsilon)_A$ (the $\Omega$-deformed cigar with $U(1)_R \subset SU(2)_R$ twist).
    \item $\TT[T]$: the genuine 2d TQFT obtained by reducing $T$ on the A-type $S^2$.
\end{itemize}

%%%%%%%%%%%%%%%%%%%%%%%%%%%%%%%%%
\section{Kinematics of the correspondence: the view from 4d}\label{sec:kinematics}

In this section we consider compactifications of four-dimensional ${\cal N}=2$ SCFTs  on $S^2\times\Sigma$.
Our goal is to construct off-shell supergravity backgrounds that include an $SU(2)_R$
twist on $\Sigma$
and 
 preserve four supercharges. Our main focus is on the B-type background that leads to our 2d/2d unitary/non-unitary correspondence, but as we are at it we also construct the A-type background (relevant for a different 2d/2d correspondence briefly previewed in Section \ref{sec:A-type}).
The results are then put to use to argue in favor of the correspondence and to elucidate some of its properties.

\subsection{Supersymmetric backgrounds}
\label{subsec:background}

 We would like to place a four-dimensional $\mathcal{N}=2$ theory $T$ on $S^2\times\Sigma$ while preserving some of the supersymmetry transformations. As we will see, we will preserve those parametrized by two 2d Dirac spinors on $S^2$, and the corresponding supersymmetry algebra will be $SU(2|1)$. Depending on the choice of background, this will be either $SU(2|1)_A$ or $SU(2|1)_B$.

\subsubsection{Coupling to background supergravity}

We follow the approach of Festuccia--Seiberg \cite{Festuccia:2011ws} of coupling the theory to a suitable supergravity multiplet, which we then freeze so that the metric is that of the curved space we are interested in and in such a way that we preserve some of the supersymmetries. As it is customary, one sets the gravitino to zero, which is consistent only if its variation under supersymmetry transformations is also zero. This leads to Killing spinor equations for the spinors that parametrize the supersymmetry transformations. If the Killing spinor equations for a fixed choice of the background supergravity fields admit a non-trivial solution for some of the spinors, then the background preserves the corresponding supersymmetries. We focus here on 4d $\mathcal{N}=2$ theories that possess at least an $SU(2)_R$ R-symmetry. For the case of the background preserving $SU(2|1)_A$ the theory does not need to have a $U(1)_r$ R-symmetry, while the one preserving $SU(2|1)_B$ exists only for 4d $\mathcal{N}=2$ SCFTs that exhibit the full $SU(2)_R\times U(1)_r$.

When looking for a supersymmetric background for 4d $\mathcal{N}=2$ theories, one needs to couple the theory to 4d $\mathcal{N}=2$ conformal supergravity with an additional compensator field to gauge fix the conformal symmetries. Some canonical choices of compensator are the vector multiplet and the tensor multiplet compensator. However, for constructing the $S^2\times\Sigma$ background of our interest we will need to follow the formalism of \cite{Butter:2015tra} of considering a generic real multiplet compensator $\Omega$ of dimension two. The Killing spinor equations derived in \cite{Butter:2015tra} with this choice of compensator in Euclidean signature are
\begin{align}\label{eq:4dKSE}
    D_m\xi^i_\alpha&=-2iG^n(\sigma_{nm}\xi^i)_\alpha-G^{ni}{}_j(\sigma_n\bar{\sigma}_m\xi^j)_\alpha-\frac{i}{2}\bar{S}^{i}{}_j(\sigma_m\bar{\xi}^j)_\alpha-\frac{i}{2}\bar{Z}_{mn}(\sigma^n\bar{\xi}^i)_\alpha\,,\nn\\
    D_m\bar{\xi}^{\dot{\alpha}i}&=2iG^n(\bar{\sigma}_{nm}\bar{\xi}^i)^{\dot{\alpha}}-G^{ni}{}_j(\bar{\sigma}_n\sigma_m\bar{\xi}^j)^{\dot{\alpha}}+\frac{i}{2}S^{i}{}_j(\bar{\sigma}_m\xi^j)^{\dot{\alpha}}+\frac{i}{2}Z_{mn}(\bar{\sigma}^n\xi^i)^{\dot{\alpha}}\,,
\end{align}
where $m,n=1,\cdots,4$ are the spacetime indices, $i,j=1,2$ are the $SU(2)_R$ indices, and $\alpha=\pm$ and $\dot{\alpha}=\dot{\pm}$ are the spinor indices. The covariant derivative contains the spin connection $\omega_m^{ab}$, the $SU(2)_R$ gauge field $V_m{}^i{}_j$ and the $U(1)_r$ gauge field $V_m$,\footnote{We work in the $K$-gauge where the dilatation connection vanishes.} however for our purposes we will always set the latter to zero.

Since we are working in Euclidean, the spinors $\xi^i_\alpha$ and $\bar{\xi}^i_{\dot{\alpha}}$ are independent. They transform in the $\bf 2$ of $SU(2)_R$ and have charges under $U(1)_r$
\begin{equation}\label{eq:4dspinorRcharges}
    r(\xi^i_\alpha)=1\,,\qquad r(\bar{\xi}^i_{\dot{\alpha}})=-1\,.
\end{equation}
The other fields on the r.h.s.~of \eqref{eq:4dKSE} are combinations of the components of the Weyl multiplet and the compensator $\Omega$. Their transformation properties under $SU(2)_R$ are obvious from the index structure, namely $G_m^{ij}$, $S^{ij}$, $\bar{S}^{ij}$ are triplets while $G_m$, $Z_{mn}$ are scalars. Their charges under $U(1)_r$ are easily determined from those of the spinors $\xi^i_\alpha$ and $\bar{\xi}^i_{\dot{\alpha}}$ by consistency with the Killing spinor equation \eqref{eq:4dKSE}
\begin{equation}
    r(G_m)=r(G_m^{ij})=0\,,\qquad r(S^{ij})=r(Z_{mn})=-2\,.
\end{equation}
Some canonical choices of compensators, like the vector multiplet or the tensor multiplet compensators, can be recovered by specializing appropriately these background fields. However, for our purposes we will need to work with the more general form of the compensator used in \cite{Butter:2015tra}.

Requiring reflection positivity for the theory in Euclidean signature implies various reality conditions on the spinors and on the background supergravity fields \cite{Butter:2015tra}. In particular, the spinors are Majorana--Weyl
\begin{equation}
    (\xi^i_\alpha)^*=\xi_i^\alpha\,,\qquad (\bar{\xi}^i_{\dot{\alpha}})^*=\bar{\xi}_i^{\dot{\alpha}}\,,
\end{equation}
while the background fields satisfy
\begin{equation}\label{eq:4dreality}
    (G_m)^*=-G_m\,,\quad (G^{ij}_m)^*=G_{mij}\,,\quad (S^{ij})^*=-S_{ij}\,,\quad (Z_{mn})^*=-Z_{mn}\,,
\end{equation}
and similarly for $\bar{S}^{ij}$ and $\bar{Z}_{mn}$. However, as discussed in \cite{Festuccia:2011ws}, in order to preserve supersymmetry on certain Euclidean manifolds one is forced to set the background fields to more general complex values, which might violate reflection positivity. This is completely fine if the goal is to compute the partition function on such manifolds using supersymmetric localization. The only price to pay is that the partition function will in general be complex rather than real. As we will see, this will be the case for the background $S^2\times\Sigma$ preserving $SU(2|1)_B$ on $S^2$ that we are primarily interested in.

Our goal is to find solutions to these equations where the background is $S^2\times\Sigma$. These lie beyond the classification of \cite{Butter:2015tra}, in that they only preserve half of the supersymmetries. We will take $m=1,2$ to correspond to the $S^2$ directions and $m=3,4$ to the $\Sigma$ directions, and accordingly decompose the various fields. In the following we will focus on the behavior on $\Sigma$ and on $S^2$ in turn, where in particular on $S^2$ we will show how to realize $SU(2|1)_A$ or $SU(2|1)_B$.

\subsubsection{Twist on $\Sigma$}

On $\Sigma$ we perform a topological twist\footnote{Following standard usage, we use the word ``topological'' here even if the theory retains dependence on the conformal structure moduli of $\Sigma$.  } by the Cartan $U(1)_R$ of the $SU(2)_R$ symmetry. Specifically, we consider the $U(1)$ isometry of $\Sigma$, which is given by a diagonal combination of the Cartan generators of the four-dimensional $SU(2)_1\times SU(2)_2$ Lorentz symmetry
\begin{equation}
    J=j_1+j_2\,,
\end{equation}
where $J$ denotes the spin on $\Sigma$, while $j_1$, $j_2$ are the 4d spins. We then mix the isometry of $\Sigma$ with  $U(1)_R$ so that half of the supercharges $Q^i_\alpha$, $\bar{Q}^i_{\dot{\alpha}}$ are scalars. These supercharges then survive the reduction on $\Sigma$ and remain as supercharges of the effective theory on $S^2$. Denoting the new spin on $\Sigma$ after the topological twist by\footnote{We normalize the $U(1)_R$ Cartan of $SU(2)_R$ so that the 4d supercharges have R-charges $\pm1$. We remind that we are also normalizing $U(1)_r$ such that the supercharges have R-charge $\pm1$.}
\begin{equation}\label{eq:newspinSigma}
    s=\frac{1}{2}R+J\,,
\end{equation}
where $R$ is the $U(1)_R$ charge, the preserved supercharges are
\begin{equation}
    Q^1_-\,,\,\, Q^2_+\,,\,\,\bar{Q}^1_{\dot{-}}\,,\,\,\bar{Q}^2_{\dot{+}}\,.
\end{equation}
This gives two chiral and two anti-chiral supercharges, so that after the reduction on $\Sigma$ we get an effective 2d $(2,2)$ theory. These are the theories of class $\FF$, which we dub $\FF[T;\Sigma;\dots]$ to stress that they depend on the 4d $\mathcal{N}=2$ SCFT $T$ with started with, on the choice of Riemann surface $\Sigma$ and on additional discrete data at the punctures encoded in the dots.

For this class of theories, the $U(1)_R$ Cartan of $SU(2)_R$ is naturally identified with the 2d vector symmetry, while $U(1)_r$ is identified with the axial symmetry
\begin{equation}\label{eq:4d2dRsymm}
    U(1)_V=U(1)_R\,,\qquad U(1)_A=U(1)_r\,.
\end{equation}
The preserved supercharges then have charges in our conventions
\begin{table}[h]
  \centering
  \begin{tabular}{c|cccc}
    & $Q^1_-$ & $Q^2_+$ & $\bar{Q}^1_{\dot{-}}$ & $\bar{Q}^2_{\dot{+}}$ \\\hline
    $U(1)_V$ & 1 & $-1$ & 1 & $-1$ \\
    $U(1)_A$ & 1 & 1 & $-1$ & $-1$
  \end{tabular}
\end{table}

\noindent In terms of the 2d right and left moving R-symmetries, which we denote by $U(1)_+$ and $U(1)_-$, we have instead
\begin{equation}\label{eq:2dRsymmLR}
    U(1)_+=\frac{1}{2}\left(U(1)_R+U(1)_r\right)\,,\qquad U(1)_-=\frac{1}{2}\left(U(1)_R-U(1)_r\right)\,,
\end{equation}
so $Q^1_-$, $\bar{Q}^2_{\dot{+}}$ are the right-moving supercharges while $Q^2_+$, $\bar{Q}^1_{\dot{-}}$ are the left-moving ones.

At the level of the spinors parametrizing the supersymmetry transformations, as we will see momentarily, we will make a choice for the background fields such that the r.h.s.~of the Killing spinor equations \eqref{eq:4dKSE} along $\Sigma$ are equal to zero for the surviving spinors
\begin{equation}\label{eq:4dKSESigma}
    D_m\xi^1_-=D_m\xi^2_+=D_m\bar{\xi}^1_{\dot{-}}=D_m\bar{\xi}^2_{\dot{+}}=0\,,\qquad m=3,4\,.
\end{equation}
Recall that the covariant derivative acts on a generic unbarred Weyl spinor as
\begin{equation}
    D_m\xi^i_\alpha=\partial_m\xi^i_\alpha-\frac{1}{2}\omega_m^{ab}(\sigma_{ab})^\beta{}_\alpha\xi^i_\beta-iV_m{}^i{}_j\xi^j_\alpha\,.
\end{equation}
Let us focus on the $\Sigma$ directions $m=3,4$ and denote by $\omega_m^{(\Sigma)}=\omega_m^{34}$ its spin connection. Specializing the $SU(2)_R$ background gauge field along the $U(1)_R$ Cartan and taking it to be proportional to the spin connection
\begin{equation}\label{eq:SU2RgaugeSigma}
    V_m{}^i{}_j=\frac{1}{2}\omega_m^{(\Sigma)}(\tau^3)^i{}_j\,,
\end{equation}
then the covariant derivative along $\Sigma$ becomes (we also use $\sigma_{34}=\tfrac{i}{2}\sigma_3$)
\begin{equation}
    D_m\xi^i_\alpha=\partial_m\xi^i_\alpha-\frac{i}{2}\omega_m^{(\Sigma)}\left((\sigma_{3})^\beta{}_\alpha\delta^i{}_j+\delta^\beta{}_\alpha(\tau_3)^i{}_j\right)\xi^j_\beta\,,
\end{equation}
where in the previous expressions we denoted by $(\tau_3)^i{}_j$ the third Pauli matrix acting on $SU(2)_R$ indices to distinguish it from the one acting on the spinor indices. We immediately see that for the spinors $\xi^1_-$, $\xi^2_+$ which are in eigenspaces of $\sigma_3$ and $\tau_3$ of opposite eigenvalues, the bracket vanishes and the covariant derivative acts just as a standard derivative. By repeating a similar analysis for the barred spinors, one then finds that the Killing spinor equations \eqref{eq:4dKSESigma} on $\Sigma$ reduce to
\begin{equation}
    \partial_m\xi^1_-=\partial_m\xi^2_+=\partial_m\bar{\xi}^1_{\dot{-}}=\partial_m\bar{\xi}^2_{\dot{+}}=0\,,\qquad m=3,4\,,
\end{equation}
which can be easily solved by taking them to be constant along $\Sigma$. On the other hand, the remaining spinors have to be set to zero to find a solution to the Killing spinor equations. Again we come to the conclusion that we preserve $(2,2)$ supersymmetry in the two directions orthogonal to $\Sigma$.

We also notice that the choice of the $SU(2)_R$ background gauge field \eqref{eq:SU2RgaugeSigma} implies that there is a non-trivial flux for the $U(1)_R$ Cartan along $\Sigma$
\begin{equation}
    c_1(R)=\frac{1}{4\pi}\int_\Sigma \mathrm{d}\omega^{(\Sigma)}=\frac{1}{2}\chi(\Sigma)=1-g\,.
\end{equation}

\subsubsection{$S^2\times \Sigma$ backgrounds}\label{subsubsec:S2Sigmabackgrounds}

It is known that one can place a 2d $(2,2)$ theory on $S^2$ preserving an $SU(2|1)$ supersymmetry algebra, which is enough to compute the partition function with supersymmetric localization \cite{Benini:2012ui,Doroud:2012xw}. Here we would like to show that the 4d Killing spinor equations \eqref{eq:4dKSE} can be solved so to have the same superalgebra on $S^2$ for the $(2,2)$ class $\FF$ theory obtained by dimensional reduction along $\Sigma$. In particular, we will show that the 4d Killing spinor equations on the $S^2$ directions reduce to the 2d ones studied in \cite{Benini:2012ui,Doroud:2012xw} when a suitable choice of the background fields is made. We start discussing the case that preserves $SU(2|1)_B$ on $S^2$ which is most relevant for us, but later we will also consider a background preserving $SU(2|1)_A$.

\subsubsection*{Preserving $SU(2\vert 1)_B$ on $S^2$}
To obtain the background preserving $SU(2|1)_B$, we turn off all the fields $G_n$, $S^{ij}$, $\bar{S}^{ij}$, $Z_{mn}$, $\bar{Z}_{mn}$ and turn on only $G^{mi}{}_j$ with the following profile:
\begin{equation}\label{eq:Gbkg}
    G^1=G^2=0\,,\qquad G^3=\frac{1}{4R_{S^2}}\tau_2\,,\qquad G^4=\frac{1}{4R_{S^2}}\tau_1\,,
\end{equation}
where $R_{S^2}$ denotes the radius of $S^2$. In particular, this field is turned on along $\Sigma$ but not $S^2$. We also point out that this background explicitly breaks also the $U(1)_R$ Cartan of $SU(2)_R$. However, it preserves $U(1)_r$ since the only field that has been activated is uncharged under it. Remembering the identification \eqref{eq:4d2dRsymm} of $U(1)_R$ with the vector symmetry and $U(1)_r$ with the axial symmetry, this is the first indication that we are preserving $SU(2|1)_B$ on $S^2$, since this superalgebra contains the axial symmetry but not the vector symmetry.

It is also important to notice that this expectation value for the $G^{mi}_j$ field satisfies the modified reality condition
\begin{equation}
    (G^{ij}_m)^*=-G_{mij}\,.
\end{equation}
In particular, since the condition in \eqref{eq:4dreality} is violated, this means that reflection positivity is not preserved in this background and that the resulting partition function will generically be complex. By the correspondence, the partition function on $S^2\times\Sigma$ coincides with the correlator of the 2d CFT $\CC[T]$ on $\Sigma$. We thus conclude that $\CC[T]$ should be non-unitary, since its correlation functions are complex. This is in agreement with our general expectation that the 2d CFT $\CC[T]$ should have as its chiral algebra the VOA $\VV[T]$ of the parent 4d $\mathcal{N}=2$ SCFT $T$, which is known to be non-unitary \cite{Beem:2013sza}.

With the choice \eqref{eq:Gbkg} we can first of all show that the Killing spinor equations on $\Sigma$ are
\begin{equation}
    D_m\xi^i_\alpha=D_m\bar{\xi}^i_{\dot{\alpha}}=0\,,\qquad m=3,4\,.
\end{equation}
One can thus preserve half of the spinors with a topological twist by $SU(2)_R$ on $\Sigma$, as previously explained. Setting the killed spinors to zero and focusing on the Killing spinor equations for the surviving ones in the directions $m=1,2$, one finds for the unbarred spinors
\begin{align}
D_{1}\xi^{2}_{+} &= \frac{i}{2R_{S^2}}\xi^{1}_{-}\,, &
D_{2}\xi^{2}_{+} &= \frac{1}{2R_{S^2}}\xi^{1}_{-}\,,\nn\\
D_{1}\xi^{1}_{-} &= \frac{i}{2R_{S^2}}\xi^{2}_{+}\,, &
D_{2}\xi^{1}_{-} &= -\frac{1}{2R_{S^2}}\xi^{2}_{+}\,,
\end{align}
and similarly for the barred spinors
\begin{align}
D_{1}\bar{\xi}^2_{\dot{+}} &= \frac{i}{2R_{S^2}}\bar{\xi}^1_{\dot{-}}\,, &
D_{2}\bar{\xi}^2_{\dot{+}} &= \frac{1}{2R_{S^2}}\bar{\xi}^1_{\dot{-}}\,,\nn\\
D_{1}\bar{\xi}^1_{\dot{-}} &= \frac{i}{2R_{S^2}}\bar{\xi}^2_{\dot{+}}\,, &
D_{2}\bar{\xi}^1_{\dot{-}} &= -\frac{1}{2R_{S^2}}\bar{\xi}^2_{\dot{+}}\,,
\end{align}
As in \cite{Benini:2012ui,Doroud:2012xw}, we do not turn on any R-symmetry gauge field on $S^2$. Moreover, we repackage the spinors into 2d Dirac spinors as
\begin{equation}\label{eq:2dKS}
    \epsilon=\begin{pmatrix}
        \xi^2_+ \\ \xi^1_-
    \end{pmatrix}\,,\qquad \bar{\epsilon}=\begin{pmatrix}
        \bar{\xi}^2_{\dot{+}} \\ \bar{\xi}^1_{\dot{-}}
    \end{pmatrix}\,.
\end{equation}
We end up with the Killing spinor equations preserving $SU(2|1)_B$ on $S^2$ (see e.g.~\cite{Benini:2012ui,Doroud:2012xw,Closset:2014pda,Gerchkovitz:2014gta})
\begin{equation}\label{eq:2dKSE}
    \nabla_m\epsilon=\frac{i}{2R_{S^2}}\gamma_m\epsilon\,,\qquad    \nabla_m\bar{\epsilon}=\frac{i}{2R_{S^2}}\gamma_m\bar{\epsilon}\,,\qquad m=1,2\,,
\end{equation}
where $\gamma_m$ are the 2d gamma matrices which coincide with the Pauli matrices. The fact that these equations correspond to $SU(2|1)_B$ can be understood from the charge assignments of the spinors \eqref{eq:4dspinorRcharges} since they preserve the axial symmetry but they break the vector symmetry, in particular $\epsilon$ and $\bar{\epsilon}$ have axial charges $1$ and $-1$ respectively but not a well-defined vector charge. Solutions to these Killing spinor equations on $S^2$ are well-known \cite{Fujii:1985bg,Lu:1998nu}. This combined with the analysis we did on $\Sigma$ guarantees the existence of a solution to the 4d $\mathcal{N}=2$ Killing spinor equations \eqref{eq:4dKSE} for the background $S^2\times\Sigma$ which preserves half of the supersymmetries, and in particular it preserves $SU(2|1)_B$ on $S^2$.

\subsubsection*{Preserving $SU(2\vert 1)_A$ on $S^2$}

Interestingly, we can also find a background that preserves $SU(2|1)_A$ on $S^2$. For this purpose we turn off $G_n$, $G^{mi}{}_j$, $S^{ij}$, $\bar{S}^{ij}$ and turn on only $Z_{mn}$, $\bar{Z}_{mn}$ with the following profile:
\begin{align}
    Z=\frac{2i}{R_{S^2}}\mathrm{Vol}_{S^2}\,,\qquad \bar{Z}=-\frac{2i}{R_{S^2}}\mathrm{Vol}_{S^2}\,,
\end{align}
where $\mathrm{Vol}_{S^2}$ denotes the volume form of $S^2$. Indeed in this case the vector symmetry is preserved since $Z_{mn}$, $\bar{Z}_{mn}$ are singlets of $SU(2)_R$, while the axial symmetry is broken since $Z_{mn}$, $\bar{Z}_{mn}$ are charged under $U(1)_r$. We also point out that this choice of  the background fields satisfies the reality conditions \eqref{eq:4dreality} and so reflection positivity is preserved in this background.

Again the Killing spinor equations along $\Sigma$ have a vanishing r.h.s.~and so we can save half of the spinors with the $SU(2)_R$ topological twist. After repackaging the spinors into 2d Dirac spinors as
\begin{equation}
    \epsilon=\begin{pmatrix}
        \bar{\xi}^1_{\dot{-}} \\ -\xi^1_-
    \end{pmatrix}\,,\qquad \bar{\epsilon}=\begin{pmatrix}
        \xi^2_+ \\ \bar{\xi}^2_{\dot{+}}
    \end{pmatrix}
\end{equation}
we find again the Killing spinor equations on $S^2$
\begin{equation}
    \nabla_m\epsilon=\frac{i}{2R_{S^2}}\gamma_m\epsilon\,,\qquad \nabla_m\bar{\epsilon}=\frac{i}{2R_{S^2}}\gamma_m\bar{\epsilon}\,,\qquad m=1,2\,,
\end{equation}
However, compared to the previous case, now the $SU(2|1)_A$ algebra is preserved on $S^2$. This is because these equations are compaitble with the assignment of charges of the spinors under the vector symmetry, but not under the axial symmetry, in particular $\epsilon$ and $\bar{\epsilon}$ have vector charges $1$ and $-1$ respectively but not a well-defined axial charge.

\subsubsection{$\mathbb{R}^2_\epsilon \times \Sigma$ background}\label{subsubsec:R2Sigmabackgrounds}

It was argued in \cite{Gomis:2012wy} that the A-type (\emph{resp.}~B-type) $S^2$ partition function of 2d $(2,2)$ theories can be factorized into those on two infinitely long hemispheres $D^2$ with opposite topological twist for the vector (\emph{resp.}~axial) symmetry. We expect that the same is true for the partition function of the 4d theory in our $S^2\times\Sigma$ background, since it coincides with the $S^2$ partition function of the 2d $(2,2)$ theory obtained by topologically twisted reduction on $\Sigma$. More precisely, we would like to argue that the $S^2\times\Sigma$ partition function that preserves $SU(2|1)_A$ (\emph{resp.}~$SU(2|1)_B$) on $S^2$ factorizes into those on $D^2\times\Sigma$ with opposite topological twist on $D^2$ for $U(1)_R\subset SU(2)_R$ (\emph{resp.}~$U(1)_r$). We will focus in particular on the B-type background that is relevant for our correspondence. This is because in this case $D^2\times\Sigma$ is equivalent to $\mathbb{R}^2_\epsilon\times\Sigma$, where $\mathbb{R}^2_\epsilon$ is the $\Omega$-deformed $\mathbb{R}^2$, still with a $U(1)_r$ topological twist \cite{Dedushenko:2023cvd}. The latter background is then related to the SCFT/VOA correspondence \cite{Oh:2019bgz,Jeong:2019pzg}. In particular, the $\mathbb{R}^2_\epsilon\times\Sigma$ partition function with $U(1)_r$ twist on $\mathbb{R}^2_\epsilon$ and $SU(2)_R$ twist on $\Sigma$ of the 4d $\mathcal{N}=2$ SCFT $T$ computes the conformal blocks of the 2d CFT $\CC[T]$ living on $\Sigma$ and whose chiral algebra coincides with the VOA $\VV[T]$ of $T$. The factorization of the $S^2\times\Sigma$ partition function that we will discuss momentarily then encodes the decomposition into conformal blocks of the full 2d CFT correlator.

Following \cite{Gomis:2012wy}, we would like to deform $S^2$ into a squashed two-sphere $S^2_b$, which can be defined by the following equation in $\mathbb{R}^3$:
\begin{equation}
    \frac{x_1^2+x_2^2}{l^2}+\frac{x_3^2}{\tilde{l}^2}=1\,,
\end{equation}
where the dimensionless squashing parameter is $b=l/\tilde{l}$. The squashed two-sphere preserves only a the Cartan $U(1)\subset SU(2)$ of the isometry of the round sphere. 

The Killing spinor equations are
\begin{equation}\label{eq:2dKSEb}
    D_m\epsilon=\frac{i}{2f(\theta)}\gamma_m\epsilon\,,\qquad    D_m\bar{\epsilon}=\frac{i}{2f(\theta)}\gamma_m\bar{\epsilon}\,,\qquad m=1,2\,,
\end{equation}
where $f^2(\theta)=\tilde{l}^2\sin{^2(\theta)}+l^2\cos{^2(\theta)}$ in terms of polar coordinates $\theta$, $\varphi$. Remembering how we obtained the Killing spinor equations \eqref{eq:2dKSE} for the round $S^2$ from the 4d ones \eqref{eq:4dKSE}, it is immediate to see that we can also get those for the squashed $S^2_b$ by just changing the configuration \eqref{eq:Gbkg} for the background supergravity field $G^m_{ij}$ to
\begin{equation}\label{eq:Gbkgb}
    G^1=G^2=0\,,\qquad G^3=\frac{1}{4f(\theta)}\tau_2\,,\qquad G^4=\frac{1}{4f(\theta)}\tau_1\,.
\end{equation}
Again as explained in \cite{Gomis:2012wy}, the same solutions to the Killing spinor equations on the round $S^2$ are also solutions on the squashed $S^2_b$ provided that we turn on a non-trivial profile for the background gauge field for $U(1)_r$, which enters in the covariant derivative $D_m$
\begin{equation}\label{eq:Vprofile}
    V\Big|_{S^2_b}=\frac{1}{2}\left(1-\frac{l}{f(\theta)}\right)\mathrm{d}\varphi\,,\qquad V\Big|_{\Sigma}=0\,.
\end{equation}
This however still has vanishing flux on the sphere $\tfrac{1}{2\pi}\int_{S^2_b}F=0$, meaning that there is no topological twist for $U(1)_r$ on $S^2_b$. 

At this point it was crucial for \cite{Gomis:2012wy} that the partition function of $S^2_b$ is independent of the squashing parameter $b$, so that in particular its value is the same for the round sphere $b=1$ and in the infinite squashing limit $b\to\infty$. In the latter case, the sphere looks like two infinitely long hemispheres glued together, as in Figure \ref{fig:spheregluing}. Moreover, they showed that the gauge field $V$ in \eqref{eq:Vprofile} takes a form near the tips of the two hemispheres such that there a non-zero flux, which is however opposite in sign in the two cases so that overall on the sphere there is no flux. This was done by performing an explicit supersymmetric localization computation on $S^2_b$ and showing that the result is independent of $b$. Even though we are not performing any localization computation, we still expect the same to be true for our $S^2_b\times\Sigma$ partition function, since this should coincide with the partition function on $S^2_b$ of the 2d $(2,2)$ theory obtained by dimensional reduction on $\Sigma$. Overall, we see that the same argument of \cite{Gomis:2012wy} applies in our 4d set-up, with the result that the $S^2_b\times\Sigma$ partition function factorizes into two $D^2\times\Sigma$ partition functions with opposite $U(1)_r$ twist on the infinitely long hemispheres $D^2$. As mentioned before, the latter can then be equivalently understood as $\mathbb{R}^2_\epsilon\times\Sigma$, still with a $U(1)_r$ twist on the $\Omega$-background $\mathbb{R}^2_\epsilon$, which is in turn related to the SCFT/VOA correspondence.

\subsection{Counterterms ambiguities}
\label{subsec:counterterm}

The partition function of 2d $(2,2)$ theories on the two-sphere is known to encode information about their conformal manifold. The latter is a K\"ahler manifold that locally takes the form $\mathcal{M}_{\text{c}}\times\mathcal{M}_{\text{tc}}$, where $\mathcal{M}_{\text{c}}$ and $\mathcal{M}_{\text{tc}}$ denote the parts parametrized by chiral and twisted chiral exactly marginal deformations respectively. It was then argued in a series of papers that \cite{Jockers:2012dk,Gomis:2012wy,Gerchkovitz:2014gta}
\begin{equation}
    Z_{S^2_B}=\mathrm{e}^{-\mathcal{K}_{\text{c}}(\lambda^i,\bar{\lambda}^i)}\,,\qquad Z_{S^2_A}=\mathrm{e}^{-\mathcal{K}_{\text{tc}}(\tilde{\lambda}^i,\bar{\tilde{\lambda}}^i)}\,,
\end{equation}
where $\mathcal{K}_{\text{c}}$ and $\mathcal{K}_{\text{tc}}$ are the K\"ahler potentials of $\mathcal{M}_{\text{c}}$ and $\mathcal{M}_{\text{tc}}$ respectively. In other words, the B-type partition function depends on the chiral exactly marginal parameters $\lambda^i$, $\bar{\lambda}^i$, while the A-type partition function depends on the twisted chiral exactly marginal parameters $\tilde{\lambda}^i$, $\bar{\tilde{\lambda}}^i$.

This result implies that the partition function is affected by ambiguities that are induced by K\"ahler transformations
\begin{equation}
    \mathcal{K}_{\text{c}}(\lambda^i,\bar{\lambda}^i)\to\mathcal{K}_{\text{c}}(\lambda^i,\bar{\lambda}^i)+\mathcal{F}_{\text{c}}(\lambda^i)+\bar{\mathcal{F}}_{\text{c}}(\bar{\lambda}^i)\,,\quad \mathcal{K}_{\text{tc}}(\hat{\lambda}^i,\bar{\hat{\lambda}}^i)\to\mathcal{K}_{\text{tc}}(\tilde{\lambda}^i,\bar{\tilde{\lambda}}^i)+\mathcal{F}_{\text{tc}}(\tilde{\lambda}^i)+\bar{\mathcal{F}}_{\text{tc}}(\bar{\tilde{\lambda}}^i)
\end{equation}
and so it is only defined up to holomorphically factorized prefactors that depend on the chiral exactly marginal operators for the B-type and the twisted chiral exactly marginal operators for the A-type
\begin{equation}\label{eq:S2Zambiguities}
    Z_{S^2_B}\simeq Z_{S^2_B}\mathrm{e}^{\mathcal{F}_{\text{c}}(\lambda^i)+\bar{\mathcal{F}}_{\text{c}}(\bar{\lambda}^i)}\,,\qquad Z_{S^2_A}\simeq Z_{S^2_A}\mathrm{e}^{\mathcal{F}_{\text{tc}}(\tilde{\lambda}^i)+\bar{\mathcal{F}}_{\text{tc}}(\bar{\tilde{\lambda}}^i)}\,.
\end{equation}

As explained in \cite{Gerchkovitz:2014gta}, these ambiguities can be understood as counterterms ambiguities. The exactly marginal couplings can indeed be promoted to background chiral and twisted chiral multiplets, more precisely they correspond to the expectation values for the bottom scalar components of such multiplets. One can then write down local counterterms that involve these background fields as well as some of the supergravity fields that are turned on in the $S^2$ background, whose contribution to the partition function precisely matches the holomorphically factorized prefactors in \eqref{eq:S2Zambiguities}. Key for this is that the two-sphere partition functions of $(2,2)$ theories, unlike many other partition functions computable with supersymmetric localization, depend on the chiral superpotential. More precisely, the B-type partition function depends on the chiral superpotential $\mathcal{W}$ as $\mathrm{e}^{-\mathcal{W}-\bar{\mathcal{W}}}$, while the A-type partion function depends on the twisted chiral superpotential $\tilde{\mathcal{W}}$ as $\mathrm{e}^{-\tilde{\mathcal{W}}-\bar{\tilde{\mathcal{W}}}}$. 

It is then a natural question whether also our $S^2\times\Sigma_{g,n}$ partition function of 4d $\mathcal{N}=2$ SCFTs is affected by similar ambiguities. We have just seen that for the background preserving the B-type $S^2$ the resulting partition function should depend on the 2d chiral exactly marginal parameters, while for the A-type background it depends on the twisted chiral ones. Hence, we first need to know how these parameters arise from the 4d perspective. As usual with dimensional reduction of quantum field theories, it is often possible to predict at least part of the spectrum of the lower dimensional theory from the knowledge of the spectrum of its higher dimensional parent. In our setup, as we will explain more in details in Section~\ref{subsec:multred}, we can extract information on certain protected operators of the 2d $(2,2)$ class $\FF$ theory from the knowledge of the BPS spectrum of the original 4d $\mathcal{N}=2$ SCFT after reduction on $\Sigma$ with the topological twist. In the following we will summarize the result of this analysis for the case of 2d exactly marginal operators. 

\newcolumntype{L}{>{\centering\arraybackslash}m{.3\textwidth}}
\begin{table}
\begin{center}
    \begin{tabular}{L|L|L}
    Partition function & Parameters & 4d origin \\\toprule
    $Z_{S^2_B}$ & chiral exactly marginal parameters & complex moduli of $\Sigma_{g,n}$ \\\hline 
    $Z_{S^2_A}$ & twisted chiral exactly marginal parameters & 4d exactly marginal deformations \\\bottomrule
    \end{tabular}
    \caption{Summary of the parameters of the class $\FF$ theories on which the A- and B-type $S^2$ partition functions depend and their 4d origin.}\label{tab:countertermpar}
    \end{center}
\end{table}

Let us consider first the B-type case that is of our main interest. Here chiral exactly marginal operators arise from complex structure moduli of the Riemann surface $\Sigma_{g,n}$. Hence, the partition function in the $S^2\times\Sigma_{g,n}$ background that preserves $SU(2|1)_B$ on $S^2$ will depend on such complex structure moduli, see the first row of Table \ref{tab:countertermpar}. This gives us an important entry of the dictionary of our correspondence. Indeed, on the CFT side the correlation function on $\Sigma_{g,n}$ of $\CC[T]$ will depend on such complex structure moduli as well. For example, in the case of a sphere with $n$ punctures there are $n-3$ moduli, which correspond to the positions of the operators after having fixed three of them at 0, 1 and $\infty$ using conformal transformations. Thus, the chiral exactly marginal deformations of the $(2,2)$ class $\FF$ theory on which the B-type $S^2$ partition function depends map to the positions of the operators in the CFT correlator. More generally for a Riemann surface of genus $g$ with $n$ punctures we have $3(g-1)+n$ moduli, which correspond to chiral exactly marginal deformations on the unitary 2d $(2,2)$ class $\FF$ side and to parameters of the correlator on the 2d non-unitary CFT $\CC[T]$ side. Moreover, the 4d partition function on the B-type $S^2\times\Sigma$ does not suffer from any ambiguity, since there is no local counterterm that we can write in 4d for these parameters as they are purely geometrical.

On the other hand, twisted chiral exactly marginal operators descend from 4d exactly marginal operators (see Appendix \ref{app:4dto2dmultiplets}). We have seen before that there exists also an $S^2\times\Sigma_{g,n}$ background preserving $SU(2|1)_A$ on $S^2$. The partition function in this background will then carry a dependence on the exactly marginal deformations in 4d, see the second row of Table \ref{tab:countertermpar}. As a consequence of this, we expect that given a 4d $\mathcal{N}=2$ SCFT $T$ with no conformal manifold, the resulting class $\FF$ theory $\FF[T;\Sigma_{g,n}]$ will not possess any twisted chiral exactly marginal deformation and thus its A-type $S^2$ partition function will be trivial. In Sections \ref{sec:dimredA1A2} and \ref{sec:dimredA1A4} we will see that this is indeed the case for the $(A_1,A_2)$ and the $(A_1,A_4)$ Argyres--Douglas theories (and more generally for any Argyres--Douglas theory with the VOA of a non-unitary RCFT), which all lack exactly marginal deformations. Moreover, we can also argue that this 4d partition function does not suffer from any ambiguity as well. Indeed, the 4d exactly marginal parameters can be promoted to background $\mathcal{N}=2$ chiral multiplets of $U(1)_r$ charge zero. One can then try to write down local counterterms involving these fields as well as the supergravity fields. Such counterterms should be of the form
\begin{equation}\label{eq:4dcounterterm}
    \int\mathrm{d}^4x\,\mathrm{d}^4\theta\,\mathcal{E}F(X^I)+\int\mathrm{d}^4x\,\mathrm{d}^4\bar{\theta}\,\bar{\mathcal{E}}\bar{F}(\bar{X}^I)\,,
\end{equation}
where $\mathcal{E}$ is the appropriate superspace measure, while $F$ is a holomorphic function of the chiral multiplets $X^I$ of the theory. When the background preserves $U(1)_r$, $F$ should have charge 2 under it. For example, in \cite{Gomis:2014woa} it was shown that on $S^4$ one can write down a counterterm that also involves the vector multiplet compensator that is used to preserve supersymmetry in this background. Such vector multiplet can be equivalently understood as a chiral of $U(1)_r$ charge 1 and so it is possible to use it to write down a consistent counterterm \eqref{eq:4dcounterterm} together with the chirals associated with the exactly marginal deformations. This recovers the ambiguity of the $S^4$ partition function related to the fact that it computes the K\"ahler potential of the 4d $\mathcal{N}=2$ conformal manifold \cite{Gerchkovitz:2014gta}. For our background instead the compensator is a generic real multiplet. Hence, the only chirals at our disposal are those encoding the exactly marginal deformations, which have trivial $U(1)_r$ charge and so are not sufficient to write down a holomorphic prepotential $F$ of charge 2. We thus conclude that the $S^2\times\Sigma_{g,n}$ partition function of 4d $\mathcal{N}=2$ theories does not suffer from any ambiguity, both for the B-type and the A-type backgrounds.

In \cite{Tachikawa:2017aux} it was shown in the context of 4d class $\SS$ theories how the ambiguity of the $S^4$ partition function of \cite{Gerchkovitz:2014gta} can be understood as an anomaly in the space of couplings which descends from the anomaly polynomial of the parent 6d $\mathcal{N}=(2,0)$ theory upon compactification on the Riemann surface. We similarly expect that the ambiguities of the $S^2$ partition functions \eqref{eq:S2Zambiguities} for the class $\FF$ theories could be derived from some anomaly of the parent 4d $\mathcal{N}=2$ SCFT. 

\subsection{Examples in free theories}
\label{subsec:free}

In this subsection we consider the free theories of a hypermultiplet and a vector multiplet in the $S^2\times \Sigma$ background preserving $SU(2|1)_B$ on $S^2$. We will show that the effective theories on $\Sigma$ after KK reduction on $S^2$ coincide with the natural non-chiral CFTs associated to their VOAs, namely the symplectic boson and the symplectic fermion CFTs respectively. 

We have seen that the background preserving $SU(2|1)_B$ on $S^2$ is characterized by a non-trivial expectation value for the $G^m_{ij}$ supergravity field. As it can be understood from the Killing spinor equations \eqref{eq:4dKSE}, this field has scaling dimension $\Delta=1$, trivial $U(1)_r$ charge $r=0$, and transforms in the triplet of $SU(2)_R$. As such, to linear level it couples to the $SU(2)_R$ R-symmetry current and, in case the theory possesses a Coulomb branch, to the K\"ahler potential $K$ \cite{Butter:2015tra},\footnote{As usual, the non-trivial metric $g_{mn}$ couples to the stress-energy tensor $T^{mn}$, with the effect of introducing a spin connection term in all covariant derivatives.}
\begin{equation}
    \Delta\mathcal{L}=G^m_{ij} J_m^{ij}+G^m_{ij}G^{ij}_mK\,.
\end{equation}
We proceed to specialize this background to the cases of a free hypermultiplet and a free vector multiplet.

\subsubsection{Free hypermultiplet}

A 4d $\mathcal{N}=2$ hypermultiplet consists of an $SU(2)_R$ doublet of complex scalars and two Weyl fermions $\psi_\alpha$, $\bar{\psi}_{\dot{\alpha}}$. We will focus for the moment on the scalars and for convenience think of them as four real scalars $q_{ia}$, so to make manifest their $SU(2)_F$ flavor symmetry corresponding to the index $a=1,2$. In the following we will often omit the contraction of the $SU(2)_F$ indices for brevity, for example $q^iq_i=q^{ia}q_{ia}$. The on-shell Lagrangian density for the scalars in our $S^2\times\Sigma$ background is \cite{Butter:2015tra}
\begin{align}\label{eq:4dhyperlagfull}
    \mathcal{L}^{\text{scal}}_{\text{hyp}}=D_mq^i D^mq_i
    +2G^m_{ij}q^iD_mq^j\,,
\end{align}
where we recall that we have a non-trivial $SU(2)_R$ gauge field \eqref{eq:SU2RgaugeSigma} on $\Sigma$ and that $G^m_{ij}$ is turned on only along $\Sigma$ with the profile \eqref{eq:Gbkg}.

We would like to show that after KK reduction on $S^2$ we obtain the non-chiral symplectic boson CFT on $\Sigma$, a.k.a.~$\beta\gamma$ system of dimension $\tfrac{1}{2}$. This is characterized by the Lagrangian density
\begin{equation}\label{eq:lagSB}
    \mathcal{L}_{\text{SB}}=\beta\bar{\partial}\gamma+\tilde{\beta}\partial\tilde{\gamma}\,.
\end{equation}
In this expression we introduced holomorphic and anti-holomorphic coordinates on $\Sigma$
\begin{equation}\label{eq:holocoord}
    z=x^3+ix^4\,,\quad \bar{z}=x^3-ix^4\quad \Rightarrow\quad \partial_3=\partial+\bar{\partial}\,,\quad \partial_4=i(\partial-\bar{\partial})
\end{equation}
with respect to which the bosons have dimensions\footnote{Notice that $\beta$ and $\gamma$ are bosons of half-integer spin. The violation of spin-statistics is related to the theory being non-unitary.}
\begin{align}
    &h(\beta)=h(\gamma)=\frac{1}{2},\qquad \bar{h}(\beta)=\bar{h}(\gamma)=0\,,\nn\\
    &h(\tilde{\beta})=h(\tilde{\gamma})=0\,,\qquad\bar{h}(\tilde{\beta})=\bar{h}(\tilde{\gamma})=\frac{1}{2}\,.
\end{align}
We will argue that the 2d symplectic bosons descend from the reduction on $S^2$ of the scalars $q_{ia}$, while the fermions $\psi_\alpha$, $\bar{\psi}_{\dot{\alpha}}$ do not survive the reduction. 

Due to the fact that in our background $G^m_{ij}$ is turned on only along $\Sigma$, we have that the Lagrangian \eqref{eq:lagSB} only contains the kinetic terms in the $S^2$ directions. Hence, we can perform a standard KK reduction on $S^2$, which results in the bosonic zero modes of $q_{ia}$ being massless. The scalars $q_{ia}$ have scaling dimension $\Delta=1$ in 4d. Their massless zero modes are usually related to their 4d parents by a factor of $R_{S^2}$ and so they have $\Delta_{\text{2d}}=0$ in 2d. However, this is no longer the case after the topological twist on $\Sigma$. In particular, following \eqref{eq:newspinSigma}, $q_{1a}$ acquires spin $s=\tfrac{1}{2}$ while $q_{2a}$ gets $s=-\tfrac{1}{2}$.\footnote{To lighten the notation, we still denote by $q_{ia}$ the massless zero modes of the 4d scalars that remain in the 2d EFT on $\Sigma$ after reduction on $S^2$.} Via the standard relations 
\begin{equation}
  \Delta_{\text{2d}}=h+\bar{h}\,,\qquad   s=h-\bar{h}\,,
\end{equation}
this means that their holomorphic and anti-holomorphic scaling dimensions have become
\begin{align}\label{eq:scal}
    h(q_{1a})=\frac{1}{2}\,,\quad h(q_{2a})=0\,,\quad\bar{h}(q_{1a})=0\,,\quad \bar{h}(q_{2a})=\frac{1}{2}
\end{align}
and they both have $\Delta_{\text{2d}}=\tfrac{1}{2}$, indicating that the correct rescaling factor between the 4d scalars and their zero modes in 2d is actually $R_{S^2}^{\frac{1}{2}}$. From these scaling dimensions, we already see that the bosonic zero modes are naturally identified with the symplectic bosons. Namely, the right moving symplectic bosons $\beta$, $\gamma$ are related to the two $SU(2)_F$ components of $q_{1a}$, while the left moving $\tilde{\beta}$, $\tilde{\gamma}$ are related to $q_{2a}$
\begin{equation}\label{eq:idSB}
    \beta=q_{11}\,,\qquad \gamma=q_{12}\,,\qquad \tilde{\beta}=q_{21}\,,\qquad \tilde{\gamma}=q_{22}\,.
\end{equation}

We can also recover the symplectic boson Lagrangian \eqref{eq:lagSB} from the dimensional reduction on $S^2$ of the 4d Lagrangian \eqref{eq:4dhyperlagfull}. First of all, we would like to argue that the kinetic term can be neglected. Indeed, with the new scaling dimensions \eqref{eq:scal} after the topological twist the kinetic term has dimension $\Delta_{\text{2d}}=3$ and thus it should be multiplied by a factor of $R_{S^2}$. However, we expect no dependence on the sizes of both $S^2$ and $\Sigma$, so in particular the result should be the same as in the limit $R_{S^2}\to 0$ where the kinetic term becomes negligible. On the other hand, the interaction term induced by $G^m_{ij}$ has dimension $\Delta_{\text{2d}}=2$ and so it should not be accompanied by any factor of $R_{S^2}$. This gives a non-trivial contribution involving only one derivative of the fields, which turns out to be proportional precisely to the symplectic boson Lagrangian density \eqref{eq:lagSB} up to the identification \eqref{eq:idSB}.

Finally, let us comments on the fermions of the hypermultiplet. Since the extra terms in the Lagrangian induced by the background field $G^m_{ij}$ only involve the scalars, this means that for the fermions we have only the ordinary kinetic term. Since the Dirac operator on $S^2$ has no vanishing eigenvalue \cite{Abrikosov:2002jr}, there is no massless zero mode surviving in the EFT on $\Sigma$ after the reduction.

\subsubsection{Free vector multiplet}

We now consider a 4d $\mathcal{N}=2$ free $U(1)$ vector multiplet. We focus on its fermionic components $\lambda^i_\alpha$, $\bar{\lambda}^i_{\dot{\alpha}}$ which are a triplet of $SU(2)_R$ but have opposite chirality. Their Lagrangian density in our $S^2\times \Sigma$ background is \cite{Butter:2015tra}
\begin{equation}\label{eq:4dveclagferm}
    \mathcal{L}_{\text{vec}}^{\text{ferm}}=\frac{1}{2}\bar{\lambda}_i\bar{\sigma}^m D_m\lambda^i+G^m_{ij}\bar{\lambda}^i\bar{\sigma}_m\lambda^j\,.
\end{equation}

We would like to show that after reduction on $S^2$ we obtain the non-chiral symplectic fermion CFT. This is characterized by the Lagrangian density
\begin{equation}\label{eq:SFlag}
    \mathcal{L}_{\text{SF}}=\epsilon_{ij}(\partial\theta_i\bar{\partial}\theta_j+\partial\bar{\theta}_i\bar{\partial}\bar{\theta}_j)\,,
\end{equation}
where the fermions have dimensions\footnote{Again the violation of spin-statistics reflects the non-unitarity of the theory.}
\begin{equation}
    h(\theta_i)=h(\bar{\theta}_i)=\bar{h}(\theta_i)=\bar{h}(\bar{\theta}_i)=0\,.
\end{equation}

After the topological twist we find that the fermions from 4d have the following dimensions:
\begin{align}\label{eq:scalferm}
    &h(\lambda^1_+)=h(\bar{\lambda}^1_{\dot{+}})=1\,,\quad \bar{h}(\lambda^1_+)=\bar{h}(\bar{\lambda}^1_{\dot{+}})=0\,,\nn\\
    &h(\lambda^2_-)=h(\bar{\lambda}^2_{\dot{-}})=0\,,\quad \bar{h}(\lambda^2_-)=\bar{h}(\bar{\lambda}^2_{\dot{-}})=1\,,
\end{align}
while $\lambda^1_-$, $\lambda^2_+$, $\bar{\lambda}^1_{\dot{-}}$, $\bar{\lambda}^2_{\dot{+}}$ have all trivial dimensions. This suggests the identification
\begin{equation}\label{id2}
    \theta_1=\lambda^1_-\,,\quad \theta_2=\bar{\lambda}^2_{\dot{+}}\,,\quad \bar{\theta}_1=\lambda^2_+\,,\quad \bar{\theta}_2=\bar{\lambda}^1_{\dot{-}}\,.
\end{equation}

Let us see how to recover the symplectic fermions Lagrangian \eqref{eq:SFlag} from the dimensional reduction on $S^2$ of the 4d Lagrangian \eqref{eq:4dveclagferm}. The idea is that, even though naively there are no fermionic massless zero modes after reduction on $S^2$, the interaction induced by $G^m_{ij}$ provides a shift of the masses in such a way that precisely the components of the gauginos appearing in \eqref{id2} actually have a massless zero mode. To see this, we first decompose the kinetic term into the part along $S^2$ and the one along $\Sigma$. We then notice that the part along $S^2$ can be combined with the interaction term induced by $G^m_{ij}$ to give
\begin{equation}
    \frac{1}{2}\bar{\lambda}^i\bar{\sigma}^m(D_m\delta_{ij}+G^n_{ij}\sigma_n\bar{\sigma}_m)\lambda^j\,,
\end{equation}
where we used that $\bar{\sigma}^m\sigma_n\bar{\sigma}_m=2\bar{\sigma}_n$ and we are focusing on the terms with $m=1,2$. This is exactly the same combination that appears in the Killing spinor equation \eqref{eq:4dKSE} (recall that in our background only $G^m_{ij}$ is turned on). Hence, we know that for the components $\lambda^1_-$ and $\lambda^2_+$ there exists a configuration for which\footnote{More explicitly, taking $G^m_{ij}$ as in \eqref{eq:Gbkg} results in the interaction induced by it being $-\tfrac{1}{2R_{S^2}}(\bar{\lambda}^1_{\dot{-}}\lambda^1_-+\bar{\lambda}^2_{\dot{+}}\lambda^2_+)$. Recalling that the eigenvalues of the Dirac operator on $S^2$ are of the form $\pm\frac{l+1}{R_{S^2}}$ with $l\in\mathbb{N}_0$ where the sign depends on the chirality \cite{Abrikosov:2002jr}, we see that for the fermions $\lambda^1_-$, $\lambda^2_+$, $\bar{\lambda}^1_{\dot{-}}$, $\bar{\lambda}^2_{\dot{+}}$ there exists one mode which is massless.}
\begin{equation}
    (D_m\delta_{ij}+G^n_{ij}\sigma_n\bar{\sigma}_m)\lambda^j=0\,,
\end{equation}
which corresponds to the aforementioned massless zero mode. Similarly by integrating by parts the original Lagrangian we can show that also $\bar{\lambda}^1_{\dot{-}}$ and $\bar{\lambda}^2_{\dot{+}}$ have a massless zero mode.

The remaining spinors instead are not involved in the interaction induced by $G^m_{ij}$, so from the kinetic term in 4d we obtain after reduction on $S^2$ mass terms for all of their KK modes, which are schematically of the form\footnote{For each KK mode this should be multiplied by the corresponding eigenvalue of the Dirac operator.}
\begin{equation}\label{eq:massferm}
    \bar{\lambda}^1_{\dot{+}}\lambda^2_-+\bar{\lambda}^2_{\dot{-}}\lambda^1_+\,.
\end{equation}
These should be combined with the kinetic terms along $\Sigma$, which after switching to the holomorphic coordinates \eqref{eq:holocoord} become\footnote{This is the leftover interaction for the massless zero modes of $\lambda^1_-$, $\lambda^2_+$, $\bar{\lambda}^1_{\dot{-}}$, $\bar{\lambda}^2_{\dot{+}}$ and for the corresponding modes of $\lambda^1_+$, $\lambda^2_-$, $\bar{\lambda}^1_{\dot{+}}$, $\bar{\lambda}^2_{\dot{-}}$ that couple to them. The other KK modes are massive and integrated out, leaving no leftover interaction in the EFT on $\Sigma$.}
\begin{equation}\label{eq:kinfermS2}
    -\bar{\lambda}^1_{\dot{+}}\bar{\partial}\lambda^2_++\bar{\lambda}^1_{\dot{-}}\partial\lambda^2_-+\bar{\lambda}^2_{\dot{+}}\bar{\partial}\lambda^1_+-\bar{\lambda}^2_{\dot{-}}\partial\lambda^1_-\,.
\end{equation}
From the sum of \eqref{eq:massferm} and \eqref{eq:kinfermS2} we can see that we can integrate out the massive $\lambda^1_+$, $\lambda^2_-$, $\bar{\lambda}^1_{\dot{+}}$, $\bar{\lambda}^2_{\dot{-}}$ and obtain for the massless fields (after integrating by parts)
\begin{equation}
    \partial\lambda^1_-\bar{\partial}\bar{\lambda}^2_{\dot{+}}+\bar{\partial}\lambda^2_+\bar{\partial}\lambda^1_{\dot{-}}\,.
\end{equation}
This precisely coincides with the symplectic fermions Lagrangian \eqref{eq:SFlag} with the identification of fields \eqref{id2}. In particular we point out that these terms have dimension $(h,\bar{h})=(1,1)$ and so they do not carry any prefactor of $R_{S^2}$, indicating that they survive in the EFT on $\Sigma$ after dimensional reduction on $S^2$.

Let us briefly comment on what happens for the bosonic fields of the vector multiplet. The complex scalar $\phi$ has Lagrangian density \cite{Butter:2015tra}
\begin{equation}\label{eq:4dveclagscal}
    \mathcal{L}_{\text{vec}}^{\text{scal}}=D_m\bar{\phi}D^m\phi+2G^{m}_{ij}G^{ij}_mK\,,
\end{equation}
where $K=\bar{\phi}\tfrac{\partial F}{\partial{\phi}}-\phi\tfrac{\partial \bar{F}}{\partial{\bar{\phi}}}$ is the K\"ahler potential and $F$ is the holomorphic prepotential. Since for a free $U(1)$ gauge field the holomorphic prepotential is simply $F=\tfrac{1}{2}\tau X^2$, we see that the extra term induced by $G^m_{ij}$ is just a mass term, which implies that there is no massless scalar zero mode in the EFT on $\Sigma$ after reducing on $S^2$. Let us comment that for a generic Lagrangian SCFT, which has multiple vector multiplets forming the adjoint representation of a gauge group $G$, the holomorphic prepotential is still quadratic, so we reach the same conclusion that the complex scalars have no massless zero modes.

The 4d gauge field $A_m$ decomposes into a scalar and a gauge field on $\Sigma$. Since $A_m$ appears just with the kinetic term in the Lagrangian, a standard KK reduction gives only a massless zero mode for the gauge field. This however will appear in the EFT Lagrangian on $\Sigma$ only through its kinetic term, with no interaction with the symplectic fermions.

%%%%%%%%%%%%%%%%%%%%%%%%%%%%%%%%%
\section{Kinematics of the correspondence: the view from 2d}
\label{sec:gencomp}

In this section we derive various general properties of the 2d theories obtained from dimensional reduction of the 4d $\mathcal{N}=2$ SCFT on either $S^2$ or $\Sigma$. These properties will be useful to us in the upcoming sections.

\subsection{Unitary side}
\label{subsec:unitary}

In this subsection we consider various aspects of the dimensional reduction of 4d $\mathcal{N}=2$ SCFTs on a Riemann surface $\Sigma$ with a holomorphic twist. This can be conveniently described as introducing a non-trivial R-symmetry connection that cancels the spin connection for some of the supercharges, rendering them covariantly constant. In 4d $\mathcal{N}=2$ SCFTs the R-symmetry is $SU(2)_R \times U(1)_r$, and there are two distinct ways to do this, depending on which factor of the R-symmetry one uses. For example, if we twist by $U(1)_r$ the resulting theory would have $\mathcal{N}=(0,4)$ supersymmetry in 2d \cite{Putrov:2015jpa,Gadde:2015wta,Bobev:2017uzs,Nawata:2023aoq}. However, as we mentioned in the Introduction and explained in more detail in Section~\ref{subsec:background}, for the proposed 2d/2d correspondence we actually consider twisting by the Cartan of $SU(2)_R$, and the resulting theory has $(2,2)$ supersymmetry. 

Next we shall briefly summarize several aspects of this type of dimensional reduction, in particular we will explain how to extract various properties of the 2d $(2,2)$ class $\FF$ theories from properties of the parent 4d $\mathcal{N}=2$ SCFTs and the details of the compactification. These predictions from 4d will constitute non-trivial consistency checks for the explicit 2d $(2,2)$ UV Lagrangians for the class $\FF$ theories that we will consider in the next section.

\subsubsection{Anomalies}
\label{eq:subsecanom}

A useful relation in dimensional reduction is that the 't Hooft anomalies of symmetries of the higher-dimensional theory transfer to the lower-dimensional theory. For the case of continuous symmetries, the 't Hooft anomalies can be encoded in an anomaly polynomial. The precise relation then is that the anomaly polynomial of the lower-dimensional theory\footnote{More precisely, the part of the anomaly polynomial containing only symmetries that originate as symmetries of the higher-dimensional theory. The anomalies of accidental symmetries, for instance, cannot be recovered in this way.} is given by the integration of the anomaly polynomial of the higher-dimensional theory on the Riemann surface \cite{Benini:2009mz,Bonelli:2009zp,Alday:2009qq}\footnote{In the case of finite symmetries, a similar relation holds for the anomaly theories \cite{Sacchi:2023omn}.}
\be
I_{d} = \int_{\Sigma} I_{d+2} \,.
\ee
We note that here it is important that $\Sigma$ has no punctures. These carry their own degrees of freedom and can give additional contributions to the anomalies, which we shall discuss momentarily. We will sometimes refer to the contribution to anomalies obtained by integrating the anomaly polynomials as ``geometric contribution", so to distinguish it from the ``punctures contribution".

For the case at hand, this relation allows for the determination of the anomalies of the 2d theories from those of the 4d theory and the compactification data. The explicit computation in the case of no punctures was performed abstractly based on the generic anomaly polynomial of 4d $\mathcal{N}=2$ SCFTs in \cite{Bobev:2017uzs}, and
here we shall borrow the results and adapt them to our case. Specifically, the anomaly polynomial of the 2d $(2,2)$ class $\FF$ theory takes the form
\be\label{eq:anpolgeom}
I_4 = \chi (2a-c) [ C^2_1 (+) - C^2_1 (-) ] \,, \label{2dAP}
\ee
where
\begin{itemize}
 \item $\chi$ is the Euler characteristic of the Riemann surface $\Sigma$. For surfaces with no punctures, this is related to the genus by $\chi = 2(1-g)$.
 \item $a$ and $c$ are the central charges of the 4d SCFT.
 \item $C_1 (+)$ and $C_1 (-)$ stand for the first Chern class of the right and left $U(1)$ R-symmetries defined in \eqref{eq:2dRsymmLR}. In terms of the central charges, \eqref{2dAP} is equivalent to
 \begin{equation}
     c_+=c_-=2\chi (2a-c)\,,
 \end{equation}
 where in our conventions the central charges are computed as $c_\pm=\mathrm{Tr}\,\gamma^3U(1)_{\pm}^2$.
\end{itemize}

The anomaly polynomial $I_4$ of a generic 2d theory might contain additional terms to those in \eqref{eq:anpolgeom} associated with flavor symmetries. In our setup, flavor symmetries can either descend from flavor symmetries of the 4d SCFT or emerge accidentally in the dimensional reduction. For the latter it is not possible to give a prediction of the anomalies from 4d, while for the former one finds that, for punctureless Riemann surfaces, the anomalies always vanish after performing a twist by the Cartan of $SU(2)_R$ twist. Hence, anomalies involving flavor symmetries that descend from 4d can only arise from inflow on the punctures. In any case, in all the examples we consider in the present paper the 4d theory has no continuous flavor symmetry and so this will not be of concern to us.

As mentioned, \eqref{eq:anpolgeom} is the anomaly polynomial of the 2d theories arising from compactification on $\Sigma$ with $SU(2)_R$ twist only if the surface $\Sigma$ has no punctures. The presence of punctures, in addition to changing the Euler characteristic to $\chi = 2-2g-n$ where $n$ is the number of punctures, will introduce additional contributions coming from the degrees of freedom localized on them \cite{Razamat:2022gpm}. As such, more generally the 2d central charges take the form
\begin{equation}\label{eq:4dc}
     c_+=c_-=2\chi (2a-c)+\sum_i \Delta c_i n_i\,,
 \end{equation}
where the index $i$ runs on the types of puncture, $\Delta c_i$ denotes the contribution of one puncture of type $i$ to the central charge and $n_i$ is the number of punctures of type $i$, so $\sum_in_i=n$. The desire to take this contribution into account stirs us to the topic of how the punctures can be described.

\subsubsection{Punctures}
\label{sec:punct}

There are various ways to try to approach punctures in dimensional reduction. Here we shall adopt the one where punctures are described as boundary conditions of the higher-dimensional theory reduced on a circle (see \cite{Razamat:2022gpm} for a review in the context of the 4d compactification of 6d SCFTs). The idea is that we can elongate the region around the puncture so that it resembles a long tube ending at the puncture. Reducing the theory on the circle direction of the tube, we see that the puncture can be thought of as a boundary for the resulting theory. The puncture can then be described by the boundary conditions imposed on the various fields. In our setup, the dimensional reduction on a circle of the 4d $\mathcal{N}=2$ SCFT will lead to a 3d $\mathcal{N}=4$ theory. As we want to preserve $(2,2)$ supersymmetry in 2d, the boundary conditions for the three-dimensional theory that we consider are restricted by that requirement. 

Whenever the 3d $\mathcal{N}=4$ theory admits a Lagrangian description, the picture of the punctures in terms of boundary conditions allows us to infer two important pieces of data:

\begin{itemize}   
 \item The representation of the punctures by boundary conditions allows us to compute the puncture specific contribution to the anomalies, which is given by half the contribution to the anomalies of the fields receiving Neumann boundary conditions. This is motivated by the thought experiment where we dimensionally reduce a fermion on an interval. The low-energy theory would then contain the fermion components that receive Neumann boundary conditions on both ends, and so each boundary should contribute half of the anomaly.  
 \item Given two identical punctures we can glue them together, which in the field theory involves the introduction of additional degrees of freedom. This appears in the above picture as some fields were given Dirichlet boundary conditions and so are killed at the boundary, but if we glue the two punctures, then these should be allowed to propagate freely. This is achieved precisely by reintroducing these fields. As such, we see that gluing two punctures should involve the addition of the fields that were given Dirichlet boundary conditions. 
\end{itemize} 

To exemplify the above general discussion, let us consider the $(A_1,A_{2k})$ Argyres--Douglas 4d $\mathcal{N}=2$ SCFTs \cite{Xie:2012hs}, which will be the focus of the next sections. Although no manifestly $\mathcal{N}=2$ Lagrangian is known for these theories, upon dimensional reduction on a circle they flow at low energies to free theories consisting of $k$ 3d $\mathcal{N}=4$ twisted hypermultiplets. Close to the boundary, we can then decompose each twisted hyper to two 2d $(2,2)$ twisted chirals.\footnote{This follows from the fact that the twisted hyper should be charged only under the part of the R-symmetry coming from the 4d $U(1)_r$. The relation between the 4d and 2d R-symmetries \eqref{eq:2dRsymmLR} then suggests that these should be thought of as twisted chirals in 2d which have opposite right and left moving R-charges.} Boundary conditions that preserve $(2,2)$ supersymmetry consist of giving Dirichlet boundary condition  to one chiral and Neumann for the other, leading us to two choices for each twisted hyper.\footnote{If we consider the Lagrangian of the 3d theory near the boundary, then we can decompose it to the kinetic term on the boundary and the one in the orthogonal direction. The former is simply described by a 2d kinetic term, but the latter needs to be described by a (twisted) superpotential. The F-term conditions of the superpotential enforce that if one twisted chiral receives Dirichlet boundary conditions then the other must receive Neumann boundary conditions, again see for instance \cite{Razamat:2022gpm} and references within for further details.} Given a puncture with one boundary condition, we can switch to the other boundary condition by flipping the twisted chiral receiving the Neumann boundary conditions.\footnote{By this we mean that the different 2d $(2,2)$ that can be obtained by different choices of boundary conditions are related to each other by a standard ``flipping procedure". Flipping an operator $\mathcal{O}$ means that we introduce a gauge singlet chiral fields $F$ that couples to it via a superpotential term $F\mathcal{O}$, so that the equations of motion of $F$ set $\mathcal{O}$ to zero in the chiral ring.} As such, this discrete option is usually viewed as different ``signs" or ``colors" of the same puncture (again, see \cite{Razamat:2022gpm}). 

While this is morally correct, the situation is actually more involved.\footnote{We are particularly grateful to Chris Beem for pointing this out.} When we consider the 4d theory on a circle with finite radius, one gets a sigma model with target space the moduli space of Higgs bundles on the Riemann surface with irregular singularity that defines the 4d AD theory \cite{Gaiotto:2009hg}. For the $(A_1,A_{2k})$ AD theories, the latter is known to have $k+1$ singular points, where the low energy effective theory is always that of $k$ free twisted hypers, but their charge under the 4d $U(1)_r$ is different \cite{Fredrickson:2017yka,Fredrickson:2017jcf}. In the zero radius limit, such singular points become infinitely separated and one has to pick one of them to zoom in, isolating the associated low energy effective theory. Curiously, in \cite{Fredrickson:2017yka,Fredrickson:2017jcf} it was proposed that each singular point was in one-to-one correspondence with a primary operator of the $\mathcal{M}(2,2k+3)$ minimal model that is the VOA of the $(A_1,A_{2k})$ theory \cite{Cordova:2015nma}. As previously noted, we expect the insertion of punctures on the surface to be dual to the insertion of operators in the correlator. We thus propose that in these models we have $k+1$ different types of punctures, one for each primary operator of $\mathcal{M}(2,2k+3)$, and that to define them we should consider the boundary conditions of the low energy effective theory at the corresponding singular point of the moduli space of the 4d theory on a circle of finite radius. The resulting punctures, while all consisting of pairs of twisted chirals, one with Dirichlet and one Neumann boundary conditions, differ by the charges of the twisted chirals under the $U(1)_r$ R-symmetry. As such, their contribution $\Delta c_i$ to the central charges will be different. We shall later see that this picture works remarkably well in the cases considered here.

Let us consider the simplest case of $(A_1,A_2)$ in greater detail. As mentioned, the 4d theory on a finite radius circle has two singular points, corresponding to the two basic operators of the Lee--Yang $\mathcal{M}(2,5)$ minimal model: the identity $I$ and the single non-trivial primary $\phi_{(2,1)}$ (see Appendix \ref{app:MqpCG} for a review). The physics locally around each point is that of a free twisted hyper, which close to the boundary can be decomposed to two twisted chirals. The two points differ by the charges of the chirals under the 4d $U(1)_r$ R-symmetry as shown in \cite{Fredrickson:2017yka,Fredrickson:2017jcf}, which in turn leads to them carrying different charges under the 2d R-symmetry. Specifically, using the identification between the 4d and 2d R-symmetries \eqref{eq:2dRsymmLR} and the fact that these fields are singlets under $SU(2)_R$, we find that the charges of the $(2,2)$ twisted chirals under the right moving R-symmetry are
\begin{equation}
    I:\,\,\left\{\frac{6}{5},-\frac{1}{5}\right\}\,,\qquad \phi_{(2,1)}:\,\,\left\{\frac{3}{5},\frac{2}{5}\right\}\,,
\end{equation}
while the left moving R-charges are the opposite of this, in accordance with the fact that the chirals are twisted.

For each singular point we can then define a puncture by giving Dirichlet boundary conditions to one twisted chiral and Neumann to the other. Here we find it more convenient to give Dirichlet boundary conditions to the twisted chiral with the smaller R-charge, however as mentioned before different choices are simply related to this by a flip of the puncture. As such, the $I$ puncture can be described by the boundary conditions on the $I$ point where the twisted chiral with R-charge $-\tfrac{1}{5}$ receives Dirichlet boundary conditions, while the one with R-charge $\tfrac{6}{5}$ receives Neumann boundary conditions. This allows us to compute the puncture specific contribution to the central charge
\begin{equation}
    \Delta c_I = \frac{1}{2}\left(1-2\times\frac{6}{5}\right) = -\frac{7}{10}\,,
\end{equation}
where we used that the contribution to the central charge from a chiral of R-charge $r$ is $1-2r$. We point out that
\begin{equation}
    \Delta c_I=-2(2a-c)\,,
\end{equation}
where we have used the fact that for the $(A_1,A_2)$ AD theory $a=\tfrac{43}{120}$ and $c=\tfrac{11}{30}$. This implies that if we have a surface with $n_I$ punctures of type $I$, the central charge should be
\begin{equation}
    c_\pm=4 (2a-c)\left(g-1+\frac{n_I}{2}\right) - \frac{7}{10}n_I = 4 (2a-c)(g-1)\,,
\end{equation}
In other words, the total effect of the $I$ puncture on the central charge is zero. This is consistent with the fact that inserting the identity should not change the correlator of the CFT, so the compactification on surfaces should give the same result regardless of the number of $I$ punctures. As such, we can treat the $I$ puncture has having no puncture, though as we shall see it is still useful to consider it in certain cases. Moreover, gluing two $I$ punctures should involve the introduction of a twisted chiral with R-charge $-\tfrac{1}{5}$, which is the field that received Dirichlet boundary conditions. We will see in Section \ref{sec:EGTQFT} that this picture is indeed correct at the level of the elliptic genus of the 2d $(2,2)$ theory.

We can define a $\phi_{(2,1)}$ puncture in a similar way, but by using the other singular point instead. Now we give Dirichlet boundary conditions to the chiral with R-charge $\tfrac{2}{5}$ and Neumann boundary conditions to the one with R-charge $\tfrac{3}{5}$. This allows us to compute the contribution to the anomaly specific to $\phi_{(2,1)}$
\begin{equation}\label{eq:cpunctA1A2}
    \Delta c_{\phi_{(2,1)}} = \frac{1}{2}\left(1-2\times\frac{3}{5}\right) = -\frac{1}{10}\,.
\end{equation}
Furthermore, we see that gluing two $\phi_{(2,1)}$ punctures should involve the introduction of a twisted chiral with R-charge $\tfrac{2}{5}$.

As a further example we shall consider the case of $(A_1,A_4)$, that will also play a role later on. Now the chiral algebra is the $\mathcal{M}(2,7)$ minimal model, also known as tricritical Lee--Yang model. As such, the 4d theory on the circle has three singular points corresponding to the three primaries of $\mathcal{M}(2,7)$: $I$, $\phi_{(2,1)}$ and $\phi_{(3,1)}$. The 3d physics around each point is that of two free twisted hypers, with the difference given by the R-charges of the fields when expressed in terms of the 4d R-charge. Translated in terms of the 2d R-charges of their twisted chiral components, we have \cite{Fredrickson:2017yka}
\begin{equation}
    I:\,\,\left\{\frac{8}{7},-\frac{1}{7};\frac{10}{7},-\frac{3}{7}\right\}\,,\qquad \phi_{(2,1)}:\,\,\left\{\frac{8}{7},-\frac{1}{7};\frac{5}{7},\frac{2}{7}\right\}\,,\qquad \phi_{(3,1)}:\,\,\left\{\frac{4}{7},\frac{3}{7};\frac{5}{7},\frac{2}{7}\right\}\,,
\end{equation}
where we separated by a ``;" the chirals coming from different twisted hypers and by a ``," those belonging to the same twisted hyper. For each point we can now define a corresponding puncture by giving supersymmetry preserving boundary conditions. The convenient punctures in our case turned out again to be giving Dirichlet boundary conditions to the chirals with smaller R-charges for each twisted hyper. As before, the fields given Dirichlet boundary conditions are the fields one needs to add in order to glue the puncture. Additionally, the contribution to the anomaly specific to the puncture can be evaluated from the fields receiving Neumann boundary conditions. Explicitly, we find
\begin{align}\label{eq:cpunctA1A4}
&\Delta c_I = \frac{1}{2}\left(2-2\times\frac{8}{7} - 2\times\frac{10}{7}\right) = -\frac{11}{7}\,,\nn\\
&\Delta c_{\phi_{(2,1)}} = \frac{1}{2}\left(2-2\times\frac{8}{7} - 2\times\frac{5}{7}\right) = -\frac{6}{7}\,,\nn\\
&\Delta c_{\phi_{(3,1)}} = \frac{1}{2}\left(2-2\times\frac{4}{7} - 2\times\frac{5}{7}\right) = -\frac{2}{7}\,.
\end{align}
Again we observe that $\Delta c_I=-2(2a-c)$, where for $(A_1,A_4)$ we have $a=\tfrac{67}{84}$ and $c=\tfrac{17}{21}$. Hence, the total contribution of the $I$ puncture to the central charge vanishes as expected.  

The above analysis can be readily generalized to the $(A_1,A_{2k})$ AD theory. As mentioned before, the associated VOA is the $\mathcal{M}(2,2k+3)$ minimal model, which has $k+1$ primary operators $\phi_{(a,1)}$ with $a=1,\cdots,k+1$, where $\phi_{(1,1)}=I$. In \cite{Fredrickson:2017yka} it was found that the theory compactified on a finite radius circle has a moduli space with $k+1$ singular points, one for each primary, where the low energy effective theory is that of $k$ free twisted hypers whose $U(1)_r$ R-charges depend on the choice of singular point
\begin{equation}
    \phi_{(a,1)}:\,\,\left\{\frac{2(k+i+1)}{2k+3},-\frac{2i-1}{2k+3}\right\}_{i=1,\cdots,k-a+1}\cup\left\{\frac{2i+1}{2k+3},\frac{2(k-i+1))}{2k+3}\right\}_{i=k-a+2,\cdots,k}\,.
\end{equation}
As before, for each twisted hyper we assign Dirichlet boundary condition to the chiral with smallest R-charge and Neumann to the other. From this data, for each type of puncture we can both compute the central charge contribution and deduce which fields we should reintroduce when gluing. For example, for the identity puncture the chirals to which we give Neumann boundary conditions have R-charges exactly equal to the dimensions of the Coulomb branch operators of the $(A_1,A_{2k})$ SCFT
\begin{equation}
    \Delta_i=\frac{2(k+i+1)}{2k+3}
\end{equation}
and so we once again find
\begin{equation}
    \Delta c_I=\frac{1}{2} \sum_i (1-2\Delta_i) = -2(2a-c)\,,
\end{equation}
where we used the Shapere--Tachikawa formula \cite{Shapere:2008zf} to relate the dimensions of the Coulomb branch operators to the $a$ and $c$ central charges. The total contribution of this puncture to the central charge would then be zero so this puncture has the right properties to be the $I$ puncture.

\subsubsection{Superconformal multiplets and predictions for the elliptic genus}
\label{subsec:multred}

Besides the anomalies, the higher-dimensional picture allows predictions regarding the BPS operator spectrum of the lower-dimensional theory. Here we shall provide a brief summary, referring the reader to Appendix \ref{app:4dto2dmultiplets} and \cite{BRZtoapp} for further details. The general idea is that given a BPS multiplet in the 4d theory, we expect it to lead to a BPS multiplet in 2d. The reason here is that a BPS multiplet contains components that are annihilated by the supercharges, and as the 2d supercharges are a subset of the 4d ones, it is reasonable that the resulting 2d multiplet will then contain components annihilated by the 2d supercharges. However, counting the full spectrum of BPS operators is in general difficult, as it can change under continuous deformations when BPS multiplets merge to form long ones. Nevertheless, it is possible to build an invariant quantity from the spectrum of BPS operators: the superconformal index. It turns out that one can make predictions for terms in the superconformal index of the lower-dimensional theory, the elliptic genus in the 2d case \cite{Gadde:2013wq,Gadde:2013dda,Benini:2013nda,Benini:2013xpa}, from the presence of certain BPS operators in the original higher-dimensional theory, which for us is the 4d $\mathcal{N}=2$ SCFT, and topological compactification data. In the following, we will consider the elliptic genus in the NSNS sector and use the conventions of \cite{Gadde:2013dda}, which we will review in more details in Section \ref{sec:EGTQFT}.

Similarly to before, we consider the reduction of the 4d $\mathcal{N}=2$ SCFTs on a Riemann surface of genus $g$ with a topological twist in the $SU(2)_R$ symmetry so that $(2,2)$ supersymmetry is preserved in 2d. Here we shall take the Riemann surface to be without punctures, as it is not known how to account for the contribution of punctures to the 2d elliptic genus. In this case, the existence of the aforementioned BPS multiplets implies the presence of certain terms in the 2d elliptic genus. We can then consider the contributions to the ellpitic genus of the 2d class $\FF$ theories coming from the main types of multiplets that can be present in the protected spectrum of 4d $\mathcal{N}=2$ SCFTs (we use the conventions of \cite{Cordova:2016emh} to denote the supermultiplets):
\begin{itemize}
    \item Every 4d $\mathcal{N}=2$ SCFT possesses a stress-energy tensor multiplet $A_2 \bar{A}_2 [0;0]^{(0;0)}_{2}$. Upon reduction to 2d it leads to $3(g-1)$ exactly marginal chiral operators, which are associated to the complex structure moduli of the Riemann surface as anticipated in Section~\ref{subsec:background}. Their contribution to the elliptic genus is
    \begin{equation}
        \mathcal{I}\supset 3(g-1)y\sqrt{q}\,.
    \end{equation}
    \item When the 4d $\mathcal{N}=2$ SCFT has a Coulomb branch, we will have associated Coulomb branch operators that reside in $L \bar{B}_1 [0;0]^{(0;r)}_{\frac{r}{2}}$ multiplets, which satisfy the relation $\Delta=\tfrac{r}{2}$ between their scaling dimension and $U(1)_r$ charge. Such a multiplet leads in 2d to $1-g$ pairs of twisted chiral operators of right moving R-charge $\Delta$ and $\Delta-1$ respectively.\footnote{The decomposition of the 4d Coulomb branch multiplet into 2d $(2,2)$ multiplets has appeared also in \cite{Gukov:2017zao}.} For each 4d Coulomb branch operator, we then get the elliptic genus contribution
    \begin{equation}
        \mathcal{I}\supset (1-g) \left(y^{1-\Delta} q^{\frac{\Delta-1}{2}}+y^{-\Delta} q^{\frac{\Delta}{2}}\right)\,.
    \end{equation}
    \item When the 4d $\mathcal{N}=2$ SCFT has a Higgs branch, we will have associated Higgs branch operators that reside in $B_1 \bar{B}_1 [0;0]^{(R;0)}_{R}$ multiplets, where $R$ is twice the $SU(2)_R$ spin. This yields $(R-1)(g-1)$ chiral operators in 2d with right moving R-charge $\frac{R}{2}$.\footnote{Turning on a flux $m$ for a $U(1)$ subgroup of the 4d flavor symmetry changes the multiplicy to $(R-1)(g-1)+mq$, where $q$ is the charge of the operator under such $U(1)$.} For each 4d Higgs branch operator, we then get the elliptic genus contribution
    \begin{equation}
        \mathcal{I}\supset (R-1)(g-1)y^{\frac{R}{2}} q^{\frac{R}{4}}\chi_{\mathcal{R}}(\vec{x})\,,
    \end{equation}
    where $\chi_{\mathcal{R}}(\vec{x})$ is the character of the representation $\mathcal{R}$ of the 4d flavor symmetry under which the Higgs branch can potentially transform, written in terms of flavor fugacities $\vec{x}$. Notice in particular that conserved current multiplets in 4d, which have $R=2$, lead to additional exactly marginal chiral operators in 2d.
   \end{itemize}

To illustrate the results, we consider the example of the $(A_1,A_2)$ and $(A_1,A_4)$ Argyres--Douglas theories. The $(A_1,A_2)$ theory has only one Coulomb branch operator of dimension $\Delta=\tfrac{6}{5}$. As such, we expect the 2d $(2,2)$ theory $\FF[(A_1,A_2),\Sigma_{g,0}]$ obtained by compactifying $(A_1,A_2)$ on a Riemann surface of genus $g$ with no punctures to contain the terms\footnote{The plethystic exponential of a function $f(x)$ is defined as $\mathrm{PE}[f(x)]=\sum_{n=1}^\infty\tfrac{1}{n}f(x^n)$.}
\begin{equation}\label{eq:EGA1A2g0n4prediction}
    \mathcal{I}_{g,n=0} \supset \mathrm{PE}\left[(1-g)y^{-\frac{1}{5}}q^{\frac{1}{10}} + 3(g-1) y\sqrt{q} + (1-g)y^{-\frac{6}{5}}q^{\frac{3}{5}}\right] \,.
\end{equation} 
where the first and the last term are the twisted chirals of dimensions $\tfrac{1}{5}$ and $\tfrac{6}{5}$ coming from the Coulomb branch operator, while the middle term comes from the stress-energy tensor. The plethystic exponential then accounts for all the products one can make from these operators.

The $(A_1,A_4)$ theory has instead two Coulomb branch operators, one of dimension $\Delta=\tfrac{8}{7}$ and the other of dimension $\Delta=\tfrac{10}{7}$. We then expect to have the following terms in the elliptic genus of $\FF[(A_1,A_a),\Sigma_{g,0}]$:
\begin{equation} \label{A1A4EGexp}
\mathcal{I}_{g,n=0} \supset \mathrm{PE}\left[(1-g) y^{-\frac{1}{7}} q^{\frac{1}{14}} + (1-g) y^{-\frac{3}{7}} q^{\frac{3}{14}} + 3(g-1) y q^{\frac{1}{2}} + (1-g) y^{-\frac{8}{7}} q^{\frac{4}{7}} + (1-g) y^{-\frac{10}{7}} q^{\frac{5}{7}}\right] \,,
\end{equation}

Finally, we briefly remark on a few subtleties regarding the above results. First, when writing the elliptic genus we need to make a choice of $(2,2)$ R-symmetry.  Here the chosen R-symmetry is the one inherited from the 4d $\mathcal{N}=2$ R-symmetry defined in \eqref{eq:2dRsymmLR}, although we stress that it may not be the superconformal one. We should also mention that the predictions presented here are not rigid as there are various subtleties in the arguments. One notable subtlety is that 2d BPS operators may also arise from extended operators in 4d wrapping the Riemann surface, rather than just local ones. As such, there could be deviations from these predictions, which have indeed been encountered in other types of compactifications, such as those from 6d to 4d and from 5d to 3d. Nevertheless, it has been observed that these predictions become more accurate as the genus increases and usually become exact for sufficiently high genus. We shall see that a similar behavior is exhibited by the theories considered here.

\subsection{Non-unitary side}
\label{subsec:kin2dnonun}

Let us next turn to the non-unitary side of the correspondence. We would like to argue that reducing the parent 4d $\mathcal{N}=2$ SCFT $T$ on $S^2$ using the background from Section \ref{subsubsec:S2Sigmabackgrounds}, we end up with a CFT $\CC[T]$ living on $\Sigma$ obtained by gluing the chiral algebra $\VV[T]$ of $T$ with its anti-chiral counterpart. Whenever the Riemann surface has punctures, these are interpreted as the insertion of primary operators in the path integral of the CFT. 

To do this, we make use of an alternative way of obtaining the chiral algebra of the 4d $\mathcal{N}=2$ SCFT from the original one of \cite{Beem:2013sza}. As discussed in \cite{Oh:2019bgz,Jeong:2019pzg} (see also \cite{Dedushenko:2023cvd}), this consists of placing the theory on $\mathbb{R}^2_\epsilon\times\mathbb{C}$ with a holomorphic-topological twist, where $\mathbb{R}^2_\epsilon$ is the $\Omega$-background. More precisely, one performs a topological twist by $U(1)_r$ on $\mathbb{R}^2_\epsilon$ which results in a 2d $\mathcal{N}=(0,4)$ theory in the orthogonal directions. Then one performs a holomorphic twist of this 2d theory so as to end up with the chiral algebra living on $\mathbb{C}$. 

Using this result, all we are left to do is to show that $S^2$ with the background preserving $SU(2|1)_B$ is equivalent to gluing two copies of $\mathbb{R}^2_\epsilon$ with opposite $U(1)_r$ topological twist, corresponding to gluing the chiral algebra with its antichiral counterpart. This fact was actually already shown in \cite{Gomis:2012wy}. Specifically, they first showed that the partition function of a 2d $(2,2)$ theory on the squashed sphere $S^2_b$ with the A-type background is actually independent of the squashing parameter $b$ and thus coincides with the one for the round sphere $S^2$ corresponding to $b=1$. One can then vary $b$ at will and in particular consider the limit $b\to0$ in which the squashed sphere degenerates into two hemispheres glued together via an infinitely long tube. In this set-up, although the background gauge field for the vector symmetry vanishes on the tube, it is non-vanishing at the tips of the two hemispheres, where it is such that it has a non-zero flux of opposite signs so that overall there is no total flux through the two-sphere. By the $\mathbb{Z}_2$ mirror automorphism of the 2d $(2,2)$ superconformal algebra, one has that also the $S^2$ partition function with the background preserving $SU(2|1)_B$ is equivalent to the gluing of two hemispheres with opposite topological twist for the axial symmetry. Since the axial symmetry descends from $U(1)_r$ in 4d via \eqref{eq:4d2dRsymm} we recover our set-up where we think of $\mathbb{R}^2_\epsilon$ as an infinitely elongated hemisphere \cite{Dedushenko:2023cvd}.

At the level of the observables computed in these backgrounds, we recover the statements we have already anticipated in the Introduction, that the B-type $S^2$ partition function of the 2d $(2,2)$ theory $\FF[T;\Sigma]$ matches with the correlator of the CFT $\CC[T]$ on $\Sigma$, while the hemisphere (or equivalently vortex) partition function matches the conformal blocks. All of these observables can be obtained from a four-dimensional partition function of the theory $T$, which is the one on $S^2\times\Sigma$ for the former and on $\mathbb{R}^2_\epsilon\times\Sigma$ for the latter. We have seen in Section \ref{subsubsec:R2Sigmabackgrounds} that the factorization property of the two-dimensional partition function uplifts to a factorization of the four-dimensional partition function $S^2\times\Sigma$ in two copies of the one on $\mathbb{R}^2_\epsilon\times\Sigma$ with opposite $U(1)_r$ twist on $\mathbb{R}^2_\epsilon$. From the CFT perspective, such a factorization of the partition function reflects the conformal blocks decomposition of the CFT correlator. Of course given a CFT correlator there will be multiple conformal blocks associated to it. These correspond to different choices of boundary conditions on the boundary of the hemisphere, and when performing the gluing to a sphere we have to sum over a complete basis for them in the same way that in the CFT we sum over a basis of conformal blocks. We will see this more explicitly in the examples of Sections \ref{sec:dimredA1A2} and \ref{sec:dimredA1A4}.

Before moving on, let us give a slightly different perspective on why dimensionally reducing the 4d $\mathcal{N}=2$ SCFT $T$ on $S^2$ leads to a non-unitary CFT built on the chiral algebra $\VV[T]$. The idea is to carry out the $S^2$ reduction in stages. One first thinks about $S^2$ as an $S^1$ fibration over an interval $I$. Upon dimensionally reducing along the $S^1$ fibers, one goes over to a 3d $\mathcal{N}=4$ SCFT, which we denote as $T_{S^1}$, on a background with topology $I\times \Sigma_{g,n}$.\footnote{As emphasized in \cite{Dedushenko:2018bpp,Dedushenko:2023cvd}, for theories $T$ which possess operators with fractional $U(1)_r$ charges, such as Argyres-Douglas theories, one must impose boundary conditions around the $S^1$ which are twisted by a holonomy for the $\mathbb{Z}_N=\langle e^{2\pi i r}\rangle$ symmetry.} Since the $SU(2)_R$ symmetry of $T$ (which we are using to twist the Riemann surface $\Sigma_{g,n}$) becomes identified with the $SU(2)_H$ R-symmetry of $T_{S^1}$, the $S^2\times \Sigma_{g,n}$ background in four dimensions is recapitulated in three dimensions away from the endpoints of the interval $I$ as a standard topological twist by $SU(2)_H$. As described in \cite{Dedushenko:2023cvd}, at the endpoints of the interval, the background implements the construction in \cite{Costello:2018fnz} which associates holomorphic/anti-holomorphic boundary conditions to $SU(2)_H$-twisted 3d $\mathcal{N}=4$ SCFTs. The construction is designed so that these boundary conditions in three dimensions reproduce the protected VOA $\VV[T]$ ($\overline{\VV[T]}$) of the four-dimensional parent theory $T$. 

\begin{figure}
    \begin{center}

\tikzset{every picture/.style={line width=0.75pt}} %set default line width to 0.75pt        

\begin{tikzpicture}[x=0.75pt,y=0.75pt,yscale=-1,xscale=1,scale=.85]

%Shape: Ellipse [id:dp676680496328376] 
\draw  [color={rgb, 255:red, 255; green, 0; blue, 33 }  ,draw opacity=1 ] (78.58,126.02) .. controls (78.58,95.21) and (81.36,70.24) .. (84.78,70.24) .. controls (88.2,70.24) and (90.98,95.21) .. (90.98,126.02) .. controls (90.98,156.82) and (88.2,181.8) .. (84.78,181.8) .. controls (81.36,181.8) and (78.58,156.82) .. (78.58,126.02) -- cycle ;
%Shape: Ellipse [id:dp8186054139921475] 
\draw  [color={rgb, 255:red, 255; green, 0; blue, 33 }  ,draw opacity=1 ] (99.41,126.1) .. controls (99.41,102.55) and (102.18,83.46) .. (105.61,83.46) .. controls (109.03,83.46) and (111.8,102.55) .. (111.8,126.1) .. controls (111.8,149.65) and (109.03,168.74) .. (105.61,168.74) .. controls (102.18,168.74) and (99.41,149.65) .. (99.41,126.1) -- cycle ;
%Shape: Ellipse [id:dp22355698056149897] 
\draw  [color={rgb, 255:red, 255; green, 0; blue, 33 }  ,draw opacity=1 ] (17.76,126.43) .. controls (17.76,102.88) and (20.54,83.79) .. (23.96,83.79) .. controls (27.38,83.79) and (30.16,102.88) .. (30.16,126.43) .. controls (30.16,149.98) and (27.38,169.07) .. (23.96,169.07) .. controls (20.54,169.07) and (17.76,149.98) .. (17.76,126.43) -- cycle ;
%Shape: Ellipse [id:dp8216798038636913] 
\draw  [color={rgb, 255:red, 255; green, 0; blue, 33 }  ,draw opacity=1 ] (38.59,126.35) .. controls (38.59,95.54) and (41.36,70.57) .. (44.78,70.57) .. controls (48.21,70.57) and (50.98,95.54) .. (50.98,126.35) .. controls (50.98,157.16) and (48.21,182.13) .. (44.78,182.13) .. controls (41.36,182.13) and (38.59,157.16) .. (38.59,126.35) -- cycle ;
%Shape: Ellipse [id:dp13005719066081178] 
\draw  [color={rgb, 255:red, 255; green, 0; blue, 33 }  ,draw opacity=1 ] (58.5,126.1) .. controls (58.5,93.24) and (61.28,66.6) .. (64.7,66.6) .. controls (68.12,66.6) and (70.9,93.24) .. (70.9,126.1) .. controls (70.9,158.96) and (68.12,185.6) .. (64.7,185.6) .. controls (61.28,185.6) and (58.5,158.96) .. (58.5,126.1) -- cycle ;
%Shape: Circle [id:dp7792617270899874] 
\draw   (5.2,126.1) .. controls (5.2,93.24) and (31.84,66.6) .. (64.7,66.6) .. controls (97.56,66.6) and (124.2,93.24) .. (124.2,126.1) .. controls (124.2,158.96) and (97.56,185.6) .. (64.7,185.6) .. controls (31.84,185.6) and (5.2,158.96) .. (5.2,126.1) -- cycle ;
%Shape: Polygon Curved [id:ds6390811640156623] 
\draw   (160.89,62.25) .. controls (183.96,50.47) and (209.42,53.85) .. (221.03,61.96) .. controls (232.63,70.07) and (228.07,82.93) .. (235.13,108.57) .. controls (242.19,134.2) and (265.18,104.92) .. (280.37,138.69) .. controls (295.57,172.46) and (241.4,213.55) .. (205.41,185.72) .. controls (169.43,157.89) and (212.13,136.68) .. (175.05,122.86) .. controls (137.97,109.03) and (137.81,74.02) .. (160.89,62.25) -- cycle ;
%Curve Lines [id:da07073666220409303] 
\draw    (179.07,92.03) .. controls (196.14,81.22) and (201.95,87.02) .. (208.24,97.03) ;
%Curve Lines [id:da7770461835804894] 
\draw    (204.05,92.38) .. controls (195.78,103.15) and (185.57,96.67) .. (183.69,90.14) ;
%Curve Lines [id:da9443483208835115] 
\draw    (216.29,152.39) .. controls (233.36,141.58) and (239.17,147.39) .. (245.46,157.39) ;
%Curve Lines [id:da8165019365174868] 
\draw    (241.27,152.74) .. controls (233,163.52) and (222.79,157.03) .. (220.91,150.51) ;
%Curve Lines [id:da6296085284241615] 
\draw    (5.37,126.27) .. controls (37.43,152.38) and (94.62,151.06) .. (124.2,126.1) ;
%Shape: Polygon Curved [id:ds8765794587365223] 
\draw   (529.69,62.85) .. controls (552.76,51.07) and (578.22,54.45) .. (589.83,62.56) .. controls (601.43,70.67) and (596.87,83.53) .. (603.93,109.17) .. controls (610.99,134.8) and (633.98,105.52) .. (649.17,139.29) .. controls (664.37,173.06) and (610.2,214.15) .. (574.21,186.32) .. controls (538.23,158.49) and (580.93,137.28) .. (543.85,123.46) .. controls (506.77,109.63) and (506.61,74.62) .. (529.69,62.85) -- cycle ;
%Curve Lines [id:da38268664085430815] 
\draw    (547.87,92.63) .. controls (564.94,81.82) and (570.75,87.62) .. (577.04,97.63) ;
%Curve Lines [id:da8250074771416602] 
\draw    (572.85,92.98) .. controls (564.58,103.75) and (554.37,97.27) .. (552.49,90.74) ;
%Curve Lines [id:da9766906838855153] 
\draw    (585.09,152.99) .. controls (602.16,142.18) and (607.97,147.99) .. (614.26,157.99) ;
%Curve Lines [id:da00866832352008906] 
\draw    (610.07,153.34) .. controls (601.8,164.12) and (591.59,157.63) .. (589.71,151.11) ;
%Straight Lines [id:da7359093169430158] 
\draw    (402.4,134.4) -- (489,134.4) ;
%Shape: Circle [id:dp24019050307985212] 
\draw  [color={rgb, 255:red, 74; green, 144; blue, 226 }  ,draw opacity=1 ][fill={rgb, 255:red, 74; green, 144; blue, 226 }  ,fill opacity=1 ] (397.6,134.4) .. controls (397.6,133.07) and (398.67,132) .. (400,132) .. controls (401.33,132) and (402.4,133.07) .. (402.4,134.4) .. controls (402.4,135.73) and (401.33,136.8) .. (400,136.8) .. controls (398.67,136.8) and (397.6,135.73) .. (397.6,134.4) -- cycle ;
%Shape: Circle [id:dp20496577654044812] 
\draw  [color={rgb, 255:red, 74; green, 144; blue, 226 }  ,draw opacity=1 ][fill={rgb, 255:red, 74; green, 144; blue, 226 }  ,fill opacity=1 ] (486.6,134.4) .. controls (486.6,133.07) and (487.67,132) .. (489,132) .. controls (490.33,132) and (491.4,133.07) .. (491.4,134.4) .. controls (491.4,135.73) and (490.33,136.8) .. (489,136.8) .. controls (487.67,136.8) and (486.6,135.73) .. (486.6,134.4) -- cycle ;
%Straight Lines [id:da25017708080894396] 
\draw    (307.6,118) -- (371.4,118) ;
\draw [shift={(373.4,118)}, rotate = 180] [color={rgb, 255:red, 0; green, 0; blue, 0 }  ][line width=0.75]    (10.93,-3.29) .. controls (6.95,-1.4) and (3.31,-0.3) .. (0,0) .. controls (3.31,0.3) and (6.95,1.4) .. (10.93,3.29)   ;

% Text Node
\draw (133.2,125) node [anchor=north west][inner sep=0.75pt]    {$\times $};
% Text Node
\draw (505.2,119.2) node [anchor=north west][inner sep=0.75pt]    {$\times $};
% Text Node
\draw (158.4,65.8) node [anchor=north west][inner sep=0.75pt]    {$\Sigma _{g,n}$};
% Text Node
\draw (10,58) node [anchor=north west][inner sep=0.75pt]    {$S^{2}_B$};
% Text Node
\draw (526.8,66.6) node [anchor=north west][inner sep=0.75pt]    {$\Sigma _{g,n}$};
% Text Node
\draw (296.4,95.6) node [anchor=north west][inner sep=0.75pt]   [align=left] {$\displaystyle S^{1}$ reduction};
% Text Node
\draw (438.8,112.4) node [anchor=north west][inner sep=0.75pt]    {$I$};
% Text Node
\draw (383.2,141.6) node [anchor=north west][inner sep=0.75pt]    {$\VV[ T]$};
% Text Node
\draw (468.8,139.2) node [anchor=north west][inner sep=0.75pt]    {$\overline{\VV[ T]}$};
% Text Node
\draw (70.4,20) node [anchor=north west][inner sep=0.75pt]   [align=left] {4d $\displaystyle \mathcal{N} =2$ SCFT $\displaystyle T$};
% Text Node
\draw (394.8,11) node [anchor=north west][inner sep=0.75pt]   [align=left] {\begin{minipage}[lt]{127.26pt}\setlength\topsep{0pt}
\begin{center}
$\displaystyle SU( 2)_{H}$-twisted 3d $\displaystyle \mathcal{N} =4$ \\SCFT with VOA boundaries
\end{center}

\end{minipage}};

\end{tikzpicture}
\caption{Thinking of $S^2_B$ as an $S^1$ fibration of the interval $I$ and reducing along the red $S^1$ fibers produces a 3d $\mathcal{N}=4$ SCFT subjected to an $SU(2)_H$ topological twist. The endpoints of the interval support holomorphic and anti-holomorphic boundary conditions.}\label{fig:KS}
    \end{center}
\end{figure}
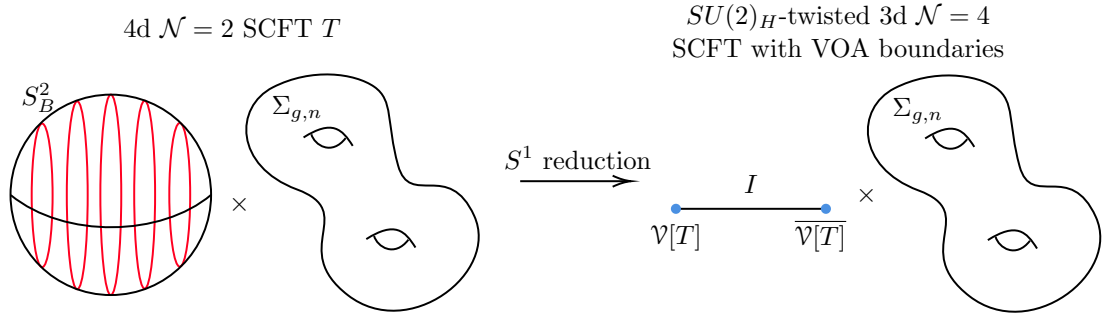

To summarize the discussion so far, after reducing $T$ on $S^2\times \Sigma_{g,n}$ along the $S^1$ fibers, one obtains a 3d TQFT on $I\times \Sigma_{g,n}$ with holomorphic/anti-holomorphic boundary conditions at the endpoints of the interval described by a left- or right-moving copy of the protected VOA $\VV[T]$ of the 4d SCFT $T$, see Figure \ref{fig:KS}. Remarkably, this configuration is essentially identical to Kapustin and Saulina's interpretation \cite{Kapustin:2010if} of the celebrated construction of 2d rational CFTs in \cite{Felder:1999mq,Fuchs:2002cm,Fjelstad:2005ua}. Indeed, the upshot of op.\ cit.\ is that any 2d rational CFT can be realized as a 3d topological QFT compactified on an interval, with chiral algebra boundary conditions imposed at the endpoints of the interval. The main difference with the present situation is that the VOA $\VV[T]$ is generally not rational and, correspondingly, the bulk TQFT is generally not semi-simple.\footnote{However, we note that there has been recent work \cite{Hofer:2025oif} on generalizing the techniques from \cite{Kapustin:2010if,Felder:1999mq,Fuchs:2002cm,Fjelstad:2005ua} to the setting of $C_2$-cofinite VOAs, which are expected to describe the protected VOAs of 4d $\mathcal{N}=2$ SCFTs $T$ without a Higgs branch \cite{Beem:2017ooy}.} Nevertheless, one expects the basic picture to go through, and one concludes that the 3d TQFT on $I\times \Sigma_{g,n}$, after reducing on $I$, yields a 2d CFT with $\VV[T]$ as part of its chiral algebra.

The 2d CFT engineered by dimensionally reducing $T$ on $S^2$ can be thought of as a kind of non-rational analog of a ``diagonal'' CFT built on $\VV[T]$. One nice feature of the perspective that Figure \ref{fig:KS} leads to is that it suggests how to engineer other modular invariant CFTs having $\VV[T]$ as part of their chiral algebra. Indeed, in the Kapustin--Saulina picture, the different consistent CFTs which can be built on top of a (strongly) rational VOA $\VV$ are labeled (sometimes redundantly) by topological surface operators in the bulk 3d TQFT: the CFT corresponding to a surface $S$ is obtained by placing $S$ at the midpoint of the interval $I$ before dimensionally reducing. This suggests that, even in the more general quasi-lisse setting of the SCFT/VOA correspondence, one may be able to obtain other CFTs having $\VV[T]$ as part of their chiral algebra by incorporating surface operators of $T_{S^1}$ which are compatible with the $SU(2)_H$-twist and become topological after twisting. Presumably, such surfaces of $T_{S^1}$ originate in the four-dimensional parent theory $T$ from supersymmetric codimension-1 membranes. The higher genus correlators of these more general non-unitary 2d CFTs built on $\VV[T]$ would be captured on the unitary side of the 2d/2d correspondence by calculating $S^2_B$ partition functions of theories of class $\FF$ decorated by codimension-1 line operators placed at the equator. We leave a more detailed analysis of this possibility to the future.

For concreteness, let us illustrate some of these ideas in the case that $T$ is taken to be the $(A_1,A_2)$ minimal Argyres-Douglas theory, which is the main example we study in Section \ref{sec:dimredA1A2}. It is conjectured \cite{Cordova:2015nma} that the VOA $\VV[(A_1,A_2)]$ is the (simple quotient of the) Virasoro vertex operator algebra at central charge $c=-22/5$, so we expect that the $S^2_B$ reduction of $(A_1,A_2)$ recovers the full Lee--Yang 2d non-unitary CFT. To see this in more detail, we first note that \cite{Dedushenko:2023cvd,ArabiArdehali:2024ysy} argued that $T_{S^1}$ --- the dimensional reduction of $T$ on $S^1$, taking care to incorporate a holonomy for $\mathbb{Z}_5=\langle e^{2\pi i r}\rangle$ --- coincides with a 3d $\mathcal{N}=4$ SCFT sometimes referred to as the Gang--Yamazaki theory, which was first studied in \cite{Gang:2018huc} as an example of IR supersymmetry enhancement of a 3d $\mathcal{N}=2$ Lagrangian theory. The $SU(2)_H$ topological twist of this SCFT is a semi-simple non-unitary 3d TQFT described by the Lee--Yang modular tensor category \cite{Dedushenko:2018bpp,Gang:2021hrd}. Furthermore, half-index calculations support the fact that $T_{S^1}$ has boundary conditions which, after carrying out the procedure of \cite{Costello:2018fnz}, lead to the obvious holomorphic boundary condition of the Lee--Yang TQFT, i.e.\ the boundary condition described by the simple quotient of the Virasoro VOA at $c=-22/5$.\footnote{By now this story has been worked out for several 4d $\mathcal{N}=2$ SCFTs. In the case of Argyres--Douglas theories, 3d $\mathcal{N}=2$ Lagrangians with supersymmetry enhancement for their $U(1)_r$-twisted circle reduction have been found in~\cite{Gang:2023rei,Gaiotto:2024ioj,ArabiArdehali:2024vli,Kim:2024dxu,Go:2025ixu,Kim:2025klh,Kim:2025rog,Nishinaka:2025ytu}.} There is only one modular invariant way to glue left- and right-movers for this VOA. (Equivalently, in the Kapustin--Saulina picture, the Lee-Yang TQFT does not have any non-trivial topological surface operators.) Thus, upon dimensional reduction on $I$, we go over to the 2d non-unitary Lee--Yang CFT.

\subsection{Remarks on the A-type 2d/2d correspondence}
\label{sec:A-type}
Before we turn to the main example studied in this paper, we pause to briefly contemplate an alternative 2d/2d correspondence obtained by starting in four dimensions with the $(\mathbb{R}^2_{\epsilon})_A\times \Sigma_{g,n}$ background, rather than  $(\mathbb{R}^2_\epsilon)_B\times \Sigma_{g,n}$. 
As we will see, this alternative ``A-type'' 2d/2d correspondence
makes direct contact with the work of Nekrasov and Shatashvili \cite{Nekrasov:2009rc,Nekrasov:2009ui,Nekrasov:2009uh,Nekrasov:2014xaa}. 

Our first observation is that, in both the $(\mathbb{R}^2_\epsilon)_A\times\Sigma_{g,n}$ and $(\mathbb{R}^2_\epsilon)_B\times \Sigma_{g,n}$ backgrounds, we are performing the same twist on $\Sigma_{g,n}$, using the Cartan of the $SU(2)_R$ symmetry of $T$. In particular, this means that in both cases, when we dimensionally reduce on $\Sigma_{g,n}$, we will obtain  the same $(2,2)$ theory of class $\FF$. However, for the A-type 2d/2d correspondence, we are required to calculate the A-type vortex partition functions of class $\FF$ theories, rather than B-type vortex partition functions, 
\begin{align}
    Z_{(\mathbb{R}^2_\epsilon)_A}[\FF[T,\Sigma_{g,n};\dots]].
\end{align}The A-type vortex partition function is sensitive to the \emph{twisted chiral} conformal manifold of $\FF[T;\Sigma_{g,n};\dots]$, rather than its chiral conformal manifold. This implies it depends on the exactly marginal couplings of $T$, but not the complex structure of the Riemann surface $\Sigma_{g,n}$. In particular, we expect the other side of the correspondence, i.e.\ the 2d theory on $\Sigma_{g,n}$, to be topological.

Indeed, following the lead of Nekrasov and Shatashvili \cite{Nekrasov:2009rc}, we may think of the 4d $\mathcal{N}=2$ SCFT $T$ on $(\mathbb{R}^2_\epsilon)_A\times \Sigma_{g,n}$ as an effective theory $\BB[T]$ with $\mathcal{N}=(2,2)$ supersymmetry on $\Sigma_{g,n}$. The twist performed on $\Sigma_{g,n}$ in 4d using the Cartan of the $SU(2)_R$ symmetry can be thought of as a 2d A-type topological twist (i.e.\ using the $U(1)_V$ vector R-symmetry) of this effective theory $\BB[T]$. The topologically twisted theory is characterized by an effective twisted superpotential ${\mathscr{W}}$ and a so-called effective dilaton term $\Omega$ (see e.g.\ \cite[Equation (2.8)]{Closset:2017zgf}) which depend in particular on $\epsilon$ and the 4d exactly marginal couplings. The states of the topologically twisted theory correspond to Bethe vacua, i.e.\ solutions to the so-called Bethe equations \cite{Nekrasov:2009uh}. The data which must be specified at the punctures of $\Sigma_{g,n}$ are insertions of local operators $\mathscr{O}_i$ residing in the twisted chiral ring, so that the observable on this side of the correspondence is 
\begin{align}\label{eqn:NSTQFT}
    \langle \mathscr{O}_1\cdots\mathscr{O}_n\rangle^{\BB[T]}_{\Sigma_{g,n}}.
\end{align}
Remarkably, there is a closed-form expression \cite{Nekrasov:2014xaa} for these topologically-twisted $n$-point genus-$g$ correlators as a sum over Bethe vacua involving $\mathscr{W}$ and $\Omega$.

The main prediction of the A-type 2d/2d correspondence is then that the Nekrasov-Shatashvili TQFT in Equation \eqref{eqn:NSTQFT} is computed by the A-type vortex partition function of theories of class $\FF$, 
\begin{align}
    Z_{(\mathbb{R}^2_\epsilon)_A}[\FF[T,\Sigma_{g,n};\dots]]\sim  \langle \mathscr{O}_1\cdots\mathscr{O}_n\rangle^{\BB[T]}_{\Sigma_{g,n}}.
\end{align}

What happens if one replaces $(\mathbb{R}^2_\epsilon)_A$ with $S^2_A$? First, we note that the TQFT defined by Equation \eqref{eqn:NSTQFT} should be more properly thought of as a \emph{relative} TQFT, i.e.\ a 2d TQFT which resides at the boundary of a non-trivial 3d bulk. Indeed, thinking of $(\mathbb{R}^2_\epsilon)_A$ as an $S^1$ fibration over $[0,\infty)$, and dimensionally reducing along the $S^1$ fibers, we obtain a $\mathcal{N}=(2,2)$ boundary condition of the $S^1$ reduction $T_{S^1}$ of $T$, which becomes a topological boundary $\BB[T]$ after subjecting the theory to the $SU(2)_H$-twist.\footnote{Note that, unlike for the B-type 2d/2d correspondence, here we can use the ordinary $S^1$ reduction of $T$ without imposing any additional holonomies around the circle.} This is analogous to the fact that the B-type vortex partition functions of theories of class $\FF$ compute conformal blocks of the VOA $\VV[T]$, which does not define an intrinsically 2d theory, but rather a holomorphic boundary of a 3d TQFT. 

Just as using $S^2_B$ glues together a left- and right-moving copy of $\VV[T]$ to obtain a full modular-invariant CFT $\CC[T]$, using $S^2_A$ glues together the topological boundary condition $\BB[T]$ with a copy of its orientation reversal to obtain a genuine 2d TQFT which we will call $\TT[T]$. In other words, Figure \ref{fig:KS} is modified by replacing $S^2_B$ with $S^2_A$ and the holomorphic boundary condition $\VV[T]$  by the topological boundary condition $\BB[T]$. In total, we claim that the dimensional reduction of $T$ on $S^2_A$ recovers the genuine 2d TQFT $\TT[T]$, whose higher genus correlators are computed by $S^2_A$ partition functions of theories of class $\FF$.

\begin{center}
\tikzset{every picture/.style={line width=0.75pt}} %set default line width to 0.75pt        
\begin{tikzpicture}[x=0.75pt,y=0.75pt,yscale=-1,xscale=1,scale=.9]
%uncomment if require: \path (0,219); %set diagram left start at 0, and has height of 219
\path[use as bounding box] (20,10) rectangle (660,180);

%Straight Lines [id:da055433214918889084] 
\draw [line width=1]    (264+30,56) -- (116.66+30,132.62) ;
\draw [shift={(114+30,134)}, rotate = 332.53] [color={rgb, 255:red, 0; green, 0; blue, 0 }  ][line width=1.5]    (14.21,-4.28) .. controls (9.04,-1.82) and (4.3,-0.39) .. (0,0) .. controls (4.3,0.39) and (9.04,1.82) .. (14.21,4.28)   ;
%Straight Lines [id:da6657060278054253] 
\draw [line width=1]    (365+30,55) -- (501.39+30,132.52) ;
\draw [shift={(504+30,134)}, rotate = 209.61] [color={rgb, 255:red, 0; green, 0; blue, 0 }  ][line width=1.5]    (14.21,-4.28) .. controls (9.04,-1.82) and (4.3,-0.39) .. (0,0) .. controls (4.3,0.39) and (9.04,1.82) .. (14.21,4.28)   ;

% Text Node
\draw (240,29) node [anchor=north west][inner sep=0.75pt]   [align=left] {4d $\displaystyle \mathcal{N} =2$ SCFT $T$ on $\displaystyle S^{2}_A \times \Sigma _{g,n}$};
% Text Node
\draw (20,142) node [anchor=north west][inner sep=0.75pt]   [align=left] {\begin{minipage}[lt]{171.41pt}\setlength\topsep{0pt}
~~~~~~~$S^{2}_A$ partition function of \\2d   $(2,2)$
 unitary SCFT $\FF[T;\Sigma_{g,n};\dots]$

\end{minipage}};
% Text Node
\draw (425,142) node [anchor=north west][inner sep=0.75pt]   [align=left] {\begin{minipage}[lt]{207.42pt}\setlength\topsep{0pt}

 ~~~~~~$\Sigma_{g,n}$ correlator of 2d
 \\TQFT $\TT[T]$ from gluing NS limits
\end{minipage}};
% Text Node
\draw (68+30,74) node [anchor=north west][inner sep=0.75pt]   [align=left] {reduce on $\displaystyle \Sigma _{g,n}$};
% Text Node
\draw (451+30,74) node [anchor=north west][inner sep=0.75pt]   [align=left] {reduce on $\displaystyle S^{2}$};

\draw [line width=1]    (275+15,160) -- (380+15,160) ;
\draw [shift={(380+15,160)}, rotate = 180] [color={rgb, 255:red, 0; green, 0; blue, 0 }  ][line width=1.5]    (14.21,-4.28) .. controls (9.04,-1.82) and (4.3,-0.39) .. (0,0) .. controls (4.3,0.39) and (9.04,1.82) .. (14.21,4.28)   ;
\draw [shift={(275+15,160)}, rotate = 0] [color={rgb, 255:red, 0; green, 0; blue, 0 }  ][line width=1.5]    (14.21,-4.28) .. controls (9.04,-1.82) and (4.3,-0.39) .. (0,0) .. controls (4.3,0.39) and (9.04,1.82) .. (14.21,4.28)   ;
\draw (292,132) node [anchor=north west][inner sep=0.75pt]   [align=left] {correspondence};
\end{tikzpicture}
\end{center}

We leave a detailed study of this correspondence to future work, and turn in the rest of this paper to illustrating the B-type 2d/2d correspondence for $T$ taken to be certain Argyres--Douglas theories. 

%%%%%%%%%%%%%%%%%%%%%%%%%%%%%%%%%
\section{A minimal example: the $(A_1,A_2)$ Argyres--Douglas theory}
\label{sec:dimredA1A2}

The simplest interacting 4d $\mathcal{N}=2$ SCFT is probably the so called $(A_1,A_2)$ Argyres--Douglas theory, sometimes also referred to as $H_0$ theory \cite{Argyres:1995jj,Argyres:1995xn}. Indeed this theory has no Higgs branch and has a single Coulomb branch operator of dimension $\Delta=\tfrac{6}{5}$. Moreover, it has the smallest $c$ central charge of any interacting 4d $\mathcal{N}=2$ SCFT. In this section we investigate our correspondence for this theory. 

We will follow a bottom-up approach. It is known that the VOA associated to the $(A_1,A_2)$ theory is that of the Lee--Yang minimal model \cite{Cordova:2015nma}. Exploiting some known expressions for the sphere $n$-point function of the Lee--Yang CFT, we will be able to reverse engineer some 2d $(2,2)$ gauged linear sigma models (GLSMs) of twisted vector and chiral fields, whose B-type $S^2$ partition functions coincide with the CFT correlators. Via the correspondence we then expect these GLSMs to flow at lowe energies to the $\FF[(A_1,A_2),\Sigma_{0,n}]$ theories, namely the theories arising from the compactification of $(A_1,A_2)$ on an $n$-punctured sphere with $SU(2)_R$ topological twist. We will provide evidence of this by performing tests based on anomalies and, in the next section, the elliptic genera that we reviewed in Section \ref{sec:gencomp}.

Our methods can be applied to any 4d $\mathcal{N}=2$ SCFT whose VOA is a non-unitary RCFT. In Section \ref{sec:dimredA1A4} we will discuss the case of $(A_1,A_4)$.

\subsection{Four-punctured sphere}
\label{subsec:4ptLY}

The Lee--Yang model has only one non-trivial primary operator $\phi_{(2,1)}$ of dimension $h=\bar{h}=-\tfrac{1}{5}$. Its unique 4-point function takes the form (see e.g.~\cite{Leitner:2018iyf})
\begin{align}\label{eq:Mpq4pt}
    \langle\phi_{(2,1)}(0)\phi_{(2,1)}(1)\phi_{(2,1)}(\infty)\phi_{(2,1)}(z,\bar{z})\rangle&=|z|^{\frac{4}{5}}|1-z|^{\frac{4}{5}}\Big|\dFu{\frac{3}{5},\frac{4}{5};\frac{6}{5};z}\Big|^2\nn\\
    &+C_{\phi_{(2,1)}\phi_{(2,1)}\phi_{(2,1)}}^2|z|^{\frac{2}{5}}|1-z|^{\frac{4}{5}}\Big|\dFu{\frac{2}{5},\frac{3}{5};\frac{4}{5};z}\Big|^2\,.
\end{align}
The structure constant is given by
\begin{equation}\label{eq:YLCppp}
    C_{\phi_{(2,1)}\phi_{(2,1)}\phi_{(2,1)}}^2=-\frac{\G{\frac{1}{5}}\G{\frac{2}{5}}\G{\frac{6}{5}}^2}{\G{\frac{4}{5}}^3\G{\frac{3}{5}}}=\frac{\G{\frac{1}{5}}^2\G{\frac{2}{5}}\G{\frac{6}{5}}}{\G{\frac{4}{5}}^2\G{\frac{3}{5}}\G{-\frac{1}{5}}}\,,
\end{equation}
where we used $\G{x+1}=x\G{x}$. Our goal is to find a 2d $(2,2)$ theory whose B-type $S^2$ partition function reproduces this correlator, possibly up to an holomorphically factorized prefactor. 

Let us consider the GLSM with $U(1)$ gauge group, $N_f$ chirals $\Phi_i$ of charge $+1$ and $N_a$ chirals $\tilde{\Phi}_j$ of charge $-1$. Its A-type $S^2$ partition function has been computed via supersymmetric localization and it is given by the integral \cite{Benini:2012ui,Doroud:2012xw}
\begin{align}\label{eq:S2pfGLSMCB}
    Z(z,\bar{z};\tau_i,\tilde{\tau}_j)&=\sum_{m\in\mathbb{Z}}\int_{-i\infty}^{i\infty}\frac{\mathrm{d}s}{2\pi i}z^{s+\frac{m}{2}}\bar{z}^{s-\frac{m}{2}}\prod_{i=1}^{N_f}\frac{\G{-i\tau_i+s-\frac{m}{2}}}{\G{1+i\tau_i-s-\frac{m}{2}}}\prod_{j=1}^{N_a}\frac{\G{-i\tilde{\tau}_j-s+\frac{m}{2}}}{\G{1+i\tilde{\tau}_j+s+\frac{m}{2}}}\,.
\end{align}
In this expression, $z=\mathrm{e}^{-2\pi \xi}$ with $\xi$ the renormalized complexified FI parameter, while $\tau_i=M_i+iq_i$ is the holomorphic combination of the twisted mass $M_i$ and the R-charge\footnote{Here and in the rest of the paper, we use ``R-charge" to refer to the charge under the right moving R-symmetry $U(1)_+$, unless otherwise specified. The charge under the left moving R-symmetry $U(1)_-$ is equal to it for chiral fields and opposite for twisted chiral fields, so we often do not specify it.} $q_i$ of the $i$-th chiral of positive charge and analogously $\tilde{\tau}_j$ for the negatively charged chirals. Exploting the $\mathbb{Z}_2$ mirror isomorphism of the $(2,2)$ superalgebra, this expression can be equivalently understood as a B-type $S^2$ partition function where the vector and chiral fields are twisted fields. This is the perspective that we will take.

This integral form of the partition function is obtained using the so-called ``Coulomb branch localization". The reason for this name is that the localizing action is taken to be such that the path integral of the partition functions localizes only on Coulomb vacua labelled by a continuous parameters $s$ and a discrete one $m$, where the former is one of the scalars in the vector multiplet while the latter labels the magnetic flux through $S^2$. However, it is possible to get a different expression by taking a localizing action such that the path integral localizes on the finitely many Higgs vacua that are not lifted by the mass deformations $\tau_i$, $\tilde{\tau}_j$. This is known as ``Higgs branch localization" and for the theory we are considering it results in an expression of the form \cite{Benini:2012ui,Doroud:2012xw}
\begin{align}\label{eq:S2pfGLSMHB}
Z(z,\bar{z};\tau_i,\tilde{\tau}_j)=\sum_{i=1}^{N_f}Z_{\text{cl}}^{(i)}(z,\bar{z})Z_{\text{1-loop}}^{(i)}(\tau_i,\tilde{\tau}_j)Z_{\text{v}}^{(i)}(z;\tau_i,\tilde{\tau}_j)Z_{\text{av}}^{(i)}(\bar{z};\tau_i,\tilde{\tau}_j)\,,
\end{align}
where $Z_{\text{cl}}^{(i)}$ and $Z_{\text{1-loop}}^{(i)}$ are the classical and 1-loop contributions, while $Z_{\text{v}}^{(i)}$ and  $Z_{\text{av}}^{(i)}$ are the contributions of vortices and anti-vortices at the north and south pole that are admitted at each Higgs vacuum, which can be also understood as hemisphere partition functions \cite{Sugishita:2013jca,Honda:2013uca,Hori:2013ika}. Their explicit expressions can be obtained either by direct localization computation or by evaluating the residues of the integral \eqref{eq:S2pfGLSMCB} at the poles coming from the positively charged chirals
\begin{align}
    Z_{\text{cl}}^{(i)}(z,\bar{z})&=|z|^{-i\tau_i}\,,\nn\\
    Z_{\text{1-loop}}^{(i)}(\tau_i,\tilde{\tau}_i)&=\prod_{k\neq i}^{N_f}\frac{\G{-i(\tau_k-\tau_i)}}{\G{1+i(\tau_k-\tau_i)}}\prod_{j=1}^{N_a}\frac{\G{-i(\tilde{\tau}_j+\tau_i)}}{\G{1+i(\tilde{\tau}_j+\tau_i)}}\,,\nn\\
    Z_{\text{v}}^{(i)}(z;\tau_i,\tilde{\tau}_j)&={}_{N_a}F_{N_f-1}\left(\left\{-i(\tilde{\tau}_j+\tau_i)\right\}_{j=1}^{N_a};\left\{1+i(\tau_k-\tau_i)\right\}_{k\neq i}^{N_f};(-1)^{N_f}z\right)\,,\nn\\
    Z_{\text{av}}^{(i)}(\bar{z};\tau_i,\tilde{\tau}_j)&={}_{N_a}F_{N_f-1}\left(\left\{-i(\tilde{\tau}_j+\tau_i)\right\}_{j=1}^{N_a};\left\{1+i(\tau_k-\tau_i)\right\}_{k\neq i}^{N_f};(-1)^{N_a}\bar{z}\right)\,,
\end{align}
where ${}_pF_q$ is the generalized hypergeometric function.

The 4-point correlator of Lee--Yang \eqref{eq:Mpq4pt} can be reproduced by the Higgs branch expression up to an holomorphically factorized prefactor taking
\begin{equation}\label{eq:spec4ptLY}
    N_f=N_a=2\,,\quad \tau_1=\tilde{\tau}_1=\frac{i}{5}\,,\quad \tau_2=\tilde{\tau}_2=\frac{2}{5}i\,.
\end{equation}
Indeed, with such specialization we get
\begin{align}\label{eq:S2pfGLSMLY}
    Z(z,\bar{z})&=\sum_{m\in\mathbb{Z}}\int_{-i\infty}^{i\infty}\frac{\mathrm{d}s}{2\pi i}z^s\bar{z}^s\frac{\G{\frac{1}{5}\pm(s-\frac{m}{2})}\G{\frac{2}{5}\pm(s-\frac{m}{2})}}{\G{\frac{4}{5}\mp(s+\frac{m}{2})}\G{\frac{3}{5}\mp(s+\frac{m}{2})}}\nn\\
    &=\frac{\G{\frac{1}{5}}}{\G{\frac{4}{5}}}|z|^{\frac{2}{5}}\Big|\dFu{\frac{2}{5},\frac{3}{5};\frac{4}{5};z}\Big|^2+\frac{\G{-\frac{1}{5}}\G{\frac{3}{5}}\G{\frac{4}{5}}}{\G{\frac{6}{5}}\G{\frac{2}{5}}\G{\frac{1}{5}}}|z|^{\frac{4}{5}}\Big|\dFu{\frac{3}{5},\frac{4}{5};\frac{6}{5};z}\Big|^2\,.
\end{align}
The prefactor consists of the $|1-z|^{\frac{4}{5}}$ in the correlator \eqref{eq:Mpq4pt} which is not reproduced by the $S^2$ partition function expression, and of a ratio of gamma functions since one can check that
\begin{equation}
    \frac{Z_{\text{1-loop}}^{(1)}}{Z_{\text{1-loop}}^{(2)}}=\frac{\G{\frac{1}{5}}^2\G{\frac{2}{5}}\G{\frac{6}{5}}}{\G{\frac{4}{5}}^2\G{\frac{3}{5}}\G{-\frac{1}{5}}}=C_{\phi_{(2,1)}\phi_{(2,1)}\phi_{(2,1)}}^2\,.
\end{equation}
We point out that our identification of the Lee--Yang 4-point function with the $S^2$ partition function of a GLSM is very similar to the result of \cite{Doroud:2012xw}, where it was shown that the partition function of SQED with $N$ flavors, i.e.~pairs of chirals with charge $\pm1$, can be matched with a certain four-point correlation function of the $A_{N-1}$ Toda CFT.

The specialization \eqref{eq:spec4ptLY} means that the $(2,2)$ theory whose $S^2$ partition function reproduces the Lee--Yang 4-point correlator is a $U(1)$ gauge theory with two pairs of chirals of charge $\pm1$, the first pair with R-charge $\tfrac{1}{5}$ and the second one $\tfrac{2}{5}$. Moreover, since all the mass parameters are set to zero the theory has no flavor symmetry. These properties are achieved by turning on a suitable superpotential interaction. We propose that the full model is actually given by the following matter content:
\begin{table}[h]
  \centering
  \begin{tabular}{c|cccccc}
    & $\Phi_1$ & $\Phi_2$ & $\tilde{\Phi}_1$ & $\tilde{\Phi}_2$ & $F_1$ & $F_2$ \\\hline
    $U(1)_{\text{gauge}}$ & 1 & 1 & $-1$ & $-1$ & 0 & 0 \\
    $U(1)_+$ & $\frac{1}{5}$ & $\frac{2}{5}$ & $\frac{1}{5}$ & $\frac{2}{5}$ & $\frac{3}{5}$ & $\frac{1}{5}$
  \end{tabular}
\end{table}

\noindent with the superpotential
\begin{equation}\label{eq:W4ptLY}
    \mathcal{W}=(\Phi_1)^2\tilde{\Phi}_1\tilde{\Phi}_2+\Phi_1\Phi_2(\tilde{\Phi}_1)^2+F_1\Phi_1\tilde{\Phi}_1+F_2\Phi_2\tilde{\Phi}_2+(F_2)^5\,.
\end{equation}
The chiral fields $F_1$, $F_2$ are singlets under the gauge symmetry and thus their contribution to the partition function is just an overall prefactor. However, their role is crucial in fixing the R-charges of the fields and breaking the flavor symmetries as required by \eqref{eq:spec4ptLY}.\footnote{Actually the superpotential \eqref{eq:W4ptLY} only fixes the R-charges of all the fields up to one parameter, which can however be reabsorbed by a gauge transformation. This is realized at the level of the matrix integral \eqref{eq:S2pfGLSMCB} by shifting the integration variable $s$ by a constant $s\to s+c$, with the only effect of producing an overall $|z|^c$ which is irrelevant since it is holomoprhically factorized.} 

We interpret this 2d $(2,2)$ GLSM as describing the $\FF[(A_1,A_2),\Sigma_{0,4}]$ theory, namely the reduction of the 4d $\mathcal{N}=2$ $(A_1,A_2)$ SCFT on a sphere with four punctures labelled by the CFT primary $\phi_{(2,1)}$, which we introduced in Section~\ref{sec:punct}. As mentioned previously, all the fields in this GLSM should be taken to be twisted fields and the interaction \eqref{eq:W4ptLY} should be understood as a twisted superpotential, so that the $S^2$ partition function that reproduces the CFT correlator is the B-type one. In a sense, this means that the GLSM description that we have found here for the compactification of the $(A_1,A_2)$ theory on a 4-punctured sphere is a mirror dual description in the sense of Hori--Vafa mirror symmetry \cite{Hori:2000kt}. We will elaborate more on this in Section~\ref{subsec:higherptLY}.

The single complex structure modulus of the 4-punctured sphere is realized in the GLSM by the FI parameter $z$, which in the Lagrangian appears multiplying the field strength multiplet. Since all the fields of the GLSM are twisted, this is an exactly marginal chiral deformation that descends from the 4d stress-energy tensor multiplet. Compatibly, the FI parameter $z$ in the $S^2$ partition function corresponds to the position of the fourth operator in the CFT correlator after using conformal invariance to place the other operators at 0, 1 and $\infty$.

As another consistency check of our proposal, we can verify that the central charge of the GLSM reproduces the one expected from 4d. From our general discussion in Section \ref{sec:gencomp}, we expect the central charges
\begin{equation} \label{eq:2dcfrom4d}
    c_\pm=\frac{7}{5}+4\times\left(-\frac{1}{10}\right)=1\,,
\end{equation}
where the first term is the geometric contribution while the second is the contribution of the four $\phi_{(2,1)}$ punctures, which are computed using that the 4d central charges of $(A_1,A_2)$ are $a=\tfrac{43}{120}$ and $c=\tfrac{11}{30}$ and that the contribution of $\phi_{(2,1)}$ puncture is as in \eqref{eq:cpunctA1A2}. These precisely match with the central charges of our GLSM
\begin{equation}\label{eq:4ptLYc}
    c_{\pm}=3\left(1-2\times\frac{1}{5}\right)+2\left(1-2\times\frac{2}{5}\right)+\left(1-2\times\frac{3}{5}\right)-1=1\,,
\end{equation}
where we used that a (twisted) chiral of R-charge $r$ contributes $1-2r$ to the central charge while a vector multiplet contributes $-1$. The contribution of the singlet fields $F_1$, $F_2$ is crucial for this matching to work. We stress that the central charge matching the 4d expectation is not necessarily the one that the 2d model has in the IR, since the 2d R-symmetry that one can embed inside the 4d one is not necessarily the IR superconformal R-symmetry. This is due to the possible emergence of symmetries at low energies that can mix with the R-symmetry. We expect this phenomenon to occur in most of our models, at least those for low genus and number of punctures, since many of the values of the central charges that we obtain from 4d are unphysical.

Finally, since this is a GLSM of twisted vector and chiral fields, its A-type $S^2$ partition function is trivial. This is consistent with the fact that the parent 4d $(A_1,A_2)$ theory has no exactly marginal deformation, as we explained in Section~\ref{subsec:background}.

The fact that this GLSM is the result of the compactification of a 4d $\mathcal{N}=2$ SCFT on a Riemann surface with a topological twist implies that it enjoys interesting dualities. Indeed, different pair-of-pants decompositions would lead to apparently different theories which are however equivalent due to the twist on the Riemann surface. The same thing happens in the 4d $\mathcal{N}=2$ class $\SS$ theories, which can be obtained by compactifying a 6d $\mathcal{N}=(2,0)$ SCFT on a Riemann surface with a topological twist. In that case, the complex structure moduli of the Riemann surface are encoded in exactly marginal gauge couplings, so the dualities that follow from considering different degeneration limits of the surface map non-trivially such gauge couplings and can thus be understood as a generalization of S-duality. In our case instead,  the FI parameters of the GLSM descend from the complex structure moduli of the surface and are mapped non-trivially under the duality. From the perspective of the CFT, different pair-of-pants decompositions correspond to different channels for the conformal block expansion of the correlator and the duality is nothing but crossing symmetry.

For the case of the four-point function \eqref{eq:Mpq4pt} of Lee--Yang, all operators are identical and so the correlator is fully invariant under any crossing symmetry transformations. The generators of such transformations can be taken to be
\begin{equation}
    z\to\frac{1}{z}\,,\qquad z\to \frac{z}{z-1}\,.
\end{equation}
At the level of the 2d $(2,2)$ theory, this translates to the statement that the GLSM is invariant under such transformations of the FI parameter, or in other words that it is self-dual. 

The self-duality associated with $z\to\tfrac{1}{z}$ is easily understood as the property of the theory of being charge conjugation invariant, thanks to the fact that we have the same number of charge $+1$ and charge $-1$ fields and that these have the same quantum numbers under the global symmetries, including the R-symmetry. Indeed, this transformation amounts to changing the sign of the FI parameter $\xi\to-\xi$, which can be compensated by redefining the vector multiplet $V\to-V$. Under such field redefinition the rest of the Lagrangian is left invariant (up to swapping the $\Phi_i$ and the $\tilde{\Phi}_i$ fields). At the level of the matrix integral \eqref{eq:S2pfGLSMLY} of the $S^2$ partition function, we can easily see that the transformation $z\to\tfrac{1}{z}$ can be undone with the change of variables $s\to-s$, $m\to-m$, so the partition function is invariant. 

The second transformation is instead less trivial and one can try to perform some tests for the validity of this duality. The invariance of the $S^2$ partition function is automatically ensured by the fact that the CFT correlator is crossing symmetric. In general for testing a duality one could also match central charges and the elliptic genus, however these tests are trivial in this case since the theory is self-dual and these quantities are insensitive to the FI parameter. In Appendix \ref{appsub:duality4pt} we provide evidence of the duality by studying in detail the classical moduli space of the GLSM, as a stepping stone to compute the Hilbert series of the twisted chiral ring. We will also observe that its structure turns out to be very similar to that of the elliptic genus that we will compute in Section~\ref{subsec:buildblocksA1A2}. We will instead see in Section \ref{sec:dimredA1A4} for the compactifications of $(A_1,A_4)$ an example of a duality between GLSMs that have a genuinely different field content thanks to the fact that the CFT in this case has more than one primary operator. In such case, we will see that the central charges and the elliptic genus agree in a non-trivial way.

\subsection{Five-punctured sphere}
\label{subsec:5ptLY}

The same bottom-up logic can be extended to higher-point functions of the Lee--Yang minimal model. It is indeed possible to derive Coulomb gas integral expressions for the corresponding conformal blocks  \cite{DiFrancesco:1997nk}, as we review in Appendix \ref{app:MqpCG}. Coulomb gas integrals of CFTs can often be manipulated to the form of the partition function of a GLSM, as observed for Toda CFT in \cite{Nedelin:2017nsb,Pasquetti:2019uop,Pasquetti:2019tix} inspired by analogous observations in the $q$-deformed case \cite{Nieri:2013yra,Aganagic:2013tta,Aganagic:2014oia}. We will be able to do the same for Lee--Yang, thus obtaining GLSM descriptions for $\FF[(A_1,A_2),\Sigma_{0,n}]$, namely the compactification of $(A_1,A_2)$ on punctured spheres with $SU(2)_R$ twist.

Since we will be looking at conformal blocks, the correct partition function that we should compare them to is the one on the hemisphere \cite{Sugishita:2013jca,Honda:2013uca,Hori:2013ika}. As mentioned previously, the full $S^2$ partition function can be obtained by gluing two copies of these, which in the CFT corresponds to the conformal block decomposition of the correlation function. Such a decomposition can in general be done with respect to a basis of conformal blocks. These are in one-to-on correspondence with the independent boundary conditions for the $(2,2)$ model on the boundary of the hemisphere. Since we are only interested in finding a GLSM prescription of such $(2,2)$ model and not in the classification of its boundary conditions, it will suffice for us to look at a single conformal block and find a GLSM whose hemisphere partition function matches the Coulomb gas integral for some choice of boundary conditions.

One of the 5-point conformal blocks of the Lee--Yang model admits the following Coulomb gas integral form with a single screening charge operator (see Appendix \ref{app:MqpCG} for more details):
\begin{equation}\label{eq:CGint5ptLY}
    F^{(5)}(z_1,z_2)=(z_1-z_2)^{\frac{1}{5}}\prod_{i=1}^2z_i^{\frac{1}{5}}(z_i-1)^{\frac{1}{5}}\int_0^1\mathrm{d}u\,u^{-{\frac{2}{5}}}(1-u)^{-{\frac{2}{5}}}\prod_{i=1}^2(1-z_iu)^{-{\frac{2}{5}}}\,.
\end{equation}
Using the identity \eqref{eq:IDCoulombGLSM}, one can show that this integral is equal to
\begin{align}\label{eq:5ptLYCG}
    \frac{F^{(5)}(z_1,z_2)}{(z_1-z_2)^{\frac{1}{5}}\prod_{i=1}^2z_i^{\frac{1}{5}}(z_i-1)^{\frac{1}{5}}}&=\frac{\G{\frac{3}{5}}}{\G{\frac{2}{5}}^2}\int_{-i\infty}^{+i\infty}\frac{\mathrm{d}s_1}{2\pi i}\left(\frac{z_1}{z_2}\right)^{s_1}\G{-s_1}\G{\frac{2}{5}+s_1}\nn\\
    &\times\int_{-i\infty}^{+i\infty}\frac{\mathrm{d}s_2}{2\pi i}\left(-z_2\right)^{s_2}\frac{\G{-s_2+s_1}\G{\frac{2}{5}+s_2-s_1}\G{\frac{3}{5}+s_2}}{\G{\frac{6}{5}+s_2}}\,.
\end{align}
This takes exactly the form of the hemisphere partition function of a GLSM whose gauge group is $U(1)^2$ and whose matter content is summarized in the following table:\footnote{To get the R-charge assignment in the table one has to perform the harmless change of variables $s_1\to s_1-\tfrac{1}{5}$, $s_2\to s_2-\tfrac{2}{5}$.}
\begin{table}[h]
  \centering
  \begin{tabular}{c|cccccccc}
      & $\Phi_1$ & $\tilde{\Phi}_1$ & $\Phi_2$ & $\tilde{\Phi}_2$ & $\Phi_3$ & $\tilde{\Phi}_3$ & $F_1$ & $F_{2,3,4}$\\\hline
      $U(1)_1$ & 1 & $-1$ & $-1$ & 1 & 0 & 0 & 0 & 0 \\
      $U(1)_2$ & 0 & 0 & 1 & $-1$ & $-1$ & 1 & 0 & 0 \\
      $U(1)_+$ & $\frac{1}{5}$ & $\frac{1}{5}$ & $\frac{1}{5}$ & $\frac{1}{5}$ & $\frac{1}{5}$ & $\frac{1}{5}$ & $\frac{1}{5}$ & $\frac{3}{5}$
  \end{tabular}
\end{table}

\noindent One can easily find a superpotential that fixes all the R-charges as desired up to gauge transformations (see Appendix \ref{appsub:duality5pt}). We note that the expression \eqref{eq:5ptLYCG} is very similar to the matrix integral obtained in the Coulomb branch localization of the $S^2$ partition function, however each chiral contributes with only one gamma function rather than a ration of two gamma functions. This is due to the fact that some of the fields inside the chiral receive Dirichlet boundary conditions and thus do not contribute to the partition function. In particular, in \eqref{eq:5ptLYCG} for the fields charged under the gauge symmetry we have that $\tilde{\Phi}_3$ has Dirichlet boundary conditions and thus contributes with a gamma at the denominator, while all the other fields have Neumann boundary conditions and thus contribute with a gamma each at the numerator.

As in the case of the four-point function, such a GLSM should be intended as composed of twisted vector and twisted chiral fields. In this way the FI parameters of the two $U(1)$ gauge groups are chiral exactly marginal deformations, which are mapped under the correspondence to combinations of the operators positions $z_1$, $z_2$ in the conformal block. These both correspond to the two complex structure moduli of the sphere with five punctures.

Another consistency check is that the central charges of the GLSM match with those expected from 4d for the compactification of $(A_1,A_2)$ on a sphere with five $\phi_{(2,1)}$ punctures
\begin{equation}
    c_{\pm}=7\left(1-2\times\frac{1}{5}\right)+3\left(1-2\times\frac{3}{5}\right)=\frac{7}{5}\times\frac{3}{2}+5\times\left(-\frac{1}{10}\right)=\frac{8}{5}\,.
\end{equation}
Again, the singlet fields turn out to be crucial for this matching.

Finally, since this is a GLSM of twisted vector and chiral fields, its A-type $S^2$ partition function is again trivial, in agreement with the fact that $(A_1,A_2)$ has no exactly marginal deformation in 4d.

\subsection{Higher number of punctures}
\label{subsec:higherptLY}

In order to make progress in extending our strategy to higher point functions, it is useful to take a different perspective on the derivation of the identities between CFT correlators or conformal blocks and partition functions of GLSMs that we have found so far for the 4- and 5-point functions. It was indeed pointed out in \cite{Nedelin:2017nsb,Pasquetti:2019uop,Pasquetti:2019tix} that this type of identities admits a three-dimensional origin. Specifically, they can be obtained as a limit of the identity of partition functions of 3d theories related by mirror symmetry \cite{Intriligator:1996ex,deBoer:1996mp,Hanany:1996ie,Aharony:1997bx}. The relevant partition function is the one on $S^2\times S^1$, also known as the supersymmetric index \cite{Bhattacharya:2008zy,Kim:2009wb,Imamura:2011su,Kapustin:2011jm,Dimofte:2011py}, for the full correlator and the one on $D^2\times S^1$, also known as the holomorphic block \cite{Pasquetti:2011fj,Beem:2012mb,Yoshida:2014ssa}, for the conformal block. The limit corresponds to shrinking the $S^1$ to zero size, which is implemented by taking $q\to 1$ where $q=\mathrm{e}^{-\beta}$ is related to the radius $\beta$ of $S^1$. In such a limit, one has to specify how the various parameters on which the 3d partition function depends, notably the FI and mass parameters, scale with $q$. Because of how these parameters are mapped under mirror symmetry, it turns out that starting from an identity of partition functions between quiver gauge theories in 3d, one ends up with a 2d GLSM partition function on one side and a Coulomb gas integral on the other. For example, for the 5-point function of Lee--Yang one starts from the 3d abelian duality between SQED with three flavors and the affine $A_2$ quiver. The limit of the SQED partition function, for a choice of scaling of the parameters, gives the Coulomb gas integral \eqref{eq:CGint5ptLY}, while the limit for the affine $A_2$ quiver gives the partition function of a similar gauge theory in 2d \eqref{eq:5ptLYCG}. A systematic discussion on how to obtain identities between 2d GLSMs and Coulomb gas integrals from 3d mirror identities, as well as their connection to 2d mirror symmetry, will be presented in \cite{ChenFaddaSacchi:WIP}.

This perspective can be used also in the case of a generic $(k+3)$-point function of Lee--Yang. It is convenient to distinguish between $k$ even and odd. When $k=2\kappa$ is even, the Coulomb gas integral is
\begin{equation}\label{eq:CGLYkeven}
    F^{(2\kappa+3)}(z_i)\propto\oint\prod_{a=1}^{\kappa}\mathrm{d}x_a\prod_{a<b}^{\kappa}(x_a-x_b)^{\frac{4}{5}}\prod_{a=1}^{\kappa}x_a^{-{\frac{2}{5}}}(x_a-1)^{-{\frac{2}{5}}}\prod_{i=1}^{2\kappa}(x_a-z_i)^{-{\frac{2}{5}}}\,,
\end{equation}
where we stripped off any prefactor depending only on the positions $z_i$ and not on the positions of the screening charges $x_a$. We are also not specifying the integration contour, which corresponds to a particular conformal block of the correlation function, since this will not be important for our following discussion. As explained in \cite{Nedelin:2017nsb,Pasquetti:2019uop,Pasquetti:2019tix}, such a Coulomb gas integral can be obtained as the $q\to1$ limit of the holomorphic block of a 3d theory whose content in $\mathcal{N}=2$ language is the following:
\begin{itemize}
    \item the contribution $\prod_{a<b}^{\kappa}(x_a-x_b)^{\frac{4}{5}}$ descends from the 1-loop contribution of a $U(\kappa)$ vector multiplet together with an adjoint chiral multiplet of R-charge $\tfrac{4}{5}$;
    \item the contribution $\prod_{a=1}^{\kappa}x_a^{-{\frac{2}{5}}}$ comes from the classical contribution of the FI parameter;
    \item the contribution $\prod_{a=1}^{\kappa}(x_a-1)^{-{\frac{2}{5}}}\prod_{i=1}^{2\kappa}(x_a-z_i)^{-{\frac{2}{5}}}$ comes from the 1-loop contribution of $2\kappa+1$ fundamental flavors, i.e.~pairs of fundamental/antifundamental chiral multiplets, of R-charge $\tfrac{3}{5}$.
\end{itemize}
The R-charges of the fields are compatible with an interaction between the adjoint and the fundamental chirals, which we denote by $A$ and $Q$, $\tilde{Q}$ respectively omitting all color and flavor indices for brevity\footnote{In 3d $\mathcal{N}=2$ the superpotential must have R-charge 2.}
\begin{equation}\label{eq:3dN4superpot}
    \mathcal{W}\supset AQ\tilde{Q}\,,
\end{equation}
which is the standard interaction of 3d $\mathcal{N}=4$ theories. The full superpotential should also contain additional terms that break the axial and the topological symmetry so that the R-charges of the chirals and the FI parameter are fixed to specific values. This is usually achieved by monopole superpotentials (see e.g.~\cite{Aharony:2013dha,Benini:2017dud}), however these will not play an important role in our discussion. The full superpotential preserves the $SU(2\kappa+1)$ vector-like symmetry acting on the chirals, whose corresponding mass parameters lead to the $z_i$ in \eqref{eq:CGLYkeven}.

We can find a mirror dual description to this 3d $\mathcal{N}=2$ theory by starting from the known mirror of the 3d $\mathcal{N}=4$ $U(\kappa)$ SQCD with $2\kappa+1$ flavors and turning on a deformation that breaks supersymmetry to $\mathcal{N}=2$ and that maps to the monopole superpotential on the SQCD side. Again we do not need the explicit form of this superpotential but just how this fixes the R-charges of the fields in the mirror dual theory, which can be worked out by matching the gauge invariant operators on the two sides of the mirror duality, for example using the superconformal index. This procedure gives us a 2d $(2,2)$ non-abelian GLSM whose hemisphere partition function coincides, up to prefactors that would be holomorphically factorized on $S^2$, with the Coulomb gas integral \eqref{eq:CGLYkeven}. Its content can be summarized with the following quiver:\footnote{The singlet fields are fixed by requiring the central charge of the theory to match with the 4d expectations, as we will comment momentarily, and by compatibility with the result for the 5-point function that we have found previously. In particular, the singlets contribute to the order $k^0$ term of the central charge, while higher order terms are only determined by the quiver.}
\begin{equation}
    \begin{tikzpicture}[scale=1.1,every node/.style={scale=1.2},font=\scriptsize]
        \node[gauge] (p1) at (0,0) {$1$};
        \node[gauge] (p2) at (1,0) {$2$};
        \node[gauge] (p3) at (2,0) {$3$};
        \node[] (p4) at (3,0) {$\cdots$};
        \node[gauge] (p5) at (4.25,0) {$\kappa-1$};
        \node[gauge] (p6) at (5.5,0) {$\kappa$};
        \node[gauge] (p7) at (6.5,0) {$\kappa$};
        \node[gauge] (p8) at (7.75,0) {$\kappa-1$};
        \node[] (p9) at (9,0) {$\cdots$};
        \node[gauge] (p10) at (10,0) {$3$};
        \node[gauge] (p11) at (11,0) {$2$};
        \node[gauge] (p12) at (12,0) {$1$};
        \node[flavor] (f1) at (5.5,-1) {$1$};
        \node[flavor] (f2) at (6.5,-1) {$1$};

        \draw[black,solid] (p2) edge [out=12+45,in=78+45,loop,looseness=4.5]  (p2);
        \draw[black,solid] (p3) edge [out=12+45,in=78+45,loop,looseness=4.5]  (p3);
        \draw[black,solid] (p5) edge [out=12+45,in=78+45,loop,looseness=4.5]  (p5);
        \draw[black,solid] (p6) edge [out=12+45,in=78+45,loop,looseness=4.5]  (p6);
        \draw[black,solid] (p7) edge [out=12+45,in=78+45,loop,looseness=4.5]  (p7);
        \draw[black,solid] (p8) edge [out=12+45,in=78+45,loop,looseness=4.5]  (p8);
        \draw[black,solid] (p10) edge [out=12+45,in=78+45,loop,looseness=4.5]  (p10);
        \draw[black,solid] (p11) edge [out=12+45,in=78+45,loop,looseness=4.5]  (p11);

        \node[cobalt] at (0.5,-0.3) {\tiny$\tfrac{1}{5}$};
        \node[cobalt] at (1.35,0.5) {\tiny$\tfrac{3}{5}$};
        \node[cobalt] at (5.2,-0.45) {\tiny$\tfrac{1}{5}$};
        \node[cobalt] at (6.8,-0.45) {\tiny$\tfrac{1}{5}$};
        \node[] at (9.5,-1) {$+$ singlets $\left\{{\color{cobalt}\tfrac{1}{5}},{\color{cobalt}\tfrac{3}{5}},{\color{cobalt}\tfrac{3}{5}},{\color{cobalt}\tfrac{3}{5}}\right\}$};

        \draw[hyper] (p1)--(p2)--(p3)--(p4)--(p5)--(p6)--(p7)--(p8)--(p9)--(p10)--(p11)--(p12);
        \draw[hyper] (p6)--(f1);
        \draw[hyper] (p7)--(f2);
        \end{tikzpicture}
\end{equation}
We use the standard quiver notation. Circles denote gauge groups, which in our case are of unitary type and with rank specified by the number inside the circle. Lines denote chiral fields, with double lines being pair of chirals in conjugate representations and arcs being adjoint chirals. In particular, a double line between two gauge nodes indicates a pair of chirals in the bifundamental representation and its conjugate, while a double line connected to a single gauge node and to a square node with a 1 inside indicates one pair of chirals in the fundamental representation and its conjugate. The flavor symmetries that are usually encoded in the square nodes are actually broken in this model by superpotential interactions, which we do not write explicitly. Such interactions also fix the R-charges of all the chirals, which we specify in blue in the drawing, in particular the bifundamental and fundamental fields have R-charge $\tfrac{1}{5}$ while the adjoint chirals have R-charge $\tfrac{3}{5}$. Finally, we are specifying the additional gauge singlet chiral fields that should be added with their R-charges.

We can similarly deal with the case of $k=2\kappa+1$ odd.\footnote{The case $k=1$ should be treated separately.} Here the Coulomb gas integral is
\begin{equation}\label{eq:CGLYkodd}
    F^{(2\kappa+2)}(z_i)\propto\oint\prod_{a=1}^{\kappa}\mathrm{d}x_a\prod_{a<b}^{\kappa}(x_a-x_b)^{\frac{4}{5}}\prod_{a=1}^{\kappa}x_a^{-{\frac{2}{5}}}(x_a-1)^{-{\frac{4}{5}}}\prod_{i=1}^{2\kappa}(x_a-z_i)^{-{\frac{2}{5}}}\,.
\end{equation}
This still descends from a 3d $\mathcal{N}=2$ $U(\kappa)$ gauge theory with one adjoint chiral of R-charge $\tfrac{4}{5}$, however this time we only have $2\kappa$ fundamental flavors of which $2\kappa-1$ have R-charge $\tfrac{3}{5}$ and one has R-charge $\tfrac{1}{5}$. As a result, the perturbative part of the superpotential, i.e.~excluding monopole operators, is not just the $\mathcal{N}=4$ one \eqref{eq:3dN4superpot}. Denoting by $Q$, $\tilde{Q}$ the $2\kappa-1$ flavors of R-charge $\tfrac{3}{5}$ and $P$, $\tilde{P}$ that of R-charge $\tfrac{1}{5}$, we have
\begin{equation}
    \mathcal{W}\supset AQ\tilde{Q}+A^2P\tilde{P}\,.
\end{equation}

Ignoring for the moment the monopole superpotential terms, as explained in \cite{Hwang:2020wpd} such a theory can be obtained from the 3d $\mathcal{N}=4$ $U(\kappa)$ SQCD with $2\kappa+1$ flavors by turning on a nilpotent mass labelled by the partition $[2,1^{2\kappa-1}]$ of $2\kappa+1$. The mirror dual of such $\mathcal{N}=4$ SQCD theory is known to be given by the following quiver:
\begin{equation}\label{SQCDmirror}
    \begin{tikzpicture}[scale=1.1,every node/.style={scale=1.2},font=\scriptsize]
        \node[gauge] (p1) at (0,0) {$1$};
        \node[gauge] (p2) at (1,0) {$2$};
        \node[gauge] (p3) at (2,0) {$3$};
        \node[] (p4) at (3,0) {$\cdots$};
        \node[gauge] (p5) at (4,0) {$\kappa$};
        \node[gauge] (p6) at (5.5,0) {$\kappa+1$};
        \node[gauge] (p7) at (7,0) {$\kappa+1$};
        \node[gauge] (p8) at (8.5,0) {$\kappa$};
        \node[] (p9) at (9.5,0) {$\cdots$};
        \node[gauge] (p10) at (10.5,0) {$3$};
        \node[gauge] (p11) at (11.5,0) {$2$};
        \node[gauge] (p12) at (12.5,0) {$1$};
        \node[flavor] (f1) at (5.5,-1) {$1$};
        \node[flavor] (f2) at (7,-1) {$1$};

        \draw[black,solid] (p1) edge [out=12+45,in=78+45,loop,looseness=4.5]  (p1);
        \draw[black,solid] (p2) edge [out=12+45,in=78+45,loop,looseness=4.5]  (p2);
        \draw[black,solid] (p3) edge [out=12+45,in=78+45,loop,looseness=4.5]  (p3);
        \draw[black,solid] (p5) edge [out=12+45,in=78+45,loop,looseness=4.5]  (p5);
        \draw[black,solid] (p6) edge [out=12+45,in=78+45,loop,looseness=4.5]  (p6);
        \draw[black,solid] (p7) edge [out=12+45,in=78+45,loop,looseness=4.5]  (p7);
        \draw[black,solid] (p8) edge [out=12+45,in=78+45,loop,looseness=4.5]  (p8);
        \draw[black,solid] (p10) edge [out=12+45,in=78+45,loop,looseness=4.5]  (p10);
        \draw[black,solid] (p11) edge [out=12+45,in=78+45,loop,looseness=4.5]  (p11);
        \draw[black,solid] (p12) edge [out=12+45,in=78+45,loop,looseness=4.5]  (p12);

        \draw[hyper] (p1)--(p2)--(p3)--(p4)--(p5)--(p6)--(p7)--(p8)--(p9)--(p10)--(p11)--(p12);
        \draw[hyper] (p6)--(f1);
        \draw[hyper] (p7)--(f2);
        \end{tikzpicture}
\end{equation}
The nilpotent mass maps in this mirror dual theory into a monopole superpotential for one of the two $U(1)$ gauge nodes, which then confines as it can be seen from the monopole duality of \cite{Benini:2017dud}. The monopole superpotential in the original SQCD that we initially ignored has the effect of fixing the R-charges of the remaining fields to some particular values, as in the case of $k$ even. Overall, we find the following quiver summarizing the 2d $(2,2)$ GLSM whose hemisphere partition function matches the Coulomb gas integral \eqref{eq:CGLYkodd}:\footnote{For $k=1$ one can repeat an identical analysis to recover the GLSM we found in Section~\ref{subsec:4ptLY}.}
\begin{equation}\label{GLSMkodd}
    \begin{tikzpicture}[scale=1.1,every node/.style={scale=1.2},font=\scriptsize]
        \node[gauge] (p2) at (1,0) {$2$};
        \node[gauge] (p3) at (2,0) {$3$};
        \node[] (p4) at (3,0) {$\cdots$};
        \node[gauge] (p5) at (4,0) {$\kappa$};
        \node[gauge] (p6) at (5.5,0) {$\kappa+1$};
        \node[gauge] (p7) at (7,0) {$\kappa+1$};
        \node[gauge] (p8) at (8.5,0) {$\kappa$};
        \node[] (p9) at (9.5,0) {$\cdots$};
        \node[gauge] (p10) at (10.5,0) {$3$};
        \node[gauge] (p11) at (11.5,0) {$2$};
        \node[gauge] (p12) at (12.5,0) {$1$};
        \node[flavor] (f1) at (5.5,-1) {$1$};
        \node[flavor] (f2) at (7,-1) {$1$};

        \draw[black,solid] (p3) edge [out=12+45,in=78+45,loop,looseness=4.5]  (p3);
        \draw[black,solid] (p5) edge [out=12+45,in=78+45,loop,looseness=4.5]  (p5);
        \draw[black,solid] (p6) edge [out=12+45,in=78+45,loop,looseness=4.5]  (p6);
        \draw[black,solid] (p7) edge [out=12+45,in=78+45,loop,looseness=4.5]  (p7);
        \draw[black,solid] (p8) edge [out=12+45,in=78+45,loop,looseness=4.5]  (p8);
        \draw[black,solid] (p10) edge [out=12+45,in=78+45,loop,looseness=4.5]  (p10);
        \draw[black,solid] (p11) edge [out=12+45,in=78+45,loop,looseness=4.5]  (p11);

        \node[cobalt] at (1.5,-0.3) {\tiny$\tfrac{1}{5}$};
        \node[cobalt] at (2.35,0.5) {\tiny$\tfrac{3}{5}$};
        \node[cobalt] at (5.1,-0.5) {\tiny$\tfrac{1}{5}$};
        \node[cobalt] at (7.4,-0.5) {\tiny$\tfrac{1}{5}$};
        \node[] at (10,-1) {$+$ singlets $\left\{{\color{cobalt}\tfrac{1}{5}},{\color{cobalt}\tfrac{3}{5}},{\color{cobalt}\tfrac{3}{5}},{\color{cobalt}\tfrac{3}{5}}\right\}$};

        \draw[hyper] (p2)--(p3)--(p4)--(p5)--(p6)--(p7)--(p8)--(p9)--(p10)--(p11)--(p12);
        \draw[hyper] (p6)--(f1);
        \draw[hyper] (p7)--(f2);
        \end{tikzpicture}
\end{equation}
We point out that the adjoint chiral of the $U(2)$ gauge node has been removed as a consequence of the application of the monopole duality of \cite{Benini:2017dud} to the leftmost $U(1)$ gauge node in \eqref{SQCDmirror}.

The proposed GLSMs for both $k$ even and odd pass the usual consistency checks. They possess $k$ FI parameters $\eta_i$, one for each gauge node of the quivers. These are related to the positions $z_i$ of the operators in the CFT correlator via
\begin{equation}
    \eta_i=\begin{cases}
        z_1 & i=1 \\
        \frac{z_i}{z_{i-1}} & i=2,\cdots,k
    \end{cases}\,.
\end{equation}
Moreover, the central charges of the GLSMs match with those expected from 4d for the compactification of $(A_1,A_2)$ on a sphere with $k+3$ punctures of type $\phi_{(2,1)}$
\begin{equation}
    c_{\pm}=\frac{3k+2}{5}=\frac{7}{5}\left(\frac{k+3}{2}-1\right)-\frac{1}{10}(k+3)\,.
\end{equation}

%%%%%%%%%%%%%%%%%%%%%%%%%%%%%%%%%
\section{TQFT structure of the elliptic genus}
\label{sec:EGTQFT}

The 2d $(2,2)$ class $\FF$ theories obtained by compactifying a 4d $\mathcal{N}=2$ SCFT $T$ on $\Sigma$ with $SU(2)_R$ twist are expected to possess interesting properties that are inherited from the Riemann surface picture. In particular, gluing two Riemann surfaces $\Sigma_1$ and $\Sigma_2$ along a puncture we expect to be able to obtain a new theory associated with the resulting surface $\Sigma$. One might then hope that there exists some prescription to implement the gluing at the level of the field theories, namely that there is some operation that allows us to combine the two theories $\FF[T;\Sigma_1]$ and $\FF[T;\Sigma_2]$ to get the theory $\FF[T;\Sigma]$. This is indeed what happens for the 4d $\mathcal{N}=2$ class $\SS$ theories \cite{Gaiotto:2009we}, where the gluing is performed by simply gauging the flavor symmetry carried by the puncture. In this section we show that this gluing operation is particularly simple when performed at the level of the elliptic genus of the 2d $(2,2)$ theories. 

As we have argued in the Introduction, considering the four-dimensional theory $T$ on $T^2\times\Sigma$ should lead to another 2d/2d correspondence that relates, on the one hand, the elliptic genus of the theory $\FF[T;\Sigma]$ and, on the other hand, the correlator of a TQFT $\EE[T]$ on $\Sigma$. In accordance with this, we will see that the elliptic genus of the theory $\FF[T;\Sigma]$ associated to a generic Riemann surface $\Sigma$ can be obtained by combining in a very simple way some basic building blocks associated with the pair-of-pants in which we can decompose the surface. These building blocks correspond to the propagators and the structure constants of $\EE[T]$. This gives us a TQFT formula for the elliptic genus, which in this section we derive for the case of $(A_1,A_2)$ and in Section~\ref{subsec:A1A4EG} for $(A_1,A_4)$, although we expect it to exist more generally. This is analogous to what happens for class $\SS$ theories, whose superconformal indices were shown in \cite{Gadde:2009kb,Gadde:2011ik,Gadde:2011uv} to coincide with correlation functions of deformations of 2d Yang--Mills theory in the zero area limit and thus also possess a TQFT structure.

Interestingly, we will be able to derive all the elementary building blocks involved in the TQFT formula for the elliptic genus just from the analysis of the punctures of Section~\ref{sec:punct} and from the GLSM for the 4-point function of Lee--Yang of Section~\ref{subsec:4ptLY}. We will then be able to use these results to compute the elliptic genus of the 2d $(2,2)$ class $\FF$ theory associated to a generic Riemann surface, even for the cases with genus $g>0$ for which we currently lack a Lagrangian description. This will allow us to perform several non-trivial consistency checks of our TQFT formula. First, we will use it to compute the elliptic genus of the model associated to a sphere with five punctures and verify that it matches with the one of the GLSM we derived in Section~\ref{subsec:5ptLY}. Then, we will show that the TQFT formula in the case of the sphere with no punctures reproduces the same elliptic genus that we can obtain from the 4d $\mathcal{N}=1$ Lagrangian for $(A_1,A_2)$ of \cite{Maruyoshi:2016tqk,Maruyoshi:2016aim} after performing the reduction on the sphere with $SU(2)_R$ twist along the lines of \cite{Gadde:2015wta}. Finally, we will verify the presence of operators expected from 4d as explained in Section~\ref{subsec:multred} for the case of a surface of genus $g$ with no punctures by expanding the elliptic genus computed with the TQFT formula as a power series in $q$.

\subsection{Building blocks and their gluing}
\label{subsec:buildblocksA1A2}

We consider the elliptic genus in the NSNS sector \cite{Gadde:2013dda}
\begin{equation}
    \mathcal{I}(v_i;q,y)=\mathrm{Tr}\,(-1)^F q^{H_-}y^{R_-}\prod_iv_i^{f_i}\,,
\end{equation}
where $H_-$ is the left moving conformal dimension, $R_-$ is the left moving R-symmetry, and $f_i$ are Cartan generators of the flavor symmetry $F$ that the theory might enjoy.

For a 2d $(2,2)$ GLSM with gauge group $G$ this takes the general form
\begin{equation}\label{eq:generalEGNSNS}
    \mathcal{I}(v_i;q,y)=\frac{1}{|W_G|}\oint_{\text{JK}}\prod_{a=1}^{\text{rk }G}\frac{\mathrm{d}u_a}{2\pi iu_a}\mathcal{I}_{\text{vec}}(u_a;q,y)\mathcal{I}_{\text{chir}}(u_a,v_i;q,y)\,.
\end{equation}
The contribution of the vector multiplet is
\begin{equation}\label{eq:EGNSNSvect}
    \mathcal{I}_{\text{vec}}(u_a;q,y)=\left(\frac{\qPoc{q}{q}^2}{\qth{q^{\frac{1}{2}}y^{-1}}}\right)^{\text{rk }G}\prod_{\alpha\in\Delta_+(\mathfrak{g})}\frac{\qth{\vec{u}^\alpha}}{\qth{q^{\frac{1}{2}}y^{-1}\vec{u}^\alpha}}\,,
\end{equation}
where $\Delta_+(\mathfrak{g})$ denotes the set of positive roots of the Lie algebra $\mathfrak{g}$ associated to the gauge group $G$. The contribution of a chiral multiplet in a representation $\mathcal{R}_G$ of the gauge symmetry, $\mathcal{R}_F$ of the flavor symmetry and with R-charge $R$ is instead
\begin{equation}\label{eq:EGNSNSchir}
    \mathcal{I}_{\text{chir}}(u_a,v_i;q,y)=\prod_{\rho\in\mathcal{R}_G}\prod_{\tilde{\rho}\in\mathcal{R}_F}\frac{\qth{q^{\frac{R+1}{2}}y^{R-1}\vec{u}^\rho\vec{v}^{\tilde{\rho}}}}{\qth{q^{\frac{R}{2}}y^{R}\vec{u}^\rho\vec{v}^{\tilde{\rho}}}}
\end{equation}
where $\rho$, $\tilde{\rho}$ are the weights of the representations $\mathcal{R}_G$, $\mathcal{R}_F$. These contributions are written in terms of the $q$-Pochhammer symbol $\qPoc{x}{q}=\prod_{k=0}^\infty(1-xq^k)$ and of the theta function $\qth{x}=\qPoc{x}{a}\qPoc{qx^{-1}}{q}$, and we also introduced the shorthand notation $\vec{u}^\rho=\prod_a u_a^{\rho_a}$. Finally, in the expression \eqref{eq:generalEGNSNS} we denoted by $|W_G|$ the dimension of the Weyl group of $G$, while JK stands for the Jeffrey--Kirwan residue prescription \cite{1993alg.geom..7001J} (see also \cite{Benini:2013xpa} for a detailed discussion) that we will explain momentarily. The contributions to the elliptic genus of twisted vector and chiral multiplets are given by the same expressions \eqref{eq:EGNSNSvect} and \eqref{eq:EGNSNSchir} but with the replacement $y\to y^{-1}$.

Any Riemann surface admits a pair-of-pants decomposition. The building blocks of the TQFT formula consist of the elliptic genera for the trinion theories, which depending on the types of punctures we denote by
\begin{equation}\label{eq:EGC}
    C_{III}\,,\qquad C_{\phi_{(2,1)}II}\,,\qquad C_{\phi_{(2,1)}\phi_{(2,1)}I}\,,\qquad C_{\phi_{(2,1)}\phi_{(2,1)}\phi_{(2,1)}}\,.  
\end{equation}
Punctures of the same type can be glued using appropriate tubes, which we denote by
\begin{equation}\label{eq:EGT}
    T_{II}\,,\qquad T_{\phi_{(2,1)}\phi_{(2,1)}}\,.
\end{equation}
Our proposed TQFT formula consists of constructing the elliptic genus for a generic Riemann surface by multiplying the contributions of the corresponding building blocks in its pair-of-pants decomposition, where when we glue two punctures we should sum over all of their possible types, which for the case of $(A_1,A_2)$ are just $I$ and $\phi_{(2,1)}$.

\begin{figure}[t]
\center
\resizebox{0.9\textwidth}{!}{\begin{tikzpicture}[
    line width=1.4pt,
    line cap=round,
    line join=round,
    every node/.style={font=\Large}
]

% --- Left Diagram: Sphere with 4 holes ---
\begin{scope}[shift={(-7,0)}]
    % Main circle
    \draw (0,0) circle (2.8cm);
    
    % Holes and Labels
    % Top Left
    \draw (-1.2, 1.2) circle (0.25cm);
    \node[penred, left=0.2cm] at (-1.1, 0.9) {$\phi_{(2,1)}$};
    
    % Bottom Left
    \draw (-1.2, -1.2) circle (0.25cm);
    \node[penred, left=0.2cm] at (-1.1, -0.9) {$\phi_{(2,1)}$};
    
    % Top Right
    \draw (1.2, 1.2) circle (0.25cm);
    \node[penred, right=0.2cm] at (1.2, 0.9) {$\phi_{(2,1)}$};
    
    % Bottom Right
    \draw (1.2, -1.2) circle (0.25cm);
    \node[penred, right=0.2cm] at (1.2, -0.9) {$\phi_{(2,1)}$};
\end{scope}

% --- Equals Sign ---
\node at (-2.5, 0) {\Huge $=$};

% --- Right Side ---

% --- Top Row ---
\begin{scope}[shift={(0, 2.2)}]
    % Left Pants
    \begin{scope}[shift={(0,0)}]
        \coordinate (TL) at (0, 1.2);
        \coordinate (BL) at (0, -1.2);
        \coordinate (R) at (2.5, 0);
        
        % Body
        \draw (0, 1.45) to[out=0, in=180] (2.5, 0.25);
        \draw (0, -1.45) to[out=0, in=180] (2.5, -0.25);
        \draw (0, 0.95) to[out=-45, in=45, looseness=1.2] (0, -0.95);
        
        % Holes
        \draw[fill=white] (TL) circle (0.25);
        \draw[fill=white] (BL) circle (0.25);
        \draw[fill=white] (R) circle (0.25);
        
        % Labels
        \node[penred, left=0.15cm] at (TL) {$\phi_{(2,1)}$};
        \node[penred, left=0.15cm] at (BL) {$\phi_{(2,1)}$};
        \node[penred, above=0.2cm] at (R) {$\phi_{(2,1)}$};
    \end{scope}

    % Tube
    \begin{scope}[shift={(3.5,0)}]
        \coordinate (L) at (0, 0);
        \coordinate (R) at (2.5, 0);
        
        \draw (0, 0.25) -- (2.5, 0.25);
        \draw (0, -0.25) -- (2.5, -0.25);
        
        \draw[fill=white] (L) circle (0.25);
        \draw[fill=white] (R) circle (0.25);
        
        \node[penblue, below=0.2cm] at (L) {$\phi_{(2,1)}$};
        \node[penblue, below=0.2cm] at (R) {$\phi_{(2,1)}$};
    \end{scope}

    % Right Pants
    \begin{scope}[shift={(7,0)}]
        \coordinate (L) at (0, 0);
        \coordinate (TR) at (2.5, 1.2);
        \coordinate (BR) at (2.5, -1.2);
        
        % Body
        \draw (0, 0.25) to[out=0, in=180] (2.5, 1.45);
        \draw (0, -0.25) to[out=0, in=180] (2.5, -1.45);
        \draw (2.5, 0.95) to[out=225, in=135, looseness=1.2] (2.5, -0.95);
        
        % Holes
        \draw[fill=white] (L) circle (0.25);
        \draw[fill=white] (TR) circle (0.25);
        \draw[fill=white] (BR) circle (0.25);
        
        % Labels
        \node[penred, above=0.2cm] at (L) {$\phi_{(2,1)}$};
        \node[penred, right=0.15cm] at (TR) {$\phi_{(2,1)}$};
        \node[penred, right=0.15cm] at (BR) {$\phi_{(2,1)}$};
    \end{scope}
\end{scope}

% --- Plus Sign ---
\node at (4.75, 0) {\Huge $+$};

% --- Bottom Row ---
\begin{scope}[shift={(0, -2.2)}]
    % Left Pants
    \begin{scope}[shift={(0,0)}]
        \coordinate (TL) at (0, 1.2);
        \coordinate (BL) at (0, -1.2);
        \coordinate (R) at (2.5, 0);
        
        \draw (0, 1.45) to[out=0, in=180] (2.5, 0.25);
        \draw (0, -1.45) to[out=0, in=180] (2.5, -0.25);
        \draw (0, 0.95) to[out=-45, in=45, looseness=1.2] (0, -0.95);
        
        \draw[fill=white] (TL) circle (0.25);
        \draw[fill=white] (BL) circle (0.25);
        \draw[fill=white] (R) circle (0.25);
        
        \node[penred, left=0.15cm] at (TL) {$\phi_{(2,1)}$};
        \node[penred, left=0.15cm] at (BL) {$\phi_{(2,1)}$};
        \node[penred, above=0.2cm] at (R) {$I$};
    \end{scope}

    % Tube
    \begin{scope}[shift={(3.5,0)}]
        \coordinate (L) at (0, 0);
        \coordinate (R) at (2.5, 0);
        
        \draw (0, 0.25) -- (2.5, 0.25);
        \draw (0, -0.25) -- (2.5, -0.25);
        
        \draw[fill=white] (L) circle (0.25);
        \draw[fill=white] (R) circle (0.25);
        
        \node[penblue, below=0.2cm] at (L) {$I$};
        \node[penblue, below=0.2cm] at (R) {$I$};
    \end{scope}

    % Right Pants
    \begin{scope}[shift={(7,0)}]
        \coordinate (L) at (0, 0);
        \coordinate (TR) at (2.5, 1.2);
        \coordinate (BR) at (2.5, -1.2);
        
        \draw (0, 0.25) to[out=0, in=180] (2.5, 1.45);
        \draw (0, -0.25) to[out=0, in=180] (2.5, -1.45);
        \draw (2.5, 0.95) to[out=225, in=135, looseness=1.2] (2.5, -0.95);
        
        \draw[fill=white] (L) circle (0.25);
        \draw[fill=white] (TR) circle (0.25);
        \draw[fill=white] (BR) circle (0.25);
        
        \node[penred, above=0.2cm] at (L) {$I$};
        \node[penred, right=0.15cm] at (TR) {$\phi_{(2,1)}$};
        \node[penred, right=0.15cm] at (BR) {$\phi_{(2,1)}$};
    \end{scope}
\end{scope}

\end{tikzpicture}}
\caption{TQFT structure of the elliptic genus of the 2d $(2,2)$ theory $\FF[(A_1,A_2),\Sigma_{0,4}]$ obtained compactifying $(A_1,A_2)$ on a sphere with four punctures of type $\phi_{(2,1)}$. We use red and blue colors to distinguish punctures of different sign, i.e.~with opposite choice of boundary conditions.}
\label{fig:TQFT4ptLY}
\end{figure}
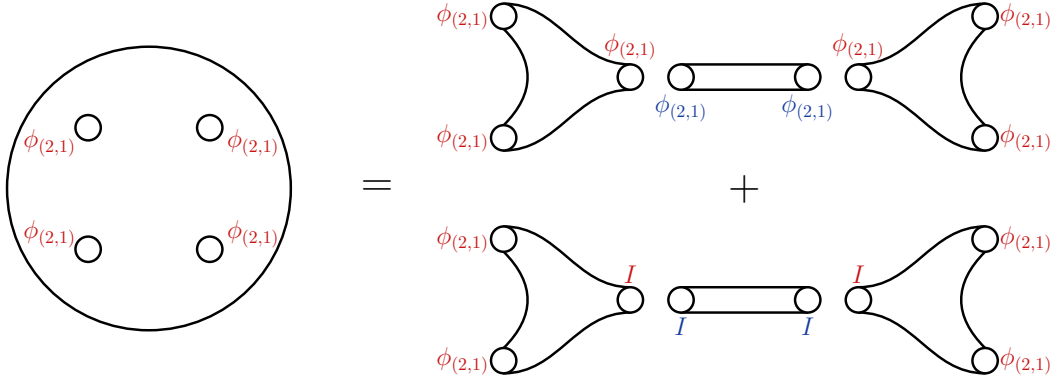

Let us consider for example the case of a sphere with four $\phi_{(2,1)}$ punctures. This can be obtained by gluing two trinions with at least two $\phi_{(2,1)}$ punctures, while the last puncture can be either $I$ or $\phi_{(2,1)}$. This gives us the sum of two contributions corresponding to each possibility, as schematically depicted in Figure \ref{fig:TQFT4ptLY}. For each term, the associated elliptic genus contribution is obtained by multiplying those of the two trinions and of the tube used in the gluing
\begin{equation}
    \mathcal{I}_{g=0,n=4}=\mathcal{I}_{C_{\phi_{(2,1)}\phi_{(2,1)}\phi_{(2,1)}}}\mathcal{I}_{T_{\phi_{(2,1)}\phi_{(2,1)}}}\mathcal{I}_{C_{\phi_{(2,1)}\phi_{(2,1)}\phi_{(2,1)}}}+\mathcal{I}_{C_{\phi_{(2,1)}\phi_{(2,1)}I}}\mathcal{I}_{T_{II}}\mathcal{I}_{C_{\phi_{(2,1)}\phi_{(2,1)}I}}\,.
\end{equation}

This procedure can be thought of as a simplified version for the gauging procedure involved when gluing punctures in class $\SS$ theories. Specifically, when gluing two punctures one generally gauges the symmetry associated with the punctures. At the level of the superconformal index, this gauging involves integrating over the gauge fugacities associated with the puncture symmetry. More generally, we can think of turning on a background flavor connection for the puncture symmetry and adding its value to the data defining the puncture. Gluing then amounts to identifying the two punctures and summing over all possible values of the background field for their symmetry (modulo gauge transformations). Our case is in a sense similar, except that the number of options is finite due to the finite number of primary operators in the Lee--Yang minimal model which are associated to distinct types of punctures, and so the integral reduces to a sum.

Overall, the problem is thus reduced to finding the elliptic genus contribution of the elementary building blocks in \eqref{eq:EGC} and \eqref{eq:EGT}. The contribution of the tubes can be argued from the analysis of the punctures of Section~\ref{sec:punct}. Specifically, we have seen that a puncture can be described as a choice of boundary conditions for the 3d $\mathcal{N}=4$ theory obtained by compactifying the 4d $\mathcal{N}=2$ SCFT on the circle of the puncture. In the case of $(A_1,A_2)$ this is just a free twisted hyper, which on the boundary is decomposed to a pair of 2d $(2,2)$ twisted chirals whose R-charges depend on the type of puncture. Specifically, for the $I$ punctures the R-charges are $\left\{\tfrac{6}{5},-\tfrac{1}{5}\right\}$ while for the $\phi_{(2,1)}$ puncture they are $\left\{\tfrac{3}{5},\tfrac{2}{5}\right\}$. One then assigns Neumann boundary conditions to one of the chirals and Dirichlet to the other. Gluing two punctures has the effect of removing the puncture. We should accordingly undo the boundary conditions, which is achieved by reintroducing the field that was given Dirichlet boundary conditions. With the choice of boundary conditions we made in Section~\ref{sec:punct}, we then find that the elliptic genus contribution of the tube $T_{II}$ should just be that of a twisted chiral of R-charge $-\tfrac{1}{5}$, while for the tube $T_{\phi_{(2,1)}\phi_{(2,1)}}$ we have a twisted chiral of R-charge $\tfrac{2}{5}$
\begin{equation}
    \mathcal{I}_{T_{II}}=\frac{\theta\left(q^{\frac{2}{5}}y^{\frac{6}{5}}\right)}{\theta\left(q^{-\frac{1}{10}}y^{\frac{1}{5}}\right)}\,,\qquad\mathcal{I}_{T_{\phi_{(2,1)}\phi_{(2,1)}}}=\frac{\theta\left(q^{\frac{7}{10}}y^{\frac{3}{5}}\right)}{\theta\left(q^{\frac{1}{5}}y^{-\frac{2}{5}}\right)}\,.
\end{equation}

It is useful to think of the tubes as spheres with punctures where we made the opposite choice of boundary conditions for the twisted chirals compared to those we made in Section~\ref{sec:punct}. This is justified by their central charges, which agree with the one expected from 4d with such different choice of boundary conditions. Using this perspective, we can use the knowledge of the elliptic genus contributions of the tubes to immediately determine the one of the trinions $C_{III}$ and $C_{\phi_{(2,1)}\phi_{(2,1)}I}$. This is because the $I$ puncture is actually equivalent to having no puncture and so these trinions are effectively also tubes. However, the punctures of these tubes are flipped (in the sense explained in Section~\ref{sec:punct}) compared to those of $T_{II}$ and $T_{\phi_{(2,1)}\phi_{(2,1)}}$, since the choice of boundary conditions at the punctures is the opposite. Hence, the elliptic genus contribution of the trinion $C_{III}$ should just be that of a twisted chiral of R-charge $\tfrac{6}{5}$, while the contribution of the trinion $C_{\phi_{(2,1)}\phi_{(2,1)}I}$ should be that of a twisted chiral of R-charge $\tfrac{3}{5}$
\begin{equation}\label{eq:EGCT}
    \mathcal{I}_{C_{III}}=\frac{\theta\left(q^{\frac{11}{10}}y^{-\frac{1}{5}}\right)}{\theta\left(q^{\frac{3}{5}}y^{-\frac{6}{5}}\right)}\,,\qquad \mathcal{I}_{C_{\phi_{(2,1)}\phi_{(2,1)} I}}=\frac{\theta\left(q^{\frac{4}{5}}y^{\frac{2}{5}}\right)}{\theta\left(q^{\frac{3}{10}}y^{-\frac{3}{5}}\right)}\,.
\end{equation}

There is another trinion whose elliptic genus contribution we can easily determine, namely $C_{\phi_{(2,1)}II}$. Indeed, from the CFT perspective we know that the one point function on the sphere of the primary $\phi_{(2,1)}$ of Lee--Yang vanishes due to conformal invariance. Via the correspondence, this implies that the 2d $(2,2)$ trinion theory $C_{\phi_{(2,1)}II}$ is actually an empty theory and its elliptic genus should just vanish
\begin{equation}
    \mathcal{I}_{C_{\phi_{(2,1)}II}}=0\,.
\end{equation}
Note that from 4d we would expect a non-zero central charge for this model, which however is negative. We interpret this as the theory not having any stable supersymmetric vacuum, which is consistent with our claim that its elliptic genus should vanish.

At this point we are left with determining the elliptic genus contribution of the trinion $C_{\phi_{(2,1)}\phi_{(2,1)}\phi_{(2,1)}}$. We can deduce this by using the GLSM description we found in Section~\ref{subsec:4ptLY} for the $\FF[(A_1,A_2),\Sigma_{0,4}]$ theory. As we will see, the elliptic genus of this GLSM has precisely the structure depicted in Figure \ref{fig:TQFT4ptLY}, which provides a first evidence of the validity of our TQFT formula.

The elliptic genus of the GLSM is computed by the following integral:
\begin{align}
    \mathcal{I}_{g=0,n=4}&=\frac{\theta\left(q^{\frac{3}{5}}y^{\frac{4}{5}}\right)\theta\left(q^{\frac{4}{5}}y^{\frac{2}{5}}\right)}{\theta\left(q^{\frac{1}{10}}y^{-\frac{1}{5}}\right)\theta\left(q^{\frac{3}{10}}y^{-\frac{3}{5}}\right)}\frac{\qPoc{q}{q}^2}{\theta\left(q^{\frac{1}{2}}y\right)}\oint_{\text{JK}}\frac{\mathrm{d}u}{2\pi i u}\frac{\theta\left(q^{\frac{3}{5}}y^{\frac{4}{5}}u^{\pm1}\right)\theta\left(q^{\frac{7}{10}}y^{\frac{3}{5}}u^{\pm1}\right)}{\theta\left(q^{\frac{1}{10}}y^{-\frac{1}{5}}u^{\pm1}\right)\theta\left(q^{\frac{1}{5}}y^{-\frac{2}{5}}u^{\pm1}\right)}\,.
\end{align}
For rank one theories, the JK contour encloses the poles inside the fundamental domain $0<|u|<q$ coming from either the positive or the negatively charged chirals \cite{Benini:2013nda}. Considering the latter, we find the following two poles:
\begin{equation}
    u=q^{\frac{1}{10}}y^{-\frac{1}{5}}\,,\qquad u=q^{\frac{1}{5}}y^{-\frac{2}{5}}\,.
\end{equation}
The residues can be evaluated using that for $n\geq 0$
\begin{equation}
    \mathrm{Res}\Big|_{u=q^{-n}}\frac{1}{\theta(u)}=\frac{(-1)^nq^{n(n+1)}}{\qPoc{q}{q}^2}\,.
\end{equation}
We then find
\begin{align}\label{eq:EG4ptLY}
    \mathcal{I}_{g=0,n=4}&=\frac{\theta\left(q^{\frac{3}{5}}y^{\frac{4}{5}}\right)^2\theta\left(q^{\frac{4}{5}}y^{\frac{2}{5}}\right)}{\theta\left(q^{\frac{1}{10}}y^{-\frac{1}{5}}\right)^2\theta\left(q^{\frac{3}{10}}y^{-\frac{3}{5}}\right)}+\frac{\theta\left(q^{\frac{2}{5}}y^{\frac{6}{5}}\right)\theta\left(q^{\frac{4}{5}}y^{\frac{2}{5}}\right)^2}{\theta\left(q^{-\frac{1}{10}}y^{\frac{1}{5}}\right)\theta\left(q^{\frac{3}{10}}y^{-\frac{3}{5}}\right)^2}\,,
\end{align}
where we simplified the contribution of pairs of fields whose R-charges sum to 1 (and so for which a mass superpotential can be generated) using
\begin{equation}
    \frac{\theta\left(q^{\frac{r+1}{2}}y^{1-r}\right)\theta\left(q^{1-\frac{r}{2}}y^{r}\right)}{\theta\left(q^{\frac{r}{2}}y^{-r}\right)\theta\left(q^{\frac{1-r}{2}}y^{r-1}\right)}=1\,,
\end{equation}
which simply follows from the fact that $\qth{x}=\qth{qx^{-1}}$. 

From \eqref{eq:EG4ptLY} we immediately see that the elliptic genus of the GLSM has the same structure depicted in Figure \ref{fig:TQFT4ptLY}, where the contributions of the tubes $T_{II}$, $T_{\phi_{(2,1)}\phi_{(2,1)}}$ and of the trinions $C_{III}$, $C_{\phi_{(2,1)}\phi_{(2,1)}I}$ are exactly those in \eqref{eq:EGT} and \eqref{eq:EGC} respectively, while that of the remaining trinion $C_{\phi_{(2,1)}\phi_{(2,1)}\phi_{(2,1)}}$ that we were looking for is
\begin{equation}
    \mathcal{I}_{C_{\phi_{(2,1)}\phi_{(2,1)}\phi_{(2,1)}}}=\frac{\theta\left(q^{\frac{3}{5}}y^{\frac{4}{5}}\right)\theta\left(q^{\frac{4}{5}}y^{\frac{2}{5}}\right)}{\theta\left(q^{\frac{1}{10}}y^{-\frac{1}{5}}\right)\theta\left(q^{\frac{3}{10}}y^{-\frac{3}{5}}\right)}\,.
\end{equation}
In other words, the elliptic genus of the last trinion is equivalent to that of two twisted chiral fields of R-charges $\tfrac{1}{5}$ and $\tfrac{3}{5}$ respectively. We also mention that the structure of the elliptic genus of this four-punctured sphere model is very similar to that of the moduli space, which we analyze in detail in Appendix \ref{appsub:duality4pt}.

Summarizing, the elliptic genus contributions of the non-trivial building blocks involved in the TQFT formula coincide with those of some twisted chiral fields as follows:
\begin{align} \label{TQFTsummary}
&   C_{I I I}=\left\{\frac{6}{5}\right\}\,,\qquad C_{\phi_{(2,1)}\phi_{(2,1)} I}=\left\{\frac{3}{5}\right\}\,,\qquad C_{\phi_{(2,1)}\phi_{(2,1)}\phi_{(2,1)}}=\left\{\frac{1}{5},\frac{3}{5}\right\}\,, \\ \nonumber   
 &   T_{II}=\left\{-\frac{1}{5}\right\}\,,\qquad T_{\phi_{(2,1)}\phi\textbf{}}=\left\{\frac{2}{5}\right\}\,,
\end{align}
where the notation $\left\{r_1,r_2,...,r_m\right\}$ denotes $m$ chiral fields with R-charges $r_1$, $r_2$, ..., $r_m$. For each non-trivial building block, the twisted chiral fields reproduce the central charges that we expect from 4d following the discussion of Section~\ref{eq:subsecanom}. When constructing the elliptic genus of a generic Riemann surface of genus $g$ and with $n$ punctures, we will have a sum of terms consisting of the contributions of the twisted chirals of the corresponding building block components. It is possible to check that for a given Riemann surface, all non-vanishing terms have the same value of the central charge, which also matches with the one predicted from 4d. 

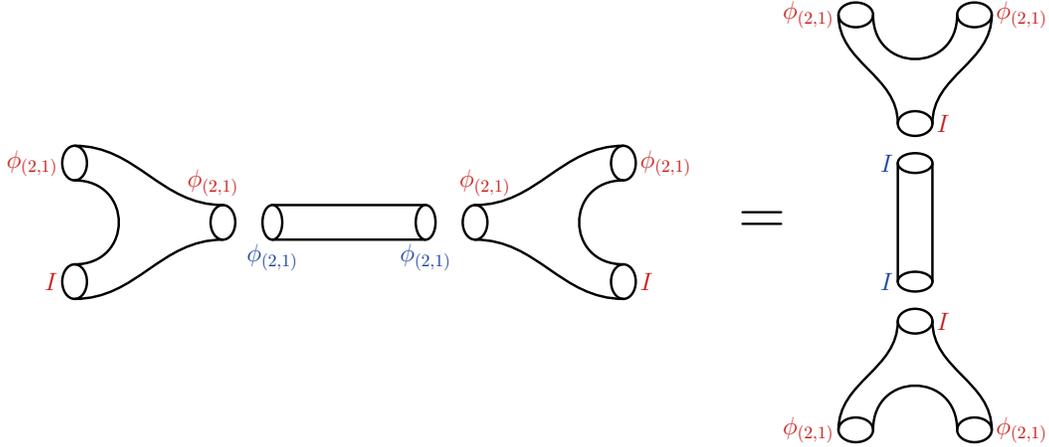
\begin{figure}[t]
\center
\resizebox{0.9\textwidth}{!}{\begin{tikzpicture}[
    line width=1.4pt,
    line cap=round,
    line join=round,
    every node/.style={font=\Large}
]

% --- LHS ---

% Left Pants
\begin{scope}[shift={(-6,0)}]
    \coordinate (TL) at (-1.5, 1.2);
    \coordinate (BL) at (-1.5, -1.2);
    \coordinate (R) at (1.5, 0);

    % Holes
    \draw[fill=white] (TL) ellipse (0.25 and 0.35);
    \draw[fill=white] (BL) ellipse (0.25 and 0.35);
    \draw[fill=white] (R) ellipse (0.25 and 0.35);

    % Body
    \draw ($(TL)+(0,0.35)$) to[out=0, in=180] ($(R)+(0,0.35)$);
    \draw ($(BL)+(0,-0.35)$) to[out=0, in=180] ($(R)+(0,-0.35)$);
    \draw ($(TL)+(0,-0.35)$) to[out=0, in=0, looseness=1.8] ($(BL)+(0,0.35)$);

    % Labels
    \node[penred, left=0.2cm] at (TL) {$\phi_{(2,1)}$};
    \node[penred, left=0.2cm] at (BL) {$I$};
    \node[penred, above=0.2cm] at ($(R)+(-0.2,0.2)$) {$\phi_{(2,1)}$};
\end{scope}

% Middle Cylinder (LHS)
\begin{scope}[shift={(-2,0)}]
    \coordinate (L) at (-1.5, 0);
    \coordinate (R) at (1.6, 0);

    % Body lines first so holes cover them if needed
    \draw ($(L)+(0,0.35)$) -- ($(R)+(0,0.35)$);
    \draw ($(L)+(0,-0.35)$) -- ($(R)+(0,-0.35)$);

    % Holes
    \draw[fill=white] (L) ellipse (0.2 and 0.35);
    \draw[fill=white] (R) ellipse (0.2 and 0.35);

    % Labels
    \node[penblue, below=0.3cm] at (L) {$\phi_{(2,1)}$};
    \node[penblue, below=0.3cm] at (R) {$\phi_{(2,1)}$};
\end{scope}

% Right Pants (LHS)
\begin{scope}[shift={(2.5-0.4,0)}]
    \coordinate (L) at (-1.5, 0);
    \coordinate (TR) at (1.5, 1.2);
    \coordinate (BR) at (1.5, -1.2);

    % Holes
    \draw[fill=white] (L) ellipse (0.25 and 0.35);
    \draw[fill=white] (TR) ellipse (0.25 and 0.35);
    \draw[fill=white] (BR) ellipse (0.25 and 0.35);

    % Body
    \draw ($(L)+(0,0.35)$) to[out=0, in=180] ($(TR)+(0,0.35)$);
    \draw ($(L)+(0,-0.35)$) to[out=0, in=180] ($(BR)+(0,-0.35)$);
    \draw ($(TR)+(0,-0.35)$) to[out=180, in=180, looseness=1.8] ($(BR)+(0,0.35)$);

    % Labels
    \node[penred, above=0.2cm] at ($(L)+(0.2,0.2)$) {$\phi_{(2,1)}$};
    \node[penred, right=0.2cm] at (TR) {$\phi_{(2,1)}$};
    \node[penred, right=0.2cm] at (BR) {$I$};
\end{scope}

% Equals Sign
\node at (6.4, 0) [scale=2.5] {=};

% --- RHS ---

\begin{scope}[shift={(9.5,0)}]
    % Top Pants
    \begin{scope}[shift={(0, 3)}]
        \coordinate (TL) at (-1.2, 1.2);
        \coordinate (TR) at (1.2, 1.2);
        \coordinate (B) at (0, -1);

        % Holes
        \draw[fill=white] (TL) ellipse (0.35 and 0.25);
        \draw[fill=white] (TR) ellipse (0.35 and 0.25);
        \draw[fill=white] (B) ellipse (0.35 and 0.25);

        % Body
        \draw ($(TL)+(-0.35,0)$) to[out=270, in=90] ($(B)+(-0.35,0)$);
        \draw ($(TR)+(0.35,0)$) to[out=270, in=90] ($(B)+(0.35,0)$);
        \draw ($(TL)+(0.35,0)$) to[out=270, in=270, looseness=1.8] ($(TR)+(-0.35,0)$);

        % Labels
        \node[penred, left=0.3cm] at (TL) {$\phi_{(2,1)}$};
        \node[penred, right=0.3cm] at (TR) {$\phi_{(2,1)}$};
        \node[penred, right=0.3cm] at (B) {$I$};
    \end{scope}

    % Middle Cylinder
    \begin{scope}[shift={(0, 0)}]
        \coordinate (T) at (0, 1.2);
        \coordinate (B) at (0, -1.2);

        % Body
        \draw ($(T)+(-0.35,0)$) -- ($(B)+(-0.35,0)$);
        \draw ($(T)+(0.35,0)$) -- ($(B)+(0.35,0)$);

        % Holes
        \draw[fill=white] (T) ellipse (0.35 and 0.2);
        \draw[fill=white] (B) ellipse (0.35 and 0.2);

        % Labels
        \node[penblue, left=0.3cm] at (T) {$I$};
        \node[penblue, left=0.3cm] at (B) {$I$};
    \end{scope}

    % Bottom Pants
    \begin{scope}[shift={(0, -3)}]
        \coordinate (T) at (0, 1);
        \coordinate (BL) at (-1.2, -1.2);
        \coordinate (BR) at (1.2, -1.2);

        % Holes
        \draw[fill=white] (T) ellipse (0.35 and 0.25);
        \draw[fill=white] (BL) ellipse (0.35 and 0.25);
        \draw[fill=white] (BR) ellipse (0.35 and 0.25);

        % Body
        \draw ($(T)+(-0.35,0)$) to[out=270, in=90] ($(BL)+(-0.35,0)$);
        \draw ($(T)+(0.35,0)$) to[out=270, in=90] ($(BR)+(0.35,0)$);
        \draw ($(BL)+(0.35,0)$) to[out=90, in=90, looseness=1.8] ($(BR)+(-0.35,0)$);

        % Labels
        \node[penred, right=0.3cm] at (T) {$I$};
        \node[penred, left=0.3cm] at (BL) {$\phi_{(2,1)}$};
        \node[penred, right=0.3cm] at (BR) {$\phi_{(2,1)}$};
    \end{scope}
\end{scope}

\end{tikzpicture}}
\caption{Check of the associativity property of the TQFT formula for the elliptic genus. We verify that in the case of a sphere with two $\phi_{(2,1)}$ and two $I$ punctures, different decompositions lead to the same answer.}
\label{fig:assoc}
\end{figure}

A first consistency check for our proposal of the TQFT structure of the elliptic genus is that it respects the associativity property. Let us consider for example a sphere with two $\phi_{(2,1)}$ and two $I$ punctures. There are two different decompositions that we consider, depending on whether we group punctures of the same type or of opposite type together, see Figure \ref{fig:assoc}. This is similar to the different channel decompositions of the CFT correlator. In the first case we have only one term in the elliptic genus corresponding to gluing with a $T_{\phi_{(2,1)}\phi_{(2,1)}}$ tube, due to the fact that the trinion $C_{\phi_{(2,1)}II}$ gives vanishing contribution. Similarly, in the second case we again have a single term, but this time corresponding to gluing with a $T_{II}$ tube. However, using the building blocks previously determined, one can easily check that the two results are identical and that the associativity property holds.

\begin{figure}[t]
\center
\resizebox{1\textwidth}{!}{\begin{tikzpicture}[
    line width=1.6pt,
    line cap=round,
    line join=round,
    % Define colors matching the image
    myred/.style={red!85!black},
    myblue/.style={blue!80!black},
    mygreen/.style={green!60!black},
    % Define symbol styles
    sym/.style={font=\small\bfseries},
    rsym/.style={sym, myred},
    bsym/.style={sym, myblue},
    % Define shape parameters
    hole/.style={fill=white, draw=black, thick, inner sep=0pt}
]

% --- Left Sphere ---
\begin{scope}[shift={(-5.5, -4.25)}]
    \draw[thick] (0,0) circle (2.2);
    % Holes inside the sphere
    \node[circle, draw, thick, inner sep=0pt, minimum size=4mm] (h1) at (0, 1.2) {};
    \node[circle, draw, thick, inner sep=0pt, minimum size=4mm] (h2) at (-1.2, 0.3) {};
    \node[circle, draw, thick, inner sep=0pt, minimum size=4mm] (h3) at (1.2, 0.3) {};
    \node[circle, draw, thick, inner sep=0pt, minimum size=4mm] (h4) at (-0.7, -1.2) {};
    \node[circle, draw, thick, inner sep=0pt, minimum size=4mm] (h5) at (0.7, -1.2) {};
    
    % Labels for holes
    \node[rsym, above=2mm] at (h1) {$\phi_{(2,1)}$};
    \node[rsym, above=2mm] at (h2) {$\phi_{(2,1)}$};
    \node[rsym, above=2mm] at (h3) {$\phi_{(2,1)}$};
    \node[rsym, above=2mm] at (h4) {$\phi_{(2,1)}$};
    \node[rsym, above=2mm] at (h5) {$\phi_{(2,1)}$};
\end{scope}

\node at (-2.2, -4.25) {\Huge =};

% --- Row 1 ---
\begin{scope}[shift={(0,1)}]
    % PantsL
    \draw[thick] (0, 0.7) to[out=0, in=180] (1.5, 0.25);
    \draw[thick] (0, -0.7) to[out=0, in=180] (1.5, -0.25);
    \draw[thick] (0, 0.25) to[out=0, in=180] (0.5, 0) to[out=0, in=180] (0, -0.25);
    \draw[thick, fill=white] (0, 0.5) ellipse (0.15 and 0.25);
    \draw[thick, fill=white] (0, -0.5) ellipse (0.15 and 0.25);
    \draw[thick, fill=white] (1.5, 0) ellipse (0.15 and 0.25);
    \node[rsym, left] at (-0.15, 0.5) {$\phi_{(2,1)}$};
    \node[rsym, left] at (-0.15, -0.5) {$\phi_{(2,1)}$};
    \node[rsym, above] at (1.5, 0.2) {$\phi_{(2,1)}$};

    % Tube 1
    \begin{scope}[shift={(2.2, 0)}]
        \draw[thick] (0, 0.25) -- (1.5, 0.25);
        \draw[thick] (0, -0.25) -- (1.5, -0.25);
        \draw[thick, fill=white] (0, 0) ellipse (0.15 and 0.25);
        \draw[thick, fill=white] (1.5, 0) ellipse (0.15 and 0.25);
        \node[bsym, below] at (0, -0.2) {$\phi_{(2,1)}$};
        \node[bsym, below] at (1.5, -0.2) {$\phi_{(2,1)}$};
    \end{scope}

    % Tee
    \begin{scope}[shift={(4.4, 0)}]
        \draw[thick] (0, -0.25) -- (1.5, -0.25);
        \draw[thick] (0, 0.25) .. controls (0.3, 0.25) .. (0.5, 0.8);
        \draw[thick] (1.5, 0.25) .. controls (1.2, 0.25) .. (1.0, 0.8);
        \draw[thick, fill=white] (0, 0) ellipse (0.15 and 0.25);
        \draw[thick, fill=white] (1.5, 0) ellipse (0.15 and 0.25);
        \draw[thick, fill=white] (0.75, 0.8) ellipse (0.25 and 0.15);
        \node[rsym, above] at (-0.1, 0.2) {$\phi_{(2,1)}$};
        \node[rsym, above] at (1.6, 0.2) {$\phi_{(2,1)}$};
        \node[rsym, above] at (0.75, 0.95) {$\phi_{(2,1)}$};
    \end{scope}

    % Tube 2
    \begin{scope}[shift={(6.6, 0)}]
        \draw[thick] (0, 0.25) -- (1.5, 0.25);
        \draw[thick] (0, -0.25) -- (1.5, -0.25);
        \draw[thick, fill=white] (0, 0) ellipse (0.15 and 0.25);
        \draw[thick, fill=white] (1.5, 0) ellipse (0.15 and 0.25);
        \node[bsym, below] at (0, -0.2) {$\phi_{(2,1)}$};
        \node[bsym, below] at (1.5, -0.2) {$\phi_{(2,1)}$};
    \end{scope}

    % PantsR
    \begin{scope}[shift={(8.8, 0)}]
        \draw[thick] (0, 0.25) to[out=0, in=180] (1.5, 0.7);
        \draw[thick] (0, -0.25) to[out=0, in=180] (1.5, -0.7);
        \draw[thick] (1.5, 0.25) to[out=180, in=0] (1.0, 0) to[out=0, in=180] (1.5, -0.25);
        \draw[thick, fill=white] (0, 0) ellipse (0.15 and 0.25);
        \draw[thick, fill=white] (1.5, 0.5) ellipse (0.15 and 0.25);
        \draw[thick, fill=white] (1.5, -0.5) ellipse (0.15 and 0.25);
        \node[rsym, above] at (0, 0.2) {$\phi_{(2,1)}$};
        \node[rsym, right] at (1.65, 0.5) {$\phi_{(2,1)}$};
        \node[rsym, right] at (1.65, -0.5) {$\phi_{(2,1)}$};
    \end{scope}
\end{scope}

\node at (5.15, -0.5) {\Huge +};

% --- Row 2 ---
\begin{scope}[shift={(0,-2.5)}]
    % PantsL
    \draw[thick] (0, 0.7) to[out=0, in=180] (1.5, 0.25);
    \draw[thick] (0, -0.7) to[out=0, in=180] (1.5, -0.25);
    \draw[thick] (0, 0.25) to[out=0, in=180] (0.5, 0) to[out=0, in=180] (0, -0.25);
    \draw[thick, fill=white] (0, 0.5) ellipse (0.15 and 0.25);
    \draw[thick, fill=white] (0, -0.5) ellipse (0.15 and 0.25);
    \draw[thick, fill=white] (1.5, 0) ellipse (0.15 and 0.25);
    \node[rsym, left] at (-0.15, 0.5) {$\phi_{(2,1)}$};
    \node[rsym, left] at (-0.15, -0.5) {$\phi_{(2,1)}$};
    \node[rsym, above] at (1.5, 0.2) {$\phi_{(2,1)}$};

    % Tube 1
    \begin{scope}[shift={(2.2, 0)}]
        \draw[thick] (0, 0.25) -- (1.5, 0.25);
        \draw[thick] (0, -0.25) -- (1.5, -0.25);
        \draw[thick, fill=white] (0, 0) ellipse (0.15 and 0.25);
        \draw[thick, fill=white] (1.5, 0) ellipse (0.15 and 0.25);
        \node[bsym, below] at (0, -0.2) {$\phi_{(2,1)}$};
        \node[bsym, below] at (1.5, -0.2) {$\phi_{(2,1)}$};
    \end{scope}

    % Tee
    \begin{scope}[shift={(4.4, 0)}]
        \draw[thick] (0, -0.25) -- (1.5, -0.25);
        \draw[thick] (0, 0.25) .. controls (0.3, 0.25) .. (0.5, 0.8);
        \draw[thick] (1.5, 0.25) .. controls (1.2, 0.25) .. (1.0, 0.8);
        \draw[thick, fill=white] (0, 0) ellipse (0.15 and 0.25);
        \draw[thick, fill=white] (1.5, 0) ellipse (0.15 and 0.25);
        \draw[thick, fill=white] (0.75, 0.8) ellipse (0.25 and 0.15);
        \node[rsym, above] at (-0.1, 0.2) {$\phi_{(2,1)}$};
        \node[rsym, above] at (1.6, 0.2) {$I$};
        \node[rsym, above] at (0.75, 0.95) {$\phi_{(2,1)}$};
    \end{scope}

    % Tube 2
    \begin{scope}[shift={(6.6, 0)}]
        \draw[thick] (0, 0.25) -- (1.5, 0.25);
        \draw[thick] (0, -0.25) -- (1.5, -0.25);
        \draw[thick, fill=white] (0, 0) ellipse (0.15 and 0.25);
        \draw[thick, fill=white] (1.5, 0) ellipse (0.15 and 0.25);
        \node[bsym, below] at (0, -0.2) {$I$};
        \node[bsym, below] at (1.5, -0.2) {$I$};
    \end{scope}

    % PantsR
    \begin{scope}[shift={(8.8, 0)}]
        \draw[thick] (0, 0.25) to[out=0, in=180] (1.5, 0.7);
        \draw[thick] (0, -0.25) to[out=0, in=180] (1.5, -0.7);
        \draw[thick] (1.5, 0.25) to[out=180, in=0] (1.0, 0) to[out=0, in=180] (1.5, -0.25);
        \draw[thick, fill=white] (0, 0) ellipse (0.15 and 0.25);
        \draw[thick, fill=white] (1.5, 0.5) ellipse (0.15 and 0.25);
        \draw[thick, fill=white] (1.5, -0.5) ellipse (0.15 and 0.25);
        \node[rsym, above] at (0, 0.2) {$I$};
        \node[rsym, right] at (1.65, 0.5) {$\phi_{(2,1)}$};
        \node[rsym, right] at (1.65, -0.5) {$\phi_{(2,1)}$};
    \end{scope}
\end{scope}

\node at (5.15, -4.25) {\Huge +};

% --- Row 3 ---
\begin{scope}[shift={(0,-6.5)}]
    % PantsL
    \draw[thick] (0, 0.7) to[out=0, in=180] (1.5, 0.25);
    \draw[thick] (0, -0.7) to[out=0, in=180] (1.5, -0.25);
    \draw[thick] (0, 0.25) to[out=0, in=180] (0.5, 0) to[out=0, in=180] (0, -0.25);
    \draw[thick, fill=white] (0, 0.5) ellipse (0.15 and 0.25);
    \draw[thick, fill=white] (0, -0.5) ellipse (0.15 and 0.25);
    \draw[thick, fill=white] (1.5, 0) ellipse (0.15 and 0.25);
    \node[rsym, left] at (-0.15, 0.5) {$\phi_{(2,1)}$};
    \node[rsym, left] at (-0.15, -0.5) {$\phi_{(2,1)}$};
    \node[rsym, above] at (1.5, 0.2) {$I$};

    % Tube 1
    \begin{scope}[shift={(2.2, 0)}]
        \draw[thick] (0, 0.25) -- (1.5, 0.25);
        \draw[thick] (0, -0.25) -- (1.5, -0.25);
        \draw[thick, fill=white] (0, 0) ellipse (0.15 and 0.25);
        \draw[thick, fill=white] (1.5, 0) ellipse (0.15 and 0.25);
        \node[bsym, below] at (0, -0.2) {$I$};
        \node[bsym, below] at (1.5, -0.2) {$I$};
    \end{scope}

    % Tee
    \begin{scope}[shift={(4.4, 0)}]
        \draw[thick] (0, -0.25) -- (1.5, -0.25);
        \draw[thick] (0, 0.25) .. controls (0.3, 0.25) .. (0.5, 0.8);
        \draw[thick] (1.5, 0.25) .. controls (1.2, 0.25) .. (1.0, 0.8);
        \draw[thick, fill=white] (0, 0) ellipse (0.15 and 0.25);
        \draw[thick, fill=white] (1.5, 0) ellipse (0.15 and 0.25);
        \draw[thick, fill=white] (0.75, 0.8) ellipse (0.25 and 0.15);
        \node[rsym, above] at (-0.1, 0.2) {$I$};
        \node[rsym, above] at (1.6, 0.2) {$\phi_{(2,1)}$};
        \node[rsym, above] at (0.75, 0.95) {$\phi_{(2,1)}$};
    \end{scope}

    % Tube 2
    \begin{scope}[shift={(6.6, 0)}]
        \draw[thick] (0, 0.25) -- (1.5, 0.25);
        \draw[thick] (0, -0.25) -- (1.5, -0.25);
        \draw[thick, fill=white] (0, 0) ellipse (0.15 and 0.25);
        \draw[thick, fill=white] (1.5, 0) ellipse (0.15 and 0.25);
        \node[bsym, below] at (0, -0.2) {$\phi_{(2,1)}$};
        \node[bsym, below] at (1.5, -0.2) {$\phi_{(2,1)}$};
    \end{scope}

    % PantsR
    \begin{scope}[shift={(8.8, 0)}]
        \draw[thick] (0, 0.25) to[out=0, in=180] (1.5, 0.7);
        \draw[thick] (0, -0.25) to[out=0, in=180] (1.5, -0.7);
        \draw[thick] (1.5, 0.25) to[out=180, in=0] (1.0, 0) to[out=0, in=180] (1.5, -0.25);
        \draw[thick, fill=white] (0, 0) ellipse (0.15 and 0.25);
        \draw[thick, fill=white] (1.5, 0.5) ellipse (0.15 and 0.25);
        \draw[thick, fill=white] (1.5, -0.5) ellipse (0.15 and 0.25);
        \node[rsym, above] at (0, 0.2) {$\phi_{(2,1)}$};
        \node[rsym, right] at (1.65, 0.5) {$\phi_{(2,1)}$};
        \node[rsym, right] at (1.65, -0.5) {$\phi_{(2,1)}$};
    \end{scope}
\end{scope}

\node at (5.15, -8) {\Huge +};

% --- Row 4 ---
\begin{scope}[shift={(0,-10)}]
    % PantsL
    \draw[thick] (0, 0.7) to[out=0, in=180] (1.5, 0.25);
    \draw[thick] (0, -0.7) to[out=0, in=180] (1.5, -0.25);
    \draw[thick] (0, 0.25) to[out=0, in=180] (0.5, 0) to[out=0, in=180] (0, -0.25);
    \draw[thick, fill=white] (0, 0.5) ellipse (0.15 and 0.25);
    \draw[thick, fill=white] (0, -0.5) ellipse (0.15 and 0.25);
    \draw[thick, fill=white] (1.5, 0) ellipse (0.15 and 0.25);
    \node[rsym, left] at (-0.15, 0.5) {$\phi_{(2,1)}$};
    \node[rsym, left] at (-0.15, -0.5) {$\phi_{(2,1)}$};
    \node[rsym, above] at (1.5, 0.2) {$I$};

    % Tube 1
    \begin{scope}[shift={(2.2, 0)}]
        \draw[thick] (0, 0.25) -- (1.5, 0.25);
        \draw[thick] (0, -0.25) -- (1.5, -0.25);
        \draw[thick, fill=white] (0, 0) ellipse (0.15 and 0.25);
        \draw[thick, fill=white] (1.5, 0) ellipse (0.15 and 0.25);
        \node[bsym, below] at (0, -0.2) {$I$};
        \node[bsym, below] at (1.5, -0.2) {$I$};
    \end{scope}

    % Tee
    \begin{scope}[shift={(4.4, 0)}]
        \draw[thick] (0, -0.25) -- (1.5, -0.25);
        \draw[thick] (0, 0.25) .. controls (0.3, 0.25) .. (0.5, 0.8);
        \draw[thick] (1.5, 0.25) .. controls (1.2, 0.25) .. (1.0, 0.8);
        \draw[thick, fill=white] (0, 0) ellipse (0.15 and 0.25);
        \draw[thick, fill=white] (1.5, 0) ellipse (0.15 and 0.25);
        \draw[thick, fill=white] (0.75, 0.8) ellipse (0.25 and 0.15);
        \node[rsym, above] at (-0.1, 0.2) {$I$};
        \node[rsym, above] at (1.6, 0.2) {$I$};
        \node[rsym, above] at (0.75, 0.95) {$\phi_{(2,1)}$};
        
        % Green cancellation
        \draw[mygreen, thick] (0.2, -0.5) -- (1.3, 1.2);
        \node[mygreen, font=\bfseries\small] at (1.7, 1.3) {$=0$};
    \end{scope}

    % Tube 2
    \begin{scope}[shift={(6.6, 0)}]
        \draw[thick] (0, 0.25) -- (1.5, 0.25);
        \draw[thick] (0, -0.25) -- (1.5, -0.25);
        \draw[thick, fill=white] (0, 0) ellipse (0.15 and 0.25);
        \draw[thick, fill=white] (1.5, 0) ellipse (0.15 and 0.25);
        \node[bsym, below] at (0, -0.2) {$I$};
        \node[bsym, below] at (1.5, -0.2) {$I$};
    \end{scope}

    % PantsR
    \begin{scope}[shift={(8.8, 0)}]
        \draw[thick] (0, 0.25) to[out=0, in=180] (1.5, 0.7);
        \draw[thick] (0, -0.25) to[out=0, in=180] (1.5, -0.7);
        \draw[thick] (1.5, 0.25) to[out=180, in=0] (1.0, 0) to[out=0, in=180] (1.5, -0.25);
        \draw[thick, fill=white] (0, 0) ellipse (0.15 and 0.25);
        \draw[thick, fill=white] (1.5, 0.5) ellipse (0.15 and 0.25);
        \draw[thick, fill=white] (1.5, -0.5) ellipse (0.15 and 0.25);
        \node[rsym, above] at (0, 0.2) {$I$};
        \node[rsym, right] at (1.65, 0.5) {$\phi_{(2,1)}$};
        \node[rsym, right] at (1.65, -0.5) {$\phi_{(2,1)}$};
    \end{scope}
\end{scope}

\end{tikzpicture}}
\caption{TQFT structure of the elliptic genus of the 2d $(2,2)$ theory $\FF[(A_1,A_2),\Sigma_{0,5}]$ obtained compactifying $(A_1,A_2)$ on a sphere with five punctures of type $\phi_{(2,1)}$.}
\label{fig:TQFT5ptLY}
\end{figure}
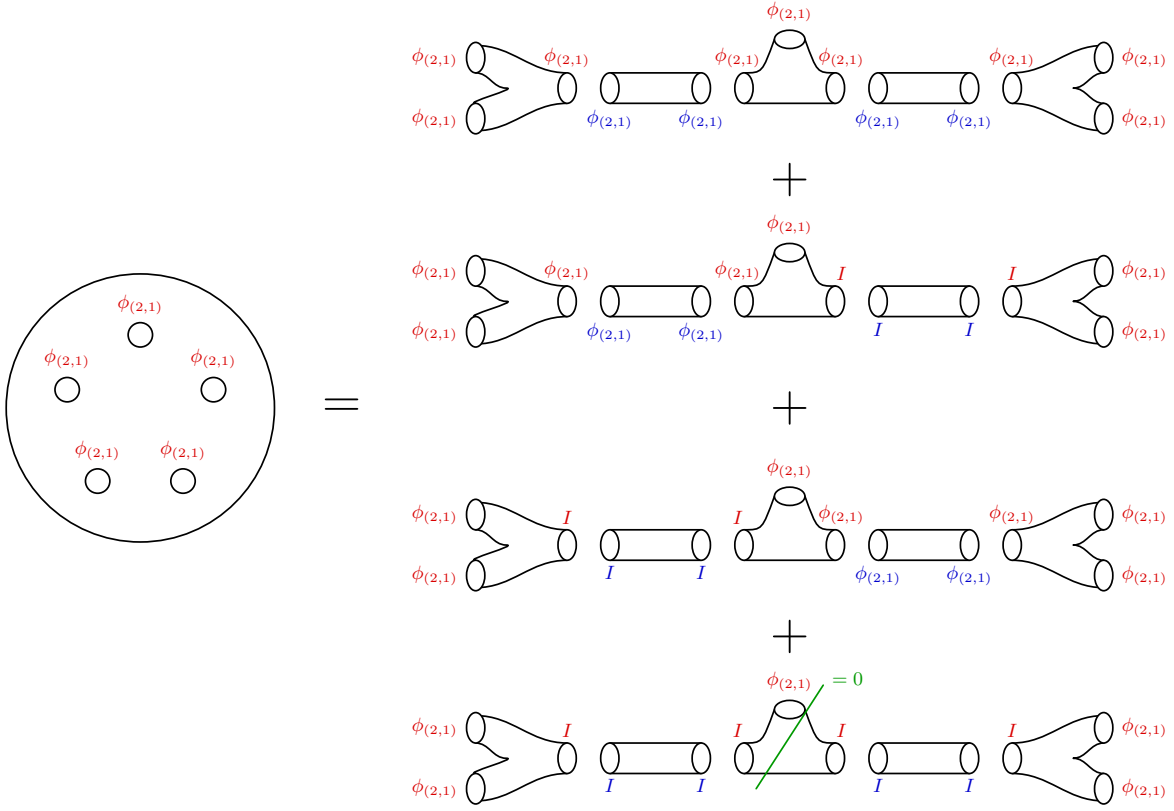

As another check, we can verify that our proposal for how to compute the elliptic genus of the theory associated to a generic Riemann surface reproduces the result that we expect from the GLSM description we found in Section~\ref{subsec:5ptLY} in the case of $\FF[(A_1,A_2),\Sigma_{0,5}]$, namely the sphere with five $\phi_{(2,1)}$ punctures. The elliptic genus of the GLSM reads
\begin{align}
    \mathcal{I}_{g=0,n=5}&=\frac{\theta\left(q^{\frac{3}{5}}y^{\frac{4}{5}}\right)\theta\left(q^{\frac{4}{5}}y^{\frac{2}{5}}\right)^3}{\theta\left(q^{\frac{1}{10}}y^{-\frac{1}{5}}\right)\theta\left(q^{\frac{3}{10}}y^{-\frac{3}{5}}\right)^3}\frac{(q)_\infty^4}{\theta\left(q^{\frac{1}{2}}y\right)^2}\nn\\
    &\times\oint_{\text{JK}}\frac{\mathrm{d}z_1}{2\pi i z_1}\frac{\mathrm{d}z_2}{2\pi i z_2}\frac{\theta\left(q^{\frac{3}{5}}y^{\frac{4}{5}}z_1^{\pm1}\right)\theta\left(q^{\frac{3}{5}}y^{\frac{4}{5}}z_1^{\mp1}z_2{\pm1}\right)\theta\left(q^{\frac{3}{5}}y^{\frac{4}{5}}z_2^{\mp1}\right)}{\theta\left(q^{\frac{1}{10}}y^{-\frac{1}{5}}z_1^{\pm1}\right)\theta\left(q^{\frac{1}{10}}y^{-\frac{1}{5}}z_1^{\mp1}z_2^{\pm1}\right)\theta\left(q^{\frac{1}{10}}y^{-\frac{1}{5}}z_2^{\mp1}\right)}\,.
\end{align}
When the gauge group is of rank greater than 1, the JK residue prescription is more involved. For rank 2 as in this case, we first need to pick a reference vector $\eta$ in $\mathbb{R}^2$ (the final result is independent of the choice of $\eta$) and then select among the vectors encoding the gauge charges of all the chirals all the pairs that span a positive cone that contains $\eta$. We should then consider the poles coming from all such pairs of chirals. Taking for example $\eta=(-1,-1)$ we find that the following three poles contribute to the integral:
\begin{align}
    &z_1=z_2=q^{\frac{1}{10}}y^{-\frac{1}{5}}\,,\nn\\
    &z_1=q^{\frac{1}{10}}y^{-\frac{1}{5}}\,,\quad z_2=q^{\frac{1}{10}}y^{-\frac{1}{5}}z_1=q^{\frac{1}{5}}y^{-\frac{2}{5}}\,,\nn\\
    &z_2=q^{\frac{1}{10}}y^{-\frac{1}{5}}\,,\quad z_1=q^{\frac{1}{10}}y^{-\frac{1}{5}}z_2=q^{\frac{1}{5}}y^{-\frac{2}{5}}\,,
\end{align}
however given the symmetry under $z_1\leftrightarrow z_2$ of the integrand the second and the third poles give the same residue contribution, so we actually only have to compute two residues. The final result of the integration is
\begin{align}\label{eq:EG5ptLY}
    \mathcal{I}_{g=0,n=5}&=\frac{\theta\left(q^{\frac{3}{5}}y^{\frac{4}{5}}\right)^3\theta\left(q^{\frac{4}{5}}y^{\frac{2}{5}}\right)}{\theta\left(q^{\frac{1}{10}}y^{-\frac{1}{5}}\right)^3\theta\left(q^{\frac{3}{10}}y^{-\frac{3}{5}}\right)}+2\frac{\theta\left(q^{\frac{2}{5}}y^{\frac{6}{5}}\right)\theta\left(q^{\frac{3}{5}}y^{\frac{4}{5}}\right)\theta\left(q^{\frac{4}{5}}y^{\frac{2}{5}}\right)^2}{\theta\left(q^{-\frac{1}{10}}y^{\frac{1}{5}}\right)\theta\left(q^{\frac{1}{10}}y^{-\frac{1}{5}}\right)\theta\left(q^{\frac{3}{10}}y^{-\frac{3}{5}}\right)^2}\,,
\end{align}
We see that this result is again compatible with a TQFT structure as shown in Figure \ref{fig:TQFT5ptLY}. Notice in particular that the presence of only three terms in the elliptic genus is compatible with the fact that the trinion $C_{\phi_{(2,1)}II}$ is trivial. Also in this case the structure of the elliptic genus seems to be identical to that of the moduli space, which for this model we analyze in detail in Appendix \ref{appsub:duality5pt}.

Before moving to other consistency checks of the TQFT formula for the elliptic genus, we would like to make a few comments on its physical interpretation. In general, the gluing procedure provides a way to build the low-energy theory associated with a given surface from those of a special set of building blocks. We noted that every term in the sum over intermediary punctures in the pair-of-pants decomposition can ultimately be interpreted as the elliptic genus of a physical 2d theory, which is some collection of chiral fields. Thus, the structure of the elliptic genus observed here is quite similar to that observed in higher dimensions. However, there are several peculiar features. The first one is that we get a finite sum of contributions. The second is that the resulting picture is insufficient to fully understand the resulting 2d theory. This is apparent when considering the $S^2$ partition function of the 4-punctured sphere \eqref{eq:S2pfGLSMLY}. The result indeed consists of two terms that can be interpreted as coming from the two collections of chirals, but it also contains the hypergeometric function coming from the vortex partition function, which cannot be reproduced from knowledge of only the free chirals and that encodes the dependence on the complex structure modulus $z$ of the sphere with four punctures. It seems that the above TQFT structure captures the low-energy theory around the singular points in the moduli space (see Appendix \ref{app:duality} for the analysis of the moduli space of the GLSM). This information is incomplete, but appears to be sufficient to reproduce the elliptic genus thanks to the fact that this does not depend on the complex structure moduli of the Riemann surface. It would be interesting to understand if this can be expanded so as to provide a complete gluing procedure of the entire field theory or at least of its $S^2$ partition function, and not just the elliptic genus. 

\subsection{Sphere compactification}
\label{subsec:sphcomp}

We have seen that the ellipic genus of the trinion $C_{I I I}$, that is the $S^2$ compactification of $(A_1,A_2)$, should be equal to that of a single chiral field of R-charge $\frac{6}{5}$, equal to the scaling dimension of the unique 4d Coulomb branch operator. This nicely agrees with the general expectation from Section~\ref{eq:subsecanom} that the central charges of the 2d $(2,2)$ theory $\FF[T;\Sigma_{0,0}]$ obtained by compactifying a 4d $\mathcal{N}=2$ SCFT $T$ with central charges $a$, $c$ on a sphere should be
\begin{equation}\label{eq:compspherec}
    c_\pm=4(2a-c)=\frac{1}{2}\sum_i(1-2\Delta_i)\,,
\end{equation}
where we used the Shapere--Tachikawa formula \cite{Shapere:2008zf} to express the central charges $a$, $c$ in terms of the scaling dimensions $\Delta_i$ of the Coulomb branch operators. These look like the central charges of a number of twisted chiral fields equal to the dimension of the 4d Coulomb branch and with R-charges identical to those of the Coulomb branch operators.

This result can be matched against the direct study of the sphere compactification of the $(A_1,A_2)$ theory using its $\mathcal{N}=1$ Lagrangian description given in \cite{Maruyoshi:2016tqk,Maruyoshi:2016aim}. Such an analysis was carried out also in \cite{Gukov:2017zao}. The 4d $\mathcal{N}=2$ Lagrangian consists of an $SU(2)$ gauge theory with one adjoint chiral $\Phi$, two fundamental chirals $q_i$ and two singlets $\alpha$, $\beta$ with the superpotential
\begin{equation}
    \mathcal{W}=\Phi q_1 q_1 + \alpha\Phi^2 + \beta\Phi q_2 q_2\,.
\end{equation}
The field content of the theory and the transformation properties under gauge and global symmetries are as follows:
\begin{table}[h]
  \centering
  \begin{tabular}{c|ccccc}
    & $\Phi$ & $q_1$ & $q_2$ & $\alpha$ & $\beta$ \\\hline
    $SU(2)$   & $\mathbf{3}$ & $\mathbf{2}$ & $\mathbf{2}$ & $\mathbf{1}$ & $\mathbf{1}$ \\
    $U(1)_t$  & $-\tfrac15$  & $\tfrac{1}{10}$ & $\tfrac{7}{10}$ & $\tfrac{2}{5}$ & $-\tfrac{6}{5}$ \\
    $U(1)_r$  & $\tfrac25$   & $\tfrac{4}{5}$ & $-\tfrac{2}{5}$ & $\tfrac{6}{5}$ & $\tfrac{12}{5}$ \\
    $U(1)_R$  & $0$   & $1$ & $1$ & $2$ & $0$
\end{tabular}
\end{table}

\noindent In the table we are reporting the charges under $U(1)_r$ and under the $U(1)_R\subset SU(2)_R$ symmetry involved in the twist, as well as under 
\begin{equation}\label{eq:U1tdef}
    U(1)_t=\frac{1}{2}\left(U(1)_R-U(1)_r\right)\,,
\end{equation}
which can be understood as a flavor symmetry from the 4d $\mathcal{N}=1$ perspective.

As explained in \cite{Kutasov:2013ffl,Gadde:2015wta}, after performing a twisted compactification by $U(1)_R$ on the sphere, each 4d chiral should lead to $1-R$ $(0,2)$ chirals in 2d if their $U(1)_R$ R-charge $R$ is smaller than 1, $R-1$ Fermi multiplets if it is bigger than 1, and no fields if it is exactly one. A 4d vector instead reduces just to a 2d $(0,2)$ vector. We then get the following 2d $\mathcal{N}=(0,2)$ Lagrangian that corresponds to the sphere compactification of the 4d $\mathcal{N}=1$ Lagrangian theory with $SU(2)_R$ twist:
\begin{table}[h]
  \centering
  \begin{tabular}{c|ccc}
    & $\Phi$ & $\alpha$ & $\beta$ \\\hline
    $SU(2)$              & $\mathbf{3}$ & $\mathbf{1}$ & $\mathbf{1}$ \\
    $U(1)_t$             & $-\tfrac15$ & $\tfrac{2}{5}$ & $-\tfrac{6}{5}$ \\
    $U(1)_{+}$ & $\tfrac15$ & $\tfrac{3}{5}$ & $\tfrac{6}{5}$
\end{tabular}
\end{table}

\noindent where $\Phi$ and $\beta$ are $(0,2)$ chirals, while $\alpha$ is a $(0,2)$ Fermi multiplet. In particular, the 4d fields $q_{1,2}$ do not survive the reduction.

We can use this 2d $(0,2)$ Lagrangian to compute the elliptic genus in the NSNS sector \cite{Gadde:2013wq}
\begin{align}
    \mathcal{I}_{C_{III}}=\frac{1}{2}\frac{(q)_\infty^2\theta\left(q^{\frac{4}{5}}t^{\frac{2}{5}}\right)}{\theta\left(q^{\frac{1}{10}}t^{-\frac{1}{5}}\right)\theta\left(q^{\frac{3}{5}}t^{-\frac{6}{5}}\right)}\oint_{\text{JK}}\frac{\mathrm{d}z}{2\pi i z}\frac{\theta\left(z^{\pm2}\right)}{\theta\left(q^{\frac{1}{10}}t^{-\frac{1}{5}}z^{\pm2}\right)}\,.
\end{align}
The evaluation of the JK residue is particularly simple since we have only two poles contributing
\begin{equation}
    z=q^{\frac{1}{20}}t^{-\frac{1}{10}}\mathrm{e}^{i\pi n}\,,\qquad n=0,1\,.
\end{equation}
The residue at these poles is equal and is given by (after some cancellations)
\begin{align}
    \mathcal{I}_{C_{III}}=\frac{\theta\left(q^{\frac{11}{10}}t^{-\frac{1}{5}}\right)}{\theta\left(q^{\frac{3}{5}}t^{-\frac{6}{5}}\right)}\,.
\end{align}
As expected, this result precisely coincides with the elliptic genus of a 2d $(2,2)$ twisted chiral of R-charge $\tfrac{6}{5}$ in \eqref{eq:EGCT} up to the identification $t=y$, which also agrees with the mapping between the 4d and 2d R-symmetry, see for instance \eqref{eq:2dRsymmLR}.

\subsection{Elliptic genus for genus $g>1$ surface with no punctures}
\label{subsec:EGgenusg}

The TQFT formula of the elliptic genus allows us to compute it for 2d class $\FF$ theories associated with arbitrary Riemann surfaces, even if we do not have a description of the low-energy theory in terms of a 2d Lagrangian theory. Here we shall use this to study the elliptic genus of the compatification on a genus $g>1$ surface without punctures, namely $\FF[(A_1,A_2),\Sigma_{g,0}]$. Our main motivation is that in this case the known BPS operator spectrum of $(A_1,A_2)$ leads to predictions regarding the terms appearing in the elliptic genus, as we explained in Section~\ref{subsec:multred}. We can then employ these predictions to test the validity of the TQFT formula. These predictions arise as 4d BPS operators can reduce to 2d BPS operators, and as such the known BPS spectrum of $(A_1,A_2)$ can be used to anticipate the BPS operators of the 2d theory (more correctly their contribution to the 2d superconformal index, that is the elliptic genus). We should mention that 2d BPS operators can also arise by other means, for instance from extended operators wrapping cycles of the compact surface. As such, there can be deviations from such predictions. Nevertheless, it is expected that these deviations occur only in special cases and that the predictions hold for sufficiently high genus. This is indeed observed in many cases of compactifications from 6d to 4d or 5d to 3d. 

Recall that we expect the expansion of the elliptic genus for the 2d $(2,2)$ theory obtained by reducing $(A_1,A_2)$ on a Riemann surface of genus $g$ and with no punctures to include the terms in \eqref{eq:EGA1A2g0n4prediction}, which we repeat here for convenience
\begin{equation}\label{eq:H04dpredictions}
    \mathcal{I}_{g,n=0} = \mathrm{PE}\left[(1-g)y^{-\frac{1}{5}}q^{\frac{1}{10}} + 3(g-1) y\sqrt{q} + (1-g)y^{-\frac{6}{5}}q^{\frac{3}{5}} +\cdots\right] \,.
\end{equation} 
We also recall that the first and the last term descend from the 4d Coulomb branch operator of dimension $\Delta=\tfrac{6}{5}$ of $(A_1,A_2)$, while the second term comes from the stress-energy tensor.\footnote{Derivatives of these operators should also contribute, so the entire expression in the $\mathrm{PE}$ should be multiplied by $(1-q)^{-1}$. However, here we shall expand the elliptic genus only up to order $q$, and so we suppresed the derivatives to simplify the expression.} The dots stand for additional operators that can arise for instance from other BPS operators of $(A_1,A_2)$. 

Next, we want to compare the above expectations against explicit calculations of the elliptic genus from the TQFT structure. Below we present the results for selected values of $g$. We begin with the case of $g=2$, where the elliptic genus to low orders reads
\begin{align}
 \mathcal{I}_{g=2,n=0} &= 1 - 2 y^{-\frac{1}{5}}q^{\frac{1}{10}} + y^{-\frac{3}{5}}q^{\frac{3}{10}} + y^{-\frac{4}{5}}q^{\frac{2}{5}} + 4 y\sqrt{q} + 2(y^{\frac{4}{5}}-y^{-\frac{6}{5}})q^{\frac{3}{5}} \\ \nonumber & - (y^{\frac{3}{5}}-y^{-\frac{7}{5}})q^{\frac{7}{10}} - 5 y^{\frac{2}{5}}q^{\frac{4}{5}} - 6 y^{\frac{1}{5}}q^{\frac{9}{10}} - 7 q + \cdots \,.
\end{align}
Up to order $q^{\frac{3}{5}}$, this can also be written as
\begin{equation}
 \mathcal{I}_{g=2,n=0} =\mathrm{PE}[- 2 y^{-\frac{1}{5}}q^{\frac{1}{10}} - y^{-\frac{2}{5}}q^{\frac{1}{5}} - y^{-\frac{3}{5}}q^{\frac{3}{10}} + 4 y\sqrt{q} + 10y^{\frac{4}{5}}q^{\frac{3}{5}} + \cdots ] \,,
\end{equation}
which completely disagrees with \eqref{eq:H04dpredictions}. However, the results approach the 4d prediction as the genus increases. Specifically, we note that the terms in the elliptic genus of the form $y^{-\frac{k}{5}}q^{\frac{k}{10}}$ appear to be consistent with \eqref{eq:H04dpredictions} if $k<g-1$. In particular, once $g\ge 3$ we indeed observe the term $3(g-1)y\sqrt{q}$. For example, for $g=4$ we get the result
\begin{align}\label{eq:EGg4}
    \mathcal{I}_{g=4,n=0}&=1-3q^{\frac{1}{10}}y^{-\frac{1}{5}}+3q^{\frac{1}{5}}y^{-\frac{2}{5}}-2q^{\frac{3}{10}}y^{-\frac{3}{5}}+9q^{\frac{1}{2}}y+3q^{\frac{1}{2}}y^{-1}-9q^{\frac{3}{5}}y^{\frac{4}{5}}+\cdots\,,
\end{align}
which can also be written as
\begin{equation}
    \mathcal{I}_{g=4,n=0}=\mathrm{PE}[- 3 y^{-\frac{1}{5}}q^{\frac{1}{10}} - y^{-\frac{3}{5}}q^{\frac{3}{10}} - 3 y^{-\frac{4}{5}}q^{\frac{2}{5}} + 9 y\sqrt{q} - 3\frac{\sqrt{q}}{y} + (18y^{\frac{4}{5}} - y^{-\frac{6}{5}})q^{\frac{3}{5}} + \cdots ] \,.
\end{equation}
This agrees with the first and second terms of \eqref{eq:H04dpredictions}, but contains additional terms of the form $y^{-\frac{k}{5}}q^{\frac{k}{10}}$, starting from $k=g-1=3$. Once we get to $g\ge 8$ the 2d result agrees with the 4d expectations, up to additional terms. For instance, for $g=8$ we find
\begin{align}
 \mathcal{I}_{g=8,n=0} &=1 - 7 y^{-\frac{1}{5}}q^{\frac{1}{10}} + 21 y^{-\frac{2}{5}}q^{\frac{1}{5}} - 35 y^{-\frac{3}{5}}q^{\frac{3}{10}} + 35 y^{-\frac{4}{5}}q^{\frac{2}{5}} + 21(y - y^{-1})\sqrt{q} \\ \nonumber & - 105 y^{\frac{4}{5}}q^{\frac{3}{5}}  + (217y^{\frac{3}{5}}+47y^{-\frac{7}{5}})q^{\frac{7}{10}} - (245 y^{\frac{2}{5}}+147 y^{-\frac{8}{5}})q^{\frac{4}{5}} \\ \nonumber & + (175 y^{\frac{1}{5}}+252 y^{-\frac{9}{5}})q^{\frac{9}{10}}   - (119 - 182 y^{-2} + 238 y^2) q + \cdots \,.
\end{align}
This can also be written as (again up to order $q^{\frac{3}{5}}$): 
\begin{equation}
    \mathcal{I}_{g=8,n=0}=\mathrm{PE}[- 7 y^{-\frac{1}{5}}q^{\frac{1}{10}} + 21 y\sqrt{q} - 7 y^{-\frac{6}{5}} q^{\frac{3}{5}} + 42 y^{\frac{4}{5}} q^{\frac{3}{5}} + \cdots ] \,,
\end{equation}
which is consistent with \eqref{eq:H04dpredictions}.

To summarize, we evaluated the elliptic genus up to $g=9$, and the results are consistent with the 4d predictions. Specifically, we observe deviations for sufficiently small genus, but the results approach the 4d predictions as the genus increases, and once we get to $g\ge 8$ we can indeed observe all three terms of \eqref{eq:H04dpredictions} in the elliptic genus, though we do observe additional terms. These do not appear to be accidental, since from the cases $g\le 9$ that we computed we observe that the behavior of the elliptic genus is asymptotic to
\begin{align}
    \mathcal{I}_{g,n=0}=\mathrm{PE}[- (g-1) y^{-\frac{1}{5}}q^{\frac{1}{10}} + 3(g-1) y\sqrt{q} - (g-1) y^{-\frac{6}{5}} q^{\frac{3}{5}} + 6(g-1) y^{\frac{4}{5}} q^{\frac{3}{5}} + \cdots ] \,,.
\end{align}
It would be interesting to understand if the additional term of $6(g-1) y^{\frac{4}{5}} q^{\frac{3}{5}}$ can also be explained from the 4d perspective, for instance in terms of additional BPS operators of $(A_1,A_2)$. Another interesting question is whether the deviations from the predictions at low genus can be better understood, especially as they appear to obey a pattern.

%%%%%%%%%%%%%%%%%%%%%%%%%%%%%%%%%
\section{Compactifications of $(A_1,A_4)$}
\label{sec:dimredA1A4}

In this section we consider the 4d $\mathcal{N}=2$ $(A_1,A_4)$ Argyres--Douglas SCFT. This is the next natural example after the $(A_1,A_2)$ theory that we have considered so far, since they both belong to the family of $(A_1,A_{2k})$ theories. Such theories all have a trivial Higgs branch, $k$ Coulomb branch operators and a VOA which is that of the $\mathcal{M}(2,2k+3)$ minimal model. In the particular case of $(A_1,A_4)$, the Coulomb branch operators have dimension $\Delta=\tfrac{8}{7}$ and $\Delta=\tfrac{10}{7}$, and the chiral algebra is that of the tricritical Lee--Yang model $\mathcal{M}(2,7)$.

We will first repeat the same analysis we did for $(A_1,A_2)$ and find GLSMs whose hemisphere partition functions match the Coulomb gas integrals for the conformal blocks of the four-point functions of the tricritical Lee--Yang model. Since this CFT possesses two non-trivial primary operators, there are several distinct four-point functions that we can consider and we will find the associated GLSMs for each of them. We will then interpret these as UV Lagrangians that flow in the IR to the 2d $(2,2)$ class $\FF$ theories arising from compactifying $(A_1,A_4)$ on a sphere with four-punctures of the type associated to the corresponding primaries in the CFT and with an $SU(2)_R$ twist. In cases in which we have both types of punctures present, considering distinct channel decompositions of the same CFT correlator will lead to appartently different GLSM. However, these are expected to actually describe the same theory in the IR since their B-type $S^2$ partition functions are both equal to the CFT correlator. Said differently, the crossing symmetry of the correlator or equivalently the different pair-of-pants decompositions of the same Riemann surface lead to a duality between the 2d $(2,2)$ theories. We will check that the central charges elliptic genera of the dual theories indeed match, which follows from a non-trivial identity of the associated JK integrals.

We will also perform the same tests that we did for the case of $(A_1,A_2)$, by verifying that the GLSMs reproduce the correct central charges predicted from 4d and that they can be used to derive the building blocks of the TQFT formula for the elliptic genus of a generic compactification of $(A_1,A_4)$ on an arbitrary Riemann surface. We will then test the validity of this TQFT formula by comparing the result for a sphere with no punctures with the one obtained by studying the compactification of the 4d $\mathcal{N}=1$ Lagrangian of $(A_1,A_4)$ \cite{Maruyoshi:2016aim} on a sphere with $SU(2)_R$ twist \cite{Gadde:2015wta}, and by verifying the presence of the operators predicted from 4d in the $q$-expansion of the elliptic genus for a surface of genus $g$ and with no punctures.

Possible generalizations consist of $(A_1,A_4)$ on a sphere with any number of punctures, as well as any Argyres--Douglas theory whose VOA is that of a non-unitary RCFT, like the $(A_1,A_{2k})$ series. Even if we will not discuss it explicitly here, the same analysis can be repeated in these cases, since Coulomb gas integral expressions for the conformal blocks are known \cite{DiFrancesco:1997nk} (see also Appendix \ref{app:MqpCG} for minimal models).

\subsection{Four-punctured spheres}
\label{subsec:A1A44pt}

The tricritical Lee--Yang model $\mathcal{M}(2,7)$ has three primary fields $I$, $\phi_{(2,1)}$, $\phi_{(3,1)}$ with dimensions repsectively 0, $-\tfrac{2}{7}$, $-\tfrac{3}{7}$. Consequently, there are many distinct four-point functions and we will consider each of them in turn.

\subsubsection*{Four $\phi_{(2,1)}$ punctures}

The four-point function of a general $\mathcal{M}(q,p)$ minimal model involving only the field $\phi_{(2,1)}$ takes the general form (see e.g.~\cite{Gerraty:2021})
\begin{align}
    \langle\phi_{(2,1)}(0)\phi_{(2,1)}(1)\phi_{(2,1)}(\infty)\phi_{(2,1)}(z,\bar{z})\rangle&=|z|^{2-3t}|1-z|^{2-3t}\Bigg|\dFu{1-t,2-3t;2-2t;z}\Bigg|^2\nn\\
    &+C_{\phi_{(2,1)}\phi_{(2,1)}\phi_{(3,1)}}^2|z|^t|1-z|^{2-3t}\Bigg|\dFu{t,1-t;2t;z}\Bigg|^2\,,
\end{align}
where we defined $t=\tfrac{q}{p}$ and we have the structure constant
\begin{align}
    C_{\phi_{(2,1)}\phi_{(2,1)}\phi_{(3,1)}}^2&=-\frac{\Gamma(2-2t)^2\Gamma(t)\Gamma(3t-1)}{\Gamma(2t)^2\Gamma(1-t)\Gamma(2-3t)}=\frac{\Gamma(2-2t)\Gamma(1-2t)\Gamma(t)\Gamma(3t-1)}{\Gamma(2t-1)\Gamma(2t)\Gamma(1-t)\Gamma(2-3t)}\,.
\end{align}
Note in particular that there are only two terms, the first one corresponding to the exchange of an intermediate $I$ operator in the conformal blocks decomposition and the second one corresponding to the exchange of $\phi_{(3,1)}$. There is no term corresponding to the exchange of $\phi_{(2,1)}$ since the structure constant $C_{\phi_{(2,1)}\phi_{(2,1)}\phi_{(2,1)}}$ vanishes unless $q=2$ and $p=5$, in which case $\phi_{(2,1)}=\phi_{(3,1)}$.

One can check that this correlator is reproduced (up to a prefactor) by the B-type $S^2$ partition function of the following GLSM of twisted fields:
\begin{table}[h]
  \centering
  \begin{tabular}{c|cccc}
    & $\Phi_1$ & $\Phi_2$ & $\tilde{\Phi}_1$  & $\tilde{\Phi}_2$ \\\hline
    $U(1)$ & 1 & 1 & $-1$ & $-1$ \\
    $U(1)_+$ & $\tfrac{t}{2}$ & $1-\tfrac{3}{2}t$ & $\tfrac{t}{2}$ & $1-\tfrac{3}{2}t$
  \end{tabular}
\end{table}

\noindent Notice that for any $q$, $p$ the GLSM consists always of a $U(1)$ gauge group with two chirals of charge $+1$ and two of charge $-1$. What changes is just the R-charges of the chirals, as well as the singlets that should be added.

Let us specialize this to the case of tricritical Lee--Yang for which $t=\tfrac{2}{7}$. We will also add a singlet field, which is needed to match the 4d expectations as we will see:
\begin{table}[H]
  \centering
  \begin{tabular}{c|ccccc}
    & $\Phi_1$ & $\Phi_2$ & $\tilde{\Phi}_1$ & $\tilde{\Phi}_2$ & $F$ \\\hline
    $U(1)$ & 1 & 1 & $-1$ & $-1$ & 0 \\
    $U(1)_+$ & $\tfrac{1}{7}$ & $\tfrac{4}{7}$ & $\tfrac{1}{7}$ & $\tfrac{4}{7}$ & $\tfrac{5}{7}$
  \end{tabular}
\end{table}

\noindent We claim that this GLSM describes the 2d $(2,2)$ class $\FF$ theory obtained by compactifying $(A_1,A_4)$ on a sphere with four punctures of type $\phi_{(2,1)}$ and $SU(2)_R$ twist. We stress once again that this model was constructed so to reproduce the $\langle\phi_{(2,1)}\phi_{(2,1)}\phi_{(2,1)}\phi_{(2,1)}\rangle$ correlation function of the tricritical Lee--Yang model, but we can check that it passes all the tests expected from its 4d origin. In particular, its 2d central charges match with the ones computed from \eqref{eq:4dc}
\begin{equation}
    c_{\pm}=2\left(1-2\times \frac{1}{7}\right)+2\left(1-2\times \frac{4}{7}\right)+\left(1-2\times \frac{5}{7}\right)-1=\frac{22}{7}+4\times\left(-\frac{6}{7}\right)=-\frac{2}{7}\,,
\end{equation}
where we used that the 4d central charges of $(A_1,A_4)$ are $a=\tfrac{67}{84}$ and $c=\tfrac{17}{21}$, and that the contribution of $\phi_{(2,1)}$ puncture is as in \eqref{eq:cpunctA1A4}. Once again, the presence of the singlet field $F$ is crucial for this matching. Moreover, the FI parameter $z$ is a chiral exactly marginal deformation coming from the single complex structure modulus of the sphere with four punctures and maps to the position of the 4th operator in the CFT correlator. Finally, since the GLSM consists only of twisted fields, its A-type $S^2$ partition function is trivial, in agreement with the fact that the $(A_1,A_4)$ theory has no exactly marginal deformation.

\subsubsection*{Three $\phi_{(2,1)}$ and one $\phi_{(3,1)}$ punctures}

In this case the correlation function $\langle\phi_{(2,1)}\phi_{(2,1)}\phi_{(2,1)}\phi_{(3,1)}\rangle$ is fully holomorphically factorized. This can be understood from the fact that in its conformal block decomposition we have only one term corresponding to the exchange on an intermediate $\phi_{(3,1)}$ operator. The exchange of $I$ and $\phi_{(2,1)}$ are indeed forbidden since the two-point function $\langle\phi_{(2,1)}\phi_{(3,1)}\rangle$ of course vanishes by conformal invariance and the structure constant $C_{\phi_{(2,1)}\phi_{(2,1)}\phi_{(2,1)}}$ vanishes as well in the tricritical Lee--Yang model. Another way to see that the correlator is fully holomorphically factorized is that the neutrality condition is satisfied without the insertion of any screening charge and so the Coulomb gas integral is trivial.

Consequently, we expect the 2d $(2,2)$ theory corresponding to this compactification to consist of just a bunch of chiral fields, without any gauge group. We propose the following field content:
\begin{table}[h]
  \centering
  \begin{tabular}{c|cc}
    & $F_1$ & $F_2$ \\\hline
    $U(1)_+$ & $\tfrac{1}{7}$ & $\tfrac{5}{7}$
  \end{tabular}
\end{table}

\noindent Indeed the central charges of these chiral fields match with those predicted from 4d
\begin{equation}
    c_{\pm}=\left(1-2\times \frac{1}{7}\right)+\left(1-2\times \frac{5}{7}\right)=\frac{22}{7}+3\times\left(-\frac{6}{7}\right)+1\times\left(-\frac{2}{7}\right)=\frac{2}{7}\,.
\end{equation}
Moreover, this proposal is consistent with the TQFT formula for the elliptic genus that we will study in the next subsection.

\subsubsection*{Two $\phi_{(2,1)}$ and two $\phi_{(3,1)}$ punctures}

This case exhibits the interesting feature that the corresponding correlation function in the CFT $\langle\phi_{(2,1)}\phi_{(2,1)}\phi_{(3,1)}\phi_{(3,1)}\rangle$ admits two distinct conformal blocks decompositions. As we will see momentarily, the GLSMs associated to each decomposition will look different, but they are expected to be IR dual as their B-type $S^2$ partition functions are equal to each other and to the CFT correlator. In other words, the crossing symmetry of the correlator or equivalently the different pair-of-pants decompositions of the same Riemann surface imply a duality of the GLSMs.

Let us first consider the channel corresponding to taking the OPE between $\phi_{(2,1)}$ and $\phi_{(3,1)}$ twice. The Coulomb gas integral for one of the conformal blocks is
\begin{equation}
    F^{(4)}(z)=z^{\frac{2}{7}}(z-1)^{\frac{6}{7}}\int_0^1\mathrm{d}u\,u^{-\frac{2}{7}}(1-u)^{-\frac{6}{7}}(1-z\,u)^{-\frac{4}{7}}\,.
\end{equation}
Applying \eqref{eq:IDCoulombGLSM} we find a representation of the conformal block in terms of a GLSM hemisphere partition function
\begin{align}
    \frac{F^{(4)}(z)}{z^{\frac{2}{7}}(z-1)^{\frac{6}{7}}}=\frac{\G{\frac{1}{7}}}{\G{\frac{4}{7}}}\int_{-i\infty}^{+i\infty}\frac{\mathrm{d}s}{2\pi i}(-z)^s\frac{\G{-s}\G{\frac{4}{7}+s}\G{\frac{5}{7}+s}}{\G{\frac{6}{7}+s}}\,.
\end{align}
With a suitable choice of singlets, the GLSM that we found is
\begin{table}[h!]
  \centering
  \begin{tabular}{c|ccccc}
    & $\Phi_1$ & $\Phi_2$ & $\tilde{\Phi}_1$ & $\tilde{\Phi}_2$ & $F$ \\\hline
    $U(1)$ & 1 & 1 & $-1$ & $-1$ & 0 \\
    $U(1)_+$ & $\tfrac{2}{7}$ & $\tfrac{3}{7}$ & $\tfrac{2}{7}$ & $\tfrac{3}{7}$ & $\tfrac{1}{7}$
  \end{tabular}
\end{table}

\noindent As usual, this reproduces the correct central charges
\begin{equation}
    c_{\pm}=2\left(1-2\times \frac{2}{7}\right)+2\left(1-2\times \frac{3}{7}\right)+\left(1-2\times \frac{1}{7}\right)-1=\frac{22}{7}+2\times\left(-\frac{6}{7}\right)+2\times\left(-\frac{2}{7}\right)=\frac{6}{7}\,,
\end{equation}
it has a single chiral exactly marginal deformation encoded in the FI parameter $z$ and it has a trivial A-type $S^2$ partition function.

Let us now consider the channel corresponding to taking the OPE between the two $\phi_{(2,1)}$ operators on the one hand and the two $\phi_{(3,1)}$ operators in the other. The Coulomb gas integral for one of the conformal blocks is
\begin{equation}
    F^{(4)}(z)=z^{\frac{2}{7}}(z-1)^{\frac{3}{7}}\int_0^1\mathrm{d}u\,u^{-\frac{2}{7}}(1-u)^{-\frac{6}{7}}(1-z\,u)^{-\frac{2}{7}}\,.
\end{equation}
Applying \eqref{eq:IDCoulombGLSM} we find a representation of the conformal block in terms of a GLSM hemisphere partition function
\begin{align}
    \frac{F^{(4)}(z)}{z^{\frac{2}{7}}(z-1)^{\frac{6}{7}}}=\frac{\G{\frac{1}{7}}}{\G{\frac{2}{7}}}\int_{-i\infty}^{+i\infty}\frac{\mathrm{d}s}{2\pi i}(-z)^s\frac{\G{-s}\G{\frac{2}{7}+s}\G{\frac{5}{7}+s}}{\G{\frac{6}{7}+s}}\,.
\end{align}
With a suitable choice of singlets, the GLSM that we found is
\begin{table}[h]
  \centering
  \begin{tabular}{c|ccccccc}
    & $\Phi_1$ & $\Phi_2$ & $\tilde{\Phi}_1$ & $\tilde{\Phi}_2$ & $F_1$ & $F_2$ & $F_3$ \\\hline
    $U(1)$ & 1 & 1 & $-1$ & $-1$ & 0 & 0 & 0 \\
    $U(1)_+$ & $\tfrac{1}{7}$ & $\tfrac{4}{7}$ & $\tfrac{1}{7}$ & $\tfrac{2}{7}$ & $\tfrac{1}{7}$ & $\tfrac{4}{7}$ & $\tfrac{5}{7}$
  \end{tabular}
\end{table}

\noindent The central charges of this GLSM coincide with those of the previous one
\begin{equation}
    c_{\pm}=2\left(1-2\times \frac{1}{7}\right)+2\times\left(1-2\times \frac{4}{7}\right)+\left(1-2\times \frac{2}{7}\right)+\left(1-2\times \frac{1}{7}\right)+\left(1-2\times \frac{5}{7}\right)-1=\frac{6}{7}\,,
\end{equation}
in agreement with our expectation that the two should be dual.

As another check of the duality, we can verify that the elliptic genera of the two GLSMs coincide. For the first one we have the following JK integral:
\begin{align}
    \mathcal{I}_1&=\frac{\theta\left(q^{\frac{4}{7}}y^{\frac{6}{7}}\right)}{\theta\left(q^{\frac{1}{14}}y^{-\frac{1}{7}}\right)}\frac{(q)_\infty^2}{\theta\left(q^{\frac{1}{2}}y\right)}\oint_{\text{JK}}\frac{\mathrm{d}z}{2\pi i z}\frac{\theta\left(q^{\frac{9}{14}}y^{\frac{5}{7}}z^{\pm1}\right)\theta\left(q^{\frac{5}{7}}y^{\frac{4}{7}}z^{\pm1}\right)}{\theta\left(q^{\frac{1}{7}}y^{-\frac{2}{7}}z^{\pm1}\right)\theta\left(q^{\frac{3}{14}}y^{-\frac{3}{7}}z^{\pm1}\right)}\,,
\end{align}
which after evaluation gives
\begin{equation}\label{eq:EG4ptM27b}
    \mathcal{I}_1=\frac{\theta\left(q^{\frac{4}{7}}y^{\frac{6}{7}}\right)^2\theta\left(q^{\frac{11}{14}}y^{\frac{3}{7}}\right)\theta\left(q^{\frac{6}{7}}y^{\frac{2}{7}}\right)}{\theta\left(q^{\frac{1}{14}}y^{-\frac{1}{7}}\right)^2\theta\left(q^{\frac{2}{7}}y^{-\frac{4}{7}}\right)\theta\left(q^{\frac{5}{14}}y^{-\frac{5}{7}}\right)}+\frac{\theta\left(q^{\frac{3}{7}}y^{\frac{8}{7}}\right)\theta\left(q^{\frac{6}{7}}y^{\frac{2}{7}}\right)}{\theta\left(q^{-\frac{1}{14}}y^{\frac{1}{7}}\right)\theta\left(q^{\frac{5}{14}}y^{-\frac{5}{7}}\right)}\,.
\end{equation}
For the second GLSM we have instead the following JK integral:
\begin{align}\label{eq:EG4ptM27c}
    \mathcal{I}_2&=\frac{\theta\left(q^{\frac{4}{7}}y^{\frac{6}{7}}\right)\theta\left(q^{\frac{11}{14}}y^{\frac{3}{7}}\right)\theta\left(q^{\frac{6}{7}}y^{\frac{2}{7}}\right)}{\theta\left(q^{\frac{1}{14}}y^{-\frac{1}{7}}\right)\theta\left(q^{\frac{2}{7}}y^{-\frac{4}{7}}\right)\theta\left(q^{\frac{5}{14}}y^{-\frac{5}{7}}\right)}\nn\\
    &\times\frac{(q)_\infty^2}{\theta\left(q^{\frac{1}{2}}y\right)}\oint_{\text{JK}}\frac{\mathrm{d}z}{2\pi i z}\frac{\theta\left(q^{\frac{4}{7}}y^{\frac{6}{7}}z^{\pm1}\right)\theta\left(q^{\frac{11}{14}}y^{\frac{3}{7}}z\right)\theta\left(q^{\frac{9}{14}}y^{\frac{5}{7}}z^{-1}\right)}{\theta\left(q^{\frac{1}{14}}y^{-\frac{1}{7}}z^{\pm1}\right)\theta\left(q^{\frac{2}{7}}y^{-\frac{4}{7}}z\right)\theta\left(q^{\frac{1}{7}}y^{-\frac{2}{7}}z^{-1}\right)}\,,
\end{align}
which after evaluation gives\footnote{In this case the chirals of gauge charge $+1$ do not have the same R-charges of those with gauge charge $-1$. Consequently, the poles and their residues coming from the former are different from those coming from the latter, which leads to apparently different expressions that however turn out to coincide, as it can be checked by expanding the result in a power series in $q$. Here we are showing the result obtained by considering the poles from the positively charged chirals.}
\begin{align}
    \mathcal{I}_2=&\frac{\theta\left(q^{\frac{4}{7}}y^{\frac{6}{7}}\right)\theta\left(q^{\frac{5}{7}}y^{\frac{4}{7}}\right)}{\theta\left(q^{\frac{1}{14}}y^{-\frac{1}{7}}\right)\theta\left(q^{\frac{3}{14}}y^{-\frac{3}{7}}\right)}+\frac{\theta\left(q^{\frac{2}{7}}y^{\frac{10}{7}}\right)\theta\left(q^{\frac{11}{14}}y^{\frac{3}{7}}\right)\theta\left(q^{\frac{6}{7}}y^{\frac{2}{7}}\right)^2}{\theta\left(q^{-\frac{3}{14}}y^{\frac{3}{7}}\right)\theta\left(q^{\frac{2}{7}}y^{-\frac{4}{7}}\right)\theta\left(q^{\frac{5}{14}}y^{-\frac{5}{7}}\right)^2}\,.
\end{align}
We have checked in a power series in $q$ that these two elliptic genera coincide
\begin{equation}
    \mathcal{I}_1=\mathcal{I}_2\,,
\end{equation}
which is a non-trivial check of the duality.

We also note that the FI parameters of the two GLSMs are mapped non-trivially under the duality. Such a map is understood in the CFT as the crossing symmetry transformation that relates the two channel decompositions of the correlator
\begin{equation}
    z\to\frac{z}{z-1}\,.
\end{equation}

\subsubsection*{One $\phi_{(2,1)}$ and three $\phi_{(3,1)}$ punctures}

In this case the correlation function $\langle\phi_{(2,1)}\phi_{(3,1)}\phi_{(3,1)}\phi_{(3,1)}\rangle$ of the tricritical Lee--Yang model exhibits the same type of OPE's in every channel. Hence, there is only one GLSM describing this compactification. The Coulomb gas integral for one of the conformal blocks is
\begin{equation}
    F^{(4)}(z)=z^{\frac{2}{7}}(z-1)^{\frac{2}{7}}\int_0^1\mathrm{d}u\,u^{-\frac{4}{7}}(1-u)^{-\frac{4}{7}}(1-z\,u)^{-\frac{2}{7}}\,.
\end{equation}
Applying \eqref{eq:IDCoulombGLSM} we find a representation of the conformal block in terms of a GLSM hemisphere partition function
\begin{align}
    \frac{F^{(4)}(z)}{z^{\frac{2}{7}}(z-1)^{\frac{6}{7}}}=\frac{\G{\frac{3}{7}}}{\G{\frac{2}{7}}}\int_{-i\infty}^{+i\infty}\frac{\mathrm{d}s}{2\pi i}(-z)^s\frac{\G{-s}\G{\frac{2}{7}+s}\G{\frac{3}{7}+s}}{\G{\frac{6}{7}+s}}\,.
\end{align}
With a suitable choice of singlets, the GLSM that we found is
\begin{table}[h!]
  \centering
  \begin{tabular}{c|ccccccc}
    & $\Phi_1$ & $\Phi_2$ & $\tilde{\Phi}_1$ & $\tilde{\Phi}_2$ & $F_1$ & $F_2$ & $F_3$ \\\hline
    $U(1)$ & 1 & 1 & $-1$ & $-1$ & 0 & 0 & 0 \\
    $U(1)_+$ & $\tfrac{1}{7}$ & $\tfrac{2}{7}$ & $\tfrac{1}{7}$ & $\tfrac{2}{7}$ & $\tfrac{1}{7}$ & $\tfrac{4}{7}$ & $\tfrac{5}{7}$
  \end{tabular}
\end{table}

\noindent As usual, this reproduces the correct central charges
\begin{align}
    c_{\pm}&=3\left(1-2\times \frac{1}{7}\right)+2\left(1-2\times \frac{2}{7}\right)+\left(1-2\times \frac{1}{7}\right)+\left(1-2\times \frac{4}{7}\right)+\left(1-2\times \frac{5}{7}\right)-1\nn\\
    &=\frac{22}{7}+1\times\left(-\frac{6}{7}\right)+3\times\left(-\frac{2}{7}\right)=\frac{10}{7}\,,
\end{align}
it has a single chiral exactly marginal deformation encoded in the FI parameter $z$ and it has a trivial A-type $S^2$ partition function.

\subsubsection*{Four $\phi_{(3,1)}$ punctures}

Again all channels are equivalent since only the same type of primary operator is involved. This time Coulomb gas integral is more involved since it is two-dimensional
\begin{equation}
    F^{(4)}(z)=z^{\frac{4}{7}}(z-1)^{\frac{6}{7}}\oint\prod_{a=1}^2\mathrm{d}x_a (x_1-x_2)^{\frac{4}{7}}\prod_{a=1}^2x_a^{-\frac{4}{7}}(x-1)^{-\frac{6}{7}}(x-z)^{-\frac{4}{7}}\,,
\end{equation}
where we have not specified the integration contour but as usual different choices correspond to different conformal blocks. Hence, in this case the identity \eqref{eq:IDCoulombGLSM} is not enough. However, we can follow the same strategy used in Section~\ref{subsec:higherptLY} for studying higher-point functions of Lee--Yang. 

The Coulomb gas integral can be obtained as the $q\to1$ limit of the holomorphic block of a 3d $\mathcal{N}=2$ $U(2)$ SQCD theory with one adjoint chiral $A$ of R-charge $\tfrac{4}{7}$, and two flavors $Q$, $\tilde{Q}$ and $P$, $\tilde{P}$ of R-charges $\tfrac{3}{7}$ and $\tfrac{1}{7}$ respectively. The perturbative part of the superpotential of this theory is
\begin{equation}
    \mathcal{W}\supset A^2Q\tilde{Q}+A^3P\tilde{P}\,.
\end{equation}
Such a theory can be obtained from the 3d $\mathcal{N}=4$ $U(2)$ SQCD with 5 flavors after a nilpotent mass deformation labelled by the partition $[3,2]$. We can then study the effect of this deformation in the mirror dual quiver of the 3d $\mathcal{N}=4$ SQCD, which corresponds to a monopole superpotential deformation that can be studied by using the monopole dualities of \cite{Benini:2017dud} several times on various gauge nodes of the dual quiver. Without presenting all the details, we obtain the following non-abelian GLSM:
\begin{table}[h!]
  \centering
  \begin{tabular}{c|cccccccc}
    & $A$ & $\Phi_1$ & $\Phi_2$ & $\tilde{\Phi}_1$ & $\tilde{\Phi}_2$ & $F_1$ & $F_2$ & $F_{3,4}$ \\\hline
    $U(2)$ & $\bf 4$ & $\bf 2$ & $\bf 2$ & $\overline{\bf 2}$ & $\overline{\bf 2}$ & 0 & 0 & 0 \\
    $U(1)_+$ & $\tfrac{2}{7}$ & $\tfrac{2}{7}$ & $\tfrac{1}{7}$ & $\tfrac{2}{7}$ & $\tfrac{1}{7}$ & $\tfrac{1}{7}$ & $\tfrac{4}{7}$ & $\tfrac{5}{7}$
  \end{tabular}
\end{table}

\noindent As usual, this reproduces the correct central charges
\begin{align}
    c_{\pm}&=8\left(1-2\times \frac{2}{7}\right)+5\left(1-2\times \frac{1}{7}\right)+\left(1-2\times \frac{4}{7}\right)+2\times\left(1-2\times \frac{5}{7}\right)-4\nn\\
    &=\frac{22}{7}+4\times\left(-\frac{2}{7}\right)=2\,,
\end{align}
it has a single chiral exactly marginal deformation encoded in the FI parameter $z$ and it has a trivial A-type $S^2$ partition function.

\subsection{Elliptic genus and TQFT structure}
\label{subsec:A1A4EG}

The TQFT structure of the elliptic genus that we observed in Section \ref{sec:EGTQFT} for the compactifications of $(A_1,A_2)$ turns out to be valid also for $(A_1,A_4)$. In this case we have more building blocks as a consequence of the fact that there is one more type of puncture, however the rules for combining them and the way they are derived are exactly as for $(A_1,A_2)$. 

The elliptic genus contribution of the trinions are the same as those of some twisted chiral fields with the following R-charges:
\begin{align}\label{eq:bluildblocksA1A4}
&   C_{I I I}=\left\{\frac{8}{7},\frac{10}{7}\right\}\,,\qquad C_{\phi_{(2,1)}\phi_{(2,1)} I}=\left\{\frac{5}{7},\frac{8}{7}\right\}\,,\qquad C_{\phi_{(3,1)}\phi_{(3,1)} I}=\left\{\frac{4}{7},\frac{5}{7}\right\}\,, \\ \nonumber 
& C_{\phi_{(3,1)}\phi_{(3,1)}\phi_{(3,1)}}=\left\{\frac{1}{7}\right\}\,,\qquad   C_{\phi_{(2,1)}\phi_{(2,1)}\phi_{(3,1)}}=\left\{\frac{5}{7}\right\} \,,\qquad C_{\phi_{(2,1)}\phi_{(3,1)}\phi_{(3,1)}}=\left\{\frac{1}{7},\frac{4}{7},\frac{5}{7}\right\}\,,
\end{align}
where all the trinions not listed here, such as $C_{\phi_{(2,1)}\phi_{(2,1)}\phi_{(2,1)}}$, have a vanishing elliptic genus. These trinions are glued together by identifying identical punctures with the following tubes:
\begin{equation}
    T_{II}=\left\{-\frac{1}{7},-\frac{3}{7}\right\}\,,\qquad T_{\phi_{(2,1)}\phi_{(2,1)}}=\left\{-\frac{1}{7},\frac{2}{7}\right\}\,,\qquad T_{\phi_{(3,1)}\phi_{(3,1)}}=\left\{\frac{2}{7},\frac{3}{7}\right\}\,.
\end{equation}
Moreover, when gluing two punctures one has to sum over all their possible types.

As a check for these building blocks, one can compute their central charges and verify that they match with those predicted from 4d. The tubes and the corresponding trinions in the first line of \eqref{eq:bluildblocksA1A4} are deduced from the analysis of the boundary conditions at the punctures of Section~\ref{subsec:multred}. The remaining trinions are instead determined by computing explicitly the elliptic genera of the GLSMs for the four-punctured spheres that we derived in the previous section. 

\begin{figure}[t]
\center
\resizebox{0.9\textwidth}{!}{\begin{tikzpicture}[
    line width=1.4pt,
    line cap=round,
    line join=round,
    every node/.style={font=\Large}
]

% --- Left Diagram: Sphere with 4 holes ---
\begin{scope}[shift={(-7,0)}]
    % Main circle
    \draw (0,0) circle (2.8cm);
    
    % Holes and Labels
    % Top Left
    \draw (-1.2, 1.2) circle (0.25cm);
    \node[penred, left=0.2cm] at (-1.1, 0.9) {$\phi_{(2,1)}$};
    
    % Bottom Left
    \draw (-1.2, -1.2) circle (0.25cm);
    \node[penred, left=0.2cm] at (-1.1, -0.9) {$\phi_{(3,1)}$};
    
    % Top Right
    \draw (1.2, 1.2) circle (0.25cm);
    \node[penred, right=0.2cm] at (1.2, 0.9) {$\phi_{(2,1)}$};
    
    % Bottom Right
    \draw (1.2, -1.2) circle (0.25cm);
    \node[penred, right=0.2cm] at (1.2, -0.9) {$\phi_{(3,1)}$};
\end{scope}

% --- Equals Sign ---
\node at (-2.5, 0) {\Huge $=$};

% --- Right Side ---

% --- Top Row ---
\begin{scope}[shift={(0, 2.2)}]
    % Left Pants
    \begin{scope}[shift={(0,0)}]
        \coordinate (TL) at (0, 1.2);
        \coordinate (BL) at (0, -1.2);
        \coordinate (R) at (2.5, 0);
        
        % Body
        \draw (0, 1.45) to[out=0, in=180] (2.5, 0.25);
        \draw (0, -1.45) to[out=0, in=180] (2.5, -0.25);
        \draw (0, 0.95) to[out=-45, in=45, looseness=1.2] (0, -0.95);
        
        % Holes
        \draw[fill=white] (TL) circle (0.25);
        \draw[fill=white] (BL) circle (0.25);
        \draw[fill=white] (R) circle (0.25);
        
        % Labels
        \node[penred, left=0.15cm] at (TL) {$\phi_{(2,1)}$};
        \node[penred, left=0.15cm] at (BL) {$\phi_{(3,1)}$};
        \node[penred, above=0.2cm] at (R) {$\phi_{(3,1)}$};
    \end{scope}

    % Tube
    \begin{scope}[shift={(3.5,0)}]
        \coordinate (L) at (0, 0);
        \coordinate (R) at (2.5, 0);
        
        \draw (0, 0.25) -- (2.5, 0.25);
        \draw (0, -0.25) -- (2.5, -0.25);
        
        \draw[fill=white] (L) circle (0.25);
        \draw[fill=white] (R) circle (0.25);
        
        \node[penblue, below=0.2cm] at (L) {$\phi_{(3,1)}$};
        \node[penblue, below=0.2cm] at (R) {$\phi_{(3,1)}$};
    \end{scope}

    % Right Pants
    \begin{scope}[shift={(7,0)}]
        \coordinate (L) at (0, 0);
        \coordinate (TR) at (2.5, 1.2);
        \coordinate (BR) at (2.5, -1.2);
        
        % Body
        \draw (0, 0.25) to[out=0, in=180] (2.5, 1.45);
        \draw (0, -0.25) to[out=0, in=180] (2.5, -1.45);
        \draw (2.5, 0.95) to[out=225, in=135, looseness=1.2] (2.5, -0.95);
        
        % Holes
        \draw[fill=white] (L) circle (0.25);
        \draw[fill=white] (TR) circle (0.25);
        \draw[fill=white] (BR) circle (0.25);
        
        % Labels
        \node[penred, above=0.2cm] at (L) {$\phi_{(3,1)}$};
        \node[penred, right=0.15cm] at (TR) {$\phi_{(2,1)}$};
        \node[penred, right=0.15cm] at (BR) {$\phi_{(3,1)}$};
    \end{scope}
\end{scope}

% --- Plus Sign ---
\node at (4.75, 0) {\Huge $+$};

% --- Bottom Row ---
\begin{scope}[shift={(0, -2.2)}]
    % Left Pants
    \begin{scope}[shift={(0,0)}]
        \coordinate (TL) at (0, 1.2);
        \coordinate (BL) at (0, -1.2);
        \coordinate (R) at (2.5, 0);
        
        \draw (0, 1.45) to[out=0, in=180] (2.5, 0.25);
        \draw (0, -1.45) to[out=0, in=180] (2.5, -0.25);
        \draw (0, 0.95) to[out=-45, in=45, looseness=1.2] (0, -0.95);
        
        \draw[fill=white] (TL) circle (0.25);
        \draw[fill=white] (BL) circle (0.25);
        \draw[fill=white] (R) circle (0.25);
        
        \node[penred, left=0.15cm] at (TL) {$\phi_{(2,1)}$};
        \node[penred, left=0.15cm] at (BL) {$\phi_{(3,1)}$};
        \node[penred, above=0.2cm] at (R) {$\phi_{(2,1)}$};
    \end{scope}

    % Tube
    \begin{scope}[shift={(3.5,0)}]
        \coordinate (L) at (0, 0);
        \coordinate (R) at (2.5, 0);
        
        \draw (0, 0.25) -- (2.5, 0.25);
        \draw (0, -0.25) -- (2.5, -0.25);
        
        \draw[fill=white] (L) circle (0.25);
        \draw[fill=white] (R) circle (0.25);
        
        \node[penblue, below=0.2cm] at (L) {$\phi_{(2,1)}$};
        \node[penblue, below=0.2cm] at (R) {$\phi_{(2,1)}$};
    \end{scope}

    % Right Pants
    \begin{scope}[shift={(7,0)}]
        \coordinate (L) at (0, 0);
        \coordinate (TR) at (2.5, 1.2);
        \coordinate (BR) at (2.5, -1.2);
        
        \draw (0, 0.25) to[out=0, in=180] (2.5, 1.45);
        \draw (0, -0.25) to[out=0, in=180] (2.5, -1.45);
        \draw (2.5, 0.95) to[out=225, in=135, looseness=1.2] (2.5, -0.95);
        
        \draw[fill=white] (L) circle (0.25);
        \draw[fill=white] (TR) circle (0.25);
        \draw[fill=white] (BR) circle (0.25);
        
        \node[penred, above=0.2cm] at (L) {$\phi_{(2,1)}$};
        \node[penred, right=0.15cm] at (TR) {$\phi_{(2,1)}$};
        \node[penred, right=0.15cm] at (BR) {$\phi_{(3,1)}$};
    \end{scope}
\end{scope}

\end{tikzpicture}}

\vspace{1cm}

\resizebox{0.9\textwidth}{!}{\begin{tikzpicture}[
    line width=1.4pt,
    line cap=round,
    line join=round,
    every node/.style={font=\Large}
]

% --- Left Diagram: Sphere with 4 holes ---
\begin{scope}[shift={(-7,0)}]
    % Main circle
    \draw (0,0) circle (2.8cm);
    
    % Holes and Labels
    % Top Left
    \draw (-1.2, 1.2) circle (0.25cm);
    \node[penred, left=0.2cm] at (-1.1, 0.9) {$\phi_{(3,1)}$};
    
    % Bottom Left
    \draw (-1.2, -1.2) circle (0.25cm);
    \node[penred, left=0.2cm] at (-1.1, -0.9) {$\phi_{(3,1)}$};
    
    % Top Right
    \draw (1.2, 1.2) circle (0.25cm);
    \node[penred, right=0.2cm] at (1.2, 0.9) {$\phi_{(2,1)}$};
    
    % Bottom Right
    \draw (1.2, -1.2) circle (0.25cm);
    \node[penred, right=0.2cm] at (1.2, -0.9) {$\phi_{(2,1)}$};
\end{scope}

% --- Equals Sign ---
\node at (-2.5, 0) {\Huge $=$};

% --- Right Side ---

% --- Top Row ---
\begin{scope}[shift={(0, 2.2)}]
    % Left Pants
    \begin{scope}[shift={(0,0)}]
        \coordinate (TL) at (0, 1.2);
        \coordinate (BL) at (0, -1.2);
        \coordinate (R) at (2.5, 0);
        
        % Body
        \draw (0, 1.45) to[out=0, in=180] (2.5, 0.25);
        \draw (0, -1.45) to[out=0, in=180] (2.5, -0.25);
        \draw (0, 0.95) to[out=-45, in=45, looseness=1.2] (0, -0.95);
        
        % Holes
        \draw[fill=white] (TL) circle (0.25);
        \draw[fill=white] (BL) circle (0.25);
        \draw[fill=white] (R) circle (0.25);
        
        % Labels
        \node[penred, left=0.15cm] at (TL) {$\phi_{(3,1)}$};
        \node[penred, left=0.15cm] at (BL) {$\phi_{(3,1)}$};
        \node[penred, above=0.2cm] at (R) {$\phi_{(3,1)}$};
    \end{scope}

    % Tube
    \begin{scope}[shift={(3.5,0)}]
        \coordinate (L) at (0, 0);
        \coordinate (R) at (2.5, 0);
        
        \draw (0, 0.25) -- (2.5, 0.25);
        \draw (0, -0.25) -- (2.5, -0.25);
        
        \draw[fill=white] (L) circle (0.25);
        \draw[fill=white] (R) circle (0.25);
        
        \node[penblue, below=0.2cm] at (L) {$\phi_{(3,1)}$};
        \node[penblue, below=0.2cm] at (R) {$\phi_{(3,1)}$};
    \end{scope}

    % Right Pants
    \begin{scope}[shift={(7,0)}]
        \coordinate (L) at (0, 0);
        \coordinate (TR) at (2.5, 1.2);
        \coordinate (BR) at (2.5, -1.2);
        
        % Body
        \draw (0, 0.25) to[out=0, in=180] (2.5, 1.45);
        \draw (0, -0.25) to[out=0, in=180] (2.5, -1.45);
        \draw (2.5, 0.95) to[out=225, in=135, looseness=1.2] (2.5, -0.95);
        
        % Holes
        \draw[fill=white] (L) circle (0.25);
        \draw[fill=white] (TR) circle (0.25);
        \draw[fill=white] (BR) circle (0.25);
        
        % Labels
        \node[penred, above=0.2cm] at (L) {$\phi_{(3,1)}$};
        \node[penred, right=0.15cm] at (TR) {$\phi_{(2,1)}$};
        \node[penred, right=0.15cm] at (BR) {$\phi_{(2,1)}$};
    \end{scope}
\end{scope}

% --- Plus Sign ---
\node at (4.75, 0) {\Huge $+$};

% --- Bottom Row ---
\begin{scope}[shift={(0, -2.2)}]
    % Left Pants
    \begin{scope}[shift={(0,0)}]
        \coordinate (TL) at (0, 1.2);
        \coordinate (BL) at (0, -1.2);
        \coordinate (R) at (2.5, 0);
        
        \draw (0, 1.45) to[out=0, in=180] (2.5, 0.25);
        \draw (0, -1.45) to[out=0, in=180] (2.5, -0.25);
        \draw (0, 0.95) to[out=-45, in=45, looseness=1.2] (0, -0.95);
        
        \draw[fill=white] (TL) circle (0.25);
        \draw[fill=white] (BL) circle (0.25);
        \draw[fill=white] (R) circle (0.25);
        
        \node[penred, left=0.15cm] at (TL) {$\phi_{(3,1)}$};
        \node[penred, left=0.15cm] at (BL) {$\phi_{(3,1)}$};
        \node[penred, above=0.2cm] at (R) {$I$};
    \end{scope}

    % Tube
    \begin{scope}[shift={(3.5,0)}]
        \coordinate (L) at (0, 0);
        \coordinate (R) at (2.5, 0);
        
        \draw (0, 0.25) -- (2.5, 0.25);
        \draw (0, -0.25) -- (2.5, -0.25);
        
        \draw[fill=white] (L) circle (0.25);
        \draw[fill=white] (R) circle (0.25);
        
        \node[penblue, below=0.2cm] at (L) {$I$};
        \node[penblue, below=0.2cm] at (R) {$I$};
    \end{scope}

    % Right Pants
    \begin{scope}[shift={(7,0)}]
        \coordinate (L) at (0, 0);
        \coordinate (TR) at (2.5, 1.2);
        \coordinate (BR) at (2.5, -1.2);
        
        \draw (0, 0.25) to[out=0, in=180] (2.5, 1.45);
        \draw (0, -0.25) to[out=0, in=180] (2.5, -1.45);
        \draw (2.5, 0.95) to[out=225, in=135, looseness=1.2] (2.5, -0.95);
        
        \draw[fill=white] (L) circle (0.25);
        \draw[fill=white] (TR) circle (0.25);
        \draw[fill=white] (BR) circle (0.25);
        
        \node[penred, above=0.2cm] at (L) {$I$};
        \node[penred, right=0.15cm] at (TR) {$\phi_{(2,1)}$};
        \node[penred, right=0.15cm] at (BR) {$\phi_{(2,1)}$};
    \end{scope}
\end{scope}

\end{tikzpicture}}
\caption{TQFT structure of the elliptic genus of the 2d $(2,2)$ theory obtained compactifying $(A_1,A_4)$ on a sphere with two punctures of type $\phi_{(2,1)}$ and two of type $\phi_{(3,1)}$ for the two possible inequivalent pair-of-pats decompositions.}
\label{fig:TQFT4ptM27b}
\end{figure}
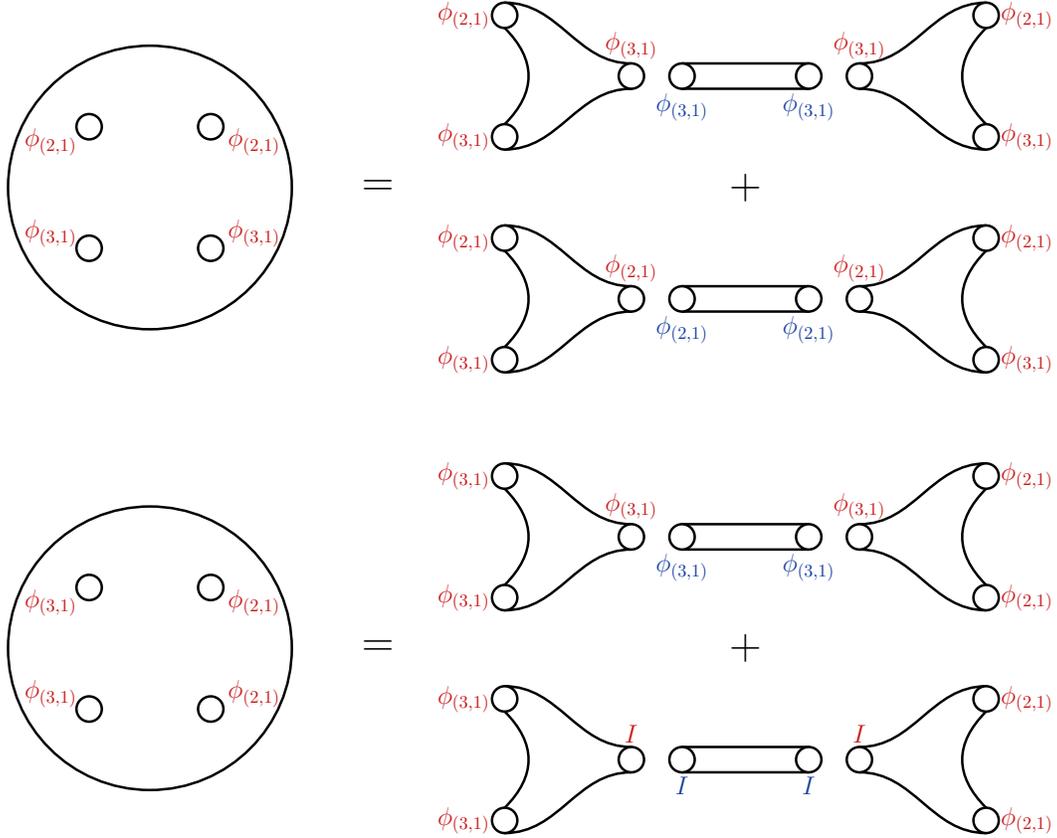

Let us consider for example the two GLSMs arising from the compactification of $(A_1,A_4)$ on a sphere with two punctures of type $\phi_{(2,1)}$ and two of type $\phi_{(3,1)}$, for which we have already computed the elliptic genera. Recall that these corresponded to two different pair-of-pant decompositions. Accordingly, the elliptic genera are expected to be computed by the diagrams depicted in Figure \ref{fig:TQFT4ptM27b}. The absence of some terms is due to the fact that some of the trinions are trivial. One can easily verify that the results coincide with \eqref{eq:EG4ptM27b} and \eqref{eq:EG4ptM27c}.

As another example, we can consider the sphere with three $\phi_{(2,1)}$ punctures and one $\phi_{(3,1)}$ puncture. Remember that the associated model actually has no gauge group and consists only of two twisted chiral fields. Accordingly, in the TQFT expansion of the elliptic genus we have only one term corresponding to the gluing of a $\phi_{(3,1)}$ puncture, since all the other terms involve trivially contributing trinions. The elliptic genus predicted from the TQFT formula then coincides with that of two twisted chirals of R-charges $\tfrac{1}{7}$ and $\tfrac{5}{7}$, in agreement with our findings from the previous section.

\begin{figure}[t]
\center
\resizebox{0.9\textwidth}{!}{\begin{tikzpicture}[
    line width=1.4pt,
    line cap=round,
    line join=round,
    every node/.style={font=\Large}
]

% --- Left Diagram: Sphere with 4 holes ---
\begin{scope}[shift={(-7,0)}]
    % Main circle
    \draw (0,0) circle (2.8cm);
    
    % Holes and Labels
    % Top Left
    \draw (-1.2, 1.2) circle (0.25cm);
    \node[penred, left=0.2cm] at (-1.1, 0.9) {$\phi_{(3,1)}$};
    
    % Bottom Left
    \draw (-1.2, -1.2) circle (0.25cm);
    \node[penred, left=0.2cm] at (-1.1, -0.9) {$\phi_{(3,1)}$};
    
    % Top Right
    \draw (1.2, 1.2) circle (0.25cm);
    \node[penred, right=0.2cm] at (1.2, 0.9) {$\phi_{(3,1)}$};
    
    % Bottom Right
    \draw (1.2, -1.2) circle (0.25cm);
    \node[penred, right=0.2cm] at (1.2, -0.9) {$\phi_{(3,1)}$};
\end{scope}

% --- Equals Sign ---
\node at (-2.5, 0) {\Huge $=$};

% --- Right Side ---

% --- Top Row ---
\begin{scope}[shift={(0, 4.4)}]
    % Left Pants
    \begin{scope}[shift={(0,0)}]
        \coordinate (TL) at (0, 1.2);
        \coordinate (BL) at (0, -1.2);
        \coordinate (R) at (2.5, 0);
        
        % Body
        \draw (0, 1.45) to[out=0, in=180] (2.5, 0.25);
        \draw (0, -1.45) to[out=0, in=180] (2.5, -0.25);
        \draw (0, 0.95) to[out=-45, in=45, looseness=1.2] (0, -0.95);
        
        % Holes
        \draw[fill=white] (TL) circle (0.25);
        \draw[fill=white] (BL) circle (0.25);
        \draw[fill=white] (R) circle (0.25);
        
        % Labels
        \node[penred, left=0.15cm] at (TL) {$\phi_{(3,1)}$};
        \node[penred, left=0.15cm] at (BL) {$\phi_{(3,1)}$};
        \node[penred, above=0.2cm] at (R) {$\phi_{(3,1)}$};
    \end{scope}

    % Tube
    \begin{scope}[shift={(3.5,0)}]
        \coordinate (L) at (0, 0);
        \coordinate (R) at (2.5, 0);
        
        \draw (0, 0.25) -- (2.5, 0.25);
        \draw (0, -0.25) -- (2.5, -0.25);
        
        \draw[fill=white] (L) circle (0.25);
        \draw[fill=white] (R) circle (0.25);
        
        \node[penblue, below=0.2cm] at (L) {$\phi_{(3,1)}$};
        \node[penblue, below=0.2cm] at (R) {$\phi_{(3,1)}$};
    \end{scope}

    % Right Pants
    \begin{scope}[shift={(7,0)}]
        \coordinate (L) at (0, 0);
        \coordinate (TR) at (2.5, 1.2);
        \coordinate (BR) at (2.5, -1.2);
        
        % Body
        \draw (0, 0.25) to[out=0, in=180] (2.5, 1.45);
        \draw (0, -0.25) to[out=0, in=180] (2.5, -1.45);
        \draw (2.5, 0.95) to[out=225, in=135, looseness=1.2] (2.5, -0.95);
        
        % Holes
        \draw[fill=white] (L) circle (0.25);
        \draw[fill=white] (TR) circle (0.25);
        \draw[fill=white] (BR) circle (0.25);
        
        % Labels
        \node[penred, above=0.2cm] at (L) {$\phi_{(3,1)}$};
        \node[penred, right=0.15cm] at (TR) {$\phi_{(3,1)}$};
        \node[penred, right=0.15cm] at (BR) {$\phi_{(3,1)}$};
    \end{scope}
\end{scope}

% --- Plus Sign ---
\node at (4.75, 2.2) {\Huge $+$};

% --- Mid Row ---
\begin{scope}[shift={(0, 0)}]
    % Left Pants
    \begin{scope}[shift={(0,0)}]
        \coordinate (TL) at (0, 1.2);
        \coordinate (BL) at (0, -1.2);
        \coordinate (R) at (2.5, 0);
        
        \draw (0, 1.45) to[out=0, in=180] (2.5, 0.25);
        \draw (0, -1.45) to[out=0, in=180] (2.5, -0.25);
        \draw (0, 0.95) to[out=-45, in=45, looseness=1.2] (0, -0.95);
        
        \draw[fill=white] (TL) circle (0.25);
        \draw[fill=white] (BL) circle (0.25);
        \draw[fill=white] (R) circle (0.25);
        
        \node[penred, left=0.15cm] at (TL) {$\phi_{(3,1)}$};
        \node[penred, left=0.15cm] at (BL) {$\phi_{(3,1)}$};
        \node[penred, above=0.2cm] at (R) {$\phi_{(2,1)}$};
    \end{scope}

    % Tube
    \begin{scope}[shift={(3.5,0)}]
        \coordinate (L) at (0, 0);
        \coordinate (R) at (2.5, 0);
        
        \draw (0, 0.25) -- (2.5, 0.25);
        \draw (0, -0.25) -- (2.5, -0.25);
        
        \draw[fill=white] (L) circle (0.25);
        \draw[fill=white] (R) circle (0.25);
        
        \node[penblue, below=0.2cm] at (L) {$\phi_{(2,1)}$};
        \node[penblue, below=0.2cm] at (R) {$\phi_{(2,1)}$};
    \end{scope}

    % Right Pants
    \begin{scope}[shift={(7,0)}]
        \coordinate (L) at (0, 0);
        \coordinate (TR) at (2.5, 1.2);
        \coordinate (BR) at (2.5, -1.2);
        
        \draw (0, 0.25) to[out=0, in=180] (2.5, 1.45);
        \draw (0, -0.25) to[out=0, in=180] (2.5, -1.45);
        \draw (2.5, 0.95) to[out=225, in=135, looseness=1.2] (2.5, -0.95);
        
        \draw[fill=white] (L) circle (0.25);
        \draw[fill=white] (TR) circle (0.25);
        \draw[fill=white] (BR) circle (0.25);
        
        \node[penred, above=0.2cm] at (L) {$\phi_{(2,1)}$};
        \node[penred, right=0.15cm] at (TR) {$\phi_{(3,1)}$};
        \node[penred, right=0.15cm] at (BR) {$\phi_{(3,1)}$};
    \end{scope}

    % --- Plus Sign ---
\node at (4.75, -2.2) {\Huge $+$};

    % --- Bottom Row ---
\begin{scope}[shift={(0,-4.4)}]
    % Left Pants
    \begin{scope}[shift={(0,0)}]
        \coordinate (TL) at (0, 1.2);
        \coordinate (BL) at (0, -1.2);
        \coordinate (R) at (2.5, 0);
        
        % Body
        \draw (0, 1.45) to[out=0, in=180] (2.5, 0.25);
        \draw (0, -1.45) to[out=0, in=180] (2.5, -0.25);
        \draw (0, 0.95) to[out=-45, in=45, looseness=1.2] (0, -0.95);
        
        % Holes
        \draw[fill=white] (TL) circle (0.25);
        \draw[fill=white] (BL) circle (0.25);
        \draw[fill=white] (R) circle (0.25);
        
        % Labels
        \node[penred, left=0.15cm] at (TL) {$\phi_{(3,1)}$};
        \node[penred, left=0.15cm] at (BL) {$\phi_{(3,1)}$};
        \node[penred, above=0.2cm] at (R) {$I$};
    \end{scope}

    % Tube
    \begin{scope}[shift={(3.5,0)}]
        \coordinate (L) at (0, 0);
        \coordinate (R) at (2.5, 0);
        
        \draw (0, 0.25) -- (2.5, 0.25);
        \draw (0, -0.25) -- (2.5, -0.25);
        
        \draw[fill=white] (L) circle (0.25);
        \draw[fill=white] (R) circle (0.25);
        
        \node[penblue, below=0.2cm] at (L) {$I$};
        \node[penblue, below=0.2cm] at (R) {$I$};
    \end{scope}

    % Right Pants
    \begin{scope}[shift={(7,0)}]
        \coordinate (L) at (0, 0);
        \coordinate (TR) at (2.5, 1.2);
        \coordinate (BR) at (2.5, -1.2);
        
        % Body
        \draw (0, 0.25) to[out=0, in=180] (2.5, 1.45);
        \draw (0, -0.25) to[out=0, in=180] (2.5, -1.45);
        \draw (2.5, 0.95) to[out=225, in=135, looseness=1.2] (2.5, -0.95);
        
        % Holes
        \draw[fill=white] (L) circle (0.25);
        \draw[fill=white] (TR) circle (0.25);
        \draw[fill=white] (BR) circle (0.25);
        
        % Labels
        \node[penred, above=0.2cm] at (L) {$I$};
        \node[penred, right=0.15cm] at (TR) {$\phi_{(3,1)}$};
        \node[penred, right=0.15cm] at (BR) {$\phi_{(3,1)}$};
    \end{scope}
\end{scope}
\end{scope}

\end{tikzpicture}}
\caption{TQFT structure of the elliptic genus of the 2d $(2,2)$ theory obtained compactifying $(A_1,A_4)$ on a sphere with four punctures of type $\phi_{(3,1)}$.}
\label{fig:TQFT4ptM27d}
\end{figure}
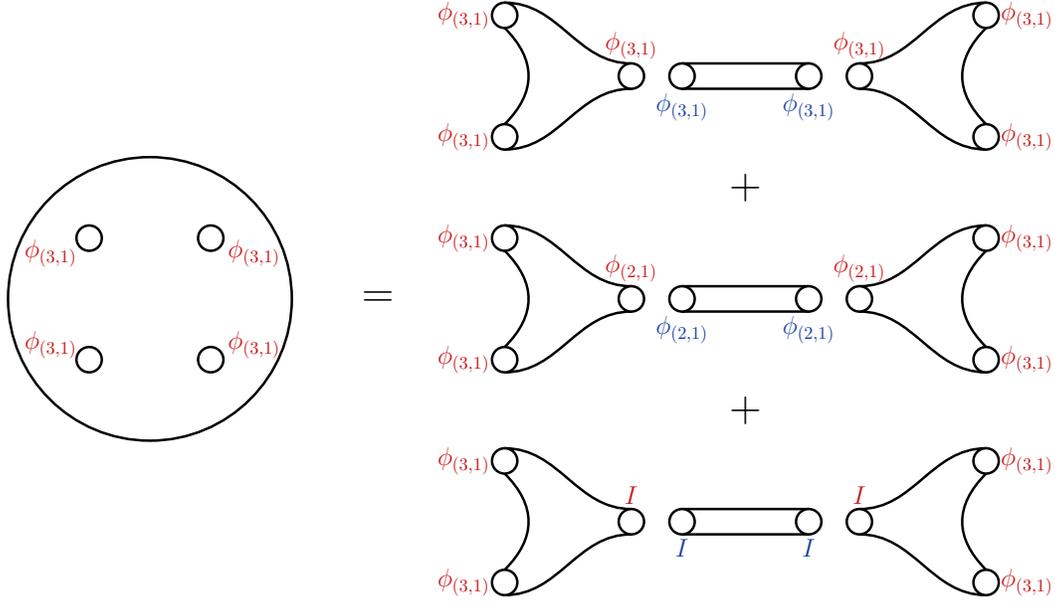

A more involved example is the sphere with four $\phi_{(3,1)}$ punctures. In this case, the TQFT expansion of the elliptic genus contains three terms, as depicted in Figure \ref{fig:TQFT4ptM27d}. This is compatible with the fact that the GLSM has a non-abelian gauge group and fields in various representations, so the structure of poles of the JK integral is more convoluted. We have checked that the evaluation of the JK integral of the GLSM contains three terms which exactly match with those expected from our TQFT formula.

In fact, in order to determine the trinions, we only need three GLSMs out of the five we found studying spheres with four punctures. Hence, the two additional GLSMs provide a consistency check of our proposal. Furthermore, we can perform the same tests that we did in the case of $(A_1,A_2)$. First, we can compare the result of the TQFT formula for a sphere with no punctures with the one obtained by studying the compactification of the 4d $\mathcal{N}=1$ Lagrangian of $(A_1,A_4)$ \cite{Maruyoshi:2016aim} on a sphere with $SU(2)_R$ twist \cite{Gadde:2015wta}. Then, we can compute the elliptic genus for a surface of genus $g$ and with no punctures as a power series in $q$ and verify the presence of the operators predicted from 4d.

We find that the compactification of $(A_1,A_4)$ on a sphere with $SU(2)_R$ twist can be described by the following 2d $\mathcal{N}=(0,2)$ Lagrangian:
\begin{table}[h]
  \centering
\begin{tabular}{c|ccccc}
    & $\Phi$ & $\alpha_1$ & $\alpha_2$ & $\beta_1$ & $\beta_2$ \\\hline
    $USp(4)$            & $\mathbf{10}$ & $\mathbf{1}$ & $\mathbf{1}$ & $\mathbf{1}$ & $\mathbf{1}$ \\
    $U(1)_t$            & $-\tfrac{1}{7}$ & $\tfrac{2}{7}$ & $\tfrac{4}{7}$ & $-\tfrac{8}{7}$ & $-\tfrac{10}{7}$ \\
    $U(1)_{+}$ & $\tfrac{1}{7}$ & $\tfrac{5}{7}$ & $\tfrac{3}{7}$ & $\tfrac{8}{7}$ & $\tfrac{10}{7}$
\end{tabular}
\end{table}

\noindent Here $\Phi$ and $\beta_i$ are chiral multiplets, while $\alpha_i$ are Fermi multiplets. We can then use this to get a JK integral expression of the elliptic genus
\begin{align}
    \mathcal{I}_{C_{III}}=\frac{1}{8}\frac{(q)_\infty^4\theta\left(q^{\frac{6}{7}}t^{\frac{2}{7}}\right)\theta\left(q^{\frac{5}{7}}t^{\frac{4}{7}}\right)}{\theta\left(q^{\frac{1}{14}}t^{-\frac{1}{7}}\right)^2\theta\left(q^{\frac{4}{7}}t^{-\frac{8}{7}}\right)\theta\left(q^{\frac{5}{7}}t^{-\frac{10}{7}}\right)}\oint_{\text{JK}}\prod_{i=1}^2\frac{\mathrm{d}z_i}{2\pi i z_i}\frac{\theta\left(z_i^{\pm2}\right)}{\theta\left(q^{\frac{1}{14}}t^{-\frac{1}{7}}z_i^{\pm2}\right)}\frac{\theta\left(z_1^{\pm1}z_2^{\pm1}\right)}{\theta\left(q^{\frac{1}{14}}t^{-\frac{1}{7}}z_1^{\pm1}z_2^{\pm1}\right)}\,.
\end{align}
which evaluates to
\begin{align}
    \mathcal{I}_{C_{III}}=\frac{\theta\left(q^{\frac{15}{14}}t^{-\frac{1}{7}}\right)\theta\left(q^{\frac{17}{14}}t^{-\frac{3}{7}}\right)}{\theta\left(q^{\frac{4}{7}}t^{-\frac{8}{7}}\right)\theta\left(q^{\frac{5}{7}}t^{-\frac{10}{7}}\right)}\,,
\end{align}
As expected this result precisely coincides with the elliptic genus of 2d $(2,2)$ twisted chirals of R-charges $\tfrac{8}{7}$ and $\tfrac{10}{7}$ in \eqref{eq:bluildblocksA1A4} up to the identification $t=y$. This agrees with the general expectation \eqref{eq:compspherec} that in the case of compactification on a sphere the central charges should be identical to those of a number of twisted chiral fields equal to the dimension of the 4d Coulomb branch and with R-charges identical to those of the Coulomb branch operators.

Finally, we discuss the last test, consisting of using the TQFT structure to compute the elliptic genus of the 2d models associated with the compactification on a genus $g>1$ surface with no punctures. As discussed in Section~\ref{subsec:multred}, the knowledge of the BPS operator spectrum of the 4d theory allows us to predict the presence of certain BPS operators in the resulting 2d theory. This prediction is most accurately formulated in terms of the apperance of certain terms in the elliptic genus of the resulting 2d theories. The prediction for the $(A_1, A_4)$ case was presented in \eqref{A1A4EGexp}, and here we shall reproduce it for convenience
\begin{equation} \label{A1A4EGexpagain}
\mathcal{I}_{g,n=0} = \mathrm{PE}\left[(1-g) y^{-\frac{1}{7}} q^{\frac{1}{14}} + (1-g) y^{-\frac{3}{7}} q^{\frac{3}{14}} + 3(g-1) y q^{\frac{1}{2}} + (1-g) y^{-\frac{8}{7}} q^{\frac{4}{7}} + (1-g) y^{-\frac{10}{7}} q^{\frac{5}{7}}+\cdots\right] \, .
\end{equation}
Here the first and fourth terms originate from the 4d Coulomb branch operator of dimension $\frac{8}{7}$, while the second and fifth terms come from the 4d Coulomb branch operator of dimension $\frac{10}{7}$. The third term is attributed to the 4d energy-momentum tensor multiplet. As before, there could be additional contributions coming from other BPS multiplets, including ones originating in extended operators of the 4d theory, which could lead to deviations from \eqref{A1A4EGexpagain}. However, we expect significant deviations to occur only for small genus, with the result becoming consistent with \eqref{A1A4EGexpagain} for sufficiently high genus.

We next put this to the test by explicitly computing the elliptic genus using the TQFT formula. This computation is more involved then the one for $(A_1, A_2)$ as there are more punctures, so we shall only carry it for $g=2,3,4$. We next summarize our result for the elliptic genus:
\bea
\mathcal{I}_{g=2,n=0} &=& 1 - y^{-\frac{1}{7}} q^{\frac{1}{14}} - 2y^{-\frac{3}{7}} q^{\frac{3}{14}} + 3y^{-\frac{4}{7}} q^{\frac{2}{7}} + y^{-\frac{5}{7}} q^{\frac{5}{14}} - 2y^{-\frac{6}{7}} q^{\frac{3}{7}} + (3y - y^{-1}) q^{\frac{1}{2}} \nonumber \\ &-& (6y^{\frac{6}{7}} + y^{-\frac{8}{7}}) q^{\frac{4}{7}} + y^{\frac{5}{7}} q^{\frac{9}{14}} + (y^{\frac{4}{7}} - y^{-\frac{10}{7}}) q^{\frac{5}{7}} + (4y^{\frac{3}{7}} + 6y^{-\frac{11}{7}}) q^{\frac{11}{14}} + (14y^{\frac{2}{7}} - y^{-\frac{12}{7}}) q^{\frac{6}{7}} \nonumber \\ &+& 22y^{\frac{1}{7}} q^{\frac{13}{14}} + (11+10y^2-y^{-2})q + \cdots \,, \\
\mathcal{I}_{g=3,n=0} &=& 1 - 2y^{-\frac{1}{7}} q^{\frac{1}{14}} + y^{-\frac{2}{7}} q^{\frac{1}{7}} - 2y^{-\frac{3}{7}} q^{\frac{3}{14}} + 4y^{-\frac{4}{7}} q^{\frac{2}{7}} - 2y^{-\frac{5}{7}} q^{\frac{5}{14}} + 2y^{-\frac{6}{7}} q^{\frac{3}{7}} + (6y - 4y^{-1}) q^{\frac{1}{2}} \nonumber \\ &-& (10y^{\frac{6}{7}} + y^{-\frac{8}{7}}) q^{\frac{4}{7}} + 8(y^{\frac{5}{7}} + y^{-\frac{9}{7}})q^{\frac{9}{14}} - (10y^{\frac{4}{7}} + 7y^{-\frac{10}{7}}) q^{\frac{5}{7}} + 2(7y^{\frac{3}{7}} + y^{-\frac{11}{7}}) q^{\frac{11}{14}} \nonumber \\ &-& 2(5y^{\frac{2}{7}} + 3y^{-\frac{12}{7}}) q^{\frac{6}{7}} + 6(y^{\frac{1}{7}} + 2y^{-\frac{13}{7}}) q^{\frac{13}{14}} + (2+21y^2-15y^{-2})q + \cdots \,, \\
\mathcal{I}_{g=4,n=0} &=& 1 - 3y^{-\frac{1}{7}} q^{\frac{1}{14}} + 3y^{-\frac{2}{7}} q^{\frac{1}{7}} - 4y^{-\frac{3}{7}} q^{\frac{3}{14}} + 9y^{-\frac{4}{7}} q^{\frac{2}{7}} - 9y^{-\frac{5}{7}} q^{\frac{5}{14}} + 6y^{-\frac{6}{7}} q^{\frac{3}{7}} + 9(y - y^{-1}) q^{\frac{1}{2}} \nonumber \\ &-& 6(4y^{\frac{6}{7}} - y^{-\frac{8}{7}}) q^{\frac{4}{7}} + (27y^{\frac{5}{7}} + 4y^{-\frac{9}{7}})q^{\frac{9}{14}} - 6(6y^{\frac{4}{7}} + y^{-\frac{10}{7}}) q^{\frac{5}{7}} + 3(17y^{\frac{3}{7}} + 6y^{-\frac{11}{7}}) q^{\frac{11}{14}} \nonumber \\ &-& 6(8y^{\frac{2}{7}} + 7y^{-\frac{12}{7}}) q^{\frac{6}{7}} + (48y^{\frac{1}{7}} + 45y^{-\frac{13}{7}})q^{\frac{13}{14}} - (69-45y^2+39y^{-2})q + \cdots \,.
\eea
The above can also be written as (up to order $q^{\frac{5}{7}}$)
\bea
\mathcal{I}_{g=2,n=0} &=& \mathrm{PE}[ - y^{-\frac{1}{7}} q^{\frac{1}{14}} - 2y^{-\frac{3}{7}} q^{\frac{3}{14}} + y^{-\frac{4}{7}} q^{\frac{2}{7}} + 2y^{-\frac{5}{7}} q^{\frac{5}{14}} - y^{-\frac{6}{7}} q^{\frac{3}{7}} + (3y + y^{-1}) q^{\frac{1}{2}} - (3y^{\frac{6}{7}} - y^{-\frac{8}{7}}) q^{\frac{4}{7}} \nn\\
&-& 2(y^{\frac{5}{7}} + 3y^{-\frac{9}{7}})q^{\frac{9}{14}} + (5y^{\frac{4}{7}} - 4y^{-\frac{10}{7}}) q^{\frac{5}{7}} + \cdots] \,,\nonumber \\
\mathcal{I}_{g=3,n=0} &=& \mathrm{PE}[ - 2y^{-\frac{1}{7}} q^{\frac{1}{14}} - 2y^{-\frac{3}{7}} q^{\frac{3}{14}} + y^{-\frac{6}{7}} q^{\frac{3}{7}} + 6y q^{\frac{1}{2}} + (2y^{\frac{6}{7}} - 3y^{-\frac{8}{7}}) q^{\frac{4}{7}} \nn\\
&+& 2(3y^{\frac{5}{7}} + 2y^{-\frac{9}{7}})q^{\frac{9}{14}} + 12y^{\frac{4}{7}} q^{\frac{5}{7}} + \cdots] \,,\nonumber \\
\mathcal{I}_{g=4,n=0} &=& \mathrm{PE}[ - 3y^{-\frac{1}{7}} q^{\frac{1}{14}} - 3y^{-\frac{3}{7}} q^{\frac{3}{14}} + 9y q^{\frac{1}{2}} + 3(y^{\frac{6}{7}} - y^{-\frac{8}{7}}) q^{\frac{4}{7}} \nn\\
&+& (9y^{\frac{5}{7}} - y^{-\frac{9}{7}})q^{\frac{9}{14}} + 3(6y^{\frac{4}{7}} - y^{-\frac{10}{7}}) q^{\frac{5}{7}} + \cdots] \,.
\eea

We can now compare the results against the general expectation \eqref{A1A4EGexpagain}. We indeed always observe the $3(g-1) y q^{\frac{1}{2}}$ contribution associated with the 4d energy-momentum tensor. Additionally, we see improved agreement with the remaining terms as the genus increases. Specifically, by $g=4$ all terms expected from \eqref{A1A4EGexpagain} appear with the expected multiplicity. For lower genus we do see some deviations, where it appears that terms of the form $y^{-\frac{k}{7}}q^{\frac{k}{14}}$ agree with the generic expectation if $k<3(g-1)$. This also suggests that the $y^{-\frac{9}{7}}q^{\frac{9}{14}}$ term at $g=4$ should vanish  at higher genus.
We expect the elliptic genus to agree with \eqref{A1A4EGexpagain} for $g>4$, although we will not pursue this here.

As before we note that the elliptic genus contains additional contributions, besides those in \eqref{A1A4EGexpagain}, that do not appear to vanish with increasing genus. Specifically, the results found so far suggests the presence of the terms $(g-1)y^{\frac{6}{7}}q^{\frac{4}{7}}$, $3(g-1)y^{\frac{5}{7}}q^{\frac{9}{14}}$ and $6(g-1)y^{\frac{4}{7}}q^{\frac{5}{7}}$ in the elliptic genus. It would be interesting understand if these can also be explained in terms of certain BPS operators of the $(A_1, A_4)$ theory.   

\section{Outlook}
\label{sec:discussion}

We have argued that the SCFT/VOA correspodence~\cite{Beem:2013sza}, which canonically associates a chiral algebra to any ${\cal N}=2$ four-dimensional SCFT, admits a natural upgrade that promotes the  chiral algebra to a full-fledged 
two-dimensional CFT. Correlation functions of the full CFT on a Riemann surface~$\Sigma$ compute (up to the expected K\"ahler ambiguities) the $S^2$ partition function of the class $\FF$  two-dimensional $(2, 2)$ SCFT obtained  
by dimensionally reducing the parent four-dimensional  theory on $\Sigma$. Our new 2d/2d correspondence is
motivated by considering the compactification of the parent 4d SCFT on $S^2 \times \Sigma$. It can 
be viewed as a  ``$2+2 =4$'' analog of the ``$4+2 = 6$'' AGT correspondence~\cite{Alday:2009aq}, with the 2d theories of class $\FF$
playing the same role as the 4d theories of class $\SS$. 

We have made this proposal precise by constructing the requisite off-shell supergravity background and establishing a detailed  dictionary 
relating the two sides of the correspondence.
The two free field theory examples of hypermultiplets and vector multiplets provide a first sanity check.
In interacting examples, it is very desirable to have an intrinsic two-dimensional presentation of the class $\FF$ SCFTs. 
We have illustrated the correspondence
in a highly non-trivial dynamical setting.
Taking the parent 4d theory to be the minimal interacting ${\cal N}=2$ SCFT, the $(A_1, A_2)$ Argyres-Douglas theory, we have conjectured  an explicit gauged linear sigma model description of its 
class $\FF$ daughters when one takes  $\Sigma$ to be an $n$-punctured sphere. The most compelling checks of our conjecture come from extracting the TQFT structure of the elliptic genus for  the whole family $\FF[(A_1, A_2), \Sigma_{g, n}]$. The elliptic genus  turns out to be compatible both with explicit localization calculations in the GLSMs and with four-dimensional expectations.

\medskip

It should be apparent that we have just scratched the surface of a very rich story.
We conclude with a partial list of several natural directions for future research:
\begin{itemize}
 \item Our prescription to obtain GLSMs from Coulomb gas integrals of minimal models is limited to correlators on the sphere, since 
 minimal model correlators on higher genus Riemann surfaces are less understood. In analogy with class $\SS$, one would like to have an algorithm to implement the gluing of punctures at the level of the $(2, 2)$ field theory. 
 While the gluing prescription that we found at the level of the elliptic genus is particularly simple, it is still unclear to us how to perform it at the level of the full $(2, 2)$ field theories or at least at the level of their $S^2$ partition functions. Understanding the gluing procedure would of course also yield new expressions for minimal model correlators at higher genus.
    \item 
    For the superconformal index of class $\SS$, the dual TQFT was identified with a certain three-parameter generalization of two-dimensional $q$-deformed Yang--Mills theory in the zero-area limit~\cite{Gadde:2009kb,Gadde:2011ik,Gadde:2011uv, Gaiotto:2012xa, Rastelli:2014jja}. It would be interesting to have an analogous understanding of the TQFT $\EE[T]$ that computes the elliptic genus of class $\FF$, starting with the cases considered in this paper, where one takes the parent theory
    to be one of the $(A_1, A_{2k})$ Argyres--Douglas theories. We have computed explicitly the structure constants  for $k=1$ and $k=2$, with the higher $k$ cases becoming more and more laborious,
    but we suspect that a more structural understanding of the TQFT would lead to  uniform expressions for any $k$.

    \item We have limited our discussion 
    to the sequence of parent 4d SCFTs  for which the associated 2d CFT is a Virasoro minimal model. We expect that our approach should rather directly 
    generalize to the many examples where the associated VOA is strongly rational. There are many interesting W-algebras arising from
    more general Argyres--Douglas theories~\cite{Xie:2019vzr} and it will be interesting to develop the 2d/2d correspondence for them.

    \item A conceptually orthogonal direction 
    is to explore four-dimensional ${\cal N}=2$ SCFTs
    whose associated VOAs are {\it not} strongly rational. 
   Barring a sparse set of strongly finite but non-rational VOAs (arising from Lagrangian SCFTs with no Higgs branch~\cite{LagrangianC2cofinite}), the vast majority 
   are quasi-lisse but not $C_2$-cofinite VOAs, arising from generic SCFTs with a non-trivial Higgs branch. In essentially all of these examples -- the  exceptions
   being (discrete gaugings of) free vector multiplets and free hypermultiplets --  constructing the full non-unitary CFT is an important open mathematical problem. We expect our correspondence to give crucial new insights. An obvious starting point are Lagrangian 4d theories, for which  the associated non-chiral CFTs can be obtained by a non-chiral version of the BRST procedure~\cite{Beem:2013sza} that computes their chiral algebra. Results in this direction will be reported elsewhere.

    \item We have barely sketched in Section~\ref{sec:A-type} a different ``A-type'' 2d/2d correspondence, closely related to the work of Nekrasov and Shatashvili~\cite{Nekrasov:2009rc,Nekrasov:2009ui,Nekrasov:2009uh,Nekrasov:2014xaa}. It will be interesting to flesh it out in concrete examples.

    \item If one takes the parent 4d theory to be a theory of  class $\SS$, our four-dimensional background uplifts to a six-dimensional backgroud, with a  (2,0) SCFT on $S^2\times\Sigma_1\times\Sigma_2$.
    The 6d SCFT has a $SO(5)$ R-symmetry consisting of two abelian Cartan factors $U(1)_1\times U(1)_2$, and we are  performing a topological twist by  $U(1)_1$ on $\Sigma_1$ and by $U(1)_2$ on $\Sigma_2$. One can then reach the 2d $(2,2)$ theory on $S^2$ either by first reducing on $\Sigma_1$ and then on $\Sigma_2$, or vice versa -- {\it if}
    the two reductions commmute (at least at the level of protected observables).
   Exchanging the order of the two reductions 
   amounts
   to switching the role of the 2d vector and axial R-symmetries, i.e.~to 2d mirror symmetry. This is nicely compatible with the fact that the $S^2_B$ partition function depends on the conformal structure moduli of the Riemann surface implicated on the 4d $\to $ 2d reduction, but does not depend on the 4d conformal manifold, which instead
   coincides with the conformal structure moduli 
   of the Riemann surface implicated in the 6d $\to$ 4d reduction. The story is exactly opposite for the $S^2_A$ partition function, as befits the fact that interchanging the role of the two Riemann surfaces amounts to 2d mirror symmetry. All in all, we would have
 \begin{align}
    Z_{S_B^2}[\FF[ \SS[ \mathfrak{g}; \Sigma_2];\Sigma_1] \sim     Z_{S_A^2}[\FF[ \SS[ \mathfrak{g}; \Sigma_1];\Sigma_2]\,.
\end{align}

\item One can also contemplate the interplay of our new correspondence with holography, starting with the paradigmatic example of  ${\cal N}=4$ 
SYM theory at large $N$. It should be possible to construct a holographic solution of Type II supergravity whose asymptotic boundary is the specific $S^2 \times \Sigma$ background that we have described, along the lines of~\cite{Bobev:2017uzs}, and explore its consequences. Finally one may fantasize about a non-chiral version of twisted holography~\cite{Bonetti:2016nma, Costello:2018zrm}.
    
\end{itemize}

%%%%
\begin{acknowledgments}

We would like to thank Chris Beem, Mykola Dedushenko, Michele Del Zotto, Pietro Ferrero, Simone Giacomelli, Noppadol Mekareeya, Alessio Miscioscia, Maria Nocchi, Sara Pasquetti, Shlomo Razamat, Palash Singh for interesting discussions. 
MS and GZ are also grateful to the organizers of the workshop ``Aspects of Supersymmetric Quantum Field Theory 2025", held at Seoul National University in November 2025, for the possibility to present the results of this work prior to publication. 
LR is partially supported by the NSF grant PHY-2513893 and by the Simons Foundation grant 681267 (Simons Investigator Award).
BR gratefully acknowledges support from DOE grant DE-SC0009988, as well as from the Sivian Fund and the Leinweber Physics Member Fund at the Institute for Advanced Study.
GZ is partially supported by the Israel Science Foundation under grant no.\ 759/23.

\end{acknowledgments}

%%%%%%%%%%%%%%%%%%%%%%%%%%%%%%%%%
\appendix
%%%%%%%%%%%%%%%%%%%%%%%%%%%%%%%%%

%%%%%%%%%%%%%%%%%%%%%%%%%%%%%%%%%
\section{From 4d to 2d superconformal multiplets}
\label{app:4dto2dmultiplets}

One interesting tool in the study of dimensional reduction of supersymmetric theories is that certain terms in the superconformal index of the lower dimensional theory can be deduced from the operator spectrum of the higher dimensional theory. We have seen some details of this in Section~\ref{subsec:multred} for the particular type of compactifications that we are studying in this paper, namely those of 4d $\mathcal{N}=2$ SCFTs on a Riemann surface with $SU(2)_R$ topological twist. The goal of this appendix is to elaborate on the derivation of the results presented in Section~\ref{subsec:multred}.

The broad idea is that given a BPS operator in higher dimensions we expect it to descend to BPS operators in lower dimensions (and potentially also to long ones). As the operators of the lower dimensional theory arise through a KK reduction, the number of such operators should be given by the zero modes of the higher dimensional operator on the compactification space. In general, the number of zero modes may be sensitive to the exact geometry of the surface, but certain differences between them are more robust, specifically those linked to topological invariants by the Atiyah--Singer index theorem. For instance, we can change the number of zero modes for a left or right handed Weyl fermion by introducing a mass, but the difference between the zero modes will not change as it is linked to the A-roof genus by the Atiyah--Singer index theorem. The idea is that this protected difference in zero modes of BPS operators is linked to a similar protected quantity in the lower dimensional theory, specifically the contribution of said operators to the superconformal index.

This idea was expanded in \cite{BRZtoapp} and applied to the case of the reduction of 6d $\mathcal{N}=(1,0)$ SCFTs on Riemann surfaces to 4d $\mathcal{N}=1$ theories. We shall specialize to the case of compactification on a genus $g>1$ Riemann surface with no flavor fluxes, that would be the main case of interest to us. The results are that the 4d superconformal index \cite{Romelsberger:2005eg,Kinney:2005ej,Dolan:2008qi} generically contains the terms
\begin{equation}
    \label{SCindex6to4d}
\mathcal{I}_{4d} = 1 + p q (g-1)(3+d_G) +\cdots\,,
\end{equation} 
where $d_G$ is the dimension of the 6d flavor symmetry (thought of as a $(1,0)$ SCFT). The term $p q$ in the index counts marginal operators minus conserved currents \cite{Beem:2012yn}, and as the result is positive for $g>1$, this usually signals the presence of $(g-1)(3+d_G)$ marginal operators.

The terms in the index originate from contributions of the 6d flavor symmetry currents (the term $(g-1)d_G$) and energy-momentum tensor (the term $3(g-1)$), which are ubiquitous in 6d SCFTs, and as such provide a generic statement. This matches similar reasoning in \cite{Benini:2009mz,Razamat:2016dpl}, who attributed these terms to marginal deformations originating from the freedom of turning on flat connections in flavor symmetries and changing the complex structure moduli of the Riemann surface. The results of \cite{BRZtoapp} then provide a different perspective on these terms, and allow for their generalization to cases with flux. We should also mention that a similar structure is seen in compactifications from other dimensions, like the reduction of 5d SCFTs on Riemann surfaces to 3d \cite{Sacchi:2021afk,Sacchi:2021wvg,Sacchi:2023rtp}.

A few comments before turning to the implication of this to the compactification of 4d SCFTs on Riemann surfaces. First, we stress that these results only provide generic expectations. There are other sources of 4d protected local operators besides just the 6d protected local operators, notably protected extended operators wrapping cycles of the surface, which can then spoil the prediction in \eqref{SCindex6to4d}. However, it is expected that the results would approach \eqref{SCindex6to4d} for sufficiently large genus. Second, the index here is evaluated using the natural $U(1)$ R-symmetry which is the Cartan of the 6d $SU(2)$ R-symmetry. This may not be the actual superconformal R-symmetry of the IR theory.\footnote{For the case considered here, compactification on a genus $g$ Riemann surface with no flavor fluxes, this R-symmetry is usually the superconformal one, baring accidental symmetries. This follows as the charge conjugation invariance of the 6d SCFT prevents mixing with other $U(1)$ flavor symmetries, as long as they originate in 6d. However, in general it will not be the superconformal R-symmetry if the charge conjugation symmetry is broken, for instance by the introduction of flavor fluxes.} Finally, the above discussion was valid for Riemann surfaces with no punctures. The proper generalization of the above statements to cases including punctures is currently unknown, and as such we shall content ourselves with working only with punctureless Riemann surfaces. 
 
Our main concern here is the generalization of these ideas to the reduction of 4d $\mathcal{N}=2$ SCFTs on Riemann surfaces with $SU(2)_R$ twist to 2d $(2,2)$ class $\FF$ theories. We recall that the right and left R-symmetries, $U(1)_+$ and $U(1)_-$, of the 2d theory are then related to the Cartans $U(1)_R$ and $U(1)_r$ of the 4d $\mathcal{N}=2$ R-symmetry by \eqref{eq:2dRsymmLR}, which we report here for convenience
\begin{equation}\label{eq:2dRsymmLRbis}
    U(1)_+=\frac{1}{2}\left(U(1)_R+U(1)_r\right)\,,\qquad U(1)_-=\frac{1}{2}\left(U(1)_R-U(1)_r\right)\,.
\end{equation}
Here we take the 4d supercharges to be in the ${\bf 2}^{-1}$ of $SU(2)_R \times U(1)_R$, and the above ensures that the charges of the 2d supercharges be $(\pm 1;0)$ and $(0;\pm 1)$ under the left and right moving R-symmetries. Additionally, the 4d Lorentz group is broken as: $SO(3,1)\rightarrow SO(1,1)\times SO(2)$, where $SO(1,1)$ is the 2d Lorentz group and the $SO(2)$ part acts on the tangent space of the Riemann surface. This fixes the embedding of the 2d $(2,2)$ supersymmetry algebra in the 4d $\mathcal{N}=2$ one.   

\subsubsection*{Stress-energy tensor and conserved current multiplets}  

We can then consider the contribution to the 2d index, that is the elliptic genus in the NSNS sector, of selected 4d $\mathcal{N}=2$ supeconformal multiplets, analogously to the 6d to 4d case discussed in \cite{BRZtoapp}. We first begin with the stress-energy tensor and conserved current multiplets, who play a pivotal role in the 6d story. These are denoted as $A_2 \bar{A}_2 [0;0]^{(0;0)}_{2}$ and $B_1 \bar{B}_1 [0;0]^{(2;0)}_{2}$, respectively, where we adopt the notation of \cite{Cordova:2016emh}. First, we can use the above embedding to decompose the states of the 4d supermultiplets into the corresponding 2d states. The charges of the states under the 4d Lorentz and R-symmetry determine their charges under the 2d Lorentz and R-symmetry, and the 4d shortening condition imply a 2d shortening condition as well, as it suggests that certain 2d supercharges annihilate the state.\footnote{Note that the conformal dimension changes in the dimensional reduction, as the dilatation symmetry only exist in the deep UV and IR. As such, we cannot use the constraint on the dimension to determine whether the multiplet will be short in 2d. Instead we determine this from the embedding of the supercharges and the 4d shortening condition.} This allows us to decompose the 4d superconformal multiples into the expected 2d superconformal ones. Performing this, one finds that both of these 4d multiplets contain the same type of 2d multiplet
\begin{equation}
    A_2 \bar{A}_2 [0;0]^{(0;0)}_{2} \supset CC[0]^{(1;1)}\,,\qquad B_1 \bar{B}_1 [0;0]^{(2;0)}_{2} \supset CC[0]^{(1;1)}\,,
\end{equation}
where for the 2d multiplets we use the convention that $[J]^{(r_+;r_-)}$ denotes a multiplet with superprimary of spin $J$, right-moving R-charge $r_+$ and left-moving R-charge $r_-$, while $C$ denotes chiral shortening conditions for either left or right moving supercharges. In other words, both the stress-energy tensor and the conserved current multiplets contain a chiral multiplet $CC[0]^{(1;1)}$ with R-charges $(1,1)$ in the decomposition. These are the charges of a chiral marginal operator in 2d, indicating that both multiplets lead to marginal operators in 2d, similarly to the 4d case. 

The multiplicity of their contribution to the index can also be found from the same reasoning as in \cite{BRZtoapp}. The computation is quite similar to the 6d case, so here we shall merely state the results. For the stress-energy tensor the multiplicity is found to be $3(g-1)$, which coincides with the dimension of the complex structure moduli space of the Riemann surface, as expected. 
For a conserved current the multiplicity is instead found to be $g-1$. Here we remind the reader that we consider the reduction on a genus $g$ surface with no punctures and no flux, besides that in the $SU(2)_R$ symmetry required to preserve the 2d $(2,2)$ supersymmetry. The multiplicity of the stress-energy tensor contribution remains unchanged even upon the addition of flavor fluxes, as it is a flavor singlet, but the multiplicity of the conserved current contribution will change if the component is charged under the flavor symmetry which is given a flux. 

Overall, this leads us to expect the following contribution to the elliptic genus due to the 4d stress-energy tensor and conserved current multiplets:
\begin{equation}
    \mathcal{I}\supset1+(g-1)(3+d_G)y\sqrt{q}\,.
\end{equation}
As in the 4d case, the above result can also be interpreted as the maginal chiral operators coming from holonomies in flavor symmetries and the complex structure moduli of the Riemann surface. 

The results for conserved currents can be generalized to arbitrary Higgs branch chiral ring operators, denoted as $B_1 \bar{B}_1 [0;0]^{(R;0)}_{R}$ in \cite{Cordova:2016emh}. These are specified by the $SU(2)_R$ representation of their primary, here of dimension $R+1$. Decomposing them into representations of the 2d superconformal algebra, we find they contain a chiral field with R-charges $(\tfrac{R}{2};\tfrac{R}{2})$.\footnote{This essentially comes from the $SU(2)_R$ highest weight of the primary, which is the one annihilated by the supercharge.}
\begin{equation}
    B_1 \bar{B}_1 [0;0]^{(R;0)}_{R} \supset CC[0]^{\left(\frac{R}{2};\frac{R}{2}\right)}\,,
\end{equation}
The multiplicity of their contribution would now be given by $(R-1)(g-1)$ so overall, we find that these would contribute to the elliptic genus as
\begin{equation}
        \mathcal{I}\supset (R-1)(g-1)y^{\frac{R}{2}} q^{\frac{R}{4}}\chi_{\mathcal{R}}(\vec{x})\,,
\end{equation}
where $\chi_{\mathcal{R}}(\vec{x})$ is the character of the representation $\mathcal{R}$ of the 4d flavor symmetry under which the Higgs branch can potentially transform, written in terms of flavor fugacities $\vec{x}$.

\subsubsection*{Coulomb branch operators}    

One important difference between the 4d $\mathcal{N}=2$ and the 6d $(1,0)$ or 5d $\mathcal{N}=1$ superconformal algebras is that Coulomb branch operators belong to short representations of the former but not of the latter. As such, once we turn to the compactification of 4d $\mathcal{N}=2$ SCFTs, we should consider what 2d multiplets would descend from the 4d Coulomb branch operators. Given the prevailing belief that any interacting 4d $\mathcal{N}=2$ SCFT has a Coulomb branch, these may be as ubiquitous as the stress-energy and conserved current multiplets. These multiplets also hold a special interest in the present discussion, the compactification of the $(A_1,A_2)$ SCFT and higher rank generalizations, as these have no Higgs branch or flavor symmetry and so the Coulomb branch operators are their only basic short representations, aside from the stress-energy tensor. 

The general idea is as before, or more generally as done in \cite{BRZtoapp}. Specifically, we consider the 4d Coulomb branch operators who sit in a $L \bar{B}_1 [0;0]^{(0;r)}_{\frac{r}{2}}$ type multiplet (and their complex conjugate though we shall assume a choice of supercharge in the index such that these are the ones that contribute). We can first decompose them into representations of the 2d $(2,2)$ algebra
\begin{equation}
    L \bar{B}_1 [0;0]^{(0;r)}_{\frac{r}{2}} \supset CA [0]^{\left(\frac{r}{2};-\frac{r}{2}\right)} \oplus CA [0]^{\left(\frac{r}{2}-1;-\frac{r}{2}+1\right)}\,,
\end{equation}
where now $A$ denotes anti-chiral shortening conditions. Hence, from the 4d Coulomb branch operator we find two twisted chiral multiplets, one $CA [0]^{\left(\frac{r}{2};-\frac{r}{2}\right)}$ with R-charges $(\tfrac{r}{2};-\tfrac{r}{2})$ and the other $CA [0]^{\left(\frac{r}{2}-1;-\frac{r}{2}+1\right)}$ with R-charges $(\tfrac{r}{2}-1;-\tfrac{r}{2}+1)$. 

Assuming both contribute to the 2d elliptic genus, we can compute the multiplicity of their contribution from the charges of their primary, as done in \cite{BRZtoapp}. We then find that the multiplicity is $1-g$ for both twisted chirals, which gives the elliptic genus contributions
\begin{equation}
        \mathcal{I}\supset (1-g) \left(y^{1-\Delta} q^{\frac{\Delta-1}{2}}+y^{-\Delta} q^{\frac{\Delta}{2}}\right)\,,
\end{equation}
where $\Delta=\tfrac{r}{2}$ is the dimension of the 4d Coulomb branch operator.

Note that when $r=2$ or $r=4$ one of these twisted chirals become marginal. This suggests that in these cases we get additional twisted chiral marginal operators in 2d. The case of $r=2$ corresponds to the free vector. In this case we expect to get a number of 2d free vectors, which then allows us to write Fayet--Iliopoulos terms which are twisted chiral marginal derfomations in 2d. The $r=4$ case is more interesting. These correspond to dimension $\Delta=2$ Coulomb branch operators, which always contain an $\mathcal{N}=2$ preserving marginal operator. Thus, we see that 4d $\mathcal{N}=2$ preserving marginal operators lead to 2d $(2,2)$ twisted marginal operators. This result is quite natural if we lift the compactification to 6d and think of it as compactifying a 6d SCFT on a product of two Riemann surfaces. In this case we expect the complex structure of both surfaces to appear as marginal 2d operators, where one is twisted and the other is not. The difference here comes from the twisting, where we embed $SO(2)_1 \times SO(2)_2$ in the 6d $SO(5)$ R-symmetry, and twist on one Riemann surface using $SO(2)_1$ and on the other using $SO(2)_2$.

%%%%%%%%%%%%%%%%%%%%%%%%%%%%%%%%%
\section{Minimal review of minimal models}
\label{app:MqpCG}

The main examples of the 2d/2d unitary/non-unitary correspondence discussed in this paper are concerned with the minimal models $\mathcal{M}(q,p)$. These are a class of 2d rational conformal field theories (RCFTs) labelled by two coprime integers $q$ and $p$. In this appendix we birefly review some of their properties, focusing in particular on the Coulomb gas integral representation of their conformal blocks. See \cite{DiFrancesco:1997nk} for a standard reference on this topic.

The minimal models have the property of having a Hilbert space made by a finite number of representations of the Virasoro algebra. For a given choice of $q$ and $p$, the finitely many primary operators are $\phi_{(r,s)}$ with the labels taking the following values:
\begin{equation}
    1\leq r< p\,,\qquad 1\leq s< q\,.
\end{equation}
The dimensions of these operators are
\begin{equation}
    h_{(r,s)}=\bar{h}_{(r,s)}=\frac{(qr-ps)^2-(q-p)^2}{4qp}\,.
\end{equation}
From these dimensions one can see that $h_{(r,s)}=h_{(p-r,q-s)}$ and similarly for the anti-holomorphic dimensions, so the operators are identified in pairs
\begin{equation}
    \phi_{(r,s)}=\phi_{(p-r,q-s)}\,.
\end{equation}
This halves the number of independent primaries, which in total are $(q-1)(p-1)/2$. One particular element in this family of operators that is present in any minimal model is the identity. This can be determined by finding the values of $r$ and $s$ for which the dimensions vanish
\begin{equation}
    I=\phi_{(1,1)}=\phi_{(p-1,q-1)}\,.
\end{equation}

Let us consider for example the case of the Lee--Yang $\mathcal{M}(2,5)$ minimal model that is the main focus of this paper. Here we have two primaries $I$ and $\phi_{(2,1)}$ and the dimension of the non-trivial primary is
\begin{equation}
    h_{(2,1)}=\bar{h}_{(2,1)}=-\frac{1}{5}\,.
\end{equation}
Another example we considered in the main text is the tricritical Lee--Yang model $\mathcal{M}(2,7)$. In this case we have three primaries $I$, $\phi_{(2,1)}$ and $\phi_{(3,1)}$ with the dimensions
\begin{equation}
    h_{(2,1)}=\bar{h}_{(2,1)}=-\frac{2}{7}\,,\qquad h_{(3,1)}=\bar{h}_{(3,1)}=-\frac{3}{7}\,.
\end{equation}

The integers $q$ and $p$ also fully specify the central charges of the minimal models
\begin{equation}
    c=1-6\frac{(q-p)^2}{6qp}\,,
\end{equation}
as well as the momenta of the primaries
\begin{equation}
    \alpha_{(r,s)}=\frac{q(1-r)-p(1-s)}{2\sqrt{qp}}\,.
\end{equation}

We are primarily interested in the computation of the conformal blocks of the minimal models using the Coulomb gas formalism. The idea is that we can re-express the correlation function of primary operators in a minimal model as that of vertex operators in a free theory with the additional insertion of screening charge operators $\mathcal{Q}_{\pm}$, which are integrated vertex operators of momenta
\begin{equation}
    \alpha_{\pm}=\pm\left(\frac{q}{p}\right)^{\pm\frac{1}{2}}\,.
\end{equation}
It is also useful to introduce the parameter
\begin{equation}
    \alpha_0=\frac{1}{2}(\alpha_++\alpha_-)=\frac{q-p}{2\sqrt{qp}}\,.
\end{equation}
More precisely, the correlation function on a sphere of primaries in the minimal model and of vertex operators in the free theory are related as
\begin{align}
    \langle\phi_{(r_1,s_1)}(z_1,\bar{z}_1)\cdots\phi_{(r_k,s_k)}(z_k,\bar{z}_k)\rangle&=\langle V_{\alpha_{(r_1,s_1)}}(z_1,\bar{z}_1)\cdots V_{\alpha_{(r_k,s_k)}}(z_k,\bar{z}_k)\mathcal{Q}_+^n\mathcal{Q}_-^m\rangle\nn\\
   \end{align}
where the numbers of screening charges $n$, $m$ is determined by the neutrality condition\footnote{The sum of the momenta of the operators in the correlator is not zero since we are on a compact surface.}
\begin{equation}\label{eq:neutrality}
    \sum_{i=1}^k\alpha_{(r_i,s_i)}+ n\alpha_++m\alpha_-=\alpha_++\alpha_-=2\alpha_0\,.
\end{equation}
The correlator of vertex operators in the free theory can be easily computed. Considering for simplicity the case $m=0$ in which we only have insertions of the positive screening charge and focusing on the holomorphic part of the correlator, i.e.~the conformal block, we have the following general expression:
\begin{equation}\label{eq:CGint}
    F^{(k)}(z_i)=\prod_{i<j}^k(z_i-z_j)^{2\alpha_{(r_i,s_i)}\alpha_{(r_j,s_j)}}\oint\prod_{a=1}^nx_a(x_a-x_b)^{2\alpha_+^2}\prod_{i=1}^k(x_a-z_i)^{2\alpha_+\alpha_{(r_i,s_i)}}\,.
\end{equation}
The contour of integration should be chosen such that the integral converges. This gives a family of independent contours that define a basis of conformal blocks for the correlation function.

As an example, let us consider the 5-point function of the $\phi_{(2,1)}$ primary of the Lee--Yang minimal model. For this model we have
\begin{equation}
    \alpha_{(2,1)}=-\frac{1}{\sqrt{10}}\,,\qquad \alpha_{(3,1)}=-\sqrt{\frac{2}{5}}\,,\qquad \alpha_+=\sqrt{\frac{2}{5}}\,,\qquad \alpha_0=-\frac{3}{2\sqrt{10}}\,.
\end{equation}
The neutrality condition \eqref{eq:neutrality} can be satisfied with a single screening charge, since
\begin{equation}
    5\alpha_{(2,1)}+\alpha_+=2\alpha_0\,.
\end{equation}
Moreover, using conformal invariance we can place three operators at 0, 1 and $\infty$, so that the correlator only depends on the positions $z_1$ and $z_2$ of the remaining two operators. The Coulomb gas integral \eqref{eq:CGint} for the conformal block in this case becomes
\begin{equation}
    F^{(5)}(z_1,z_2)=(z_1-z_2)^{\frac{1}{5}}\prod_{n=1}^2z_a^{\frac{1}{5}}(z_a-1)^{\frac{1}{5}}\oint\mathrm{d}x\,x^{-{\frac{2}{5}}}(x-1)^{-{\frac{2}{5}}}\prod_{n=1}^2(x-z_n)^{-{\frac{2}{5}}}\,.
\end{equation}
In the main text we focused on a single conformal block since this is enough to determine the GLSM associated with this correlation function.  We choose in particular the one corresponding to the contour $[1,\infty)$, which after performing the change of variable $x=u^{-1}$ is given by the following integral:
\begin{equation}
    F^{(5)}(z_1,z_2)=(z_1-z_2)^{\frac{1}{5}}\prod_{n=1}^2z_n^{\frac{1}{5}}(z_n-1)^{\frac{1}{5}}\int_0^1\mathrm{d}u\,u^{-{\frac{2}{5}}}(1-u)^{-{\frac{2}{5}}}\prod_{n=1}^2(1-z_nu)^{-{\frac{2}{5}}}\,.
\end{equation}
This is precisely the same integral we used in \eqref{eq:CGint5ptLY}.

In fact for most of the examples appearing in the main text only one screening charge turns out to be needed. In this situation, we can prove the following useful identity for the Coulomb gas integral of the conformal block of a $(k+3)$-point function:
\begin{align}\label{eq:IDCoulombGLSM}
    I_k&=\int_0^1\mathrm{d}u\,u^{-2-a-b-\sum_{n=1}^{k}c_a}(1-u)^b\prod_{n=1}^{k}(1-z_nu)^{c_n}\nn\\
    &=\frac{\G{b+1}}{\prod_{n=1}^{k}\G{-c_n}}\int_{-i\infty}^{+i\infty}\frac{\mathrm{d}s_1}{2\pi i}\left(\frac{z_1}{z_2}\right)^{s_1}\G{-s_1}\G{-c_1+s_1}\nn\\
    &\times\prod_{n=2}^{k-1}\left(\int_{-i\infty}^{+i\infty}\frac{\mathrm{d}s_n}{2\pi i}\left(\frac{z_n}{z_{n+1}}\right)^{s_n}\G{-s_n+s_{n-1}}\G{-c_n+s_n-s_{n-1}}\right)\nn\\
    &\times\int_{-i\infty}^{+i\infty}\frac{\mathrm{d}s_{k}}{2\pi i}\left(-z_{k}\right)^{s_{k}}\frac{\G{-s_{k}+s_{k-1}}\G{-c_{k}+s_{k}-s_{k-1}}\G{-a-b-1-\sum_{n=1}^{k}c_n+s_k}}{\G{-a-\sum_{n=1}^{k}c_n+s_k}}\,,
\end{align}
where the parameters $a$, $b$, $c$ should be taken so that both integrals are convergent. This is the main identity that allows us to recast the Coulomb gas integral in the form of a hemisphere partition function of an abelian GLSM.

In order to derive it, we make use of two basic well-know identities
\begin{align}\label{eq:1F0}
    &(1-u)^{-\alpha}=\frac{1}{\G{\alpha}}\int_{-i\infty}^{+i\infty}\frac{\mathrm{d}s}{2\pi i}(-u)^s\G{-s}\G{\alpha+s}\,,
\end{align}
\begin{align}\label{eq:betafunction}
    &\int_0^1\mathrm{d}u\,u^{\beta-1}(1-u)^{\gamma-1}=\frac{\G{\beta}\G{\gamma}}{\G{\beta+\gamma}}\,.
\end{align}
These are equivalent representations of the ${}_1F_0(\alpha;-u)$ hypergeometric function and of the beta-function, respectively, and they are related to each other by a Mellin transform. We first use \eqref{eq:1F0} for each of the $k$ factors $(1-z_nu)^{c_n}$ in the definition of $I_k$ and reorder the integrals
\begin{align}
    I_k&=\prod_{n=1}^{k}\left(\frac{1}{\G{-c_n}}\int_{-i\infty}^{+i\infty}\frac{\mathrm{d}s_n}{2\pi i}\left(-z_n\right)^{s_n}\G{-s_n}\G{-c_n+s_n}\right)\nn\\
    &\times\int_0^1\mathrm{d}u\,u^{-2-a-b-\sum_{n=1}^k(c_n-s_n)}(1-u)^b\,.
\end{align}
We can now remove the original integral by using \eqref{eq:betafunction}
\begin{align}
    I_k&=\frac{\G{b+1}}{\prod_{n=1}^{k}\G{-c_n}}\prod_{n=1}^{k}\left(\int_{-i\infty}^{+i\infty}\frac{\mathrm{d}s_n}{2\pi i}\left(-z_n\right)^{s_n}\G{-s_n}\G{-c_n+s_n}\right)\nn\\
    &\times\frac{\G{-a-b-1-\sum_{n=1}^k(c_n-s_n)}}{\G{-a-\sum_{n=1}^k(c_n-s_n)}}\,.
\end{align}
The result \eqref{eq:IDCoulombGLSM} is finally obtained by performing the change of variables
\begin{equation}
    s_n\to s_n-s_{n-1}\,,\qquad n=2,\cdots,k\,.
\end{equation}

The identity \eqref{eq:IDCoulombGLSM} and its derivation have an interesting physical interpretation. The identity can be obtained as the $q\to1$ limit of the holomorphic blocks \cite{Pasquetti:2011fj,Beem:2012mb} identity associated to the 3d abelian mirror duality that relates SQED with $k+1$ flavors to the affine $A_k$ quiver \cite{Intriligator:1996ex,deBoer:1996mp,Hanany:1996ie,Aharony:1997bx}. As observed in \cite{Nedelin:2017nsb,Pasquetti:2019uop,Pasquetti:2019tix}, this limit of partition functions of 3d mirror dualities generically leads to identities between Coulomb gas integrals and partition functions of 2d GLSMs (see \cite{Aharony:2017adm} for the interpretation in terms of 2d dualities). In our case, the limit of the partition function of SQED reduces to the one-dimensional Coulomb gas integral, while the limit of the partition function of the quiver mirror dual reduces to the 2d partition function of a similar GLSM which appears as the r.h.s.~of \eqref{eq:IDCoulombGLSM}. Moreover, the derivation that we provided of \eqref{eq:IDCoulombGLSM} is completely analogous to the piecewise derivation of abelian mirror symmetry of \cite{Kapustin:1999ha}. The latter indeed allows to derive the mirror duality for SQED with $k+1$ flavors by iterating the basic one for $k=0$, which relates 3d $\mathcal{N}=2$ SQED with one flavor to the XYZ model. The auxiliary identities \eqref{eq:1F0}-\eqref{eq:betafunction} can themselves be obtained as $q\to1$ limits of the 3d identity associated to the SQED/XYZ duality.

%%%%%%%%%%%%%%%%%%%%%%%%%%%%%%%%%
\section{Moduli spaces and duality from crossing symmetry}
\label{app:duality}

In this appendix we analyze in detail the structure of the classical moduli space\footnote{Of course, by the Coleman--Mermin--Wagner theorem, since we are in two dimensions there cannot be an actual vacuum moduli space of the quantum theory.} for the 2d $\mathcal{N}=(2,2)$ GLSMs that describe the compactification of $(A_1,A_2)$ on a sphere with four and five punctures of type $\phi_{(2,1)}$, which we derived in Subsections \ref{subsec:4ptLY} and \ref{subsec:5ptLY} respectively.
In the four-punctured case, we deduce the Hilbert series of the twisted chiral ring at both $\xi =0$ and $\xi = \infty$. We find that they agree,
as predicted from the the self-duality that this model is expected to enjoy from the crossing symmetric property of the associated Lee--Yang four-point function.

\subsection{Four-punctured sphere}
\label{appsub:duality4pt}

For the correlator of four identical operators located at 0, 1, $\infty$ and $z$, we have an invariance under the two transformations
\begin{equation}
    z\to\frac{1}{z}\,,\qquad z\to \frac{z}{z-1}\,,
\end{equation}
which relate the $s$-channel to the $u$ and the $t$-channels, respectively. Recall that $z$ is identified with the FI parameter of the GLSM as $z=\mathrm{e}^{-2\pi \xi}$. Hence, the GLSM is expected to be left invariant under such transformations of the FI parameter. We interpret this as a non-trivial self-duality of the theory.

As mentioned in the main text, the first symmetry transformation can be easily understood as the theory being invariant under charge conjugation. The second transformation is less trivial and we would like to provide further evidence for the corresponding self-duality of the GLSM. The $S^2$ partition function is of course invariant under $z\to \frac{z}{z-1}$, since it was engineered so to coincide the CFT correlator. The central charge and the elliptic genus are also trivially invariant since they do not depend on the FI parameter. To perform a non-trivial test of the duality we will study the moduli space of the theory. In particular, since the symmetry transformation maps $\xi=0$ to $\xi=+\infty$, we would like to compare the moduli space at these two points.

From the superpotential \eqref{eq:W4ptLY} we get the following F-term equations:
\begin{align}
& 2 \Phi_1 \tilde{\Phi}_1 \tilde{\Phi}_2 + \tilde{\Phi}_1^2 \Phi_2 + F_1 \tilde{\Phi}_1 = 0 \,, \nn\\
& \Phi_1^2 \tilde{\Phi}_2 + 2 \Phi_1 \Phi_2 \tilde{\Phi}_1 + F_1 \Phi_1 = 0 \,,\nn \\
& \tilde{\Phi}_1^2 \Phi_1 + F_2 \tilde{\Phi}_2 = 0 \,, \nn\\ 
& \Phi_1^2 \tilde{\Phi}_1 + F_2 \Phi_2 = 0 \,, \nn\\ 
& \Phi_1 \tilde{\Phi}_1 = 0 \,, \nn\\ 
& \Phi_2 \tilde{\Phi}_2 + 5 F_2^4 = 0 \,.
\end{align}
The last four equations can be simplified to $F_2 = \Phi_1 \tilde{\Phi}_1 = \Phi_2 \tilde{\Phi}_2 = 0$. This further allows us to simplify the first two equations, ending up with the following simplified version of the F-terms:
\begin{align}\label{F-terms}
& \tilde{\Phi}_1 (\tilde{\Phi}_1 \Phi_2 + F_1) = 0 \,, \nn\\
& \Phi_1 (\Phi_1 \tilde{\Phi}_2 + F_1) = 0 \,, \nn\\
& \Phi_1 \tilde{\Phi}_1 = 0 \,, \nn\\ 
& \Phi_2 \tilde{\Phi}_2 = 0 \,, \nn\\ 
& F_2 = 0 \,.
\end{align}
Furthermore, we have the D-term equation
\begin{equation}\label{D-terms}
|\Phi_1|^2 + |\Phi_2|^2 - |\tilde{\Phi}_1|^2 - |\tilde{\Phi}_2|^2 = \xi
\end{equation} 
and we should also quotient by the action of the $U(1)$ gauge group.

We begin with the case of $\xi=0$. In this case we can just focus on gauge invariants subject to the F-terms. The last three conditions in \eqref{F-terms} essentially leave us with only 3 gauge invariants: $x = F_1$ , $y = \Phi_1 \tilde{\Phi}_2$ , $z = \tilde{\Phi}_1 \Phi_2$. The first two conditions in \eqref{F-terms} further restrict them to obey:
\begin{equation}
z (z + x) = 0 \, ,\qquad y (y + x) = 0 \,.
\end{equation}
These equations have three solutions:
\begin{align}
& \text{I}:\,\, z = y = 0 , x\neq 0 \,,\nn\\
& \text{II}:\,\, z = 0 , x = -y \,, \nn\\
& \text{III}:\,\, y = 0 , x = -z \,. \end{align}

The three solutions coincide at the point $x=y=z=0$, but are otherwise disjoint and thus describe three distinct branches. Each branch is isomorphic to $\mathbb{C}$, and we thus conclude that the moduli space consists of three copies of $\mathbb{C}$ adjointed at the origin. Note that the three gauge invariants ($x$, $y$ and $z$) all have R-charge $\frac{3}{5}$ and as such the three branches appear completely symmetric.\footnote{The symmetry between branches II and III follows from charge conjugation. Classically the gauge symmetry is not broken on branch I so it is not clear if there is actually a symmetry of the theory permuting the three branches. We also note that the charge conjugation symmetry comes from the duality which becomes a symmetry at this point. It is interesting if similar symmetries exist at the other self-duality points.} We can now write the Hilbert series of the moduli space by combining the contribution of all three branches and subtracting for the triple counting of the origin
\be\label{eq:HSxi0}
\mathrm{HS}(\mathcal{M}_{\xi=0}) = \frac{3}{1-t^3} - 2 = \frac{1 + 2t^3}{1-t^3} \,,
\ee
where we are using a normalization of the fugacity $t$ such that its exponent is $5R$ with $R$ the R-charge.

\begin{figure}
\center
\includegraphics[width=0.90\textwidth]{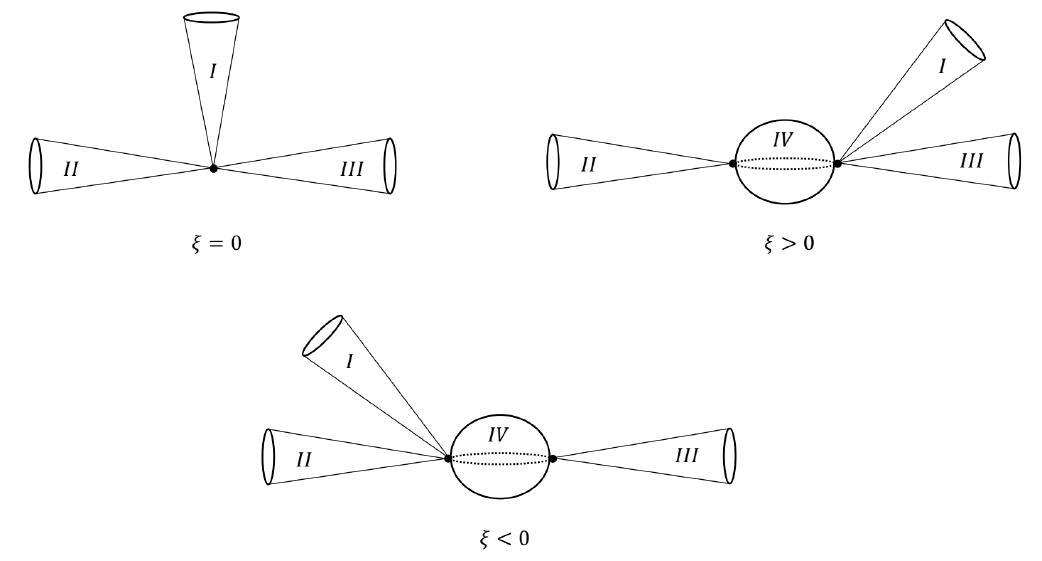} 
\label{MS}
\caption{Schematic depiction of the moduli space for the 2d $(2,2)$ GLSM obtained by $SU(2)_R$ twisted compactification $(A_1,A_2)$ on a sphere with four punctures of type $\phi_{(2,1)}$ for different values of the FI $\xi$. Branches I, II and III are isomorphic to $\mathbb{C}$ while branch IV is isomorphic to a $\mathbb{CP}^1$ of radius $\sqrt{|\xi|}$.}
\end{figure}

Next we turn to the case with non-zero $\xi$. Here we shall concentrate on the case $\xi>0$, where the negative case can then be generated by charge conjugation. The D-term equation \eqref{D-terms} can be solved by setting
\bea
& & |\Phi_1| = \sqrt{\xi} \cosh{(\nu)} \cos{(\theta)}\, , \nonumber \\
& & |\Phi_2| = \sqrt{\xi} \cosh{(\nu)} \sin{(\theta)} \,, \nonumber \\
& & |\tilde{\Phi}_1| = \sqrt{\xi} \sinh{(\nu)} \cos{(\phi)} \,, \nonumber \\
& & |\tilde{\Phi}_2| = \sqrt{\xi} \sinh{(\nu)} \sin{(\phi)} \,. 
\eea
We then need to enforce the F-term conditions \eqref{F-terms}. We find four distinct solutions
\bea
& & I: \,\,\,\nu=0\,,\,\,\, \theta=\frac{\pi}{2} \quad\Rightarrow\quad \Phi_1 = \tilde{\Phi}_1 = \tilde{\Phi}_2 = 0\,, \,\,\, \Phi_2 = \sqrt{\xi} \,,\,\,\, F_1\neq 0  \,, \nonumber \\
& & II: \,\,\,\theta=0\,,\,\,\, \phi=\frac{\pi}{2}\,,\,\,\, F_1 = -\Phi_1 \tilde{\Phi}_2 \quad\Rightarrow\quad \tilde{\Phi}_1 = \Phi_2 = 0 \,,\,\,\, |\Phi_1|^2-|\tilde{\Phi}_2|^2=\xi \,,\,\,\, F_1 = -\Phi_1 \tilde{\Phi}_2 \,, \nonumber \\
& & III: \,\,\,\theta=\frac{\pi}{2}\,,\,\,\, \phi=0\,,\,\,\, F_1 = -\Phi_2 \tilde{\Phi}_1 \quad\Rightarrow\quad \tilde{\Phi}_2 = \Phi_1 = 0 \,,\,\,\, |\Phi_2|^2-|\tilde{\Phi}_1|^2=\xi \,,\,\,\, F_1 = -\Phi_2 \tilde{\Phi}_1 \,, \nonumber \\
& & IV: \,\,\nu=0\,, F_1=0 \quad\Rightarrow\quad F_1 = \tilde{\Phi}_1 = \tilde{\Phi}_2 = 0 \,,\,\,\, |\Phi_1|^2+|\tilde{\Phi}_2|^2=\xi \,.
\eea
Solutions I, II and III are analogous to the solutions found in the $\xi=0$ case, except they no longer touch at a point. Specifically, branch I and III touch at the point $\Phi_2 = \sqrt{\xi}$ with all other fields zero, while branch II has no intersection with them. Additionally, we now have an additional branch, IV. Unlike the other branches, this one is compact, being isomorphic to $S^3/S^1 = \mathbb{CP}^1$ with radius $\sqrt{\xi}$. Note that branches I and III intersect with IV at the point $\Phi_2 = \sqrt{\xi}$ with all other fields zero, while branch II intersect with it at the point $\Phi_1 = \sqrt{\xi}$ with all other fields zero. As such, we see that turning on an FI term has the effect of partially blowing up the origin. The three branches I, II and III remain but are partly separated, with I going with branch III if $\xi>0$ and with branch II if $\xi<0$. A sketch of the moduli space in the different cases is given in Figure \ref{MS}.

Next we want to consider what happens in the limit $\xi\rightarrow \infty$. Note that in this case the distance in field space between the points $\Phi_1 = \sqrt{\xi}$ and $\Phi_2 = \sqrt{\xi}$ becomes infinite. The moduli space then appears to consists of two singular points that are completely decoupled. Specifically, we have the point $\Phi_1 \rightarrow \infty$, where we have branch II connected with the decompactified $\mathbb{CP}^1$, and the point $\Phi_2 \rightarrow \infty$, where branches I and III are connected with themselves and the decompactified $\mathbb{CP}^1$.

We can attempt to compute the Hilbert series by summing the contributions of the two disjoint regions, similarly to as done before. We begin with the region $\Phi_1 \rightarrow \infty$. We expand the original superpotential \eqref{eq:W4ptLY} around that point, finding
\be
\mathcal{W} = \xi \tilde{\Phi}_1 \tilde{\Phi}_2 + \sqrt{\xi} \tilde{\Phi}^2_1 \Phi_2 + \sqrt{\xi} F_1 \tilde{\Phi}_1 + F_2 \Phi_2 \tilde{\Phi}_2 + F^5_2 \,. 
\ee      
The F-term equations now set 
\begin{equation}
    \tilde{\Phi}_1 = F_2 = 0\,,\qquad \Phi_2 \tilde{\Phi}_2 = 0\,,\qquad F_1 = - \sqrt{\xi} \tilde{\Phi}_2\,.
\end{equation}
This leaves two solutions, corresponding to branches II and IV
\begin{align}
    &II:\,\,\,\Phi_2 = 0\,,\qquad F_1 = - \sqrt{\xi} \tilde{\Phi}_2\neq 0\,,\nn\\
    &IV:\,\,\,F_1 = \tilde{\Phi}_2 = 0\,,\qquad \Phi_2\neq 0\,.
\end{align}
We note that around this point $F_1$ and $\tilde{\Phi}_2$ have R-charge $\tfrac{3}{5}$, while $\Phi_2$ has R-charge $\tfrac{1}{5}$. The Hilbert series of this region can then be evaluated by summing the contribution of the two branches subtracting for the shared origin:
\be
\mathrm{HS}(\mathcal{M}^{(1)}_{\xi\rightarrow \infty}) = \frac{1}{1-t} + \frac{1}{1-t^3} - 1 = \frac{1+t+t^2+t^3}{1-t^3} .
\ee  

Similarly, we can expand around the second region, $\Phi_2 \rightarrow \infty$. The superpotential now reads
\be
\mathcal{W} = \Phi^2_1 \tilde{\Phi}_1 \tilde{\Phi}_2 + \sqrt{\xi} \tilde{\Phi}^2_1 \Phi_1 + F_1 \Phi_1 \tilde{\Phi}_1 + \sqrt{\xi} F_2 \tilde{\Phi}_2 + F^5_2 .
\ee 
The F-term equations now set 
\begin{equation}
   \tilde{\Phi}_2 = F_2 = 0\,,\qquad  \Phi_1 \tilde{\Phi}_1 = 0\,,\qquad F_1 \Phi_1 = 0\,,\qquad \tilde{\Phi}_1 (F_1 + \sqrt{\xi} \tilde{\Phi}_1 ) = 0 \,.
\end{equation}
There are three possible solutions, corresponding to branches I, III and IV
\begin{align}
    &I:\,\,\,\Phi_1 = \tilde{\Phi}_2 = 0\,,\qquad F_1\neq 0\,,\nn\\
    &III:\,\,\, \Phi_1 = 0\,,\qquad F_1 = - \sqrt{\xi} \tilde{\Phi}_1\neq0\,,\nn\\
    &IV:\,\,\,F_1 = \tilde{\Phi}_1 = 0\,,\qquad \Phi_1\neq 0\,.
\end{align}
We note that around this point $F_1$ and $\tilde{\Phi}_1$ have R-charge $\tfrac{3}{5}$, while $\Phi_1$ has R-charge $-\tfrac{1}{5}$. The Hilbert series of this region can then be evaluated by summing the contribution of the three branches subtracting for the shared origin
\be
\mathrm{HS}(\mathcal{M}^{(2)}_{\xi\rightarrow \infty}) = \frac{1}{1-t^{-1}} + \frac{2}{1-t^3} - 2 = \frac{t(t^2-t-1)}{1-t^3} .
\ee  

For finite $\xi$ the two regions have branch IV in common, but in the $\xi\rightarrow \infty$ limit they become decoupled, and we expect the full Hilbert series to be given by just their sum. Combining the two terms we then get
\be
\mathrm{HS}(\mathcal{M}_{\xi\rightarrow \infty}) = \mathrm{HS}(\mathcal{M}^{(1)}_{\xi\rightarrow \infty}) + \mathrm{HS}(\mathcal{M}^{(2)}_{\xi\rightarrow \infty}) = \frac{1+2t^3}{1-t^3},
\ee
which is indeed the same Hilbert series as in the $\xi=0$ case \eqref{eq:HSxi0}.

Note that even though the Hilbert series matches, the moduli space appear to be different at $\xi=0$ compared to $\xi=\infty$. This is expected, since in two dimensions the moduli spaces of dual theories are not necessarily identical, but the rings of chiral operators which are counted by the Hilbert series should. We also comment that the structure of the moduli space at $\xi\neq0$ is very similar to that of the elliptic genus \eqref{eq:EG4ptLY} of the theory. Indeed, the elliptic genus can be expressed as the sum of two contributions, each of which looks like the elliptic genus of chiral fields with fixed R-charges. These are precisely the chiral fields that we found in the low energy effective theories around each of the two singular points. It would be interesting to investigate further the relation between the moduli space and the elliptic genus, at least in the context of 2d $(2,2)$ arising from $SU(2)_R$ twisted compactifications of 4d $\mathcal{N}=2$ SCFTs.

\subsection{Five-punctured sphere}
\label{appsub:duality5pt}

We can extend the previous moduli space analysis to the case of the five-punctured sphere. Recall that the field content of the theory is two $U(1)$ gauge groups and chiral fields with charges as shown in the following table:
\begin{table}[h]
  \centering
  \begin{tabular}{c|cccccccc}
      & $\Phi_1$ & $\tilde{\Phi}_1$ & $\Phi_2$ & $\tilde{\Phi}_2$ & $\Phi_3$ & $\tilde{\Phi}_3$ & $F_1$ & $F_{2,3,4}$\\\hline
      $U(1)_1$ & 1 & $-1$ & $-1$ & 1 & 0 & 0 & 0 & 0 \\
      $U(1)_2$ & 0 & 0 & 1 & $-1$ & $-1$ & 1 & 0 & 0 \\
      $U(1)_R$ & $\frac{1}{5}$ & $\frac{1}{5}$ & $\frac{1}{5}$ & $\frac{1}{5}$ & $\frac{1}{5}$ & $\frac{1}{5}$ & $\frac{1}{5}$ & $\frac{3}{5}$
  \end{tabular}
\end{table}

\noindent To enforce the R-charges, we need to introduce a superpotential. The most general superpotential consistent with the R-charge assignment is
\begin{align}
\mathcal{W} &= a_1 F^5_1 + a_2 \Phi^2_1 \tilde{\Phi}^2_1 F_1 + a_3 \tilde{\Phi}^2_3 \Phi^2_3 F_1 + a_4 \tilde{\Phi}^2_2 \Phi^2_2 F_1 + a_5 \Phi_1 \tilde{\Phi}_1 F^3_1 + a_6 \tilde{\Phi}_3 \Phi_3 F^3_1 + a_7 \tilde{\Phi}_2 \Phi_2 F^3_1 \nonumber \\ \nonumber & + a_8 \Phi^2_1 \tilde{\Phi}_1 \Phi_3 \Phi_2 + a_9 \Phi_1 \tilde{\Phi}_3 \Phi^2_3 \Phi_2 + a_{10} \Phi_1 \tilde{\Phi}_2 \Phi_3 \Phi^2_2 + a_{11} \Phi_1 \tilde{\Phi}^2_1 \tilde{\Phi}_2 \tilde{\Phi}_3 + a_{12} \Phi_3 \tilde{\Phi}_1 \tilde{\Phi}_2 \tilde{\Phi}^2_3 \\ \nonumber & + a_{13} \Phi_2 \tilde{\Phi}_1 \tilde{\Phi}^2_2 \tilde{\Phi}_3 + a_{14} \Phi_1 \tilde{\Phi}_1 F_2 + a_{15} \tilde{\Phi}_3 \Phi_3 F_2 + a_{16} \tilde{\Phi}_2 \Phi_2 F_2 + a_{17} \Phi_1 \tilde{\Phi}_1 F_3 + a_{18} \tilde{\Phi}_3 \Phi_3 F_3 \\ \nonumber & + a_{19} \tilde{\Phi}_2 \Phi_2 F_3 + a_{20} \Phi_1 \tilde{\Phi}_1 F_4 + a_{21} \tilde{\Phi}_3 \Phi_3 F_4 + a_{22} \tilde{\Phi}_2 \Phi_2 F_4 + a_{23} \Phi_1 \Phi_3 \Phi_2 F^2_1 \\  & + a_{24} \tilde{\Phi}_1 \tilde{\Phi}_3 \tilde{\Phi}_2 F^2_1 + a_{25} F_2 F^2_1 + a_{26} F_3 F^2_1 + a_{27} F_4 F^2_1\,.
\end{align}
We will analyze the moduli space under the assumption that the coefficients in the superpotential are generic. We shall not write the F-term conditions associated with the above superpotential explicitly since these are quite complicate, but we will write their forms in the special cases of interest to us.

We begin with the case of $\xi_1 = \xi_2 =0$ (vanishing FI terms for the two $U(1)$ gauge groups), where the D-term equation can be solved by restriction to gauge invariants. The basic invariants are the 4 flip fields $F_i$ and the following combination of the charged fields:
\begin{equation}
    x = \Phi_1 \tilde{\Phi}_1 \,,\qquad y = \tilde{\Phi}_3 \Phi_3 \,,\qquad z = \tilde{\Phi}_2 \Phi_2 \,,\qquad u = \Phi_1 \Phi_2 \Phi_3 \,,\qquad v = \tilde{\Phi}_1 \tilde{\Phi}_2 \tilde{\Phi}_3\,.
\end{equation}
Note that these are not all independent, but rather obey the constraint $x y z = u v$.

It is convenient to first assume that $F_1 = 0$, and then return to the case of $F_1 \neq 0$ later. The reason is that in this case the F-terms associated with the fields $F_{1,2,3,4}$ reduce to the following conditions:
\bea
\nonumber& & a_2 x^2 + a_3 y^2 + a_4 z^2 = 0 \,, \\ \nonumber
& & a_{14} x + a_{15} y + a_{16} z = 0 \,, \\ \nonumber
& & a_{17} x + a_{18} y + a_{19} z = 0 \,, \\ 
& & a_{20} x + a_{21} y + a_{22} z = 0 \,.
\eea
Assuming the coefficients are generic, these will have the only solution $x=y=z=0$. Using this allows us to simplify the F-terms for the other fields to
\bea \label{FtermsPhi}
\nonumber & & \tilde{\Phi}_1 ( a_{14} F_2 + a_{17} F_3 + a_{20} F_4 + a_{11} v ) = 0 \,, \\ \nonumber
& & \Phi_1 ( a_{14} F_2 + a_{17} F_3 + a_{20} F_4 + a_{8} u ) = 0 \,, \\ \nonumber
& & \Phi_3 ( a_{15} F_2 + a_{18} F_3 + a_{21} F_4 + a_{9} u ) = 0 \,, \\ \nonumber
& & \tilde{\Phi}_3 ( a_{15} F_2 + a_{18} F_3 + a_{21} F_4 + a_{12} v ) = 0 \,, \\ \nonumber
& & \Phi_2 ( a_{16} F_2 + a_{19} F_3 + a_{22} F_4 + a_{10} u ) = 0 \,, \\ 
& & \tilde{\Phi}_2 ( a_{16} F_2 + a_{19} F_3 + a_{22} F_4 + a_{13} v ) = 0 \,.
\eea
These, together with the condition $x=y=z=0$ have the following set of solutions:
\begin{itemize} 
\item $u=v=0$: in this case $F_{2,3,4}$ can be non-zero and the moduli space is locally $\mathbb{C}^3$.
\item $u=0$, $v\neq 0$: this implies $\Phi_1 = \Phi_3 = \Phi_2 = 0$, which automatically satisfies half the conditions in \eqref{FtermsPhi}. The remaining three conditions can be written as the matrix equation $M F = v A$, with $F=(F_2, F_3, F_4)$. As long as the $a_i$'s are sufficiently generic $\mathrm{det}(M)\neq 0$, and there is a unique solution for $F$ in terms of $v$. As such this branch is locally $\mathbb{C}$. 
\item $v=0$, $u\neq 0$: this implies $\tilde{\Phi}_1 = \tilde{\Phi}_3 = \tilde{\Phi}_2 = 0$, which again automatically satisfies half the conditions in \eqref{FtermsPhi}. The remaining three conditions can be written as the matrix equation $M F = u A$, with $F=(F_2, F_3, F_4)$, and as long as the $a_i$'s are generic enough $\mathrm{det}(M)\neq 0$. In this case, we again have a unique solution for $F$ in terms of $u$,  and this branch is locally $\mathbb{C}$. 
\end{itemize} 

This brings us to the case where $F_1 \neq 0$. In that case, the F-term equations associated with $F_{1,2,3,4}$ imply
\begin{align}
& 5 a_1 F_1^{4}
+ 2 F_1 \left( a_{23} u + a_{24} v + a_{25} F_2 + a_{26} F_3 + a_{27} F_4 \right)
+ 3 F_1^{2} \left( a_5 x + a_6 y + a_7 z \right)
\nonumber\\
&\qquad
+\, a_2 x^{2} + a_3 y^{2} + a_4 z^{2}
= 0 \,, \nonumber \\[4pt]
& a_{25} F_1^{2} + a_{14} x + a_{15} y + a_{16} z = 0 \,, \nonumber \\[2pt]
& a_{26} F_1^{2} + a_{17} x + a_{18} y + a_{19} z = 0 \,, \nonumber \\[2pt]
& a_{27} F_1^{2} + a_{20} x + a_{21} y + a_{22} z = 0 \,.
\end{align}
The last three equations suggest that $x$, $y$ and $z$ are now non-zero and their values are fixed by the value of $F_1$ (again assuming generic $a_i$'s). The first equation can be solved to express the combination $a_{23}u + a_{24}v + a_{25}F_2 + a_{26}F_3 + a_{27}F_4$ in terms of $F_1$. We can now write the F-term conditions for the remaining fields
\bea
& & a_5 F^3_1 x + a_{14} F_2 x + a_{17} F_3 x + a_{20} F_4 x + 2a_2 F_1 x^2 + a_{11} x v + a_{23} F^2_1 u + 2 a_8 x u + a_9 y u + a_{10} z u = 0 \,, \nonumber  \\ \nonumber 
& & a_5 F^3_1 x + a_{14} F_2 x + a_{17} F_3 x + a_{20} F_4 x + 2a_2 F_1 x^2 + 2a_{11} x v + a_{24} F^2_1 v + a_8 x u + a_{12} y v + a_{13} z v = 0 \,, \\ \nonumber
& & a_6 F^3_1 y + a_{15} F_2 y + a_{18} F_3 y + a_{21} F_4 y + 2a_3 F_1 y^2 + a_{9} y u + a_{24} F^2_1 v + a_{11} x v + 2a_{12} y v + a_{13} z v = 0 \,, \\ \nonumber 
& & a_6 F^3_1 y + a_{15} F_2 y + a_{18} F_3 y + a_{21} F_4 y + 2a_3 F_1 y^2 + a_{12} y v + a_{23} F^2_1 u + a_{8} x u + 2a_{9} y u + a_{10} z u = 0 \,, \\ \nonumber
& & a_7 F^3_1 z + a_{16} F_2 z + a_{19} F_3 z + a_{22} F_4 z + 2a_4 F_1 z^2 + a_{10} z u + a_{24} F^2_1 v +  a_{11} x v + a_{12} y v + 2a_{13} z v = 0 , \\ \nonumber& & a_7 F^3_1 z + a_{16} F_2 z + a_{19} F_3 z + a_{22} F_4 z + 2a_4 F_1 z^2 + a_{13} z v + a_{23} F^2_1 u +  a_{8} x u + a_{9} y u + 2a_{10} z u = 0 \,,  \\ && 
\eea
where we used the fact that the $\Phi$'s and $\tilde{\Phi}$'s must get a non-zero value so we multiply by them to recast the F-terms using gauge invariants. 
One may note that taking the difference of the second and first equation is equal to the difference between the third and fourth equation which is equal to the difference between the fifth and sixth equation. Thus, if we add the $F_1$ F-term equation, there are effectively five equation in the five unknowns which can be solved to express $F_{2,3,4}$ and the five invariants made from the charged fields in terms of $F_1$. As such, we seem to find a unique solution. However, recall that the invariants are not independent, rather they obey $x y z = u v$. There is no reason for the solution to the F-term equations to obey this relation and so, for generic coefficients, there should actually be no solution to both the F-term and D-term equations for which $F_1 \neq 0$. 

\begin{figure}
\center
\includegraphics[width=1\textwidth]{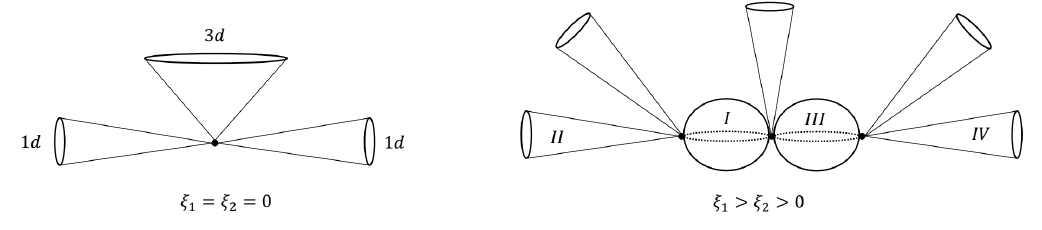} 
\caption{Schematic depiction of the moduli space for the 2d $(2,2)$ GLSM obtained by $SU(2)_R$ twisted compactification $(A_1,A_2)$ on a sphere with five punctures of type $\phi_{(2,1)}$ for $\xi_1=\xi_2=0$ and $\xi_1>\xi_2>0$. Each cone denotes a copy of $\mathbb{C}$ while the spheres denote $\mathbb{CP}^1$.}
\label{5pcSphereMS}
\end{figure}

Overall, we see that the moduli space in this case contains three branches, as shown on the left of Figure \ref{5pcSphereMS}. Two are one dimensional cones where either $u$ or $v$ acquire a non-zero value, while the last is a three dimensional space where $F_{1,2,3}$ have arbitrary values. 

We next turn to the case where the FI parameters are not zero. For simplicity, we shall assume that $\xi_1 > \xi_2 >0$. \footnote{We can exchange $\xi_1$ and $\xi_2$ with the quiver reflection symmetry, and change their signs with charge conjugation. This can be used to reach other corners of the parameter space where $\xi_1$ and $\xi_2$ have the same sign. Note that the superpotential may not be invariant under this symmetry, but this should not matter as we do not rely on the specific coefficients here, only that they are generic enough.} The FIs modify the D-terms as
\bea
& & |\Phi_1|^2 + |\tilde{\Phi}_2|^2 - |\tilde{\Phi}_1|^2 - |\Phi_2|^2 = \xi_1 , \\ \nonumber 
& & |\tilde{\Phi}_3|^2 + |\Phi_2|^2 - |\Phi_3|^2 - |\tilde{\Phi}_2|^2 = \xi_2 .
\eea 
These can be solved by setting
\bea
\nonumber& & |\Phi_1| = \sqrt{\xi_1} \cosh{(\nu)}\cos{(\theta)}\, , \\ \nonumber
& & |\tilde{\Phi}_1| = \sqrt{\xi_1} \sinh{(\nu)}\cos{(\phi)}\, , \\ \nonumber
& & |\tilde{\Phi}_3| = \sqrt{\xi_2} \cosh{(\mu)}\cos{(\psi)}\ , \\ \nonumber
& & |\Phi_3| = \sqrt{\xi_2} \sinh{(\mu)}\cos{(\omega)}\, , \\ \nonumber
& & |\tilde{\Phi}_2| = \sqrt{\xi_1} \cosh{(\nu)}\sin{(\theta)} = \sqrt{\xi_2} \sinh{(\mu)}\sin{(\omega)}\ , \\ 
& & |\Phi_2| = \sqrt{\xi_1} \sinh{(\nu)}\sin{(\phi)} = \sqrt{\xi_2} \cosh{(\mu)}\sin{(\psi)}\, ,
\eea
but note that we are going to have to the enforce the conditions of the last two lines, so $\nu$, $\mu$, $\theta$, $\phi$, $\psi$ and $\omega$ are not independent.

Next we need to solve the F-term equations. Now we can immediately set $F_1 = 0$ as we have seen that there should not be any solution to the F-terms where $F_1 \neq 0$. The solution now proceeds as before, where the F-terms for the $F_i$ fields force the condition: $x=y=z=0$. It is straightforward to see that we have the following solutions:
\begin{itemize}
\item $\nu = \psi = 0$, $\omega=\frac{\pi}{2}$: this implies $\tilde{\Phi}_1 = \Phi_3 = \Phi_2 = 0$, while $|\Phi_1| = \sqrt{\xi_1} \cos{(\theta)}$, $|\tilde{\Phi}_3| = \sqrt{\xi_2} \cosh{(\mu)}$, $|\tilde{\Phi}_2| = \sqrt{\xi_1} \sin{(\theta)} = \sqrt{\xi_2} \sinh{(\mu)}$. Given any value of $\theta$, the constraint can be solved for $\mu$. This suggests that this branch of the moduli space is spanned by the angle $\theta$ and a single phase (three fields but two phases are irrelevant due to the gauge symmetry). This suggests that this space is at least topologically $\mathbb{CP}^1$. 
\item $\phi=\psi=0$, $\theta=\omega=\frac{\pi}{2}$: this implies $\Phi_1 = \Phi_3 = \Phi_2 = 0$, while $|\tilde{\Phi}_1| = \sqrt{\xi_1} \sinh{(\nu)}$, $|\tilde{\Phi}_3| = \sqrt{\xi_2} \cosh{(\mu)}$, $|\tilde{\Phi}_2| = \sqrt{\xi_1} \cosh{(\nu)} = \sqrt{\xi_2} \sinh{(\mu)}$. Given any value of $\nu$ the constraint can be solved for $\mu$. This branch of the moduli space is then spanned by $0\leq\nu< \infty$ and a single phase, suggesting that this space is at least topologically $\mathbb{C}$. Note that on this branch we generally have that $u=0$, $v\neq 0$.
\item $\theta=\mu=0$, $\phi=\frac{\pi}{2}$: this implies $\tilde{\Phi}_1 = \Phi_3 = \tilde{\Phi}_2 = 0$, while $|\Phi_1| = \sqrt{\xi_1} \cosh{(\nu)}$, $|\tilde{\Phi}_3| = \sqrt{\xi_2} \cos{(\psi)}$, $|\Phi_2| = \sqrt{\xi_1} \sinh{(\nu)} = \sqrt{\xi_2} \sin{(\psi)}$. Given any value of $\psi$, the constraint can be solved for $\mu$. This suggests that this branch of the moduli space is spanned by the angle $\psi$ and a single phase, implying that this space is at least topologically $\mathbb{CP}^1$.
\item $\theta=\omega=0$, $\psi=\phi=\frac{\pi}{2}$: this implies $\tilde{\Phi}_1 = \tilde{\Phi}_3 = \tilde{\Phi}_2 = 0$, while $|\Phi_1| = \sqrt{\xi_1} \cosh{(\nu)}$, $|\Phi_3| = \sqrt{\xi_2} \sinh{(\mu)}$, $|\Phi_2| = \sqrt{\xi_1} \sinh{(\nu)} = \sqrt{\xi_2} \cosh{(\mu)}$. Given any value of $\mu$ the constraint can be solved for $\nu$. This suggests that this branch of the moduli space is spanned by $0\leq \mu< \infty$ and a single phase, implying that this space is at least topologically $\mathbb{C}$. Note that on this branch we generally have that $v=0$, $u\neq 0$. 
\end{itemize}

Next we need to solve \eqref{FtermsPhi} for each case. As we have seen before, if three of the $\Phi_i$ or $\tilde{\Phi}_i$ fields vanish while the other three don't, then $F_{2,3,4}$ are completely determined in terms of them. The reason is that we have three non-trivial equations, from the non-vanishing fields, that for generic coefficients give a unique solution for the three $F$ fields. As this is the generic case in all of the above options, we see that the $F_i$ fields would just be determined from the $\Phi_i$ and $\tilde{\Phi}_i$ fields and the branches are as outlined above. However, there are points where more than three $\Phi_i$, $\tilde{\Phi}_i$ fields vanish. This happens at the intersection of the branches. Particularly, branch II intersects with branch I at the point $\Phi_1 = \tilde{\Phi}_1 = \Phi_3 = \Phi_2 = 0$, $\tilde{\Phi}_2 = \sqrt{\xi_1}$, $\tilde{\Phi}_3 = \sqrt{\xi_2+\xi_1}$. Branch I intersects with branch III at the point $\tilde{\Phi}_1 = \Phi_3 = \tilde{\Phi}_2 = \Phi_2 = 0$, $\Phi_1 = \sqrt{\xi_1}$, $\tilde{\Phi}_3 = \sqrt{\xi_2}$. Finally, branch III intersects branch IV at the point $\tilde{\Phi}_1 = \tilde{\Phi}_3 = \Phi_3 = \tilde{\Phi}_2 = 0$, $\Phi_1 = \sqrt{\xi_1+\xi_2}$, $\Phi_2 = \sqrt{\xi_2}$. At each of these three points four $\Phi_i$, $\tilde{\Phi}_i$ fields vanish, and as such \eqref{FtermsPhi} reduce to just two equations in the three $F_i$ which will have a one dimensional space of solutions. Therefore, there should be an additional 1d cone emanating at each of these points. Thus, the moduli space has the schematic form shown in Figure \ref{5pcSphereMS}.  

We can next consider the behavior near the three singular points in the moduli space where the branches intersect. First we consider the point where I and II intersect. Here, only the fields $\tilde{\Phi}_3$ and $\tilde{\Phi}_2$ receive a non-zero value, which breaks the $U(1)$ gauge symmetries. As such, locally around this point the theory as an effective description in terms of chirals interacting through a superpotential. The R-charges of the chirals are 
\begin{table}[h]
  \centering
  \begin{tabular}{c|cccccc}
    & $\Phi_1$ & $\tilde{\Phi}_1$ & $\Phi_3$ & $\Phi_2$ & $F_1$ & $F_{2,3,4}$ \\\hline
    $U(1)_R$ & $-\tfrac{1}{5}$ & $\tfrac{3}{5}$ & $\tfrac{2}{5}$ & $\tfrac{2}{5}$ & $\tfrac{1}{5}$ & $\tfrac{3}{5}$
  \end{tabular}
  \label{tableI-II}
\end{table}

\noindent The R-charges allow mass terms between $\Phi_3$ and $\Phi_2$, and combinations of $\tilde{\Phi}_1$ and $F_{2,3,4}$. In fact the superpotential around this point includes such mass terms (assuming it is sufficiently generic). As such these can be integrated out, and the low-energy theory will be described by the chirals $\Phi_1$, $F_1$ and the two remaining combinations of $\tilde{\Phi}_1$ and $F_{2,3,4}$.  

We next consider the behavior near the intersection of III and IV. This point behaves similarly to the previous point, as both are related by charge conjugation. Specifically, at this point only the fields $\Phi_1$ and $\Phi_2$ have a non-zero value, which breaks the $U(1)$ gauge symmetries. The theory is again described by a collection of chiral fields with the R-charges 
\begin{table}[h]
  \centering
  \begin{tabular}{c|cccccccc}
    & $\tilde{\Phi}_3$ & $\Phi_3$ & $\tilde{\Phi}_1$ & $\tilde{\Phi}_2$ & $F_1$ & $F_2$ & $F_3$ & $F_4$ \\\hline
    $U(1)_R$ & $-\tfrac{1}{5}$ & $\tfrac{3}{5}$ & $\tfrac{2}{5}$ & $\tfrac{2}{5}$ & $\tfrac{1}{5}$ & $\tfrac{3}{5}$ & $\tfrac{3}{5}$ & $\tfrac{3}{5}$ 
  \end{tabular}
\end{table}

\noindent The R-charges allow mass terms between $\tilde{\Phi}_1$ and $\tilde{\Phi}_2$, and combinations of $\Phi_3$ and $F_{2,3,4}$, which are indeed present in the superpotential. As such these can be integrated out, and the low-energy theory will be described by the chirals $\tilde{\Phi}_3$, $F_1$ and the two remaining combinations of $\Phi_3$ and $F_{2,3,4}$.

Finally, we have the point where I and III intersect. At this point only the fields $\Phi_1$ and $\tilde{\Phi}_3$ receive a non-zero value, which breaks the $U(1)$ gauge symmetries. The theory is again described by a collection of chiral fields with the R-charges
\begin{table}[h]
  \centering
  \begin{tabular}{c|cccccccc}
    & $\tilde{\Phi}_1$ & $\Phi_3$ & $\tilde{\Phi}_2$ & $\Phi_2$ & $F_1$ & $F_2$ & $F_3$ & $F_4$ \\\hline
    $U(1)_R$ & $\tfrac{2}{5}$ & $\tfrac{2}{5}$ & $\tfrac{1}{5}$ & $\tfrac{1}{5}$ & $\tfrac{1}{5}$ & $\tfrac{3}{5}$ & $\tfrac{3}{5}$ & $\tfrac{3}{5}$ 
  \end{tabular}
\end{table}

\noindent The R-charges allow mass terms between $\tilde{\Phi}_1$ and $\Phi_3$, and combinations of $F_{2,3,4}$, which are indeed present in the superpotential. As such these can be integrated out, and the low-energy theory will be described by the chirals $\tilde{\Phi}_2$, $\Phi_2$, $F_1$ and the remaining combination of $F_{2,3,4}$.

Overall, one can verify by comparing with \eqref{eq:EG5ptLY} that the sum of the elliptic genera of the EFTs at each of the three singular points agrees with the elliptic genus of the five-punctured sphere model.

%%%%%%%%%%%%%%%%%%%%%%%%%%%%%%%%%%%%%%%%%%%%%%%%%%%%%%%%%%%
%%%%%%%%%%%%%%%%%%%%%%%%%%%%%%%%%%%%%%%%%%%%%%%%%%%%%%%%%%
\bibliographystyle{JHEP}
\bibliography{refs}

\end{document}